\documentclass[12pt,aps,epsf,onecolumn,numberedappendix,appendixfloats]{emulateapj}

\usepackage{amsmath}
\DeclareMathOperator\arctanh{arctanh}
\usepackage{graphicx}
\usepackage[FIGTOPCAP]{subfigure}
\usepackage{rotating}
\usepackage{epstopdf}
\usepackage{dcolumn}
\usepackage{bm}

\def\fun#1#2{\lower3.6pt\vbox{\baselineskip0pt\lineskip.9pt
\ialign{$\mathsurround=0pt#1\hfil##\hfil$\crcr#2\crcr\sim\crcr}}}
\def\lap{\mathrel{\mathpalette\fun <}}

\def\kms{\mathrm{km\,s}^{-1}}

\def\mass{{\cal M}}

\def\msun{{\mass_\odot}}

\def\beq{\begin{equation}}
\def\eeq{\end{equation}}
\def\barr{\begin{eqnarray}}
\def\earr{\end{eqnarray}}
\def\bsub{\begin{subequations}}
\def\esub{\end{subequations}}
\def\lbin{l_\mathrm{bin}}
\def\Lbin{L_\mathrm{bin}}
\def\lstar{l_\mathrm{star}}
\def\Lstar{L_\mathrm{star}}

\def\mh{M_{\bullet}}
\def\sbh{SBH}

\def\sbhs{SBHs}

\def\sbh{SBH}

\def\sbhs{SBHs}

\shorttitle{Binary Black Holes in Rotating Nuclei}
\shortauthors{Rasskazov and Merritt}

\begin{document}

\title{Evolution Of Binary Supermassive Black Holes In Rotating Nuclei}

\author{Alexander Rasskazov and David Merritt}
\affil{School of Physics and Astronomy and Center for Computational Relativity and Gravitation, 
Rochester Institute of Technology, Rochester, NY 14623}

\begin{abstract}
Interaction of a binary supermassive black hole with stars in a galactic nucleus can result in changes
to all the elements of the binary's orbit, including the angles that define its orientation.
If the nucleus is rotating, the orientation changes can be large, causing large changes in the binary's
orbital eccentricity as well.
We present a general treatment of this problem based on the Fokker-Planck equation for $f$, defined as
the probability distribution for the binary's orbital elements.
First- and second-order diffusion coefficients are derived for the orbital elements of the binary
using numerical scattering experiments, and analytic approximations are presented for some of these coefficients.
Solutions of the Fokker-Planck equation are then derived under various assumptions about the initial
rotational state of the nucleus and the binary hardening rate.
We find that the evolution of the orbital elements can become qualitatively different when we introduce nuclear rotation: 
1) the orientation of the binary's orbit evolves toward alignment with the plane of rotation of the nucleus; 
2) binary orbital eccentricity decreases for aligned binaries and increases for counter-aligned ones. 
We find that the diffusive (random-walk) component of a binary's evolution is small in nuclei with non-negligible rotation, 
and we derive the time-evolution equations for the semimajor axis, eccentricity and inclination in that approximation. 
The aforementioned effects could influence gravitational wave production as well as the relative orientation of host galaxies and radio jets. 
\end{abstract}


\section{\label{Section:Intro} Introduction}

According to the current paradigm, galaxies are surrounded by extensive dark matter halos, and galaxies can grow in size when they come close enough to other galaxies for the dark matter to induce a merger \citep{Mo2010}.
Many galaxies are also known to contain a supermassive black hole (\sbh) at their center, and it is commonly assumed that \sbhs\ are universally present in early-type galaxies, and in the bulges of disk galaxies, at least for galaxies above a certain mass \citep{DEGN}.
Taken together, these two hypotheses imply the formation of binary \sbhs.
The idea was first explored by \citet{Begelman1980}, who broke down the likely evolution of a massive binary into three stages:

\begin{enumerate}
\item In the early phases of the galaxy merger, the two \sbhs\ are far enough apart that they move independently in the potential of the merger remnant. Both \sbhs\ sink toward the center of the potential due to dynamical friction against the stars.
\item When they are close enough together -- roughly speaking, within their mutual spheres of gravitational influence -- the two  \sbhs\ form a bound pair.
Their two-body orbit continues to shrink due to exchange of energy and angular momentum with nearby matter:
through gravitational slingshot interactions with stars, or gravitational torques from gas.
\item If the binary separation manages to shrink to a small fraction of a parsec, emission of gravitational waves brings the two \sbhs\
even closer together, resulting ultimately in coalescence.
\end{enumerate}

The present paper focusses on the second of these three phases. Furthermore, only interactions of the massive binary with stars are considered; gaseous torques are ignored. In certain respects, this is well-trodden ground.
Using numerical scattering experiments, \citet{Mikkola1992}, \citet{Quinlan1996} and \citet{Sesana2006} derived
expressions for the rates of change of binary semimajor axis  and eccentricity, for binaries in spherical nonrotating nuclei.
\citet{Merritt2002} noted that the same interactions would induce changes also in the other elements of the binary's orbit
-- for instance, its inclination -- and he obtained expressions for the rate of change of a binary's orientation from scattering experiments.
If the nucleus is spherical and nonrotating, these changes take the form of a random walk,
similar in many ways to the ``rotational Brownian motion''
of a polar molecule that collides with other molecules in a dielectric material \citep{Debye1929}.
In both cases, evolution can be described via a Fokker-Planck equation
in which the independent variable is a quantity (angle) that defines the orientation: the orbital plane in the case of a massive
binary, the dipole moment in the case of a molecule.

$N$-body simulations of galaxy mergers suggest that the stellar nuclei of merged galaxies should be flattened and rotating \citep[e.g.][]{Milosavljevic2001,GualandrisMerritt2012}.
Since there is a preferred axis in such nuclei, it would not be surprising if the orbital plane of a massive binary evolved
in a qualitatively different manner, due to slingshot interactions, as compared with binaries in spherical and nonrotating nuclei.
Recent $N$-body work has addressed this possibility \citep{Gualandris2012,Cui2014,Wang2014}.
One finds in fact that the orbital angular momentum vector of the binary tends to align with the rotation axis of the nucleus.
There are corresponding changes in the evolution of the binary's eccentricity \citep{Sesana2011}.
Stellar encounters tend to circularize the binary if its angular momentum is in the same direction as that of the nucleus, and vice versa, while in nonrotating nuclei the eccentricity is always slowly increasing.

In the present paper, we return to a Fokker-Planck description of the evolution of a massive binary at the center of a galaxy.
As in \citet{Merritt2002}, we use scattering experiments to extract the diffusion coefficients that appear in the Fokker-Planck equation.
However we generalize the treatment in that paper in a number of ways. (i) In \citet{Merritt2002} (as in \citet{Debye1929}),
a single diffusion coefficient described changes in the orbital inclination, and this coefficient was assumed to be independent
both of the binary's instantaneous orientation and of the direction of its change. In the present work, those assumptions are
 relaxed, allowing us to describe orientation changes in the general case of a binary evolving in an anisotropic (rotating) stellar background.
(ii) Both first and second-order diffusion coefficients are calculated; the former are most important in the case of rapidly rotating
nuclei, the latter in the case of slowly rotating nuclei. (iii) Terms describing the rate of change of binary separation and
eccentricity due to gravitational wave emission are included; in this respect, our work carries the evolution of the binary
into the third of the three phases  defined by Begelman et al.

A shortcoming of this approach is that the scattering experiments assume an unchanging distribution of stars in the nucleus,
while in reality, evolution of a massive binary is likely to be accompanied by changes in the stellar density.
Exactly how these two sorts of evolution are coupled has been debated in the past.
At one extreme, it is possible for the binary to ``empty the loss cone'' corresponding to orbits that pass near the binary.
If this happened, the density of stars in the vicinity of the binary would drop drastically, and the binary would cease to
harden; or it would harden at a rate determined by collisional orbit repopulation, which is very slow in all but the smallest
galaxies. The possibility that binaries ``stall'' at parsec-scale separations was considered likely by \citet{Begelman1980},
and the term ``final-parsec problem'' was coined by \citet{Milosavljevic2003} to describe the difficulty of evolving a binary past this point.
However, recent work \citep{Khan2013,Vasiliev2014,Vasiliev2015,Gualandris2017} has made a strong case that massive binaries typically do not stall in this way.
Rather, one finds that even slight departures of a nucleus from spherical symmetry allow stars to be continually fed to a central binary, at rates that decrease slowly with time, but which can be much greater than rates due to collisional orbital repopulation. This is an especially important effect considering that the product of a galactic merger is expected to be generically triaxial \citep{GualandrisMerritt2012, Khan2016}.
We incorporate the results of this work, and in particular the study of \citet{Vasiliev2015}, into our evolution equations, and thus account in an approximate way for the back-reaction of the binary's evolution on its stellar surroundings.

This paper is organized as follows.
In \S\ref{Section:BinaryEvolve} we generalize the Fokker-Planck formalism used by \citet{Debye1929} and
\citet{Merritt2002}  to include changes in all the elements of a binary's orbit, in a stellar nucleus that has an axis of rotational symmetry.
 \S\ref{Section:DiffCoeff} describes the scattering experiments and the method for extracting diffusion coefficients.
 In \S\ref{Section:UnderstandingScattering} we present a qualitative analysis of the results of the scattering experiments
 and try to explain some of their phenomenology.
 In \S\ref{Section:GR} we estimate the influence of post-Newtonian effects.
 \S\ref{Section:DiffusionCoefficients} presents the results of numerical calculation of diffusion coefficients for all the orbital components of the binary. Finally, in \S\ref{Section:FPESolution} we use these results to solve the FPE for the distribution function of binary's orbital inclination. \S\ref{Section:Conclusions} sums up and discusses some observational implications of our results. 
 
 An important application of the results obtained here is to calculations of the stochastic gravitational wave spectrum
 produced by a cosmological population of massive binaries in merging galaxies.
This is the subject of Paper II \citep{PaperII}.
 
\section{Equations of binary evolution}
\label{Section:BinaryEvolve}
Consider a massive binary at the center of a galaxy.
The components of the binary have
masses $M_1$ and $M_2$, which are assumed to be unchanging, and $M_1\ge M_2$.
If the binary is treated as an isolated system,
its energy $E_\mathrm{bin}$ and angular momentum $L_\mathrm{bin}$
are related to its semimajor axis $a$ and eccentricity $e$ via
\begin{equation}\label{Equation:EL_bin}
E_\mathrm{bin} = - \mu\frac{GM_{12}}{2a}, \ \ \ \
L_\mathrm{bin} = \mu\sqrt{GM_{12}a\left(1-e^2\right)}
\end{equation}
where $M_{12} = M_1 + M_2$ is the binary's total mass and $\mu = M_1M_2/\left(M_1+M_2\right)$ 
its reduced mass.
For the remainder of this section, we will use $E\equiv E_\mathrm{bin}/\mu$ and 
$L\equiv L_\mathrm{bin}/\mu$ to denote the specific energy and specific angular momentum,
respectively, of the massive binary:
\beq\label{Equation:ELreduced_bin}
E = -\frac{GM_{12}}{2a},\ \ \ \ L = \sqrt{GM_{12}a(1-e^2)} \,.
\eeq

Five variables are needed to completely specify the shape and orientation of the binary's orbit.
Four of these can be taken to be $(E,\boldsymbol{L})$; the fifth variable determines the orientation of the major 
axis of the binary's orbit (in the plane determined by the direction of $\boldsymbol{L}$) and is usually 
taken to be $\omega$, the argument of periapsis.
Both $E$ and $\boldsymbol{L}$ are independent of $\omega$.
In principle, one could evaluate changes in $\omega$ due to interaction of the binary with stars
using the numerical scattering experiments described below.
We choose to ignore changes in $\omega$ in our Fokker-Planck description of the binary's evolution.
That is a valid approximation in two limiting cases: when $\omega$ does not change at all;
or when $\omega$ changes so rapidly that we can average all the other diffusion coefficients over $\omega$. 
In \S\ref{Section:D(omega)} we show the latter to be a good approximation for a wide range of possible system parameters.
Accordingly, in much of what follows, our expressions for quantities like the diffusion coefficients in
$E$ and $\boldsymbol{L}$ will be averaged over $\omega$.

\subsection{Fokker-Planck equation}
\label{Section:BinaryEvolveFPE}

The binary is assumed to interact with stars, causing changes in its orbital elements.\footnote{
Changes in the location of the binary's center of mass are ignored; these were discussed by \citet{Merritt2001}.}
In the simplest representation, the binary's orbit would evolve smoothly and deterministically
with respect to time.
We consider a slightly more complex model, in which a random, or diffusive, component to the
binary's evolution is allowed as well.

Accordingly, define $f(E,\boldsymbol{L},t)\, dE\, d\boldsymbol{L} $ to be the probability that the binary's
energy $E$ and angular momentum  $\boldsymbol{L}$ lie in the intervals
$E$ to $E+dE$ and $\boldsymbol{L}$ to $\boldsymbol{L} + d\boldsymbol{L}$, respectively, at time $t$.
Let $\Delta t$ denote an interval of time that is short compared with the time
over which the orbit of the binary changes due to encounters
with stars, but still long enough that many encounters occur.
Define the transition probability $\Psi(E,\boldsymbol{L};\Delta E,\boldsymbol{\Delta L})$
that the energy and angular momentum of the binary change by $\Delta E$ and
$\boldsymbol{\Delta}\boldsymbol{L}$, respectively, in time $\Delta t$.
Then
\begin{equation}\label{Equation:Transition}
f\left(E,\boldsymbol{L},t+\Delta t\right) = 
\int f\left(E - \Delta E, \boldsymbol{L}-\boldsymbol{\Delta L}, t\right) 
\Psi\left(E - \Delta E,\boldsymbol{L}-\boldsymbol{\Delta L};\Delta E, \boldsymbol{\Delta L}\right)
dE \boldsymbol{d\Delta L}.
\end{equation}
This equation assumes in addition that the evolution of $f$ depends only
on its instantaneous value, that is, that its previous history can be ignored
(``Markov process'').

We now expand $f(E,\boldsymbol{L},t+\Delta t)$ on the left-hand side of
Equation (\ref{Equation:Transition}) as a Taylor series in $\Delta t$,
and $f\left(E-\Delta E,\boldsymbol{L}-\boldsymbol{\Delta L}, t\right)$ and  $\Psi\left(E-\Delta E,\boldsymbol{L}-\boldsymbol{\Delta L};\Delta E, \boldsymbol{\Delta L}\right)$
on the right-hand side as Taylor series in $\Delta E$ and $\boldsymbol{\Delta L}$.
Retaining only terms up to second order, the result is
\begin{eqnarray}\label{Equation:FPxy}
\frac{\partial f}{\partial t} &=&
-\frac{\partial}{\partial L_x}\left(f\langle\Delta L_x\rangle\right)
-\frac{\partial}{\partial L_y}\left(f\langle\Delta L_y \rangle\right)
-\frac{\partial}{\partial L_z}\left(f\langle\Delta L_z \rangle\right)
-\frac{\partial}{\partial E}\left(f\langle\Delta E \rangle\right) \nonumber \\
&+& \frac12 \frac{\partial^2}{\partial L_x^2}\left(f\langle\Delta L_x^2\rangle\right)
+ \frac12\frac{\partial^2}{\partial L_y^2}\left(f\langle(\Delta L_y)^2\rangle\right)
+ \frac12\frac{\partial^2}{\partial L_z^2}\left(f\langle(\Delta L_z)^2\rangle\right)
+ \frac12\frac{\partial^2}{\partial E^2}\left(f\langle(\Delta E)^2\rangle\right)
\nonumber \\
&+& \frac{\partial^2}{\partial L_x\partial L_y}\left(f\langle\Delta L_x\Delta L_y\rangle\right)
+ \frac{\partial^2}{\partial L_x\partial L_z}\left(f\langle\Delta L_x\Delta L_z\rangle\right)
 +\frac{\partial^2}{\partial L_y\partial L_z}\left(f\langle\Delta L_y\Delta L_z\rangle\right) \nonumber \\
 &+& \frac{\partial^2}{\partial L_x\partial E}\left(f\langle\Delta L_x\Delta E\rangle\right)
+ \frac{\partial^2}{\partial L_y\partial E}\left(f\langle\Delta L_y\Delta E\rangle\right)
 +\frac{\partial^2}{\partial L_z\partial E}\left(f\langle\Delta L_z\Delta E\rangle\right) .
\end{eqnarray}
Diffusion coefficients are defined in the usual way as
\begin{eqnarray}
\langle\Delta x\rangle  &=& \frac{1}{\Delta t}\int \Psi(E,\boldsymbol{L};
\Delta E,\boldsymbol{\Delta L}) \,\Delta x \, d\Delta E\;\boldsymbol{d\Delta L} ,  \nonumber \\ 
\langle\Delta x\Delta y\rangle  &=&  \frac{1}{\Delta t}\int \Psi(E,\boldsymbol{L}; 
\Delta E, \boldsymbol{\Delta L})\,\Delta x\,\Delta y \,d\Delta E\, \boldsymbol{d\Delta L} ,
\label{Equation:DefDiff}
\end{eqnarray}
where $\{x,\,y\}$ can be any of $\{L_x,\,L_y,\,L_z,\,E\}$. 

\begin{figure}[ht]
\includegraphics[width=\textwidth]{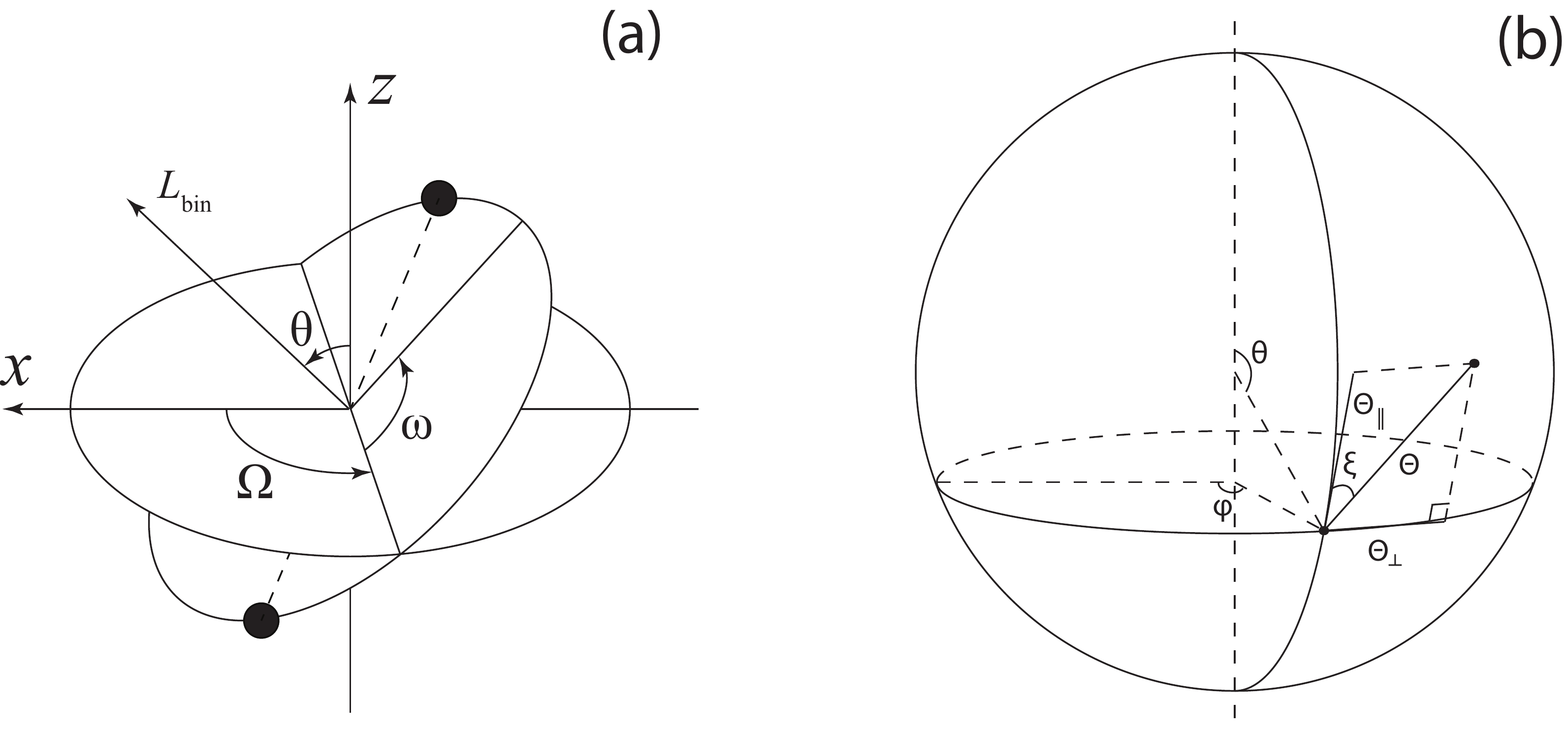}
\caption{Orbital parameters of the massive binary. (a) Angular momentum $L_\mathrm{bin}$, inclination $\theta$, longitude of ascending node $\Omega$, argument of periapsis $\omega$. 
The $z$ axis coincides with the axis of rotation of the nuclear cluster.
(b) $(\theta,\phi)$ is the direction of the binary's angular momentum vector in spherical coordinates; $(\Theta,\xi)$ or $(\Theta_\perp,\Theta_\parallel)$ denote its change after a single interaction.}
\label{Figure:binary}
\end{figure}

We will often be interested in the case of a binary that evolves in a rotating stellar nucleus.
Suppose that the nucleus is unchanging and spherical and that the center of mass of the
binary coincides with that of the nuclear star cluster.
Assume furthermore that the total angular momentum with respect to the nuclear center, 
of stars in any interval of orbital energy,
is directed along a fixed direction which we define to be the $z$ axis.
The binary's angular momentum vector may be inclined with respect to this axis,
by an angle $\theta(t)$.
In this case it is useful to express the Fokker-Planck equation in terms of angular momentum
variables for the binary that are defined with respect to the $z$-axis, for instance
\beq\label{Equation:NewVariables}
x_1 = L, \ \ \ x_2 = \mu = \cos\theta = L_z/L, \ \ \ x_3 = \phi, \ \ \ x_4 = E  .
\eeq
With the right choice of ``reference axis'' and ``reference plane,'' 
$\theta$ is equivalent to the orbital inclination of the binary, 
usually denoted by $i$, and $\phi$ is equivalent to the longitude of its ascending node,
or $\Omega$  (Figure~\ref{Figure:binary}a).

\citet[][Section 4.9]{Risken1989} shows how to transform Equation~(\ref{Equation:FPxy}) under a 
change in variables:
\begin{equation}\label{Equation:FPGeneral}
{\cal J}\frac{\partial f}{\partial t} = -\displaystyle\sum_{i=1}^4\frac{\partial}{\partial x_i}\left(\langle\Delta x_i\rangle {\cal J}f\right)
+ \displaystyle\sum_{i,j=1}^4\frac{1}{2}\frac{\partial^2}{\partial x_i\partial x_j}\left(\langle\Delta x_i\Delta x_j\rangle {\cal J}f\right)
\end{equation}
where ${\cal J}=\mathrm{Det}\{\partial L_i/\partial x_j\}$ is the Jacobian relating old
($E, L_i$) to new ($x_i$) variables,
and the new diffusion coefficients are related to the old diffusion coefficients in the following way:
\begin{subequations}\label{Equation:Delta}
\begin{eqnarray}
\langle\Delta x_i\rangle &=& \frac{\partial x_i}{\partial t} +
\displaystyle\sum_{j=1}^4 \frac{\partial x_i}{\partial L_j} \langle\Delta L_j\rangle
+ \frac{1}{2}\displaystyle\sum_{j,k=1}^4\frac{\partial^2 x_i}{\partial L_j\partial L_k} \langle\Delta L_j\Delta L_k\rangle
\label{Equation:Deltax}, \\
\langle\Delta x_i\Delta x_j\rangle &=& \displaystyle\sum_{k,l=1}^4
\frac{\partial x_i}{\partial L_k}\frac{\partial x_j}{\partial L_l}
\langle\Delta L_k\Delta L_l\rangle .
\label{Equation:DeltaxDeltay}
\end{eqnarray}
\end{subequations}
In our case, the old variables are
\beq\label{Equation:OldVariables}
L_1 = L_x, \ \ \ L_2 = L_y, \ \ \ L_3 = L_z, \ \ \ L_4 = E  . \nonumber
\eeq
Setting $(x_1, x_2, x_3, x_4) = (L,\mu,\phi,E)$ we obtain ${\cal J}=L^2$
and Equation (\ref{Equation:FPGeneral}) becomes
\begin{eqnarray}\label{Equation:FPLmuphi}
\frac{\partial g}{\partial t} &=&
- \frac{\partial}{\partial L}(g\langle\Delta L\rangle)
- \frac{\partial}{\partial \mu}(g\langle\Delta \mu\rangle)
- \frac{\partial}{\partial \phi}(g\langle\Delta \phi\rangle)
- \frac{\partial}{\partial E}(g\langle\Delta E\rangle)
\nonumber \\
&+& \frac12\frac{\partial^2}{\partial L^2}(g\langle\Delta L^2\rangle) +
\frac12\frac{\partial^2}{\partial \mu^2}(g\langle\Delta \mu^2\rangle) + \frac12\frac{\partial^2}{\partial \phi^2}(g\langle\Delta \phi^2\rangle) +
\frac12\frac{\partial^2}{\partial E^2}(g\langle\Delta E^2\rangle)
\nonumber \\
&+& \frac{\partial^2}{\partial \mu \partial L}(g\langle\Delta L\Delta \mu\rangle)
+ \frac{\partial^2}{\partial \mu \partial \phi}(g\langle\Delta \phi\Delta \mu\rangle)
+ \frac{\partial^2}{\partial L \partial \phi}(g\langle\Delta \phi\Delta L\rangle)
\nonumber \\
&+& \frac{\partial^2}{\partial E \partial L}(g\langle\Delta L\Delta E\rangle)
+ \frac{\partial^2}{\partial E \partial \mu}(g\langle\Delta \mu\Delta E\rangle)
+ \frac{\partial^2}{\partial E \partial \phi}(g\langle\Delta \phi\Delta E\rangle)
\end{eqnarray}
where $g=fL^2$.
Furthermore
\begin{eqnarray}
f(E,\boldsymbol{L}) dE\,d\boldsymbol{L} &=& f(E,{\boldsymbol L})\; dE\,dL_x\,dL_y\,dL_z
= f(E,L,\mu,\phi)\; L^2 dE\, dL\, d\mu\, d\phi = g(E,L,\mu,\phi) \; dE\, dL\, d\mu\, d\phi .
\end{eqnarray}
The new diffusion coefficients can be expressed in terms of the old ones
via Equations (\ref{Equation:Delta}); we give the explicit expressions in Appendix~\ref{Appendix:DiffCoefs}. 
Expressed in terms of any other choices for the independent variables, 
the Fokker-Planck equation would have the same form as Equation~(\ref{Equation:FPLmuphi}) but with different ${\cal J}=g/f$. 
For instance, ${\cal J} = L$ for $x_i = (L,L_z,\phi,E)$ or ${\cal J} = \left(\mu\sqrt{GM_{12}a}\right)^3 2e\sqrt{1-e^2}$ for $x_i = (e,\mu,\phi,E)$. 

When the distribution of velocities and angular momenta in the nucleus has an axis of symmetry that is unchanging with respect to time, all the diffusion coefficients are independent of $\phi$, 
and furthermore we may not be interested in the dependence of $f$ on $\phi$.
These considerations motivate the definition of the reduced probability density $\overline{f}$:
\beq
\overline{f} = \int_0^{2\pi}fd\phi
\eeq
and $\overline{g}=\overline{f} L^2$, so that
\begin{equation}
\int_{-\infty}^0 dE \int_0^\infty dL \int_{-1}^1d\mu\, \overline{g}  = 
\int_{-\infty}^0 dE \int_0^\infty dL \int_{-1}^1 d\mu\, \overline{f}L^2  = 1 .
\end{equation}
Integrating both sides of Equation (\ref{Equation:FPLmuphi}) over $\phi$
eliminates the terms containing $\partial/\partial\phi$:
\begin{eqnarray}\label{Equation:FPLmuphiav}
\frac{\partial \overline{g}}{\partial t} &=&
- \frac{\partial}{\partial L}(\overline{g}\langle\Delta L\rangle)
- \frac{\partial}{\partial \mu}(\overline{g}\langle\Delta \mu\rangle)
- \frac{\partial}{\partial E}(\overline{g}\langle\Delta E\rangle)
\nonumber\\
&+& \frac{1}{2}\frac{\partial^2}{\partial L^2}(\overline{g}\langle\Delta L^2\rangle)
+ \frac{1}{2}\frac{\partial^2}{\partial \mu^2}(\overline{g}\langle\Delta \mu^2\rangle)
+ \frac{1}{2}\frac{\partial^2}{\partial E^2}(\overline{g}\langle\Delta E^2\rangle) \nonumber\\
&+& \frac{\partial^2}{\partial \mu \partial L}(\overline{g}\langle\Delta L\Delta \mu\rangle)
+ \frac{\partial^2}{\partial E \partial L}(\overline{g}\langle\Delta L\Delta E\rangle)
+ \frac{\partial^2}{\partial E \partial \mu}(\overline{g}\langle\Delta \mu\Delta E\rangle).
\end{eqnarray}

\subsection{Evolution equation for the binary's orientation}
\label{Section:BinaryEvolveOrientation}

We also consider the case of a binary for which the energy, $E$, and the magnitude of the
angular momentum, $L$, 
change with time in some specified way: $E=E_0(t),\,L=L_0(t)$. In that case,
the reduced probability density is
\beq
\overline{g}(E,L,\mu) = \delta(L-L_0(t)) \delta(E-E_0(t)) \overline{f}(\mu) L_0(t)^2 .
\eeq
Substituting this expression into Equation~(\ref{Equation:FPLmuphiav}) and 
integrating over $E$ and $L$ leaves a Fokker-Planck equation describing the evolution
of the binary's orientation:
\begin{equation}\label{Equation:FPonlymu}
\frac{\partial \overline{f}}{\partial t} = - \frac{\partial}{\partial \mu}
\left(\overline{f}\langle\Delta \mu\rangle\right)
+ \frac{1}{2}\frac{\partial^2}{\partial \mu^2}
\left(\overline{f}\langle\Delta \mu^2\rangle\right) .
\end{equation}

This reduced problem is similar to  one considered by \citet{Debye1929},
who derived a Fokker-Planck equation describing the evolution of the orientation
of a polar molecule in an electric field, subject to collisions with other molecules.
Debye's treatment appears to be the closest existing treatment to our own, and
it is of interest to demonstrate the correspondence of his expressions
with the equations derived here.
We begin by replacing $\mu$ by $\cos\theta$:
\begin{eqnarray}
\frac{\partial}{\partial\mu} &=& -\frac{1}{\sin\theta}\frac{\partial}{\partial\theta},\ \ \ \
\frac{\partial^2}{\partial\mu^2} = \frac{1}{\sin^2\theta}\frac{\partial^2}{\partial\theta^2}
- \frac{\cos\theta}{\sin^3\theta}\frac{\partial}{\partial\theta},
\nonumber \\
\langle\Delta\mu\rangle &=& -\langle\Delta\theta\rangle\sin\theta - \frac12 \langle(\Delta\theta)^2\rangle\cos\theta,\ \ \ \
(\Delta\mu)^2= \langle(\Delta\theta)^2\rangle\sin^2\theta 
\label{Equation:DeltaThetaToDeltaMu}
\end{eqnarray}
so that Equation (\ref{Equation:FPonlymu}) becomes
\begin{equation}\label{Equation:FPonlytheta}
\sin\theta \frac{\partial \overline{f}}{\partial t} = - \frac{\partial}{\partial \theta}\left(\overline{f}\sin\theta\langle\Delta \theta\rangle\right) + \frac{1}{2}\frac{\partial^2}{\partial \theta^2}\left(\overline{f}\sin\theta\langle(\Delta\theta)^2\rangle\right) .
\end{equation}
For instance, we will show below that for a binary in a rotating nucleus, 
\begin{eqnarray}\label{Equation:thetaD1D2}
\langle(\Delta\theta)^2\rangle \approx \zeta(t) C_2 , \ \ \ \ 
\langle\Delta\theta\rangle \approx \zeta(t) \left(-C_1\sin\theta + \frac{1}{2}C_2\cot\theta\right)
\end{eqnarray}
where $C_{1,2}$ are non-negative constants and $\zeta(t)$ is some function of time; in a nonrotating nucleus, $C_1=0$.
With these forms for the diffusion coefficients, the evolution equation (\ref{Equation:FPonlytheta}) becomes
\beq\label{Equation:FPtheta}
\frac{\partial {\overline f}}{\partial \tau} = \frac{1}{\sin\theta}\frac{\partial}{\partial\theta}\left[\sin\theta\left(\alpha\frac{\partial \overline{f}}{\partial\theta} + {\overline f}\sin\theta\right)\right]
\eeq
where $d\tau = C_1\zeta(t)dt$ and $\alpha = C_2/(2C_1)$. 
This equation has exactly the same form as Equation (46) of \citet{Debye1929}. 
We can also write this equation in terms of $\mu$:
\begin{eqnarray}\label{Equation:FPmu}
\frac{\partial \overline{f}}{\partial \tau} = \frac{\partial}{\partial\mu}\left[\left(1-\mu^2\right)\left(\alpha\frac{\partial \overline{f}}{\partial\mu} - \overline{f}\right)\right]
\end{eqnarray}

In the case of no external electric field (equivalent to the case of a nonrotating nucleus in our model) 
\cite{Debye1929} made an additional simplifying assumption: that
$\Psi = \Psi(\Delta \chi)$ is a function only of the (spherical) angular
displacement between $\boldsymbol{L}$ and $\boldsymbol{L+\Delta L}$, i.e. of
\begin{equation}
\cos\left(\Delta\chi\right) =
\frac{ {\boldsymbol L}\cdot \left({\boldsymbol L}+ {\boldsymbol \Delta L}\right)}
{\left|\boldsymbol{L}\right| \left|\boldsymbol{L}+\boldsymbol{\Delta  L}\right|}.
\end{equation}
Following Debye, we now derive diffusion coefficients
$\langle\Delta\theta\rangle$ and $\langle(\Delta\theta)^2\rangle$
from this ansatz.
Figure \ref{Figure:binary}b defines a new spherical-polar coordinate system
with principal axis directed along $\boldsymbol{L}$ (not $\boldsymbol{z}$),
and surface area element $\sin\Theta\: d\Theta \: d\xi$.
(In Debye's Figure 25 these coordinates are labeled $\Theta$ and $\phi$; while
in his text, the symbol $\theta$ is used to represent the same angle labelled
$\Theta$ in his figure.
Debye uses the symbol $\vartheta$ for our $\theta$.
Note that $\Theta$ -- which is small by assumption -- is a differential angle
and so can equally well be written as $\Delta\Theta$.)
\citet{Debye1929} showed via spherical trigonometry that the differential in
(our) $\theta$ is given in terms of $(\Theta, \xi)$ by
\beq
\Delta \theta = -\Theta\cos\xi + \frac{\Theta^2}{2}\frac{\cos\theta}{\sin\theta}\sin^2\xi + \ldots .
\eeq
Thus
\begin{eqnarray}
\langle\Delta\theta\rangle &=& \int_0^\pi \sin\Theta\:  d \Theta
\int_0^{2\pi} \Delta\theta\: \Psi(\Theta,\xi) \: d\xi
\approx \int_0^\pi \int_0^{2\pi} \Psi(\Theta) \frac{\Theta^2}{2}\frac{\cos\theta}{\sin\theta} \sin^2\xi \sin\Theta\:d\Theta\: d\xi \nonumber \\
&=& \frac12\frac{\cos\theta}{\sin\theta}  \int_0^\pi
\Psi(\Theta) \Theta^2  \sin\Theta\:d\Theta\:
\int_0^{2\pi} \sin^2\xi\: d\xi
= \frac{\pi}{2}\frac{\cos\theta}{\sin\theta}  \int_0^\pi
\Psi(\Theta) \Theta^2  \sin\Theta \:d\Theta = \frac14 \frac{\cos\theta}{\sin\theta} \langle(\Delta\Theta)^2\rangle
\end{eqnarray}
where
\begin{equation}
\langle(\Delta\Theta)^2\rangle \equiv  \int_0^\pi \int_0^{2\pi} \Theta^2
\Psi(\Theta)  \sin\Theta \:d\Theta d\xi .\nonumber \\
\end{equation}
In the same way,
\begin{eqnarray}
\langle(\Delta\theta)^2\rangle &=& \int_0^\pi \sin\Theta\:  d \Theta
\int_0^{2\pi} (\Delta\theta)^2\: \Psi(\Theta,\xi) \: d\xi
\approx \int_0^\pi \int_0^{2\pi} \Psi(\Theta) \Theta^2
 \cos^2\xi \sin\Theta\:d\Theta\: d\xi \nonumber \\
&=&  \int_0^\pi
\Psi(\Theta) \Theta^2  \sin\Theta\:d\Theta\:
\int_0^{2\pi} \cos^2\xi\: d\xi = \pi\int_0^\pi
\Psi(\Theta) \Theta^2  \sin\Theta \:d\Theta =  \frac 12\langle(\Delta\Theta)^2\rangle.
\end{eqnarray}
Equation (\ref{Equation:FPonlytheta}) is then
\begin{eqnarray}\label{Equation:FPDebye}
\frac{\partial \overline{f}}{\partial t} &=&
\frac{\langle(\Delta\Theta)^2\rangle}{4\sin\theta} \frac{\partial}{\partial\theta}
\left(\sin\theta\frac{\partial \overline{f}}{\partial\theta}\right)
\end{eqnarray}
which has the same form as Debye's Equation (46) if his drift term is set to zero.

\citet{Merritt2002} evaluated $\langle(\Delta\Theta)^2\rangle$
via scattering experiments for a circular-orbit, equal-mass binary and discussed
time-dependent solutions to Equation (\ref{Equation:FPDebye}).
He used the term ``rotational Brownian motion'' to describe the evolution of a binary's
orientation in response to random encounters with stars.

Returning to the more general case described by Equations (\ref{Equation:FPonlymu})
or (\ref{Equation:FPonlytheta}): we can recast these equations also in terms of $(\Theta,\xi)$.
As illustrated in Figure \ref{Figure:binary}b, 
we define the new angles $\{\Theta_\parallel, \Theta_\perp\}$ via
\begin{equation}
\Theta_\parallel = \Theta\cos\xi, \ \ \ \ \Theta_\perp = \Theta\sin\xi .
\end{equation}
(Note the analogy with the velocity-space diffusion coefficients for a single
star, which can be expressed in terms of $\{\Delta v_\parallel, \Delta v_\perp\}$.)
The diffusion coefficients for $\theta$ are easily expressed in terms of these variables:
\begin{subequations}\label{Equation:LocalDiffusionCoefficientsInTheta}
\begin{eqnarray}
\langle\Delta\theta\rangle &=& - \langle\Theta\cos\xi\rangle + \frac{1}{2}\cot\theta\langle\Theta^2\sin^2\xi\rangle = - \langle\Delta\Theta_\parallel\rangle + \frac{1}{2}\cot\theta\langle(\Delta\Theta_\perp)^2\rangle,
\\
\langle(\Delta\theta)^2\rangle &=& \langle\Theta^2\cos^2\xi\rangle
= \langle(\Delta\Theta_\parallel)^2\rangle .
\end{eqnarray}
\end{subequations}
In Appendix \ref{Appendix:Debye}, we show that the Fokker-Planck equation for the angular part of the probability density can then
be written as
\begin{eqnarray}
\sin\theta\frac{\partial f}{\partial t} &=& - \frac{\partial}{\partial\theta} \left[f\left(-\sin\theta\langle\Delta\Theta_\parallel\rangle + \frac12\cos\theta\langle(\Delta\Theta_\perp)^2\rangle\right)\right] -
\frac{\partial}{\partial\phi}\left[f\left(\langle\Delta\Theta_\perp\rangle + \cot\theta\langle\Delta\Theta_\parallel\Delta\Theta_\perp\rangle\right)\right] +\nonumber\\
&+& \frac12\frac{\partial^2}{\partial\theta^2}\left[f\sin\theta\langle(\Delta\Theta_\parallel)^2\rangle\right] - \frac{\partial^2}{\partial\phi\partial\theta}\left[f\langle\Delta\Theta_\parallel\Delta\Theta_\perp\rangle\right] + \frac{1}{2}\frac{\partial^2}{\partial\phi^2}\left[f\frac{1}{\sin\theta}\langle(\Delta\Theta_\perp)^2\rangle\right] 
\end{eqnarray}
where for the sake of generality a possible dependence on $\phi$ has been included.
In the case of a symmetric transition probability, as considered by Debye,
$\langle\Theta\sin\xi\rangle = \langle\Theta\cos\xi\rangle = \langle\Theta^2\sin\xi\cos\xi\rangle = 0$ and $\langle\Theta^2\sin^2\xi\rangle = \langle\Theta^2\cos^2\xi\rangle = \frac{1}{2}\langle\Theta^2\rangle$.
Thus
\begin{equation}\label{Equation:SphericalRelations}
\langle\Delta\Theta_\parallel\rangle = \langle\Delta\Theta_\perp\rangle = \langle\Delta\Theta_\parallel\Delta\Theta_\perp\rangle = 0 , \ \ \ \
\langle(\Delta\Theta_\parallel)^2\rangle =
\langle(\Delta\Theta_\perp)^2\rangle =
\frac12\langle(\Delta\Theta)^2\rangle = \mathrm{const.}
\end{equation}
and the Fokker-Planck equation returns to the form of Equation (\ref{Equation:FPDebye}).

\section{Numerical evaluation of the diffusion coefficients}
\label{Section:DiffCoeff} 
\subsection{\label{Section:DiffCoeffOne} Interaction of the massive binary with a single star}

We begin by considering the interaction of the massive binary with a single, initially unbound star (``field star'').
Aside from the presence of the field star, we approximate the binary as an isolated system, 
with energy and angular momentum given by equation~(\ref{Equation:EL_bin}).
We assume that the star approaches the binary from infinitely far away, and
that after some (possibly long) time, the star either escapes from the binary along an asymptotically
linear orbit -- the ``gravitational slingshot'' -- or (with much lower probability) 
it becomes bound to one or the other of the binary's components.

We write the energy per unit mass of the field star as $\varepsilon$ and its angular momentum per unit 
mass as $l$.
Given changes in $\varepsilon$ and $l$, 
we wish to find expressions for the corresponding changes in the binary's orbital parameters.
The latter include the binary's semimajor axis $a$ and eccentricity $e$, 
but also the orbital inclination $\theta$, the longitude of the ascending node $\Omega$, and the argument of periapsis $\omega$ (Figure~\ref{Figure:binary}a).
Given such expressions, we can compute rates of change of the binary's elements via scattering
experiments.

It is convenient to work in a frame such that the center of mass of the binary-star system
is located at the origin with zero linear momentum.
Henceforth we refer to this as the ``center-of-mass'' (COM) frame.
Let $\boldsymbol{r}_\mathrm{bin}$ and $\boldsymbol{v}_\mathrm{bin}$ be the position
and velocity of the massive binary's center of mass with respect to the COM frame.
Then
\begin{equation}\label{Equation:COM}
M_{12}\boldsymbol{r}_\mathrm{bin} = - m_f\boldsymbol{r},\ \ \ \
M_{12}\boldsymbol{v}_\mathrm{bin} = - m_f\boldsymbol{v}
\end{equation}
where $m_f$, $\boldsymbol{r}$ and $\boldsymbol{v}$ are the field star's mass, position vector and velocity respectively.
Conservation of energy and angular momentum of the binary-field star system implies
\begin{subequations}\label{Equation:ConservationLawsEL}
\begin{eqnarray}
E_\mathrm{bin} + \frac12 M_{12}v^2_\mathrm{bin} + \frac{1}{2} m_fv^2 &=& \mathrm{const.},
\label{Equation:ConservationLawsE} \\
\boldsymbol{L}_\mathrm{bin} + M_{12}\boldsymbol{r}_\mathrm{bin}\times\boldsymbol{v}_\mathrm{bin} + m_f\boldsymbol{r}\times\boldsymbol{v} &=& \mathrm{const.}
\label{Equation:ConservationLawsL}
\end{eqnarray}
\end{subequations}
Expressing $\boldsymbol{v}_\mathrm{bin}$ from equation (\ref{Equation:COM}) and substituting into equation (\ref{Equation:ConservationLawsE}), we find
\begin{equation}
E_\mathrm{bin} + \left(1 + \frac{m_f}{M_{12}}\right)\cdot \frac{1}{2} m_fv^2 = E_\mathrm{bin} + \left(1 + \frac{m_f}{M_{12}}\right)\cdot m_f\varepsilon = \mathrm{const.}
\end{equation}
which allows us to express the change in the binary's energy in terms of
the change in star's energy, in a single collision, as
\begin{equation}\label{Equation:deltaE}
\delta E_\mathrm{bin} = - \left(1 + \frac{m_f}{M_{12}}\right) m_f \delta\varepsilon .
\end{equation}
In the same way, combining equations (\ref{Equation:COM}) and
(\ref{Equation:ConservationLawsL}) yields
\begin{equation}\label{Equation:deltaL}
\delta \boldsymbol{L}_\mathrm{bin} = - \left(1 + \frac{m_f}{M_{12}}\right) m_f \delta \boldsymbol{l} .
\end{equation}
Typically we will be concerned with the case $m_f/M_{12} \ll 1$.
In this limit, equations (\ref{Equation:deltaE}) and (\ref{Equation:deltaL}) imply that
the field star's effect on the binary's orbital elements ($a,e$) is almost the same as if
the binary had remained fixed in space.

Recalling equations (\ref{Equation:EL_bin}),
we can express the binary's semimajor axis and eccentricity in terms of $E_\mathrm{bin}$ and $L_\mathrm{bin}$:
\begin{equation}
\frac{1}{a} = - \frac{2E_\mathrm{bin}}{GM_{12}\mu}, \ \ \ \
e^2 = 1 + \frac{2E_\mathrm{bin}L_\mathrm{bin}^2}{GM_{12}^2\mu^3} .
\end{equation}
Since the changes in both quantities are proportional to $m_f/\mu$,
we can assume them to be small, and write
\begin{subequations}\label{Equation:deltaae}
\begin{eqnarray}
\delta\left(\frac{1}{a}\right) &=& \frac{2m_f}{\mu} \frac{\delta\varepsilon}{GM_{12}},
\label{Equation:deltaa}\\
\delta e &=& \frac{m_f}{\mu} \frac{1-e^2}{e} \left(-\frac{\delta\varepsilon}{GM_{12}/a} + \frac{\delta l_\parallel}{\sqrt{GM_{12}a(1-e^2)}}\right)
\label{Equation:deltae}
\end{eqnarray}
\end{subequations}
where $l_\parallel$ is the projection of $\boldsymbol{l}$ on $\boldsymbol{L}_\mathrm{bin}$, so that $m_f\delta l_\parallel = - \delta L_\mathrm{bin}$.

We can also derive expressions for the change in the orientation of the orbit, i.e.,
the direction of the binary's angular momentum vector $\boldsymbol{L}_\mathrm{bin}$.
In terms of the binary's orbital inclination $\theta$ and nodal angle $\Omega$:
\begin{subequations}
\begin{eqnarray}\label{Equation:deltatheta}
\delta\theta &=& \frac{\delta l_{\mathrm{bin},\theta}}{l_\mathrm{bin}} = - \frac{m_f}{\mu} \frac{\delta l_\theta}{\sqrt{GM_{12}a(1-e^2)}},
\\ \label{Equation:deltaOmega}
\delta\Omega &=& - \frac{m_f}{\mu} \frac{\delta l_\Omega}{\sqrt{GM_{12}a(1-e^2)}\sin\theta}
\end{eqnarray}
\end{subequations}
where the designations are as follows:
\begin{itemize}
\item $l_\theta$ is the projection of $\boldsymbol{l}$ onto $\boldsymbol{l}_\mathrm{bin}\times\left(\boldsymbol{l}_\mathrm{bin}\times\hat{z}\right)$ (the axis lying in the $(\hat{z},\boldsymbol{l}_\mathrm{bin})$ plane and perpendicular to $\boldsymbol{l}_\mathrm{bin})$
\item $l_\Omega$ is the projection of $\boldsymbol{l}$ onto $\hat{z}\times\boldsymbol{l}_\mathrm{bin}$ .
\end{itemize}

\subsection{\label{Section:DiffCoeffDiffCoeff}Diffusion coefficients}
\begin{figure}[h]
\includegraphics[width=0.75\textwidth]{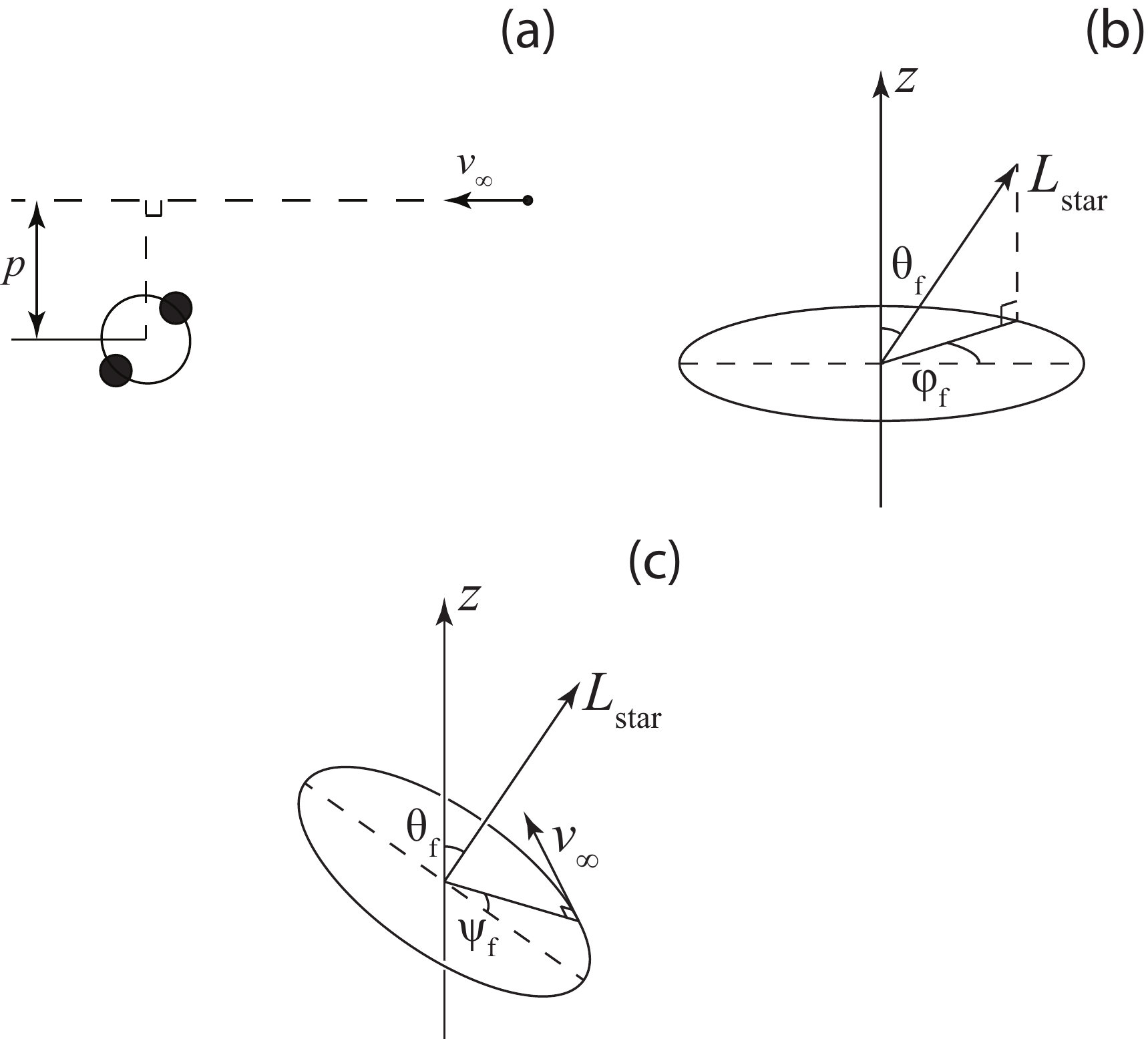}
\caption{Notations for initial stellar orbital parameters used in this paper. (a) Impact parameter $p$ and velocity at infinity $v_\infty$. (b), (c) Other parameters of a star's initial orbit: angular momentum $L_\mathrm{star}$, angles defining the direction of angular momentum $\theta_f$ and $\varphi_f$ (analogs of inclination and longitude of ascending node, respectively, for an unbound orbit), angle defining the direction of initial velocity in the orbital plane $\psi_f$ (analog of argument of periapsis for an unbound orbit).}
\label{Figure:star}
\end{figure}

We compute changes in $\varepsilon$ and $l$ via scattering experiments \citep{Hills1983}. A field star is assigned initial conditions, expressed in terms of its impact parameter $p$, velocity at infinity $v_\infty$,
and any additional parameters that are required to fully specify the initial stellar orbit 
(Figure~\ref{Figure:star}), all defined in the COM frame.
Starting from a separation much greater than the binary semimajor axis, the trajectory of the star
is integrated forward, in the time-dependent gravitational field of the rotating binary, typically until
the star has escaped again from the binary and is moving nearly rectilinearly away from it.
The orbital motion of the two components of the binary is assumed to be unaffected by the interaction;
 a valid approximation if $m_f\ll M_{12}$ \citep{Mikkola1992}.
Changes in the field-star's energy and angular momentum are then used, via the expressions derived
in the previous section, to compute changes in the orbital elements of the massive binary.

Given the results from a large number of scattering experiments,
diffusion coefficients describing changes in $Q$ associated with the
binary can then be computed as follows:
\begin{subequations}\label{Equation:DiffQ}
\begin{eqnarray}
\langle\Delta Q\rangle = \int_0^\infty \int_0^{p_\mathrm{max}} \frac{dN(p,v_\infty)}{dt}\, \overline{\delta Q}\, dp\, dv_\infty , \label{Equation:DiffQa} \\
\langle(\Delta Q)^2\rangle = \int_0^\infty \int_0^{p_\mathrm{max}} \frac{dN(p,v_\infty)}{dt}\, \overline{\delta Q^2}\, dp\, dv_\infty \label{Equation:DiffQb}
\end{eqnarray}
\end{subequations}
where $(d/dt)N(p,v_\infty)\,dp\,dv_\infty$ is the number of stars, with impact parameters $p$ to $p+dp$ and velocities at infinity $v_\infty$ to $v_\infty+dv_\infty$, that interact with the binary per unit time.
The `` $\overline{\phantom{x}}$ '' symbol here denotes an average over the
binary's initial mean anomaly, as well as over directions of the field star's initial
velocity and angular momentum. 

The scattering experiments ignore the gravitational potential from the stars;
furthermore, all stellar trajectories are initially unbound with respect to the binary,
since the initial energy of the field star is $v_\infty^2/2 > 0$.
Before proceeding, we need a scheme that relates $N(p,v_\infty)$ to the known distribution
 of orbits in the stellar nucleus.
The latter is defined in terms of the unperturbed orbits in the nuclear potential, and this potential
includes a contribution from all the stars in the nucleus.

In all of the models discussed below, the field-star distribution is 
assumed to be spherically symmetric initially.
Even if the nuclear cluster should depart from spherical symmetry (due to ejection
of stars by the massive binary, say),
the gravitational potential will continue to be dominated by the massive binary,
and so to a good approximation the total gravitational potential can be assumed to remain
spherically symmetric, at least at radii $a\lap r \lap r_\mathrm{m}$, where
$r_\mathrm{m}$ is the gravitational influence radius of the binary (defined below).
We therefore write the contribution to the gravitational potential from the stars as
$\Phi_\star(r)$.
The energy per unit mass of a single star is then
\beq
E = \frac{v^2}{2} + \Phi_\star(r) - \frac{GM_{12}}{r} \equiv \frac{v^2}{2} + \Phi(r)
\eeq
where the binary has been approximated as a point mass.
The other conserved quantity is the orbital angular momentum per unit mass, $\boldsymbol{L}$.
Let $f_\star(E,L)$ be the phase-space number density of stars in the nucleus,
and $N(E,L)\, dE\, dL$ the number of stars with orbital elements in the range $E$ to $E+dE$ and
$L$ to $L+dL$.
$N$ and $f_\star$ are related via
\beq
N(E,L)\, dE\, dL = 8 \pi^2 f_\star(E,L) P(E,L)\, L\, dE\, dL
\eeq
\citep[][Eq. 3.44]{DEGN}; here $P(E,L)$ is the radial period.

We wish to establish a one-to one correspondence between ($p$, $v_\infty$) and ($E$, $L$).
Since the trajectories in the scattering experiments are different from those in the nucleus,
there is no unique way to do this.
We are most interested in stars' interaction with the binary, and
such interactions occur mostly when the stars come close to the binary.
We therefore choose $p=p(E,L)$ and $v_\infty=v_\infty(E,L)$ in such a way that the two
representations of the orbit have the same periapsis distance, $r_p$, and the same
velocity at periapsis, $v_p$.
Having established this mapping, we can then compute the Jacobian determinant that relates the two distributions:
\begin{eqnarray}
N(p,v_\infty)\,dp\,dv_\infty = N(E,L)\,dE\,dL = 
N(E,L) \left|\frac{\partial(E,L)}{\partial(p,v_\infty)}\right| dp\,dv_\infty
\end{eqnarray}
allowing us to write
\begin{eqnarray}
 \frac{d}{dt} N(p,v_\infty) =  \left|\frac{\partial(E,L)}{\partial(p,v_\infty)}\right| \frac{d}{dt} N(E,L) 
 =  \left|\frac{\partial(E,L)}{\partial(p,v_\infty)}\right|  \frac{N(E,L)}{P(E,L)} .
\end{eqnarray}

At periapsis, the variables $E$ and $L$ are related via
\begin{equation}
\frac{L^2}{r_p^2} = 2\left[E-\Phi(r_p)\right]
\end{equation}
while in the scattering experiments,
\begin{equation}\label{Equation:r_p}
r_p \equiv r_p(p,v_\infty) = \frac{1}{GM_{12}}\frac{p^2v_\infty^2}{1+\sqrt{1+p^2v_\infty^4/G^2M_{12}^2}} ;
\end{equation}
in both expressions, we represent the binary by a point of mass $M_{12}$.
From these equations we find the desired mapping:
\begin{subequations}\label{Equation:Eofpv}
\begin{eqnarray}
E(p, v_\infty) &=& \Phi\left[r_p(p,v_\infty)\right] + \frac{p^2v^2_\infty}{2r_p^2(p,v_\infty)} =
\Phi_\mathrm{\star}\left[r_p(p,v_\infty)\right] + \frac{v_\infty^2}{2}, \label{Equation:Eofpva} \\
L(p, v_\infty) &=& pv_\infty \label{Equation:Eofpvb}
\end{eqnarray}
\end{subequations}
with Jacobian determinant
\begin{eqnarray}
\left|\frac{\partial(E,L)}{\partial(p,v_\infty)}\right| = \begin{vmatrix} \frac{\partial E}{\partial p} & \frac{\partial E}{\partial v_\infty} \\ \frac{\partial L}{\partial p} & \frac{\partial L}{\partial v_\infty} \end{vmatrix} = v_\infty^2\left(1+\frac{M_\star\left[r_p(p,v_\infty)\right]}{M_{12}\sqrt{1+p^2v_\infty^4/G^2M_{12}^2}}\right) .
\end{eqnarray}
The orbits of most interest have $r_p\lesssim a$; in the case of a hard binary, $M_\star(r_p=a)\ll M_{12}$ and the Jacobian determinant reduces to
\begin{eqnarray}
\left|\frac{\partial(E,L)}{\partial(p,v_\infty)}\right| \approx v_\infty^2 .
\end{eqnarray}

We can now rewrite equation (\ref{Equation:DiffQa}) as
\begin{eqnarray}\label{Equation:DiffCoeffinQ}
\langle\Delta Q\rangle &=& \int_0^\infty \int_0^{p_\mathrm{max}} \frac{N(E,L) }
{P(E,L)} \overline{\delta Q} \left|\frac{\partial(E,L)}{\partial(p,v_\infty)}\right|\, dp\, dv_\infty
\approx \int_0^\infty \int_0^{p_\mathrm{max}} \frac{N(E,L)}{P(E,L)}\, \overline{\delta Q} \, v_\infty^2\, dp\, dv_\infty \nonumber \\
&\approx& 8\pi^2 \int_0^\infty \int_0^{p_\mathrm{max}} Lf_\star(E,L) \,\overline{\delta Q}\,v_\infty^2  dp\, dv_\infty
\end{eqnarray}
and similarly for $\langle(\Delta Q)^2\rangle$.
In these expressions, $E$ and $L$ are understood to be functions of $p$ and $v_\infty$
via equations (\ref{Equation:Eofpv}).

Previous studies \cite[e.g.][]{Quinlan1996,Merritt2001}
have usually modelled the field star distribution as an infinite homogeneous medium with number density $n$ and isotropic velocity distribution $f_v(v)$
(which we normalize such that $4\pi\int_0^\infty f_v v^2 dv = 1$).
The corresponding expressions for the diffusion coefficients are
\begin{subequations}\label{Equation:DiffQhomo}
\begin{eqnarray}
\langle\Delta Q\rangle = \int_0^\infty \int_0^{p_\mathrm{max}} \overline{\delta Q}\times n\times 2\pi p\,dp \times v_\infty\times 4\pi f_v(v_\infty) v_\infty^2 dv_\infty , \\
\langle(\Delta Q)^2\rangle = \int_0^\infty \int_0^{p_\mathrm{max}} \overline{\delta Q^2}\times n\times 2\pi p\,dp \times v_\infty\times 4\pi f_v(v_\infty) v_\infty^2 dv_\infty .
\end{eqnarray}
\end{subequations}
Recalling that $L=pv_\infty$, we see that these are equivalent to equations
(\ref{Equation:DiffCoeffinQ}) if we assume that $f_\star(E,L)=f_\star(E)= nf_v(v)$ and
identify the unperturbed field star velocities with $v_\infty$. 
But a question then arises: in realistic galactic nuclei, density $n(r)$ and velocity dispersion $\sigma(r)$  
are functions of radius. At what radius should we evaluate $n$ and $\sigma$ in equation~(\ref{Equation:DiffQhomo})? Intuition suggests that this radius should be roughly 
the influence radius of the binary; this guess is confirmed in Appendix~\ref{Appendix:n(r)}. 

Rotation of the nuclear cluster is introduced as follows. As above, we choose the $z$-axis to be aligned with the total angular momentum of the stars.
Starting from a nonrotating cluster (i.e. $f_\star=f_\star(E,L)$), we identify stars whose angular
momentum vectors are displaced by an angle larger than $\pi/2$ with respect to the
$z$-axis.
A specified fraction ($2\eta-1$) of these ``counteraligned'' stars have their velocities reversed,
causing their angular momentum vectors also to reverse. 
This operation results in a nonzero total angular momentum of the nucleus while leaving the distribution $N(E,L)$ unchanged.  What does change is the distribution of the directions of the angular momentum
vectors, so that we now have
\bsub\label{Equation:Overline}
\begin{eqnarray}
\overline{\delta Q} &\equiv& \int_0^{2\pi}\int_0^{2\pi}\frac{d\psi_b}{2\pi}\frac{d\psi_f}{2\pi} \frac{1}{4\pi}\int_0^{2\pi}d\varphi_f \cdot
 \left(2\eta\int_0^{\pi/2}\sin\theta_f d\theta_f\,\delta Q + 2(1-\eta)\int_{\pi/2}^\pi\sin\theta_f d\theta_f\,\delta Q
 \right), \\
\overline{\delta Q^2} &\equiv& \int_0^{2\pi}\int_0^{2\pi}\frac{d\psi_b}{2\pi}\frac{d\psi_f}{2\pi} \frac{1}{4\pi}\int_0^{2\pi}d\varphi_f \cdot
 \left(2\eta\int_0^{\pi/2}\sin\theta_f d\theta_f\,\delta Q^2 + 2(1-\eta)\int_{\pi/2}^\pi\sin\theta_f d\theta_f\,\delta Q^2\right).
\end{eqnarray}
\esub
Here $\psi_b$ is binary's initial mean anomaly, $\varphi_f$ and $\theta_f$ are the spherical coordinates of the field star's angular momentum direction (with $\boldsymbol{n}$ taken as the polar axis), and $\psi_f$ determines the direction of star's initial velocity in its orbital plane (i. e. the direction from which the field star is initially approaching; see Figure~\ref{Figure:star}). 
Setting $\eta=1/2$ would correspond to a nonrotating nucleus, while $\eta=0$ or $\eta=1$ 
represents a ``maximally'' counter- or corotating nucleus.

We can see immediately from Eq.~(\ref{Equation:Overline}) that the dependence of any diffusion coefficient on the degree of corotation $\eta$ is always linear and thus completely defined just by two parameters; convenient choices 
are $\langle\Delta Q\rangle_{\eta=1/2}$ and $\langle\Delta Q\rangle_{\eta=1}$, so that the value of $\langle\Delta Q\rangle$ at some intermediate value of $\eta$ is just a linear combination of these two:
\beq\label{Equation:DeltaQ(eta)}
\langle\Delta Q\rangle (\eta) = \langle\Delta Q\rangle_{\eta=1/2} \cdot 2(1-\eta) + \langle\Delta Q\rangle_{\eta=1} \cdot 2(\eta-1/2).
\eeq

The distribution of $v_\infty$ was assumed to be Maxwellian:
\beq\label{Equation:MaxwellianDistribution}
f_v(v_\infty) = \frac{1}{(2\pi\sigma^2)^{3/2}}e^{-v_\infty^2/2\sigma^2} .
\eeq
The resulting expressions for the diffusion coefficients are obtained by combining 
equations~(\ref{Equation:DiffQhomo}), (\ref{Equation:Overline}) and (\ref{Equation:MaxwellianDistribution}):
\begin{subequations}\label{Equation:DQ}
\begin{eqnarray}
\langle\Delta Q\rangle &=& na^2\sigma\cdot2\sqrt{2\pi} \int_0^\infty \int_0^{p_\mathrm{max}} \frac{pdp}{a^2} \cdot \frac{v^3dv}{\sigma^4} e^{-v^2/2\sigma^2} \overline{\delta Q} , \\
\langle\left(\Delta Q\right)^2\rangle &=& na^2\sigma\cdot2\sqrt{2\pi} \int_0^\infty \int_0^{p_\mathrm{max}} \frac{pdp}{a^2} \cdot \frac{v^3dv}{\sigma^4} e^{-v^2/2\sigma^2} \overline{\delta Q^2} , \\
\overline{\delta Q} &\equiv& \int_0^{2\pi}\int_0^{2\pi}\frac{d\psi_b}{2\pi}\frac{d\psi_f}{2\pi} \frac{1}{4\pi}\int_0^{2\pi}d\varphi_f \cdot
 \left(2\eta\int_0^{\pi/2}\sin\theta_f d\theta_f\,\delta Q + 2(1-\eta)\int_{\pi/2}^\pi\sin\theta_f d\theta_f\,\delta Q\right) , \\
\overline{\delta Q^2} &\equiv& \int_0^{2\pi}\int_0^{2\pi}\frac{d\psi_b}{2\pi}\frac{d\psi_f}{2\pi} \frac{1}{4\pi}\int_0^{2\pi}d\varphi_f \cdot
 \left(2\eta\int_0^{\pi/2}\sin\theta_f d\theta_f\,\delta Q^2 + 2(1-\eta)\int_{\pi/2}^\pi\sin\theta_f d\theta_f\,\delta Q^2\right) .
\end{eqnarray}
\end{subequations}

Numerically, $\langle\Delta Q\rangle$ and $\langle(\Delta Q)^2\rangle$ 
were computed after replacing the integrals by summations over discrete field star-binary encounters.
The latter were computed in much the same manner as in previous studies \citep[e.g.][]{Quinlan1996,Merritt2002,Sesana2006}, by integrating the trajectories of
massless ``stars'' in the time-dependent gravitational field of the massive binary.
Integrations were carried out using ARCHAIN, an implementation of algorithmic regularization  \citep{Mikkola2008}. 
ARCHAIN was developed to treat small-$N$ systems.
We found that for three-body systems, ARCHAIN can be even faster than an algorithm which 
advances the binary orbit via Kepler's equation and integrates only 
the field star's equations of motion, as in the studies just cited.
In the case of circular binaries, the relative change in Jacobi's constant
was always less than $10^{-5}$.

Field star trajectories were assumed to be Keplerian until the star had approached
within a distance of $50a$ from the binary's center of mass, after which 
the orbit was numerically integrated until it had exited the sphere of radius $50a$ with positive total energy.
The final energy and angular momentum of the star were then recorded. Given the changes in the field-star trajectory, the changes $\delta Q$ or $\delta Q^2$ 
were computed using the expressions in \S \ref{Section:DiffCoeffOne}. 
If this did not happen after about $10^4$ binary periods, 
the star was considered to be captured by the binary, 
and it was not included when computing the diffusion coefficients. 
The fraction of captured stars was always less than $1\%$. 

Finally, the ``VEGAS'' method 
developed by \citet{Lepage1980} was used to numerically calculate the integrals.
We used the implementation in the GNU Scientific Library
\citep{GSL}.
The VEGAS algorithm is based on importance sampling: it samples points from the probability distribution described by the absolute value of the integrand, so that the points are concentrated in the regions that make the largest contribution to the integral. In practice it is not possible to sample from the exact distribution for an arbitrary function; the VEGAS algorithm approximates the exact distribution by making a number of passes over the integration region while histogramming the integrand. Each histogram is used to define a sampling distribution for the next pass. Asymptotically this procedure converges to the desired distribution.

\subsection{\label{Section:DiffCoeffBoundStars}Bound vs. unbound stars}
In the scattering experiments, all field-star orbits are initially unbound with respect to the binary.
Some of these orbits have periapsis parameters $(r_p,v_p)$ that are associated
also with bound orbits in the full galactic potential, i.e., orbits with $E<0$,
and these are the orbits that will appear in integrals like that of equation
(\ref{Equation:DiffCoeffinQ}).
However, some ($p, v_\infty$) values map onto orbits with $E>0$ in the full
galactic potential,
and there likewise exist orbits with $E<0$ having periapsis parameters that
are not matched by orbits with any ($p,v_\infty$) in the scattering experiments.
Some orbits of very negative $E$ fall into this category, since they move
effectively in the potential of the binary alone (like in the scattering experiments)
but are nevertheless bound to the binary (unlike in the scattering experiments).
Orbits such as these will {\it not} be represented in integrals like (\ref{Equation:DiffCoeffinQ})
even though they might exist in the real galaxy, and this is a potential source
of systematic error in our computation of the diffusion coefficients.

\begin{figure}[h!]
\plottwo{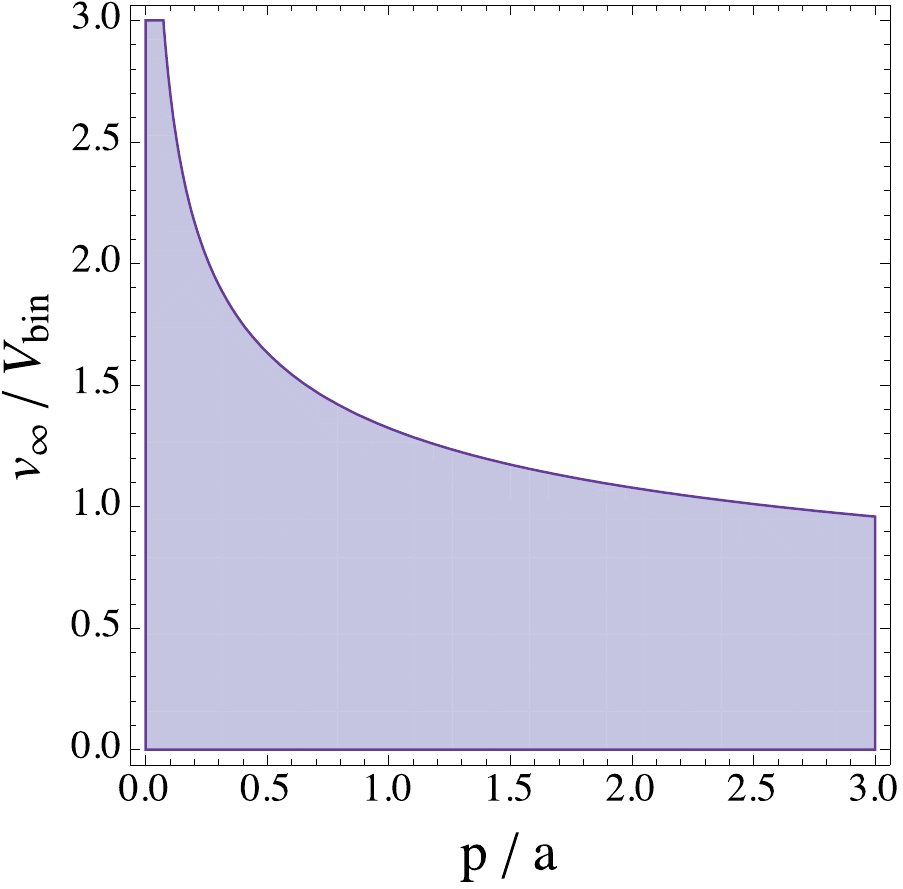}{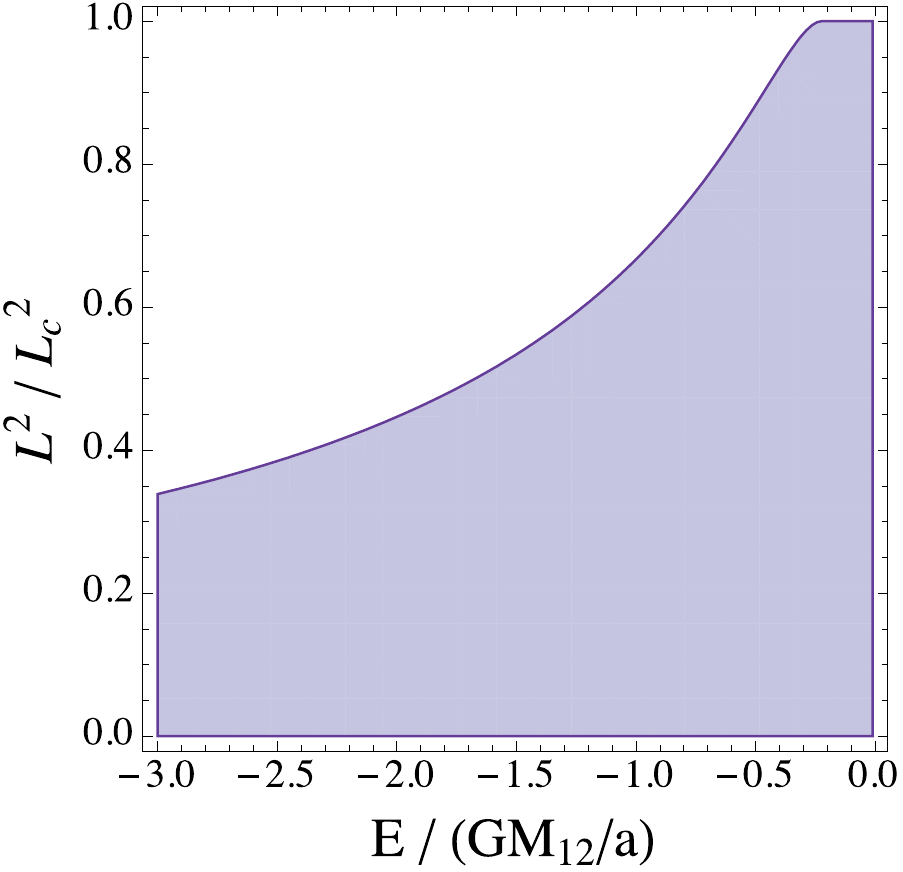}
\caption{Left: shaded (blue) area corresponds to orbits in the scattering experiments which
would be bound to the galaxy given our adopted mapping $(p,v_\infty)\Leftrightarrow (E,L)$,
assuming $\gamma=5/2$ and $S=6$.
Right: shaded area shows the region in $(E,L)$ space corresponding to orbits that would be 
included in the scattering experiments, also for $\gamma=5/2$, $S=6$.
\label{Figure:pv_region_power-law}}
\end{figure}

To get a better idea of which orbits in the galactic potential are being excluded,
we adopt a particular form for the stellar density profile:
\begin{eqnarray}
\rho(r) = \rho_0\left(\frac{r}{r_0}\right)^{-\gamma} , \ \ \ \ \gamma<3 .
\end{eqnarray}
The potential induced by the stars (excluding the case $\gamma=2$) is
\begin{eqnarray}
\Phi_\star(r) = -\Phi_0\left(\frac{r}{r_0}\right)^{2-\gamma},\ \
\Phi_0=\frac{4\pi G\rho_0r_0^2}{(3-\gamma)(\gamma-2)}
\end{eqnarray}
and the energy of a  star, expressed in terms of $p$ and $v_\infty$ via
the mapping defined above, is
\begin{eqnarray}
E = - \frac{GM_{12}}{r_p(p,v_\infty)} - \Phi_0\left(\frac{r_p(p,v_\infty)}{r_0}\right)^{2-\gamma} + \frac{p^2v^2_\infty}{2r_p^2(p,v_\infty)} .
\end{eqnarray}
The condition $E<0$ turns out to be equivalent to
\begin{eqnarray}\label{Equation:BoundCondition}
\frac{v_\infty^2}{2} < \Phi_0 \left[\frac{r_p(p,v_\infty)}{r_0}\right]^{2-\gamma} .
\end{eqnarray}
If we measure $p$ and $v_\infty$ in units of $a$ and $V_\mathrm{bin}=\sqrt{GM_{12}/a}$,
i.e.
\beq
\tilde p = \frac{p}{a}, \ \ \ \ 
\tilde{v}_\infty = \frac{v_\infty}{V_\mathrm{bin}},
\eeq
we can rewrite this condition as
\begin{eqnarray}
\frac{\tilde{v}_\infty^2}{2} < \frac{2}{S^{6-2\gamma}(\gamma-2)} 
\left(\frac{\tilde{p}^2 \tilde{v}_\infty^2}{1+\sqrt{1+\tilde{p}^2\tilde{v}_\infty^4}}\right)^{2-\gamma}
\end{eqnarray}
where $S$ is a dimensionless measure of the binary hardness:
\begin{eqnarray}
S \equiv \sqrt{\frac{r_m}{a}}
\end{eqnarray}
with $r_m$ defined as the radius containing a mass in stars equal to $2M_{12}$.
(If we, arbitrarily, replace $r_m$ in this expression with  $r_\mathrm{infl}\equiv GM_{12}/\sigma^2$,
then $S=V_\mathrm{bin}/\sigma$ which is a more common definition of binary hardness.
The two definitions are equivalent in an ``isothermal'' nucleus, i.e. $\rho\propto r^{-2}$ and
$\sigma=$ const.)

Values of $p$ and $v_\infty$ that violate the condition (\ref{Equation:BoundCondition}) correspond (via our adopted mapping) to orbits that would be unbound and hence not present in the galaxy.
Figure~\ref{Figure:pv_region_power-law}a illustrates the allowed values of ($p,v_\infty$)
for the case $\gamma=5/2$, $S=6$.

We are more interested in the values of $(E,L)$ that are not accessible,
via the mapping (\ref{Equation:Eofpv}), to any $(p,v_\infty)$.
 Figure~\ref{Figure:pv_region_power-law}b illustrates the allowed $(E,L)$ region
for the power-law model with $\gamma=5/2$.
(We chose a relatively large $\gamma$ so that the stellar gravitational potential $\Phi_\star(r)$ would not be infinitely large at infinity -- that would cause problems since most of the interacting stars come from large distances.)
We see that tightly-bound orbits, $E\rightarrow -\infty$, are representable
but only if they are very eccentric.
Orbits that are highly bound and nearly circular are excluded.

\begin{figure}[h]
\includegraphics[width=0.5\textwidth]{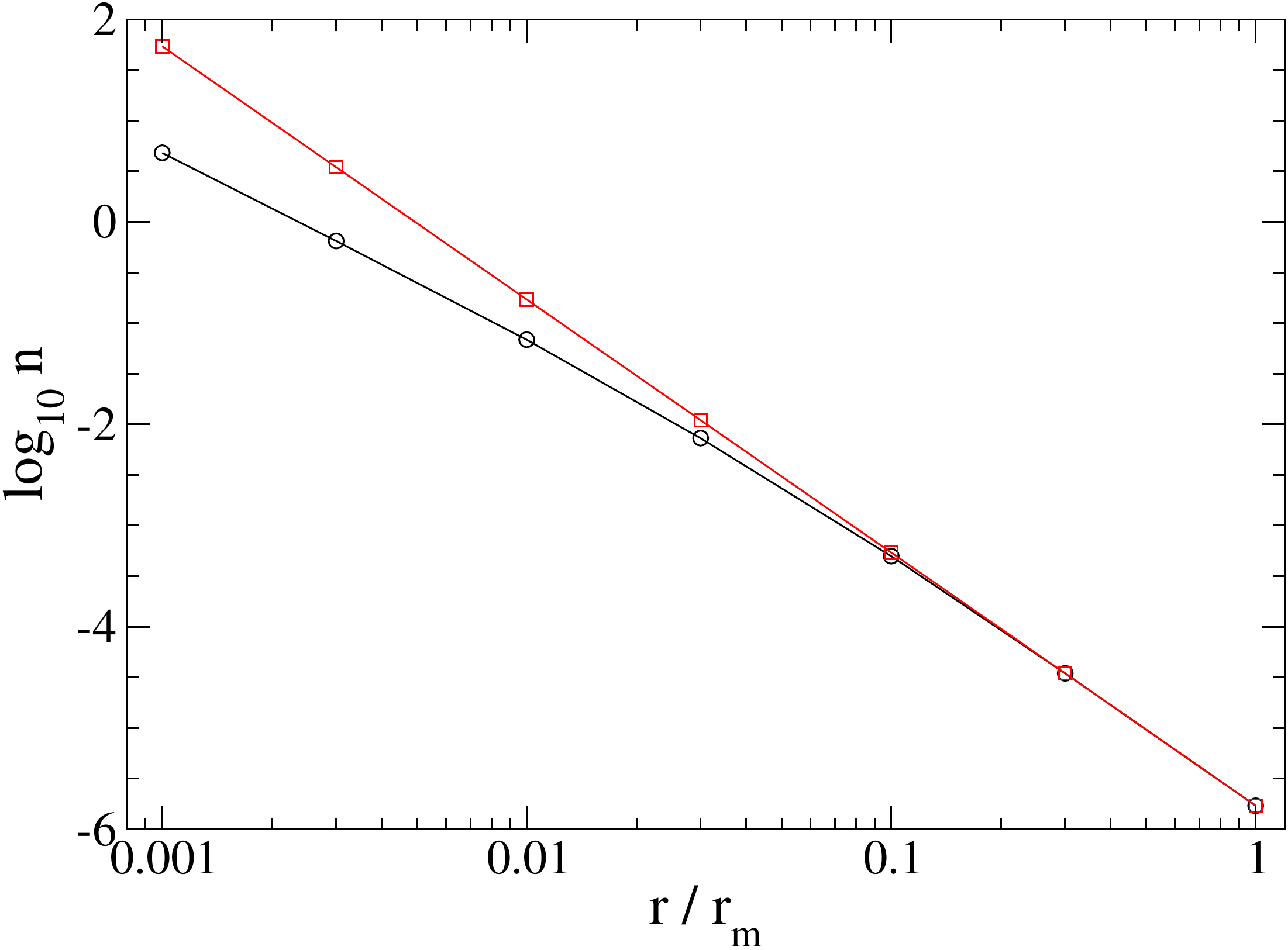}
\caption{Black: number density of stars having $(E,L)$ values that
are representable via the scattering experiments.
Red: total number density.
This figure assumes a power-law density profile, $n\propto r^{-5/2}$, and a binary hardness $S=6$.}
\label{Figure:n_unbound_power-law}
\end{figure}
By excluding certain orbits, we are in effect changing the density profile
of the stars that are allowed to interact with the binary.
Figure~\ref{Figure:n_unbound_power-law} compares the number density of all stars
in the galaxy with the density of stars that are representable via the scattering experiments,
again for $S=6$, $\gamma=5/2$.
When we carry out the same analysis for a more realistic, broken-power-law density, 
the pictures for $(E,L)$ and $(r_p,v_p)$ stay qualitatively the same,
while the region in $(p,v_\infty)$ has lost its high-velocity tail.
We would argue that this loss is not important given that, for a hard binary, 
most of the stars have initial velocities (at infinity) $\sim S^{-1} V_\mathrm{bin} \ll V_\mathrm{bin}$.

So far, we have ignored the possible effects of stars that are bound to the massive binary. 
Such stars can of course interact with the  binary and influence the evolution of its orbital parameters. 
That influence was studied by \citet{Sesana2008} and \citet{Sesana2010}, who used a ``hybrid'' code
that combined scattering experiments with an approximate representation of the dynamical evolution of the 
nucleus. 
They found that the ejection of bound stars can significantly change the binary's orbit, 
but that once such stars are ejected, essentially no stars replace them, 
and subsequent evolution of the binary is only due to the unbound stars. 
The closer the binary's mass ratio is to one, the shorter is the characteristic time for depletion
of the initially-bound stars, and for equal-mass binaries that time is only a few binary periods. 
Furthermore, in full $N$-body simulations starting from realistic (pre-merger) initial conditions \citep{Milosavljevic2001, GualandrisMerritt2012} and mass ratios close to unity, the early phase
of evolution due to bound stars is not observed; perhaps because this phase is so short that it
can not be distinguished from the phase of binary formation.

\section{Understanding the results from the scattering experiments}
\label{Section:UnderstandingScattering}
Here we discuss some systematic features arising from the scattering experiments,
particularly in regard to the direction of the field-star angular momentum changes,
and provide some quantitative interpretations.
Unless otherwise indicated, results in this section are presented in dimensionless
units such that $GM_{12}=a=1$.
All experiments in this section adopt a circular-orbit, equal-mass binary, and spherically symmetric distribution of stellar velocities and angular momenta.
\begin{figure*}[h]
	\centering
	\subfigure{\includegraphics[width=0.49\textwidth]{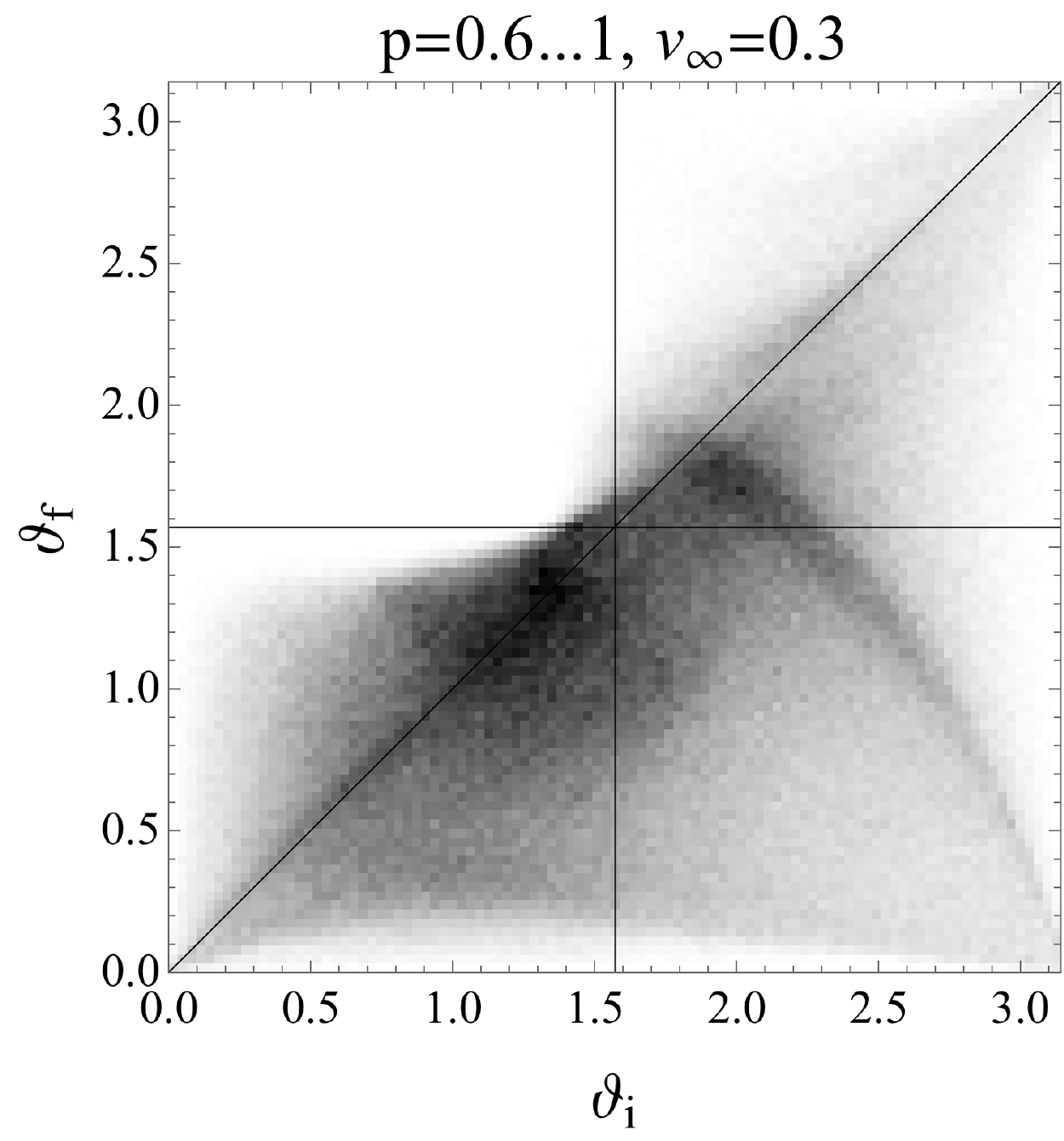}}
	\subfigure{\includegraphics[width=0.49\textwidth]{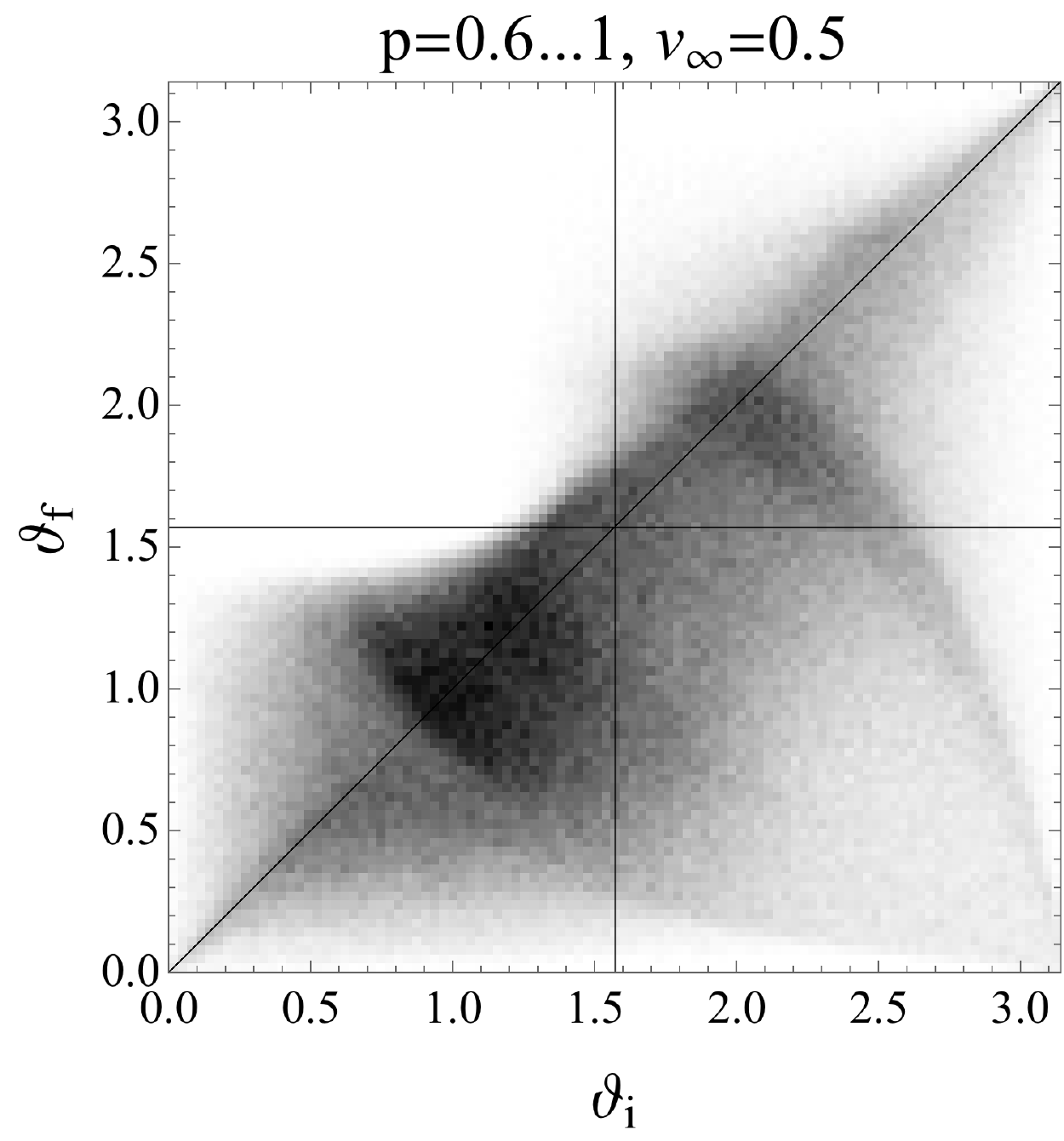}}
	\subfigure{\includegraphics[width=0.49\textwidth]{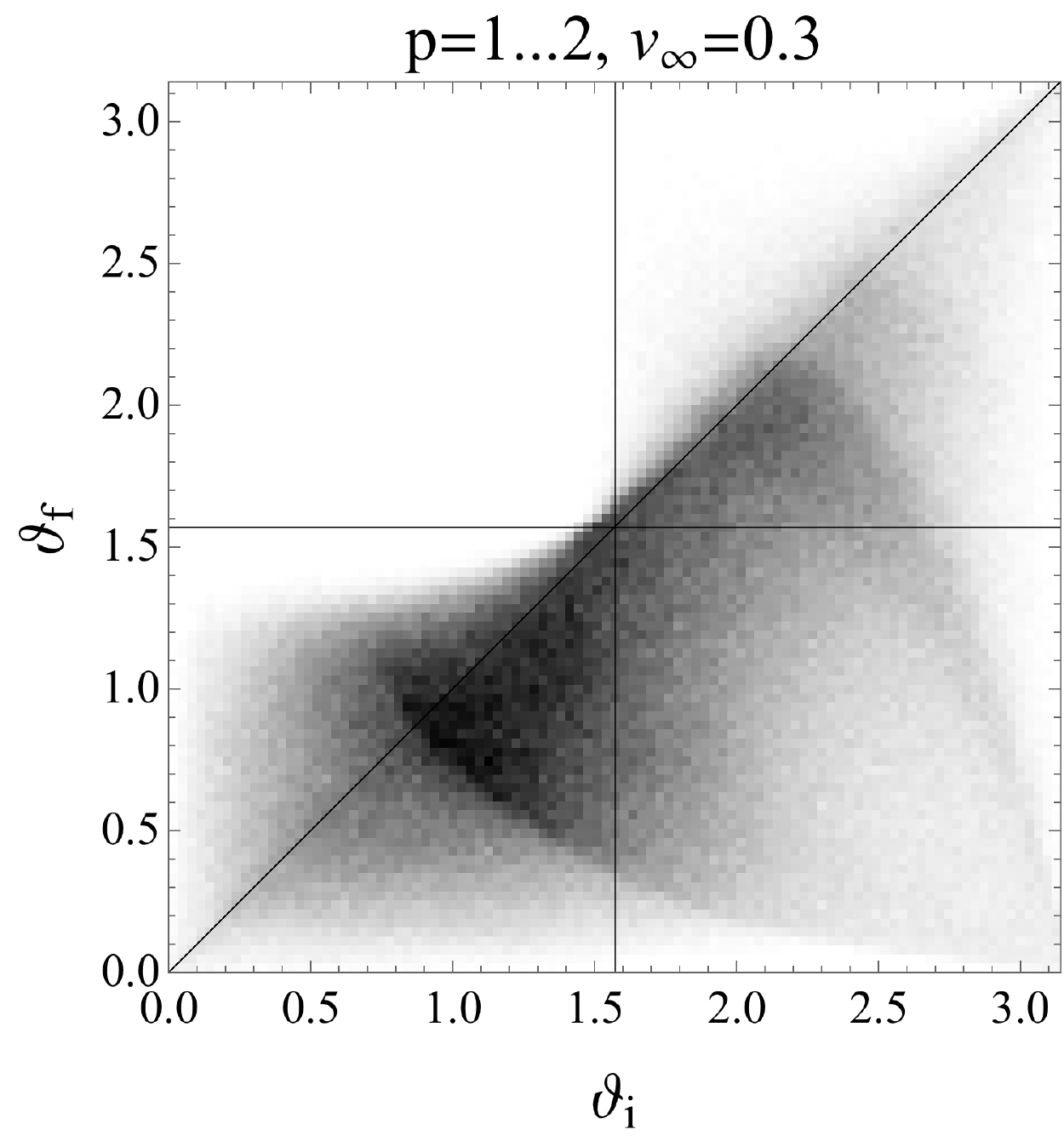}}
	\subfigure{\includegraphics[width=0.49\textwidth]{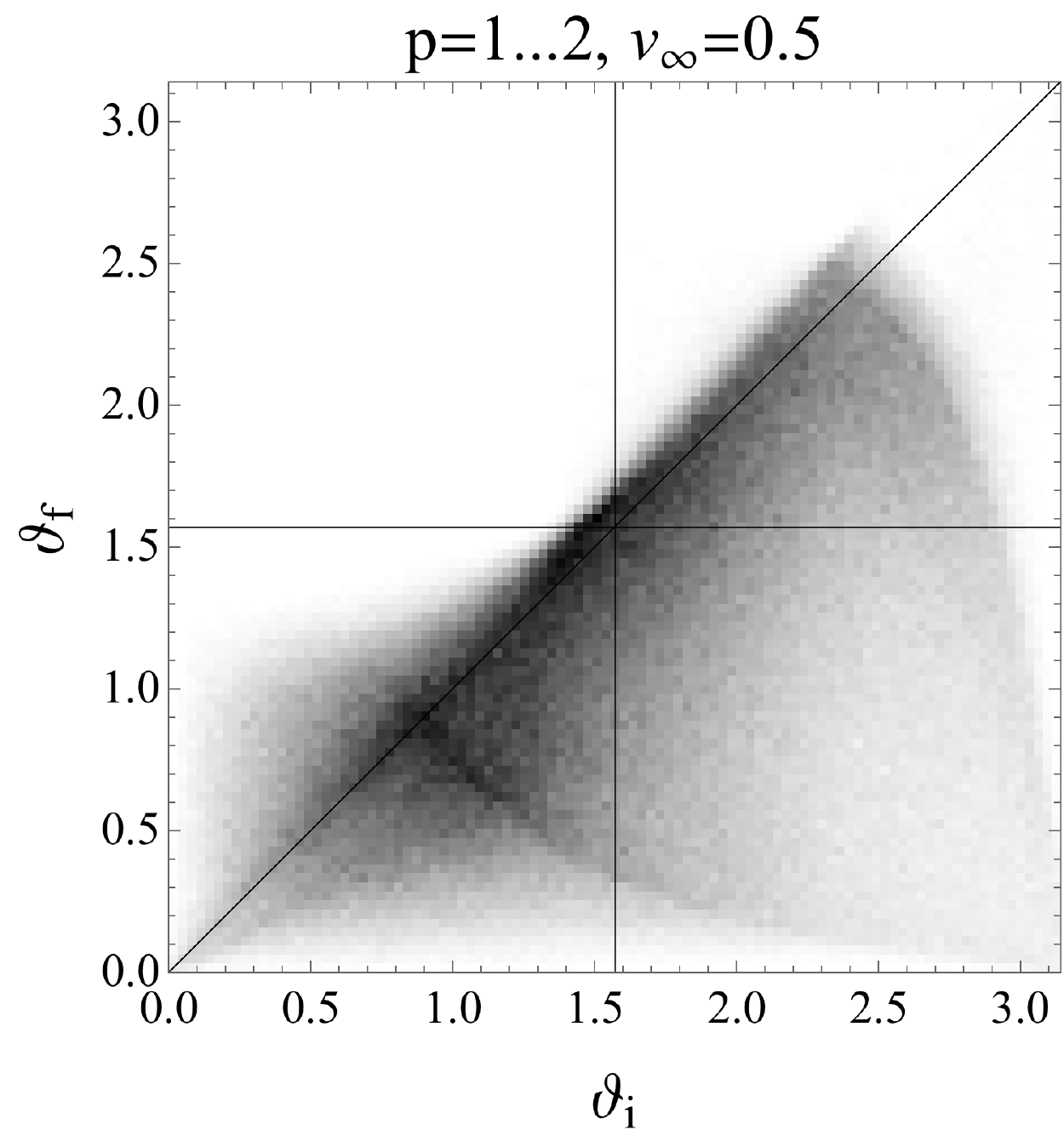}}
\caption{Density plots showing the final angle, $\theta_f$, between $\boldsymbol{l}$ and $\boldsymbol{l}_\mathrm{bin}$ versus the initial angle, $\theta_i$.
Each frame contains results from $10^6$ trajectories for different values of $p$ and $v_\infty$.
}
\label{Figure:Flipping}
\end{figure*}

A striking result from the numerical integrations is illustrated in Figure~\ref{Figure:Flipping},
which shows the relation between $\vartheta_i$ and $\vartheta_f$, the initial and final values
of the angle $\vartheta$ between $\boldsymbol{l}$ and $\boldsymbol{l}_\mathrm{bin}$.
Stars that are initially counterrotating with respect to the binary
($\pi/2\lesssim\vartheta_i\lesssim\pi$) tend to become corotating
after the interaction ($0\lesssim\vartheta_f\lesssim \pi/2$), as if their orbits had been ``flipped''.
Orbits that are initially corotating, on the other hand, tend to remain corotating.
Stated differently: stars tend to align their angular momenta with that of the binary.

Inspection of the detailed orbits of stars that undergo significant changes in their orbital parameters suggests that most of them interact with the binary in a series of
brief and close encounters (distances $\ll a$) with $M_1$ and/or $M_2$,
continuing until the star is  ejected.
Furthermore, in the case of the initially nearly corotating stars, the number of close interactions can reach a few tens, while almost all of the initially counterrotating stars
experience ejection after just one close interaction.
The probable reason is that a counterrotating star has larger velocity with respect to the binary component that it closely interacts with, making a ``capture'' less likely.

Inspection of plots like those in Figure~\ref{Figure:Flipping} reveals another
regularity in the outcomes of the scattering experiments:
values of $\{p,v_\infty\}$ that imply the same $r_p$ for the
initial orbit, equation (\ref{Equation:r_p}), tend to yield similar results
(e.g. the upper-right and lower-left panels in Figure~\ref{Figure:Flipping}).

While the interaction of a field star with the binary is typically chaotic in character,
there can be conserved quantitites associated with the star's motion,
and the existence of such quantities might help to explain
regularities like those discussed above.
In the restricted circular three-body problem (i.e. a zero-mass field star
interacting with a circular-orbit binary), the Jacobi integral $H_J$ is precisely conserved
\citep[][Eq. 8.168]{DEGN}:
\begin{subequations}\label{Equation:DefineJacobiConstant}
\begin{eqnarray}\label{Equation:JacobiConstanta}
H_\mathrm{J} &=& 2\left(\frac{GM_1}{r_1} + \frac{GM_2}{r_2}\right)
+2nl_z -\dot{x}^2 - \dot{y}^2 - \dot{z}^2 \\
&=& -2\left(E - nl_z\right)
\label{Equation:JacobiConstantb}
\end{eqnarray}
\end{subequations}
where $l_z = x\dot{y} - y\dot{x}$ is the specific angular momentum of the field
star with respect to the binary center of mass, $r_1$ and $r_2$ are the distance
of the field star from $M_1$ and $M_2$ respectively, and $n=2\pi/P$ is
the (fixed) angular velocity of the binary, whose angular momentum is aligned
with the $z$-axis.

At times either long before or long after its interaction with the binary,
the field star's Jacobi integral is
\begin{equation}\label{Equation:Jacobi}
H_\mathrm{J} \approx -v^2 + 2n\,l_z .
\end{equation}
Conservation of $H_\mathrm{J}$, in the case of a circular-orbit binary, therefore implies that
the total change in the field star's energy is related to the change in the
component of its angular momentum parallel to $\boldsymbol{l}_\mathrm{bin}$:
\begin{equation}\label{Equation:deltas}
\delta E = n\,\delta l_z .
\end{equation}
A star that escapes to infinity must have final energy $E = v^2/2 > 0$, so from equation (\ref{Equation:deltas}) it follows that a lower limit exists on $\delta l_z$:
\begin{equation}\label{boundary}
\delta l_z > -\frac{v_\infty^2}{2n}
\end{equation}
where $v_\infty$ is, as always, the field-star velocity at $t\rightarrow -\infty$.
\begin{figure}
\includegraphics[width=\textwidth]{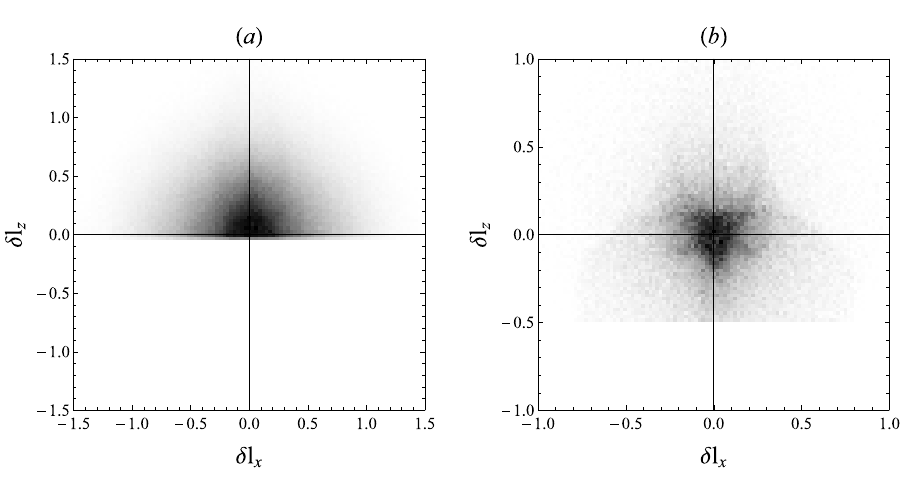}
\caption{(a) Distribution of angular momentum changes for the stars with
$p=1\ldots2, v_\infty=0.3$. (b) Distribution of angular momentum changes for the stars with
$p=0.3, v_\infty=1$. \label{Figure:dlxdlz}}
\end{figure}

Figure~\ref{Figure:dlxdlz}(a) illustrates this result, based on scattering experiments
with a circular-orbit, equal-mass binary
and an assumed isotropic distribution of field stars having impact parameters
in the range $p=1\ldots 2$ and a single velocity $v_\infty=0.3$; the sharp lower boundary
is at $-v_\infty^2/2 \approx -0.045$.
Since a typical value of $|\delta l|$ is $\sim 1$
(both the torque acting on a star during an encounter,
and the time it spends close to the binary, are of order unity),
which is much greater than $v_\infty^2/2$,
it is not surprising that $\langle\delta l_z\rangle \sim 1$,
hence $\delta E>0$, i.e. most encounters take energy from the binary.
Now, if we imagine increasing $v_\infty$, the lower bound on $\delta l_z$ becomes
smaller (more negative).
This is illustrated by Figure~\ref{Figure:dlxdlz}(b) which sets $(p, v_\infty)=(0.3, v=1)$, for which
$(\delta l_z)_\mathrm{min}\approx -0.5$.
 In this case, the average $\delta l_z$, i.e. the average energy gain, is almost zero
(even negative, if we take only the corotating stars, as discussed below).

Recall that in our adopted units, a typical field-star velocity is
$v_\infty\approx \sigma=1/S\ll 1$ for a hard binary, implying
 $v_\infty^2/2 \ll 1$, hence $\langle\delta l_z\rangle > 0$.
This leads us to the conclusion that only stars with
$v_\infty \lesssim v_\mathrm{bin}$ contribute to hardening of the binary.

\citet{VrceljKiewietdeJonge1978} found a conserved quantity, analogous to the Jacobi integral, in the non-circular restricted three-body problem; however, it contains a nonintegrable term, becoming integrable only in special cases:
\begin{equation}\label{Equation:vrcelj}
\varepsilon - \frac{n}{(1-e^2)^{1/2}}l_z + \mu n^2a^2\frac{e}{1-e^2}\frac{\delta e}{m_f} = \mathrm{const}
\end{equation}
Here $\delta e = \int_0^t \mathrm{d}e$ is the net change in binary's eccentricity, calculated in the approximation of infinitesimal field-star mass $m_f$ (which means that $\delta e\sim m_f$ and $\delta e/m_f$ doesn't depend on $m_f$).
This relation is actually equivalent to (\ref{Equation:deltae}) -- we need only recall that $n=2\pi/T=\sqrt{GM_{12}/a^3}$ and the constant on the right-hand side of equation (\ref{Equation:vrcelj}) is the initial value of the left-hand side:
\begin{equation}
\mathrm{const} = \varepsilon_i - \frac{n}{(1-e^2)^{1/2}}l_{z,i}
\end{equation}
which yields
\begin{equation}
\delta\varepsilon - \frac{\sqrt{GM/a^3}}{(1-e^2)^{1/2}}\delta l_z + \mu \frac{GM}{a}\frac{e}{1-e^2}\frac{\delta e}{m_f} = 0
\end{equation}
which is the same as (\ref{Equation:deltae}).
In the case of a circular binary, $e=0$, the last term is zero and this generalized conserved quantity turns into Jacobi constant.
In the case of a large-mass-ratio binary, the last term also becomes negligible, which
results in
\begin{equation}
\varepsilon - \frac{n}{(1-e^2)^{1/2}}l_z = \mathrm{const}
\end{equation}
This expression, very similar to the Jacobi constant, gives us a limitation for the angular momentum change, similar to that of equation (\ref{boundary}):
\begin{equation}
\delta l_z > -\frac{v_0^2(1-e^2)}{2n} .
\end{equation}
This, in turn, allows us to use the arguments analogous to those presented in the previous section to explain the net increase in the binary's angular momentum in the case of binary large mass ratio and any eccentricity.


\section{Numerical calculation of diffusion coefficients}
\label{Section:DiffusionCoefficients}
In this section we present values for the drift and diffusion coefficients that describe changes in 
the binary's orbital elements,
as computed from the scattering experiments in the manner described above 
(\S~\ref{Section:Scattering}).
Results are presented for the orbital elements $a$ (semimajor axis),
$\theta$ (orbital inclination), $e$ (eccentricity),
$\omega$ (argument of periapsis),
and $\Omega$ (longitude of ascending node).

With the exception of the diffusion coefficients for $\omega$ itself, results presented here are
averaged over $\omega$ (except in the special cases where $\omega$ is ignorable, e.g. 
$e=0$).

The diffusion coefficients are functions of the orbital elements themselves, 
as well as the following three parameters:

\begin{itemize}
\item The ratio of binary component masses, $q\equiv M_1/M_2$. 
Usually we assume $q\geq1$, but unless otherwise specified, the formulae we give stay the same when one replaces $q$ for $1/q$.
\item The degree of corotation of the stellar nucleus, $\eta$  (see \S\ref{Section:Scattering}, Equation~\ref{Equation:DQ}).
$\eta=1/2$ corresponds to a nonrotating nucleus, $\eta=1$ or $\eta=0$ to a maximally co- or counterrotating nucleus (defined with respect to the sense of rotation of the binary).
\item The upper cutoff to the impact parameter of incoming stars, $p_{\rm max}$. 
Ideally, we would want to set $p_{\rm max}=\infty$. 
We found that increasing $p_{\rm max}/a$ above $\sim 6S = 6 V_\mathrm{bin}/\sigma$ did not result in any appreciable change in any of the diffusion coefficients, so we fixed $p_{\rm max}/a$ at $6S$ in what follows.
\end{itemize}
\noindent
Aside from $p_\mathrm{max}$ and $\omega$, 
there are six parameters on which the diffusion coefficients can depend: $a$, $\theta$, $e$, $\Omega$, $q$ and $\eta$.
This is too large a number to explore fully, but in what follows, we attempt to identify the most important
dependences.

\subsection{\label{Section:D(a)}Drift and diffusion coefficients for the semimajor axis}

A standard definition of the dimensionless binary hardening rate \citep[e. g.][Section 8.1]{DEGN} is
\beq\label{Equation:DefinitionOfH}
H\equiv\frac{\sigma}{G\rho}\frac{d}{dt}\left(\frac{1}{a}\right).
\eeq
In the Fokker-Planck formalism, $da/dt$ corresponds to $\langle\Delta a\rangle$. 
Accordingly, we express the first- and second-order diffusion coefficients for $a$ in terms
of the dimensionless quantities $H$ and $H'$, as follows:

\bsub\label{Equation:HH'}
\barr
\langle\Delta a\rangle &=& - \frac{\rho G a^2}{\sigma} H\label{Equation:HtoDeltaa}, \\
\langle(\Delta a)^2\rangle &=& \frac{m_f}{M_{12}} \frac{\rho G a^3}{\sigma} H'\label{Equation:H'toDeltaa2}, \\
H &=& - \frac{4\sqrt{2\pi} S^2}{\nu} \int_0^\infty \int_0^{p_{max}/a} \,\frac{pdp}{a^2} \,dz \,z^3 e^{-(Sz)^2/2} \frac{\overline{\delta\varepsilon}}{GM_{12}/a} \label{Equation:H}, \\
H' &=& \frac{4\sqrt{2\pi} S^2}{\nu^2} \int_0^\infty \int_0^{p_{max}/a} \,\frac{pdp}{a^2} \,dz \,z^3 e^{-(Sz)^2/2} \frac{\overline{\delta\varepsilon^2}}{\left(GM_{12}/a\right)^2} \label{Equation:H'}
\earr
\esub
with $S \equiv V_\mathrm{bin}/\sigma$,  $z \equiv v/V_\mathrm{bin}$ and $\nu \equiv \mu/M_{12}=q/(1+q)^2$.
Here $\delta\varepsilon$ is the change in specific energy of the star during one interaction with the binary (see \S\ref{Section:DiffCoeffOne}, in particular Equation~\ref{Equation:deltaa}).
For convenience, we henceforth adopt the following notational convention:
\beq
\overline{\overline{\delta Q}} \equiv 4\sqrt{2\pi} S^2 \int_0^\infty \int_0^{p_{max}/a} \,\frac{pdp}{a^2} \,dz \,z^3 e^{-(Sz)^2/2} \overline{\delta Q}
\eeq
so that Equations~(\ref{Equation:H}) and (\ref{Equation:H'}) become
\beq
H = -\frac{2}{\nu}\frac{\overline{\overline{\delta\varepsilon}}}{GM_{12}/a},\ \ \ \ 
H' = \frac{4}{\nu^2}\frac{\overline{\overline{\delta\varepsilon^2}}}{(GM_{12}/a)^2}.
\eeq

\begin{figure}
	\centering
	\subfigure[]{\includegraphics[width=0.49\textwidth]{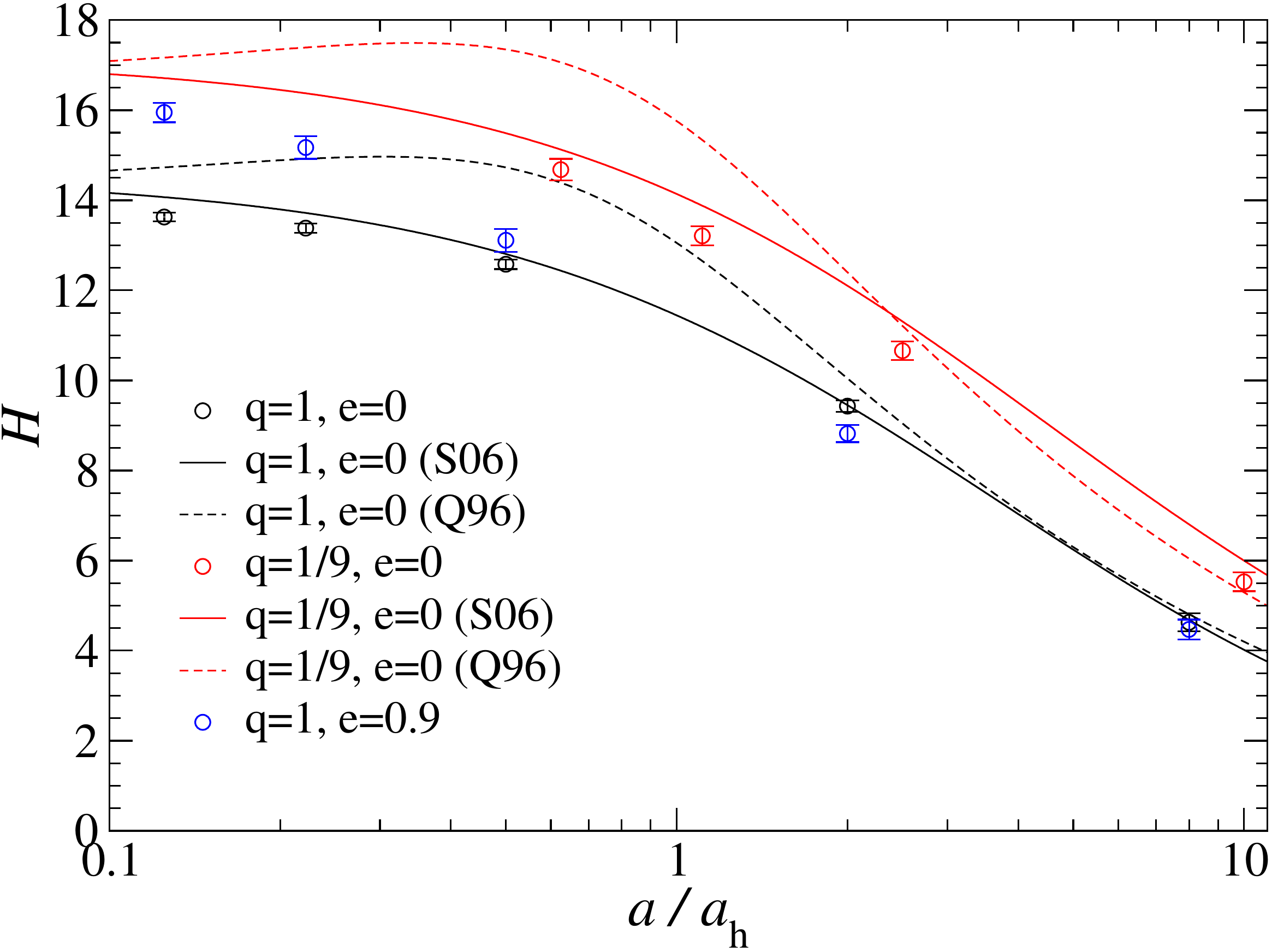}}
	\subfigure[]{\includegraphics[width=0.49\textwidth]{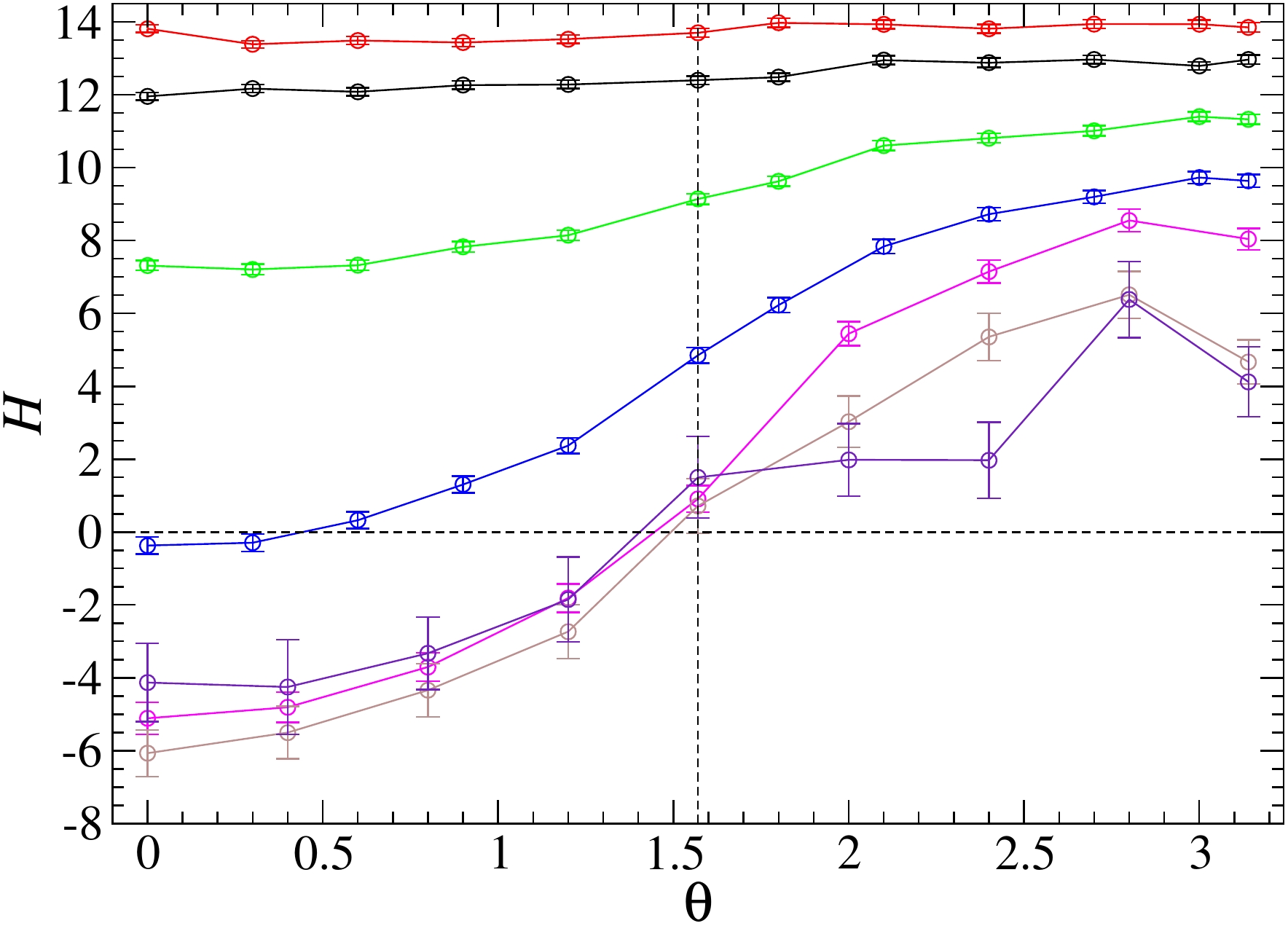}}
	\subfigure[]{\includegraphics[width=0.49\textwidth]{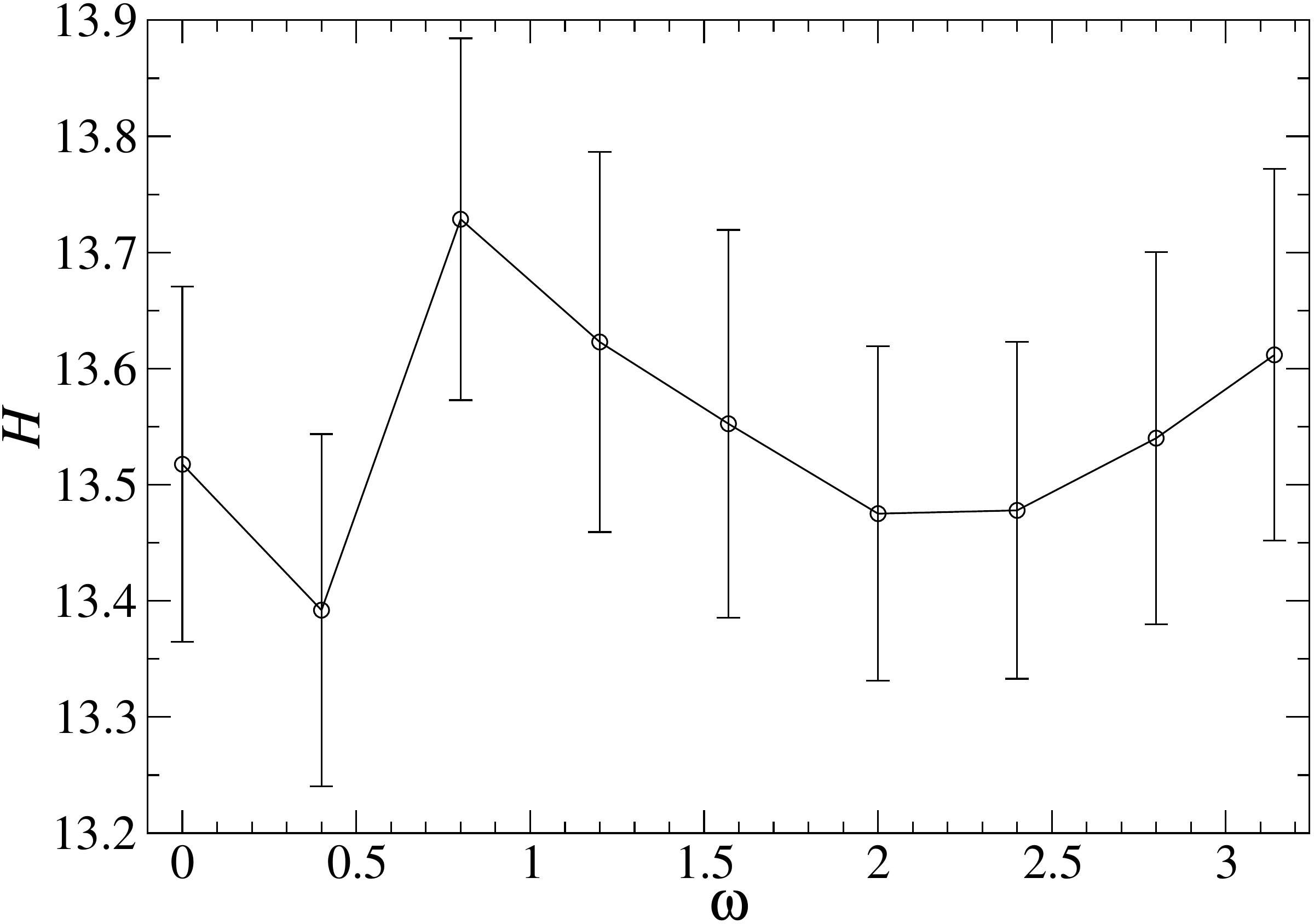}}
	\subfigure[]{\includegraphics[width=0.49\textwidth]{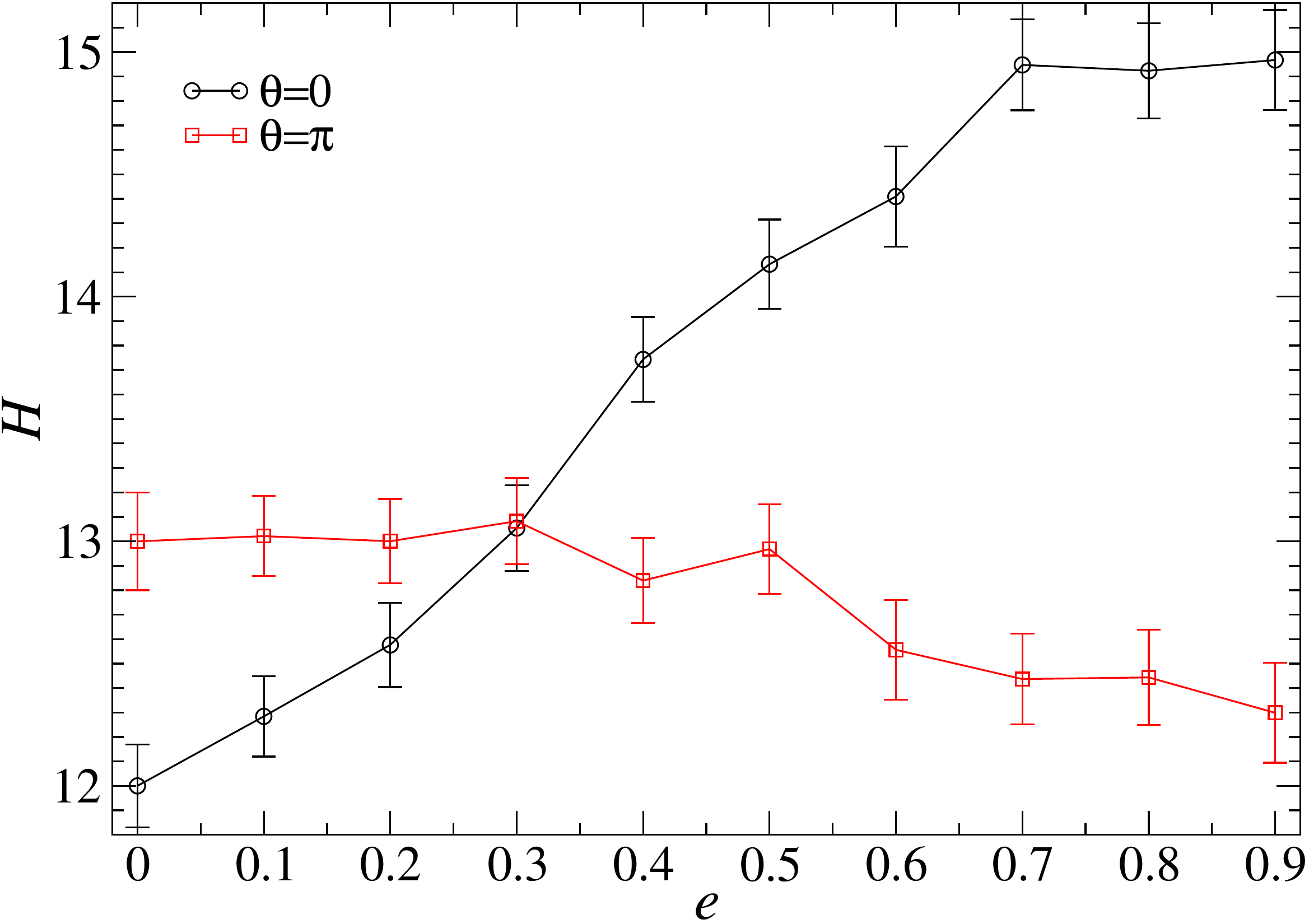}}
	\caption{
	Dimensionless hardening rate $H$ (Equation~\ref{Equation:H}) as a function of binary orbital elements
	and the parameters that define the stellar nucleus.
		(a) Dependence of $H$ on binary hardness $a/a_h$ (Equation~\ref{Equation:a_h}) for a binary in a nonrotating nucleus. Black is for $q=1,\,e=0$, red is for $q=9,\,e=0$, blue is for $q=1,\,e=0.9$.
		 Symbols are the results of scattering experiments, solid curves are Equation (16) of  
		 \citet{Sesana2006}, dashed curves are Equation (18) of \citet{Quinlan1996}. 
(b) Dependence of $H$ on binary orbital inclination $\theta$ for a circular, equal-mass binary in a maximally rotating ($\eta=1$) nucleus. Different colors are for $S=0.1,\,0.2,\,0.5,\,1,\,2,\,4,\,8$ ($a/a_h=1600,\,400,\,64,\,16,\,4,\,1,\,0.25$), with higher $S$ (lower $a/a_h$) corresponding to higher $H(\theta=\pi)$. 
(c) Dependence of $H$ on the argument of periapsis $\omega$ for an equal-mass binary in a maximally-rotating nucleus, with $S=4$, $e=0.5$, $\theta=\pi/2$. 
(d) Dependence of $H$ on binary eccentricity $e$ for an equal-mass binary in a maximally-rotating nucleus; $S=4$. Black: $\theta=0$; red: $\theta=\pi$. The values of $H$ are averaged over $\omega$, assuming a uniform distribution of $\omega$.
		}
	\label{Figure:H}
\end{figure}

In a nonrotating nucleus, the hardening rate depends only on the parameters $S$, $q$ and $e$.
\citet{Mikkola1992}, \citet{Quinlan1996} and \citet{Sesana2006} studied these dependences and derived analytical approximations for them.
\citet[Section 3]{Sesana2006} find that the dependence of $H$ on binary hardness is roughly the same for all values of $q$ and $e$ if the hardness is measured in $a/a_h$, where
\bsub\label{Equation:a_h} 
\barr
a_h \equiv \frac{G\mu}{4\sigma^2} &\approx& 2.7 \nu \frac{M_{12}}{10^8 \msun} 
\left(\frac{\sigma}{200\,\kms}\right) ^{-2} \mathrm{pc}, \\
\frac{a_h}{a} &=& \frac{\nu S^2}{4}.
\earr
\esub
Our results for the hardening rate are in good agreement with those of \citet{Sesana2006},
as shown in Figure~\ref{Figure:H}a. 

In a rotating nucleus, $da/dt$ depends also on $\eta$ and $\theta$.
Figure~\ref{Figure:H}b shows the $\theta$-dependence in maximally-rotating nuclei. 
We see that $H\approx c_1+c_2\cos\theta$ in this case, and that $H$ tends to a constant ($c_2\approx0$),
independent of $\theta$, for $S\gtrsim4$. For sufficiently {\it soft} binaries ($S\lesssim1$), the hardening rate can be negative for $\theta<\pi/2$; this is qualitatively different than the nonrotating case for which $H$ is always positive. 
Evidently, a binary in a nucleus with a high enough degree of corotation need not harden at all, 
at least in the case that the dynamical friction force fades before the three-body hardening rate becomes positive.

 This difference can be traced to the different nature of star-binary interactions in the two cases. 
 In the case of a hard binary, the initial velocity of the star is negligible compared to the escape velocity from the binary's orbit
 and the interaction is rather chaotic in nature; the final parameters of the stellar orbit are practically random and independent of the initial ones. Since the typical, final velocity is of the order of escape velocity, most of the stars gain energy as a result of the interaction, and the binary becomes harder (see also Figure~\ref{Figure:dlxdlz} and the arguments about the conservation of Jacobi constant in \S~\ref{Section:UnderstandingScattering}).
In the case of a soft binary, the star approaches the binary with a velocity much greater than the escape velocity, and interaction consists typically of only one close interaction with one of the binary components, with a 
relatively small change in the star's velocity. At the moment of that close interaction, the binary component moves (more or less) in the same (opposite) direction as the star in the corotating (counterrotating) case. 
Considering that the star is massless and the interaction is elastic, we know from classical mechanics that the star loses energy as a result of interaction in the first case, and gains energy in the second case. This explains the aforementioned dependence of hardening rate on $\theta$.

We note that \citet{Khan2015} obtained a different result by means of $N$-body simulations.
In rotating nuclei, the hardening rate was found to always be higher than in nonrotating systems regardless of the binary's orientation. 
The disagreement with our results may be related to the binary's center-of-mass motion in their models. 
In the counterrotating case, they found that the binary exhibited a random walk but with a seemingly higher amplitude than 
in nonrotating nuclei; 
while in the corotating case, the binary was observed to go into a circular orbit with a radius larger than the Brownian motion amplitude in both cases. 
As a result, the effective stellar scattering cross-section in rotating models was probably higher.
We note that the amplitude of the binary's center-of-mass motion is likely to be strongly dependent on $m_\star/M_{12}$ and
that this ratio is much larger in $N$-body models than in real galaxies.

For eccentric binaries, there are two more parameters on which $H$ could depend: argument of periapsis $\omega$ and eccentricity $e$. 
Our results suggest no dependence of $H$ on $\omega$ (Fig.~\ref{Figure:H}c) and only a weak 
dependence on $e$ (Figure~\ref{Figure:H}d), with at most $\sim25\%$ difference in $H$ between circular and eccentric binaries, similar to the nonrotating case.

\begin{figure}
	\centering
	\subfigure[]{\includegraphics[width=0.49\textwidth]{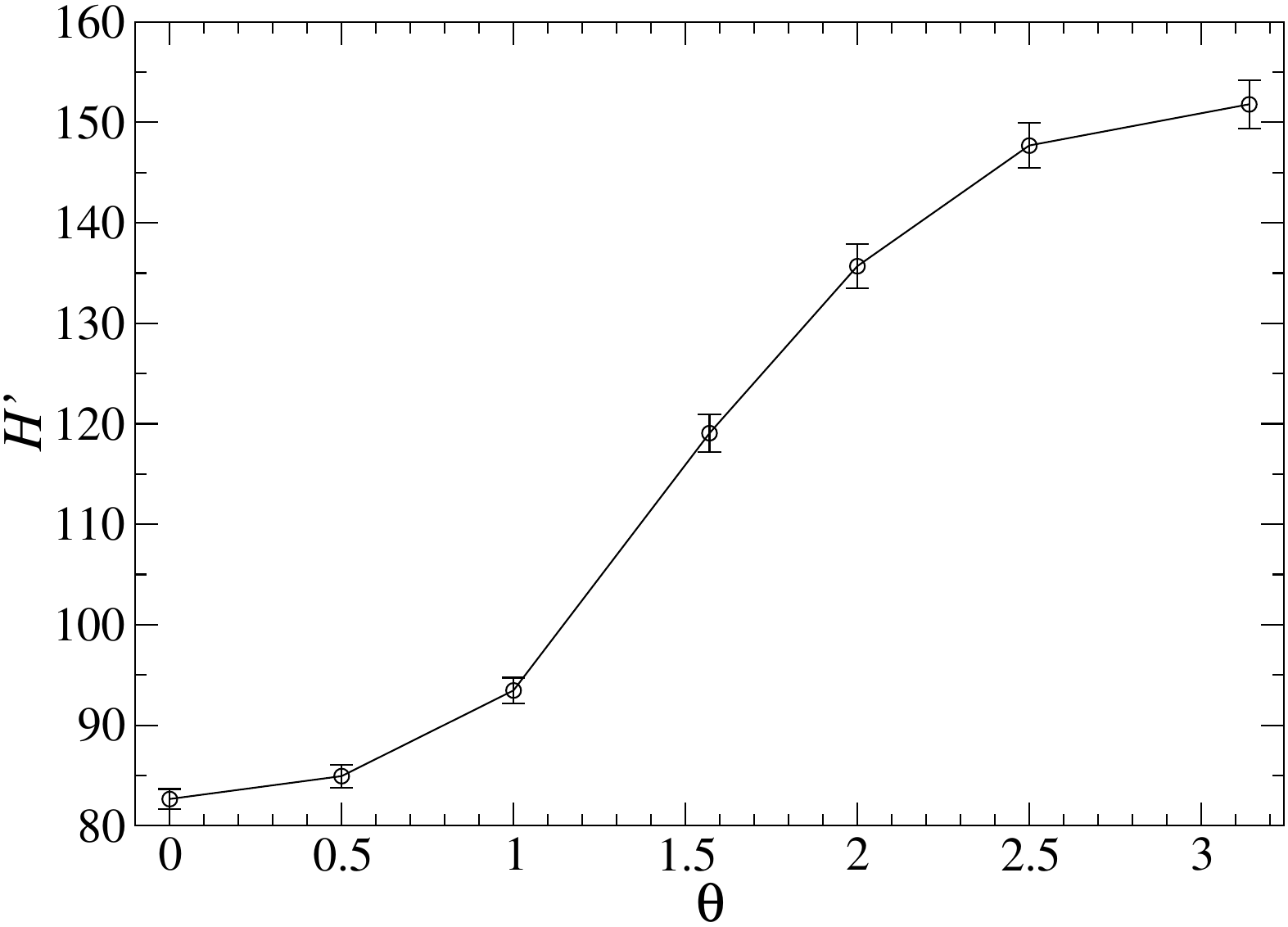}}
	\subfigure[]{\includegraphics[width=0.49\textwidth]{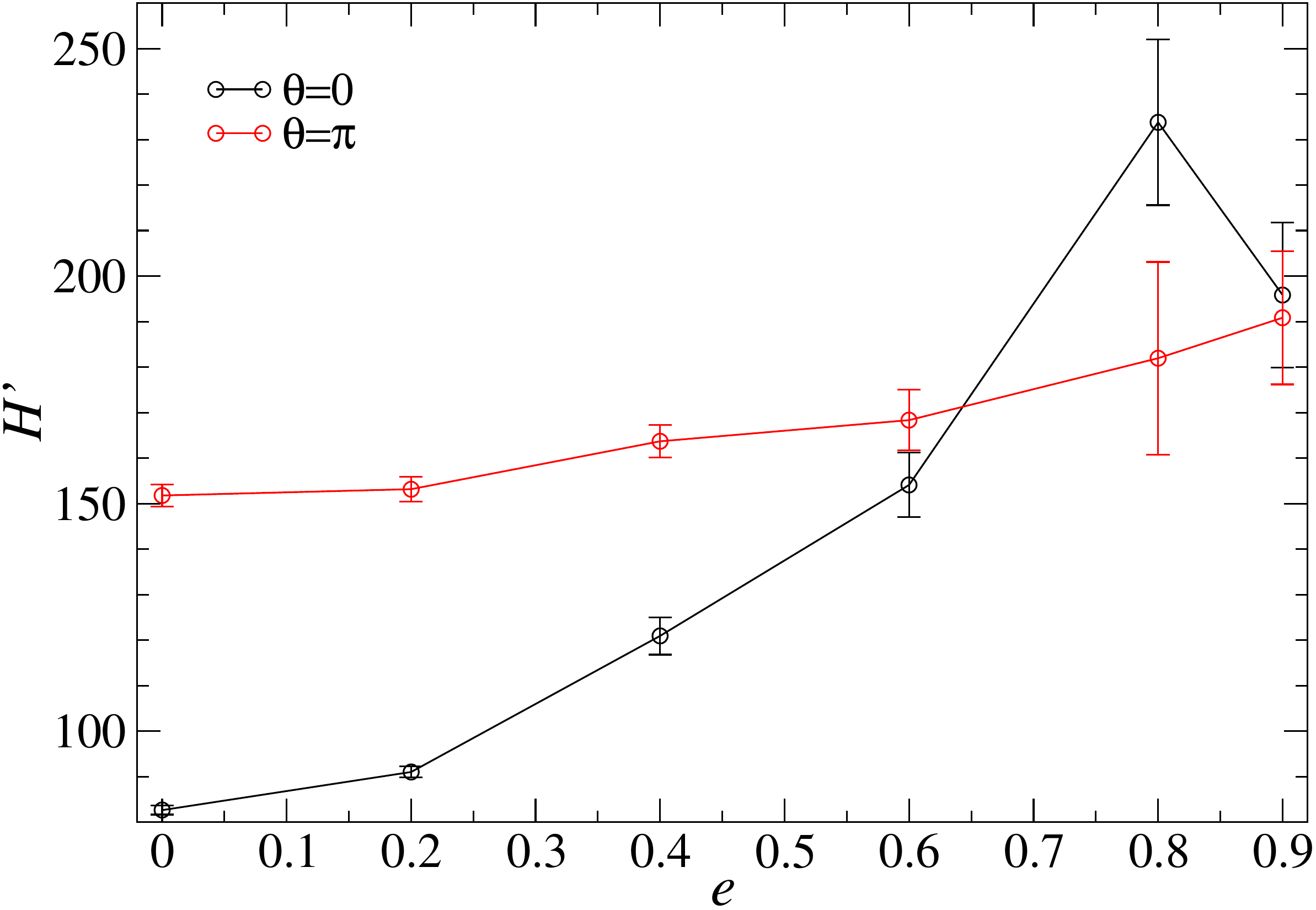}}
	\caption{
	Dimensionless diffusion coefficient in semimajor axis $H'$ (Equation~\ref{Equation:H'}) as a function of binary orbital elements for $q=1$, $\eta=1$.
(a) Dependence of $H'$ on binary orbital inclination $\theta$ for $e=0$.
(b) Dependence of $H$ on binary eccentricity $e$. Black: $\theta=0$; red: $\theta=\pi$.
		}
	\label{Figure:H'}
\end{figure}

Next we consider the dimensionless coefficient $H^\prime$ that determines the second-order diffusion coefficient
(Equation~\ref{Equation:H'}).
It turns out that $H^\prime$ is not too strongly dependent on the orbital elements or the parameters 
defining the stellar nucleus: $50\lesssim H' \lesssim 200$, i. e. $H'\sim10^2$. As shown below, such small values of $H'$ are small enough to ignore the second-order effects completely, thus we haven't studied the dependence of H on different parameters in detail. Fig.~\ref{Figure:H'} shows the dependence of $H'$ on $\theta$ and $e$.

In \S~\ref{Section:BinaryEvolve}, we derived a one-dimensional Fokker-Planck equation for binary orientation assuming that we knew a priori the time dependence of the binary's energy, i. e., semimajor axis $a$. Our finding of $H$ being approximately independent of any orbital parameters other than $a$ confirms that assumption; with $H=\rm const$, Eq.~(\ref{Equation:DefinitionOfH}) gives 
\bsub\label{Equation:a(t)}
\barr
a(t) &=& \left(\frac{G\rho Ht}{\sigma}+\frac{1}{a_0}\right)^{-1} = \frac{a_0}{1+t/t_h},\\
a_0 &=& {\rm const},\,\,t_\mathrm{hard}\equiv\frac{\sigma}{\rho GaH}\label{Equation:Definethard}
\earr
\esub
Also, the aforementioned assumption that we can replace $a$ with $a(t)$ requires the second-order
terms in $a$ to be negligible, i. e. it requires the deterministic change in $a$ in one hardening time:
\beq
(\Delta a)_1 \equiv \left|\langle\Delta a\rangle\right|\times t_\mathrm{hard} \approx a
\eeq
to be greater than the change due to diffusion:
\beq
(\Delta a)_2 \equiv \sqrt{\langle(\Delta a)^2\rangle\times t_\mathrm{hard}} 
\approx a\sqrt{\frac{m_f}{M_{12}}\frac{H'}{H}}
\eeq
yielding the criterion
\beq\label{Equation:da2<da1}
\frac{m_f}{M_{12}}\frac{H'}{H} \lesssim 1 .
\eeq
For hard binaries $H'/H\lesssim10$, and even for binaries as soft as $a/a_h=10$, $H'/H\lesssim20$. 
The largest star-binary mass ratio that is consistent with our test-mass approximation is $m_f/M_{12}\approx0.1$. Considering that in reality $m_f/M_{12}$ is usually a few orders of magnitude smaller than that, we can be sure that condition (\ref{Equation:da2<da1}) is fulfilled under all realistic parameter values.

Returning to the first-order diffusion coefficient $\langle\Delta a\rangle$: we found that
rotation of the nucleus significantly affected the hardening rate for soft binaries  ($a/a_h\gtrsim8$ (which corresponds to $S\gtrsim1$ for $q=1$) for $\eta=1$ and even softer for $\eta<1$; see Figure~\ref{Figure:H}b).
However, applying our 3-body scattering technique at such high binary separations may yield misleading 
results for the following reasons:

\begin{enumerate}
\item Dynamical friction acting on the two binary components independently may play a significant role
when $a\gtrsim a_f$, where $a_f$ is the separation at which the stellar mass within radius $a_f$ is 
$\sim 2M_2$, according to \citet{GualandrisMerritt2012}. In their simulations, $a_f\approx100a_h$.
\item At large separations, the two \sbhs~may not be bound yet (and not follow the Keplerian trajectories). 
We have analyzed the $N$-body data of \citet{GualandrisMerritt2012} and found that in their models, this is true for $a/a_h>20...30$. 
\item The hardening time may be shorter than the binary orbital period, invalidating our assumption that
the two black holes follow a Keplerian orbit. 
In the simulations of \citet{GualandrisMerritt2012}, this was the case  for $a/a_h\gtrsim10$. 
For nonrotating (or weakly rotating) nuclei we can estimate the characteristic separation, as follows.
Adopting the analytical approximation for the hardening rate form \citet{Sesana2006}, which is consistent with our results, as shown on Figure~\ref{Figure:H}a:
\beq
H \approx 15\left(1+\frac{a}{3.5a_h}\right)^{-1} \approx 53\left(\frac{a}{a_h}\right)^{-1} ,\quad a/a_h\gg1 .
\eeq
The condition $T>t_\mathrm{hard}$, where $T$ is the binary's (Keplerian) period, then yields
\beq
\frac{a}{a_h} > 0.67 \frac{\sigma^4}{G^2\rho^{2/3}M_{12}^{4/3}} = 12
\left(\frac{M_{12}}{10^8\msun}\right)^{-4/3} 
\left(\frac{\sigma}{200\,\kms}\right) ^{4}
\left(\frac{\rho}{10^3\msun\mathrm{pc}^{-3}}\right)^{-2/3} .
\eeq
Other orbital elements change too, but, as will be shown later in this section, the characteristic times for them are either comparable to or longer than the hardening time.
\item In our scattering experiments we assumed that stars approach the binary on Keplerian trajectories until they reach a separation of $50a$ from the binary; that is: we assumed that the binary dominates the gravitational potential at $r<50a$.
This may not be the case if $r_\mathrm{infl}\ll50a$. 
In addition, the derivation of the formulae which we used to calculate the diffusion coefficients (\S\ref{Section:DiffCoeffDiffCoeff}) relies on the assumption that $r_\mathrm{infl}\gg a$.
\end{enumerate} 

\subsection{\label{Section:D(e)}Drift and diffusion coefficients for the eccentricity}

\begin{figure}
	\centering
	\subfigure[]{\includegraphics[width=0.49\textwidth]{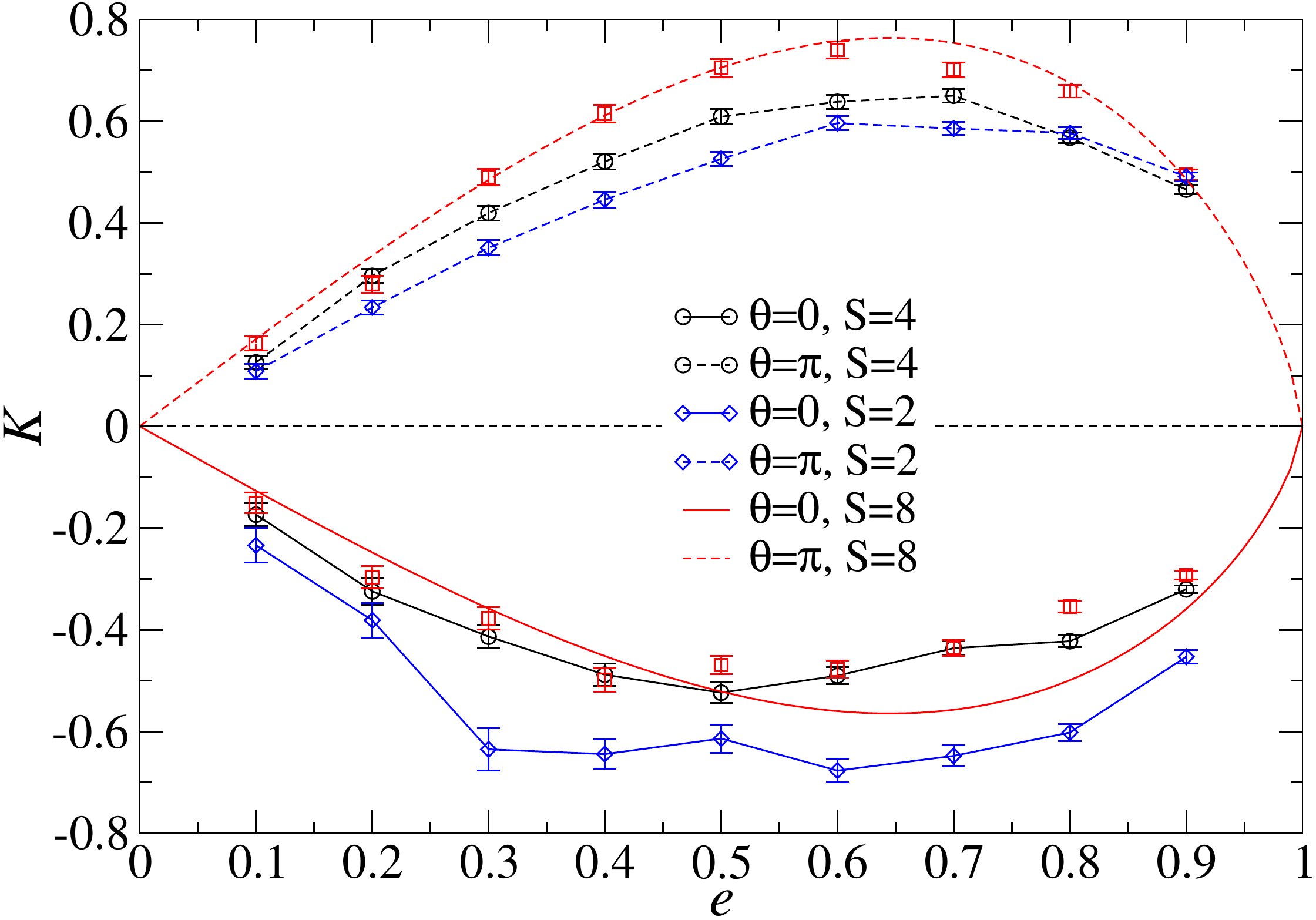}}
	\subfigure[]{\includegraphics[width=0.49\textwidth]{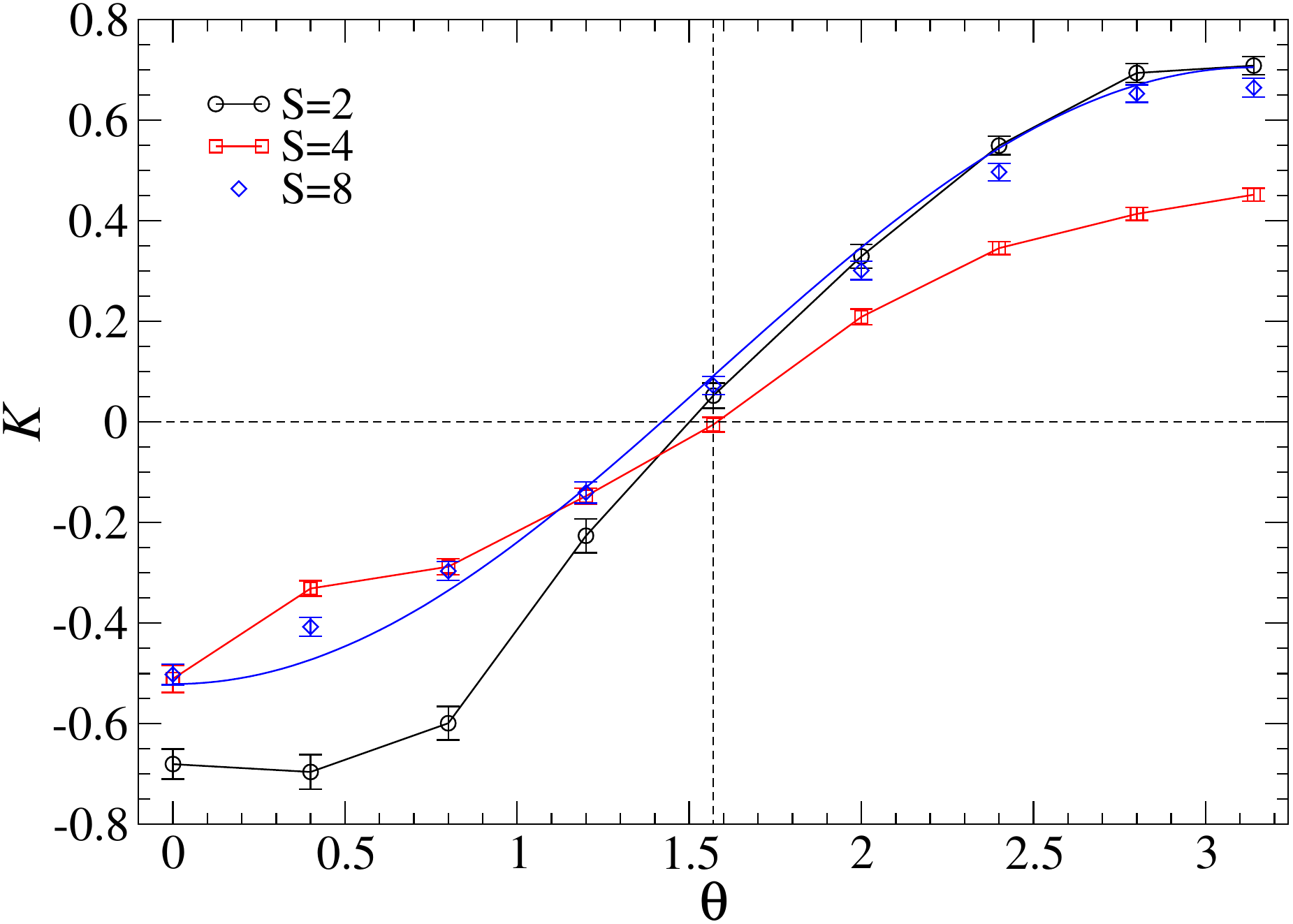}}
	\subfigure[]{\includegraphics[width=0.49\textwidth]{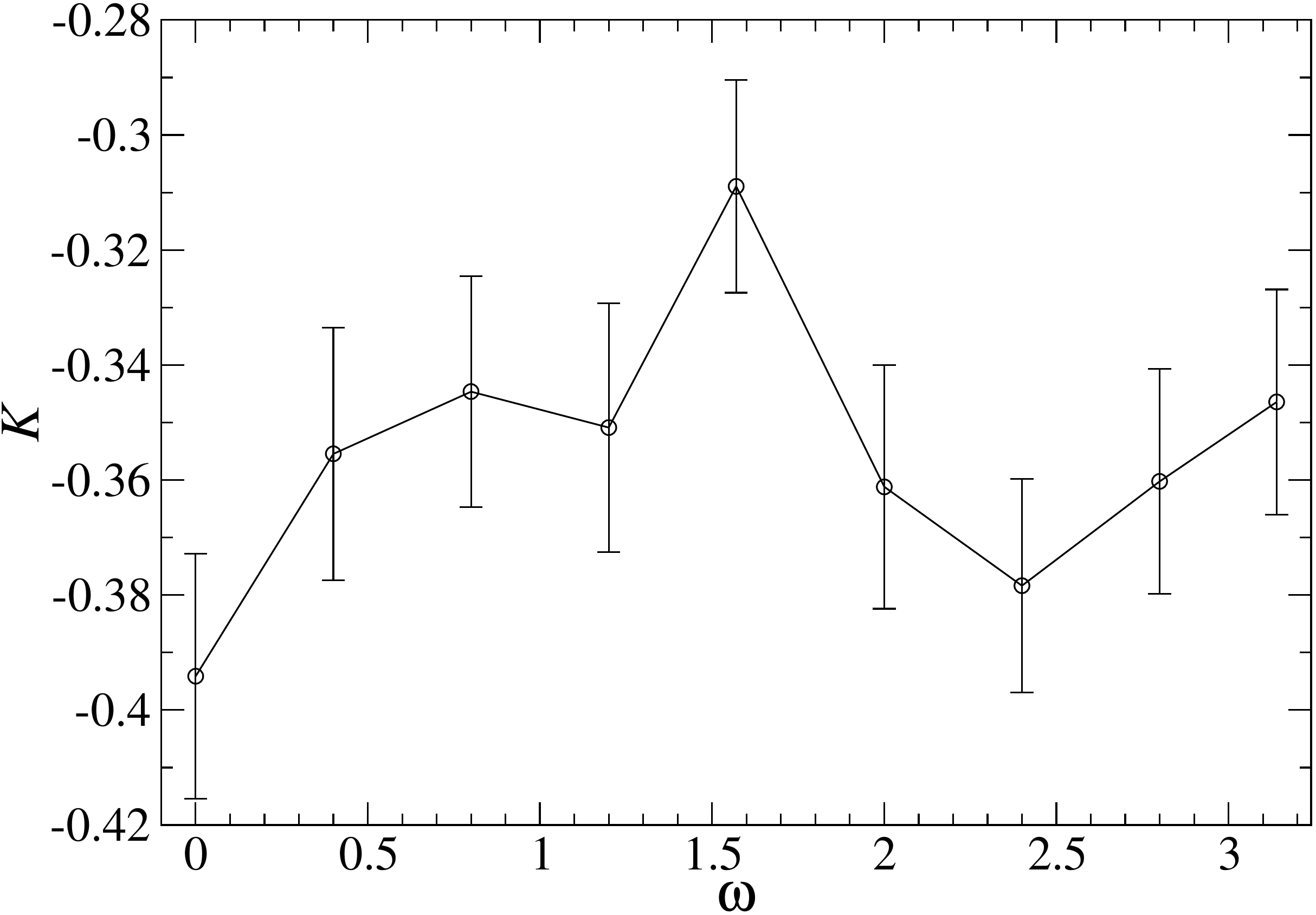}}
	\subfigure[]{\includegraphics[width=0.49\textwidth]{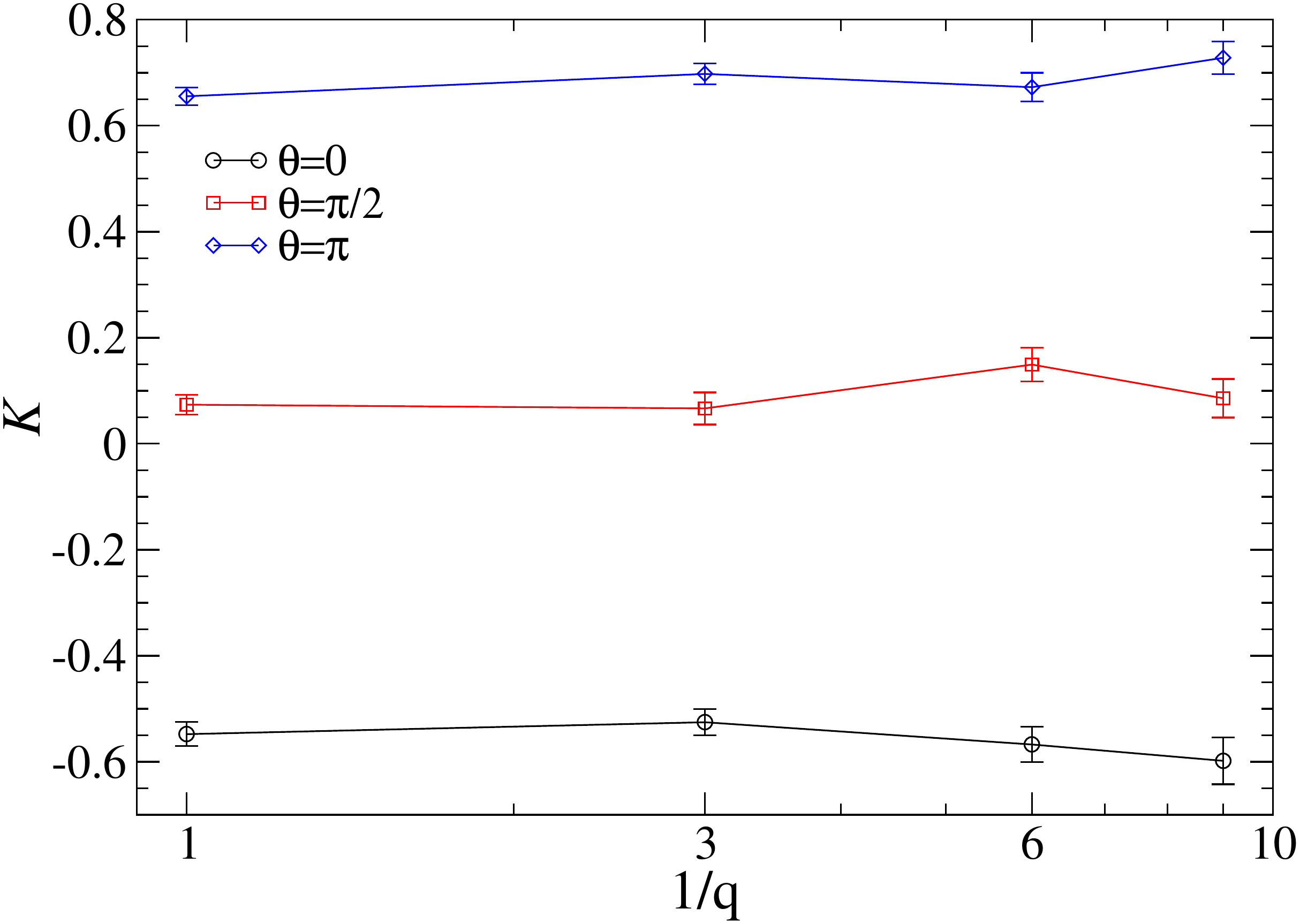}}
	\caption{Dimensionless rate of change of eccentricity $K$ (Equation~\ref{Equation:KK'}c).
		Points: the dependence of $K$ on various parameters for a maximally-rotating nucleus ($\eta=1$). Unless otherwise indicated, $q=1$, $S=4$ ($S=8$ on Fig. d), $e=0.5$, $\theta=\pi/4$. All of the figures show values of $K$ averaged over $\omega$ assuming a uniform distribution of $\omega$. Curves: fits to Equation~(\ref{Equation:K(e,theta)}).
		}
	\label{Figure:K}
\end{figure}

A standard definition for the dimensionless rate of change of binary eccentricity \citep[e. g.][Section 8.1]{DEGN}
is
\beq
K \equiv \frac{de}{d\ln{1/a}} = -\frac{de}{da}a .
\eeq
In the Fokker-Planck formalism, $K$ is related to the first-order diffusion coefficient in $e$ as
\beq
K = - \frac{\langle\Delta e\rangle}{\langle\Delta a\rangle} a .
\eeq
As in the case of semimajor axis, we define a second  dimensionless variable $K'$ such that
\beq
K' = - \frac{M_{12}}{m_f} \frac{\langle(\Delta e)^2\rangle}{\langle\Delta a\rangle} a .
\eeq
Using Equation~(\ref{Equation:deltae}),
we can then express $\langle\Delta e\rangle$ and $\langle(\Delta e)^2\rangle$ as
\bsub\label{Equation:KK'}
\barr
\langle\Delta e\rangle &=& \frac{\rho G a}{\sigma} KH\label{Equation:KtoDeltae}, \\
\langle(\Delta e)^2\rangle &=& \frac{m_f}{M_{12}} \frac{\rho G a}{\sigma} K'H\label{Equation:K'toDeltae2}, \\
K &=& \frac{1}{\nu H} \times \frac{1-e^2}{e} \left[ \frac{1}{\sqrt{1-e^2}}\frac{\overline{\overline{\delta l_\parallel}}}{aV_\mathrm{bin}} - \frac{\overline{\overline{\delta\varepsilon}}}{GM_{12}/a}\right], \\
K' &=& \frac{1}{\nu^2 H} \times \left[\frac{1-e^2}{e}\right]^2 \overline{\overline{\left[
\frac{1}{\sqrt{1-e^2}}\frac{\delta l_\parallel}{aV_\mathrm{bin}} - \frac{\delta\varepsilon}{GM_{12}/a} 
\right]^2}} .
\earr
\esub

\citet{Sesana2011} studied the evolution of eccentricity in rotating stellar environments and found that co- and counter-rotating binaries, started from $e=0.5$, quickly evolve to $e\approx0$ and $e\approx1$, respectively, in about one hardening time. Our results, shown in Figure~\ref{Figure:K}, are in good agreement:
as long as $e$ is not too close to 0 or 1 and $S\gtrsim2$, $K\approx -0.5$ for $\theta=0$ and  $K\approx 0.5$ for $\theta=\pi$. $K$ can be understood as eccentricity change per hardening time, so the agreement is not only qualitative, but quantitative as well. 

\begin{figure}[h]
\includegraphics[width=0.5\textwidth]{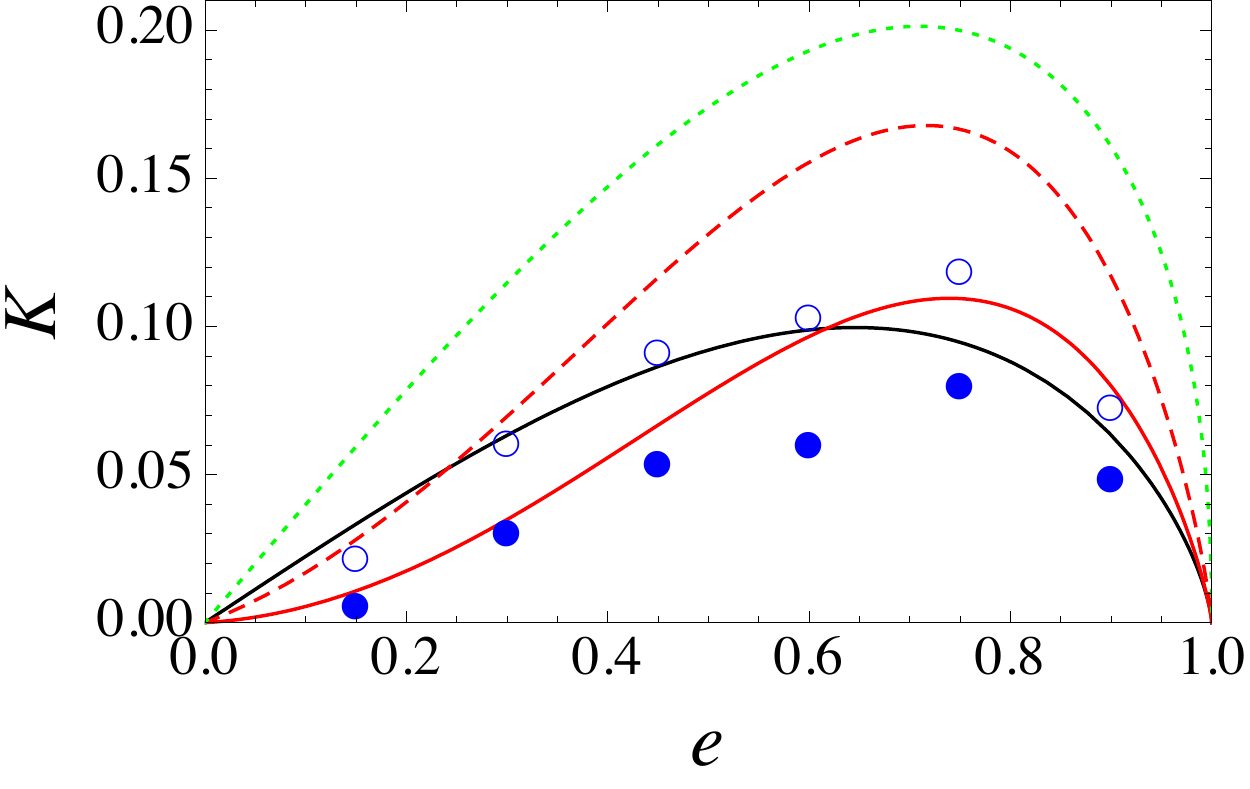}
\caption{Eccentricity growth rate $K$ for equal-mass binaries in nonrotating nuclei. Black line: our analytical approximation. Green line: the results of \citet{Mikkola1992} in $S\rightarrow\infty$ limit. Red lines: the results of \citet{Quinlan1996} for $S=10$ (solid) and $S=30$ (dashed). Blue circles: the results of \citet{Sesana2006} for $S=10$ (filled) and $S=30$ (empty).}
\label{Figure:K(e)-nonrotating}
\end{figure}

The dependence of $K$ on both $e$ and $\theta$ in the hard-binary limit can be crudely approximated as 
\beq\label{Equation:K(e,theta)}
K (e, \theta, \eta) \approx 1.5\,e\, (1-e^2)^{0.7}\, [0.15-(2\eta-1)\cos\theta] .
\eeq
Previously $K$ was calculated only for nonrotating systems \citep{Mikkola1992, Quinlan1996, Sesana2006}. 
The results of \citet{Mikkola1992} and \citet{Quinlan1996} agree well with each other, but not so well with those of \citet{Sesana2006}. Our Equation (\ref{Equation:K(e,theta)}) gives the following result for $\eta=1/2$:

\beq\label{Equation:K(e)-nonrotating}
K_{\eta=1/2} (e) \approx 0.225\,e\, (1-e^2)^{0.7} .
\eeq
We plot this function, and the earlier approximations, in Figure~\ref{Figure:K(e)-nonrotating}. 
Our expression is consistent with that of \citet{Sesana2006} in the $S\rightarrow\infty$ limit (which is almost reached at $S=30$). The discrepancy between different authors is probably due to the difficulty
of computing $K$ from scattering experiments, as emphasized by \citet{Quinlan1996}.

As Figures \ref{Figure:K}c and \ref{Figure:K}d show, $K$ is practically independent of $q$ and $\omega$. 

As it was shown in \S\ref{Section:D(a)}, for $S\lesssim1$ there are values of $\theta$ where $H\rightarrow0$, so by definition $K\rightarrow\pm\infty$ (because $KH$ is still nonzero). This just means that the definition of $K$ loses its meaning, because the binary doesn't harden, and we can't use $a$ as a proxy for time.

\subsection{\label{Section:D(theta)}Diffusion coefficients for orbital inclination}
In this section the diffusion coefficients describing changes in the binary's orbital inclination are presented. Inclination is defined here via the angle $\theta$, defined in \S\ref{Section:BinaryEvolve} and Figure~\ref{Figure:binary} as the angle between the binary's angular momentum vector and the rotation axis
of the stellar nucleus.
A number of other angular variables were defined in \S\ref{Section:BinaryEvolveFPE} and
\S\ref{Section:BinaryEvolveOrientation}; we refer the
reader to those sections, where transformation equations between the various diffusion coefficients 
describing orbital inclination are presented. 

We express the diffusion coefficients in terms of the dimensionless rates $D_{\theta,1}$ and $D_{\theta,2}$,
as follows:
\bsub\label{Equation:D12}
\barr
\langle\Delta\theta\rangle &=& - \frac{\rho G a}{\sigma} D_{\theta,1}\label{Equation:D1toDeltatheta}, \\
\langle(\Delta\theta)^2\rangle &=& \frac{m_f}{M_{12}} \frac{\rho G a}{\sigma} D_{\theta,2}\label{Equation:D2toDeltatheta2}, \\
D_{\theta,1} &=& - \frac{1}{\nu\sqrt{1-e^2}} 
\frac{\overline{\overline{\delta l_\theta}}}{aV_\mathrm{bin}}, \\
D_{\theta,2} &=& \frac{1}{\nu^2(1-e^2)}
 \frac{\overline{\overline{\delta l^2_\theta}}}{\left(aV_\mathrm{bin}\right)^2} .
\earr
\esub
These expressions were obtained from Equations~(\ref{Equation:deltatheta}) and~(\ref{Equation:DiffQhomo})  assuming a Maxwellian velocity distribution (Equation~\ref{Equation:MaxwellianDistribution}). 
The expression for $\langle(\Delta\theta)^2\rangle$ is similar to Equation~(20) of \citet{Merritt2002}. 
In the simplest case of a circular equal-mass binary in a spherically symmetric nucleus, $D_{\theta,1} = 0$ and $D_{\theta,2}$ depends on two parameters only:
\begin{eqnarray}
S \equiv \frac{V_\mathrm{bin}}{\sigma},\, R \equiv \frac{p_{max}\sigma^2}{GM_{12}} .
\end{eqnarray}
Figure~3 of \citet{Merritt2002} suggests that setting $R=6$ is acceptable for any hardness $S\gtrsim1$ and
we adopt that value in what follows.

\begin{figure}[h]
	\subfigure[]{\includegraphics[width=0.49\textwidth]{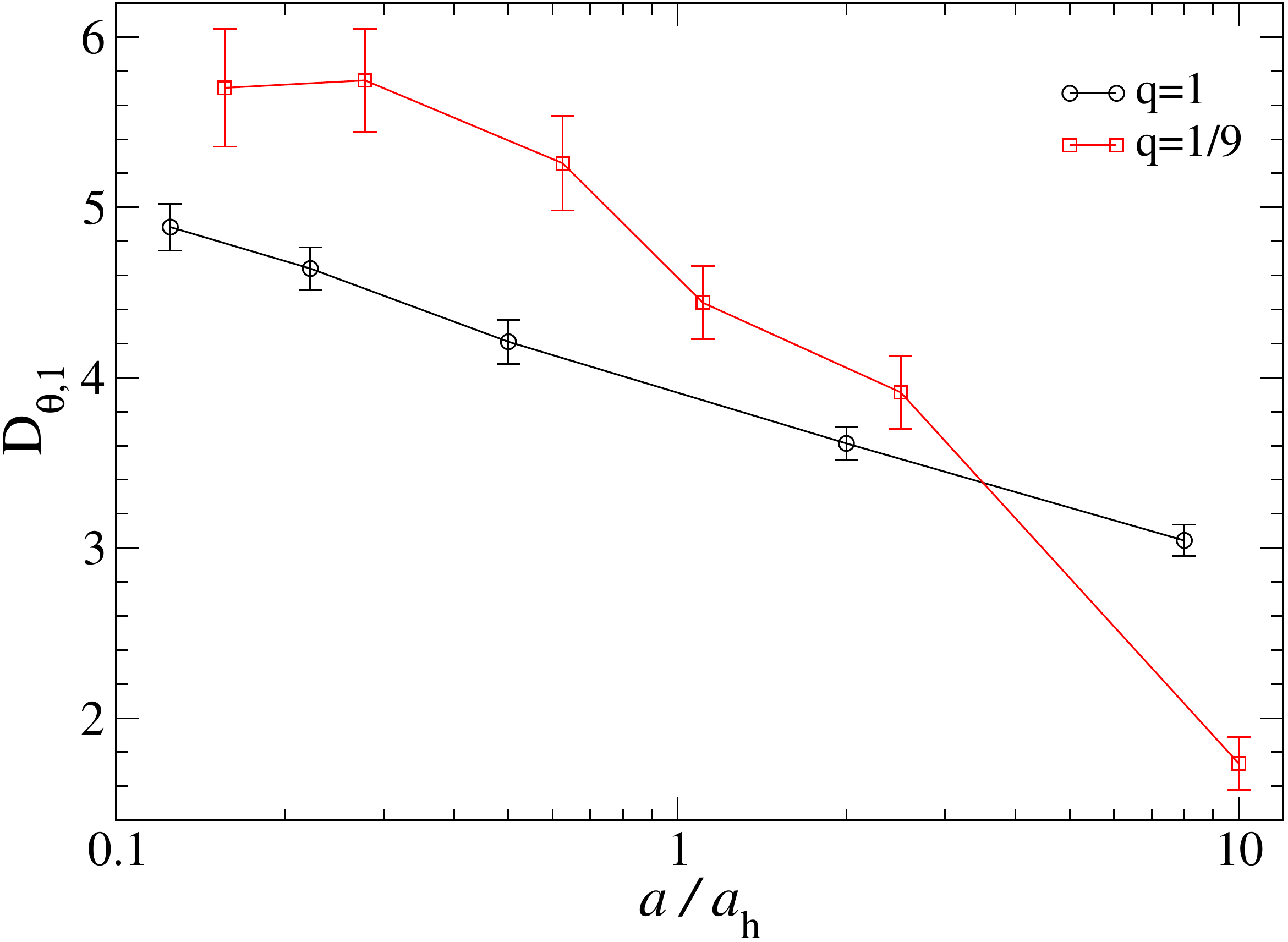}}
	\subfigure[]{\includegraphics[width=0.49\textwidth]{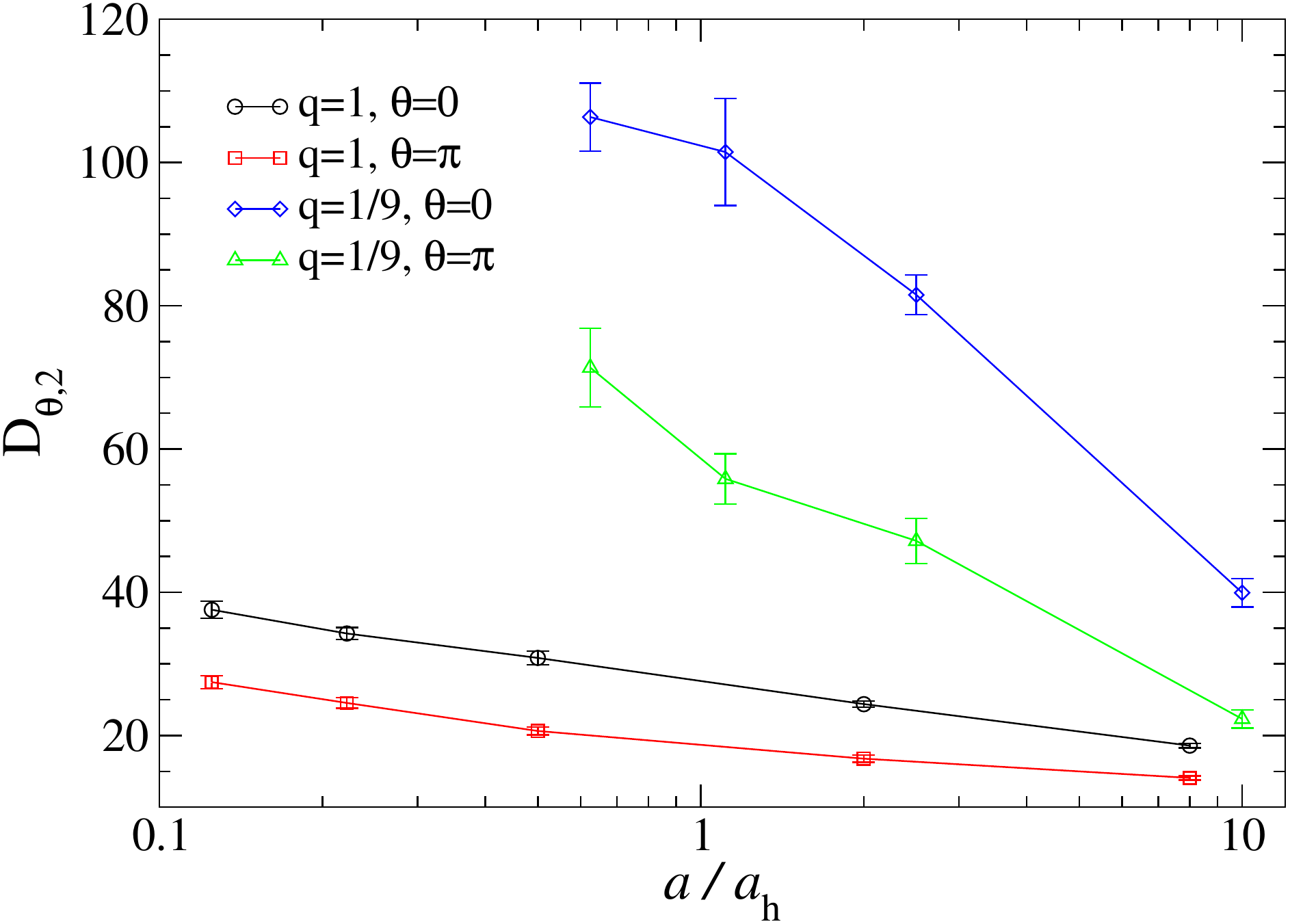}}
	\subfigure[]{\includegraphics[width=0.49\textwidth]{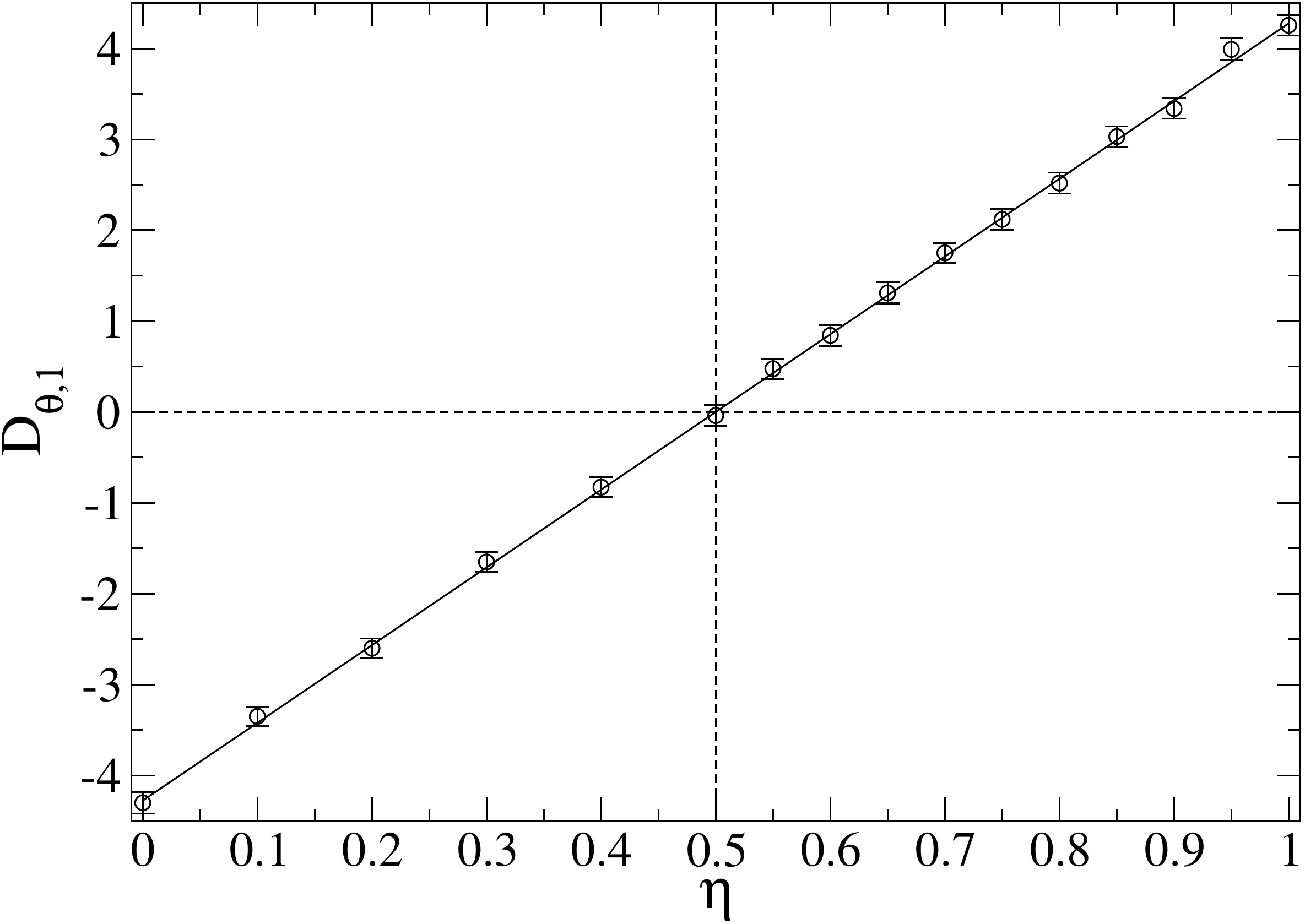}}
	\subfigure[]{\includegraphics[width=0.49\textwidth]{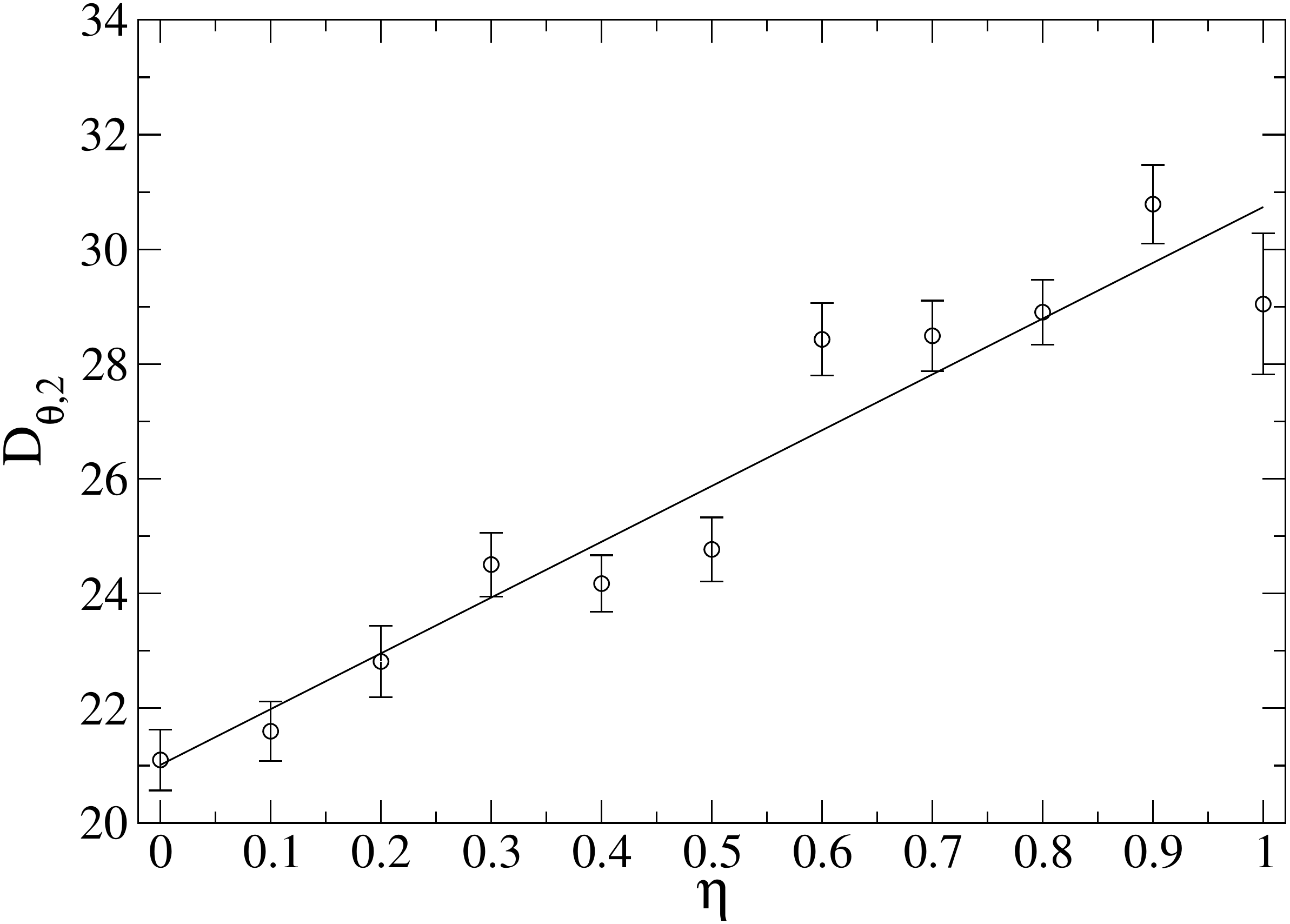}}
	\subfigure[]{\includegraphics[width=0.49\textwidth]{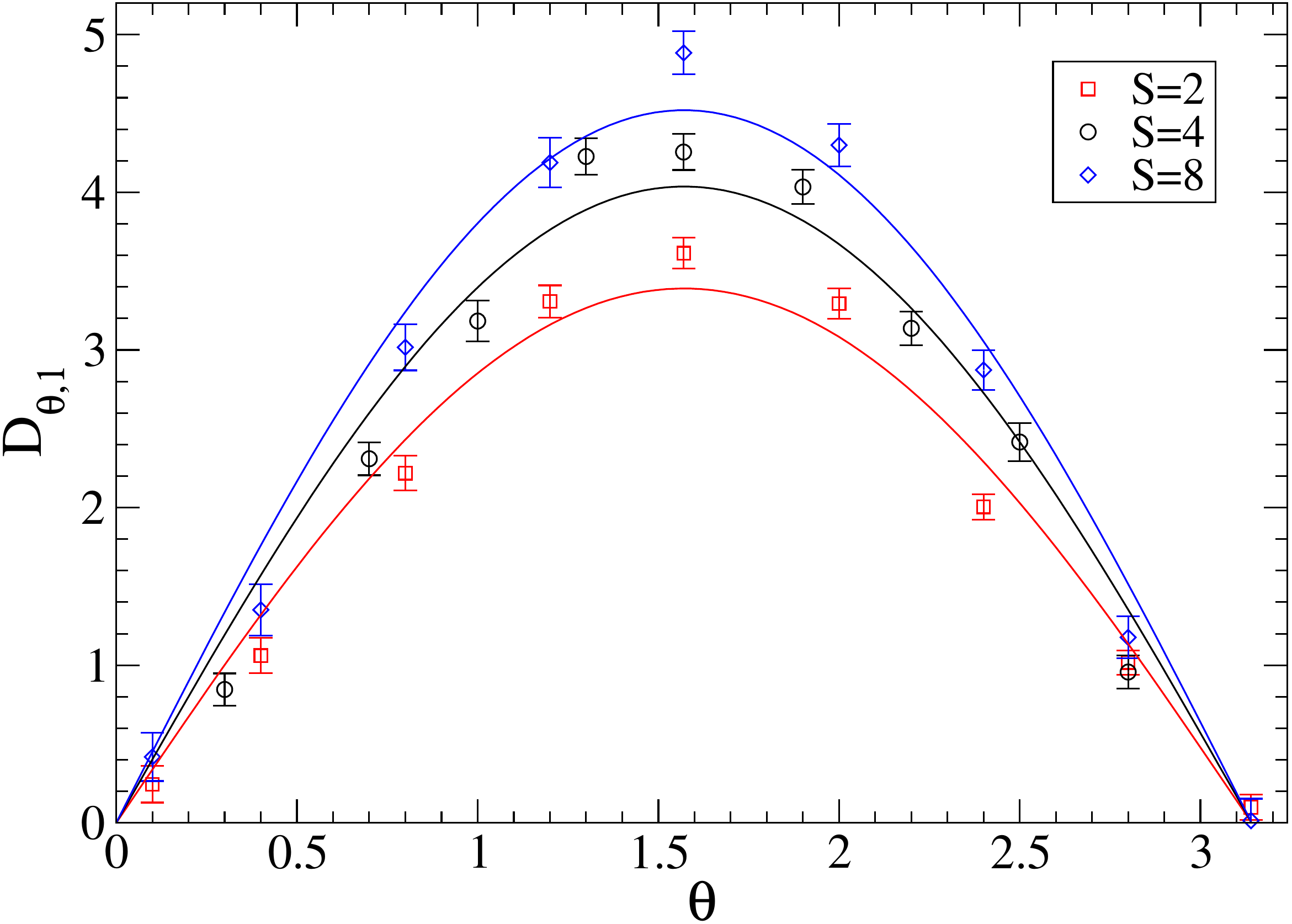}}
	\subfigure[]{\includegraphics[width=0.49\textwidth]{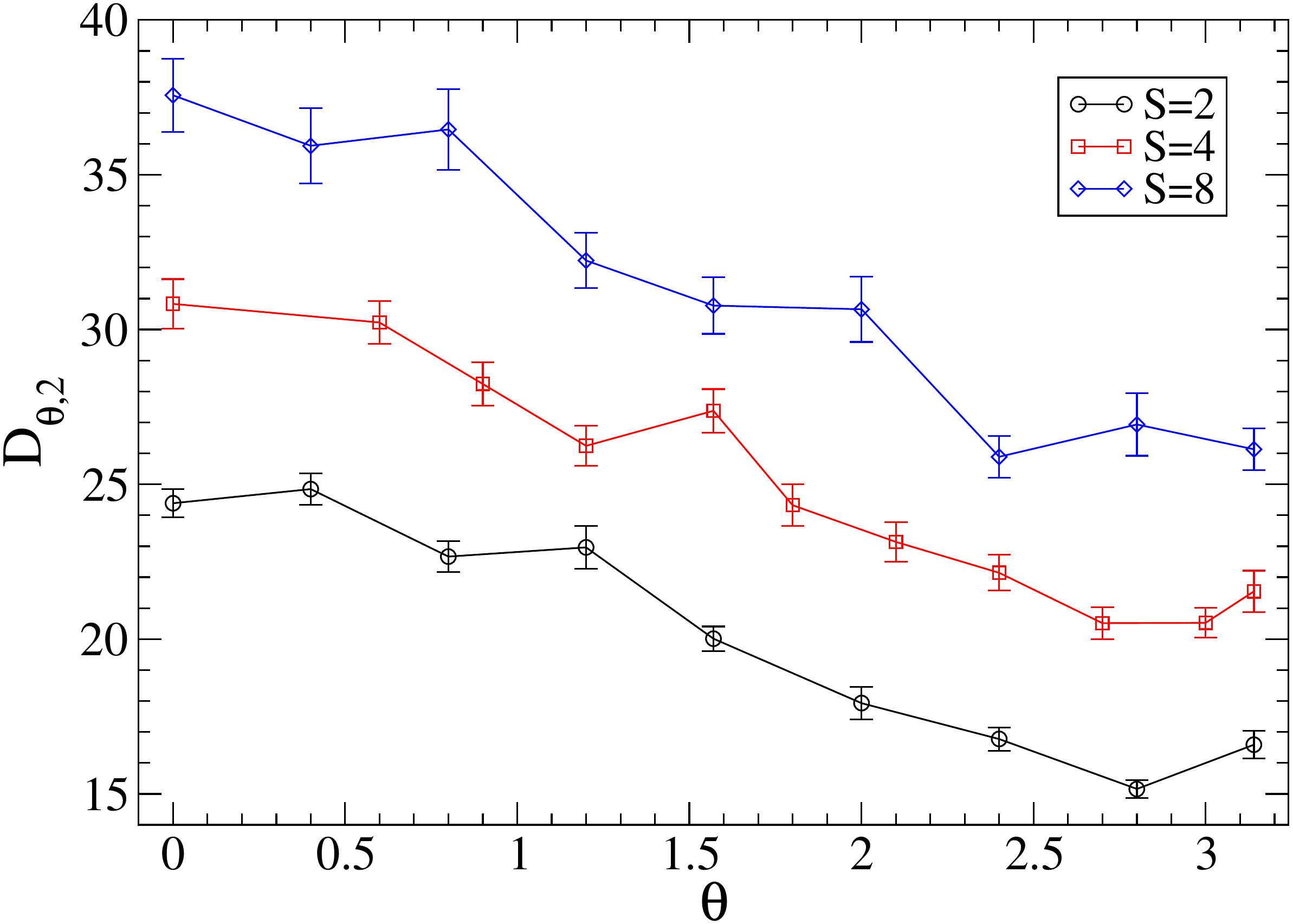}}
\caption{Dependence of $D_{\theta,1}$ and $D_{\theta,2}$ on various parameters: semimajor axis $a$ expressed in units of $a_h$ (defined in Equation~\ref{Equation:a_h}), degree of corotation $\eta$, binary inclination $\theta$, eccentricity $e$, binary mass ratio $q$ and argument of periapsis $\omega$. If not stated otherwise, $S=4$, $\eta=1$, $e=0$ and $q=1$ for both $D_{\theta,1}$  and $D_{\theta,2}$, $\theta=\pi/2$ for $D_{\theta,1}$ and $\theta=0$ for $D_{\theta,2}$.}
\label{Figure:D(a,eta,theta)}
\end{figure}

\begin{figure}[h!]
	\subfigure[]{\includegraphics[width=0.49\textwidth]{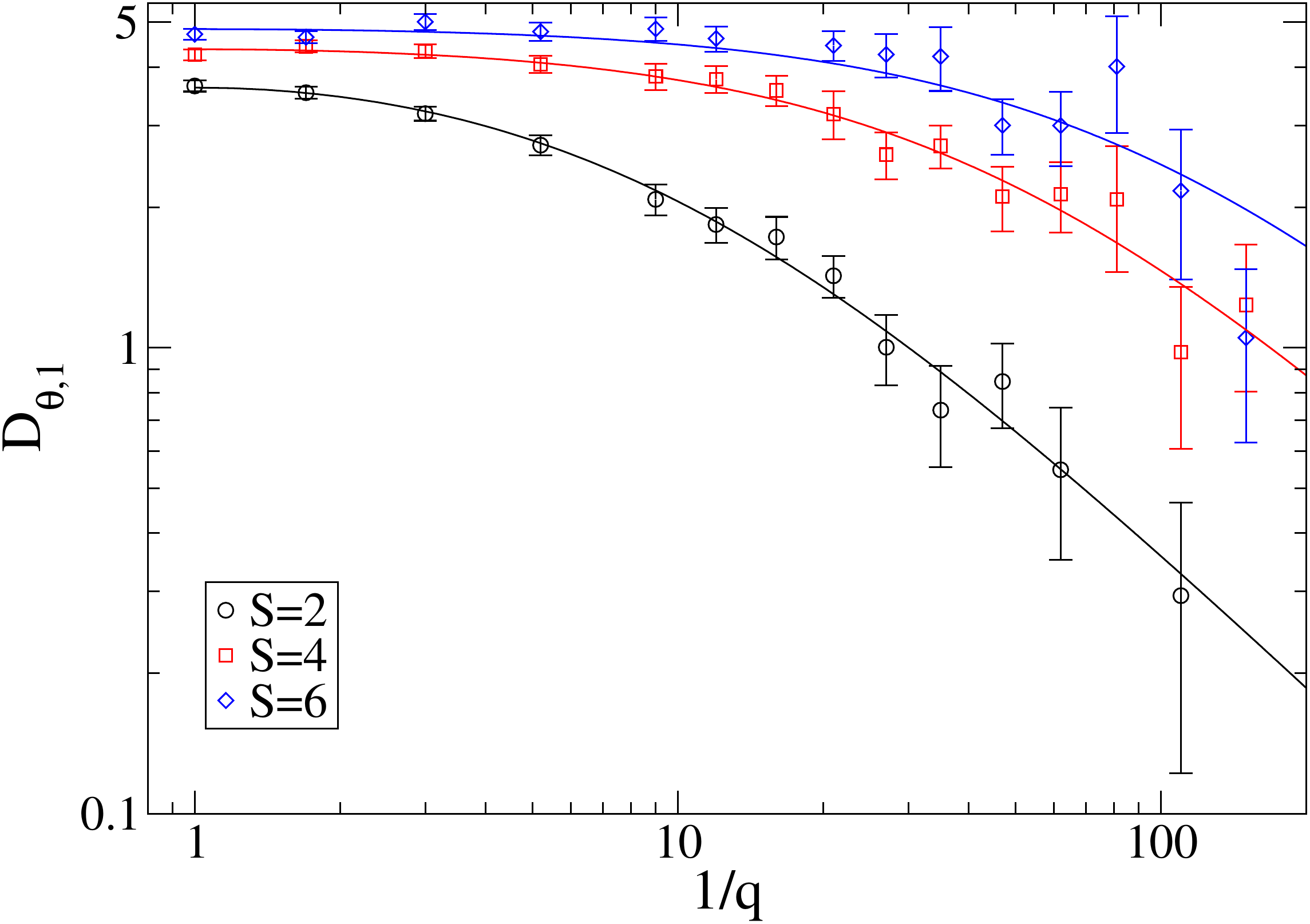}}
	\subfigure[]{\includegraphics[width=0.49\textwidth]{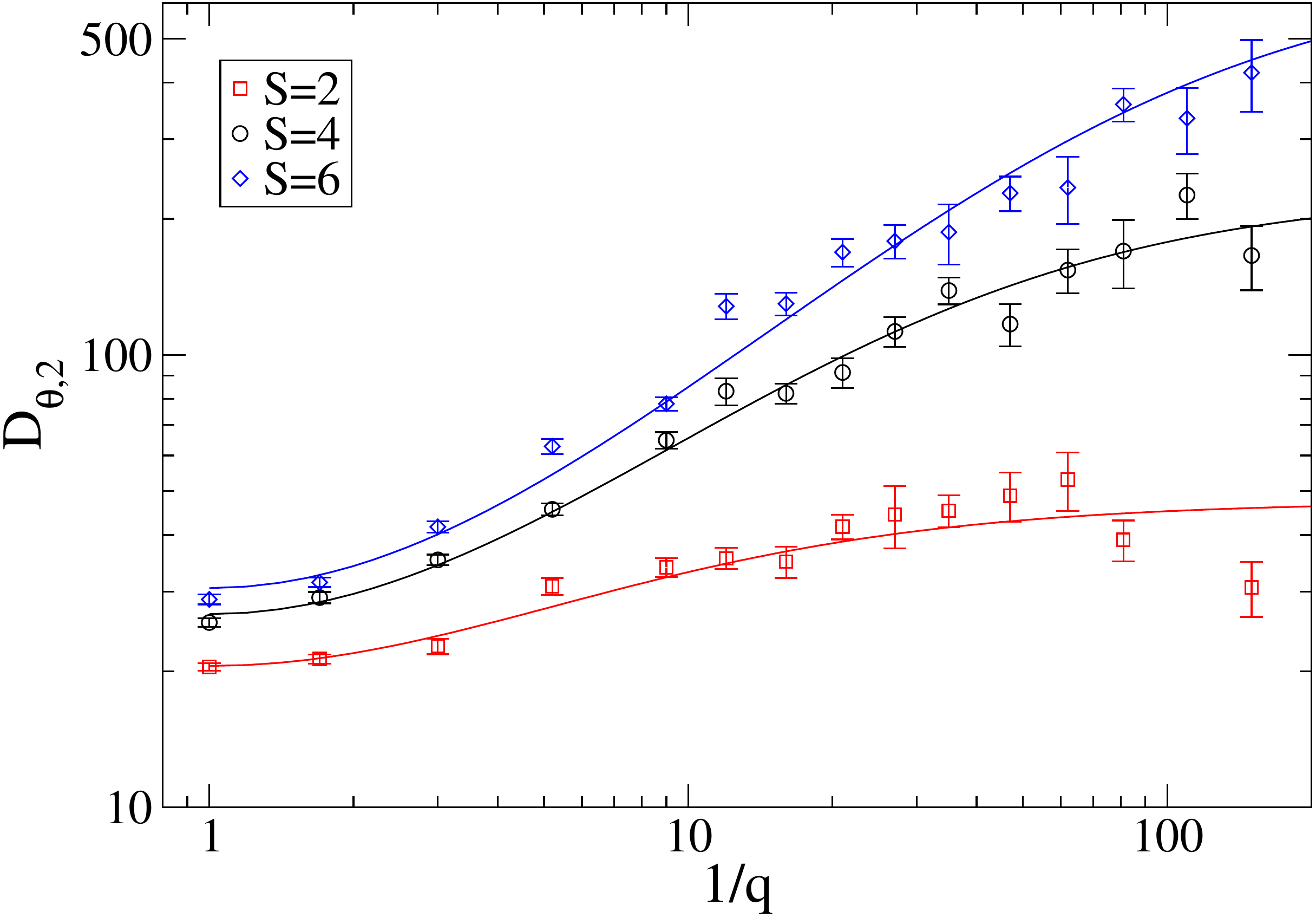}}
	\subfigure[]{\includegraphics[width=0.49\textwidth]{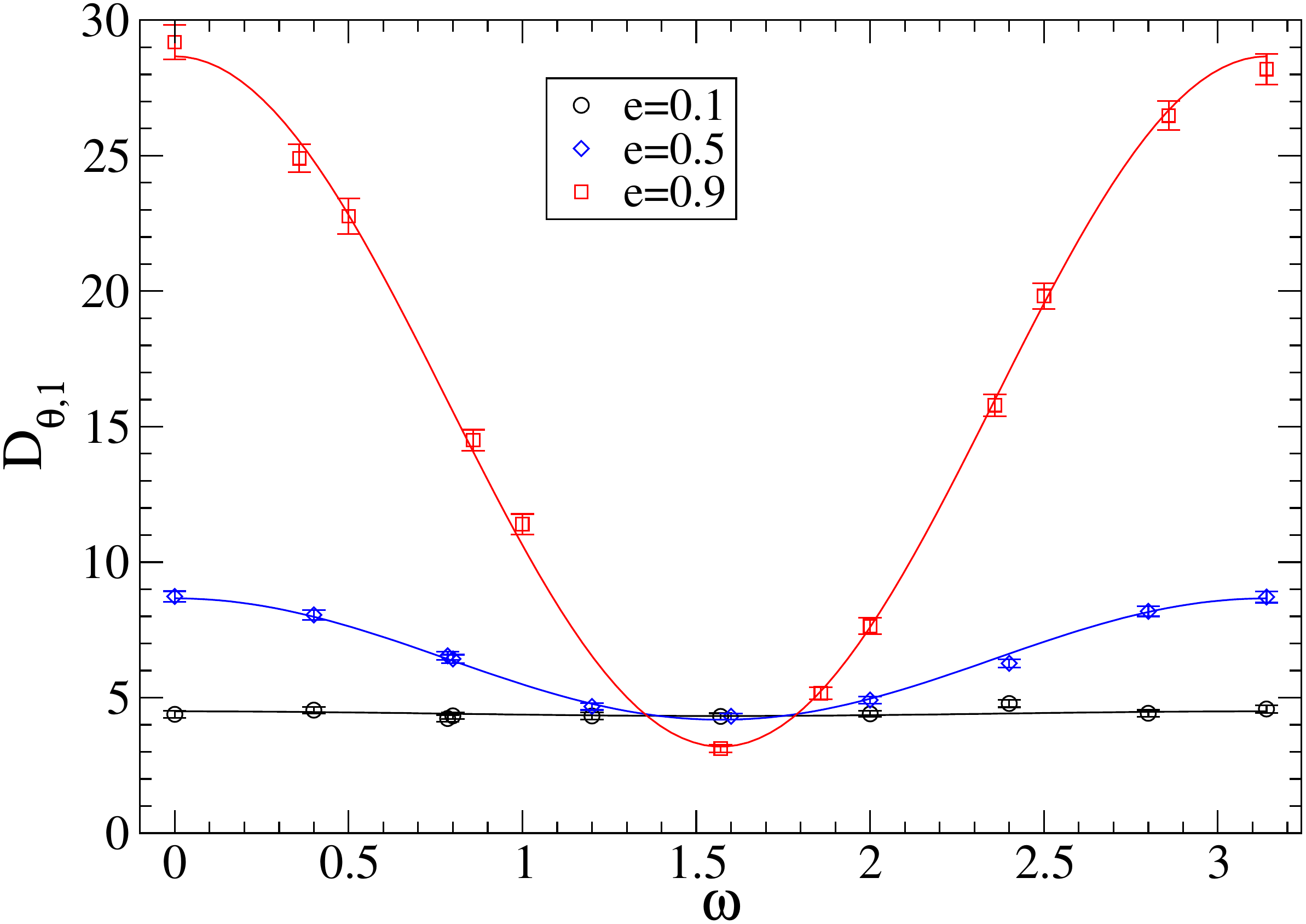}}
	\subfigure[]{\includegraphics[width=0.49\textwidth]{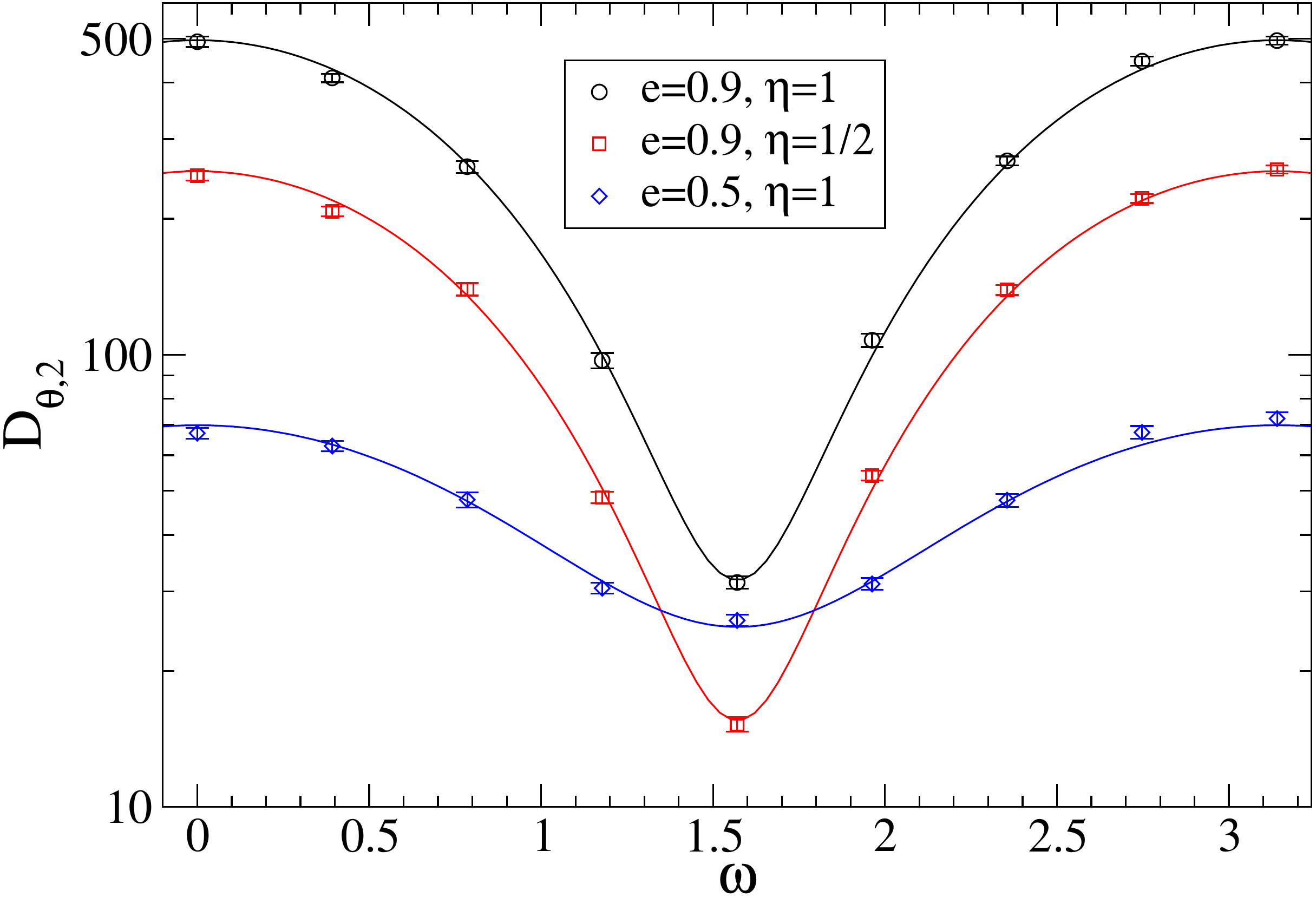}}
	\subfigure[]{\includegraphics[width=0.49\textwidth]{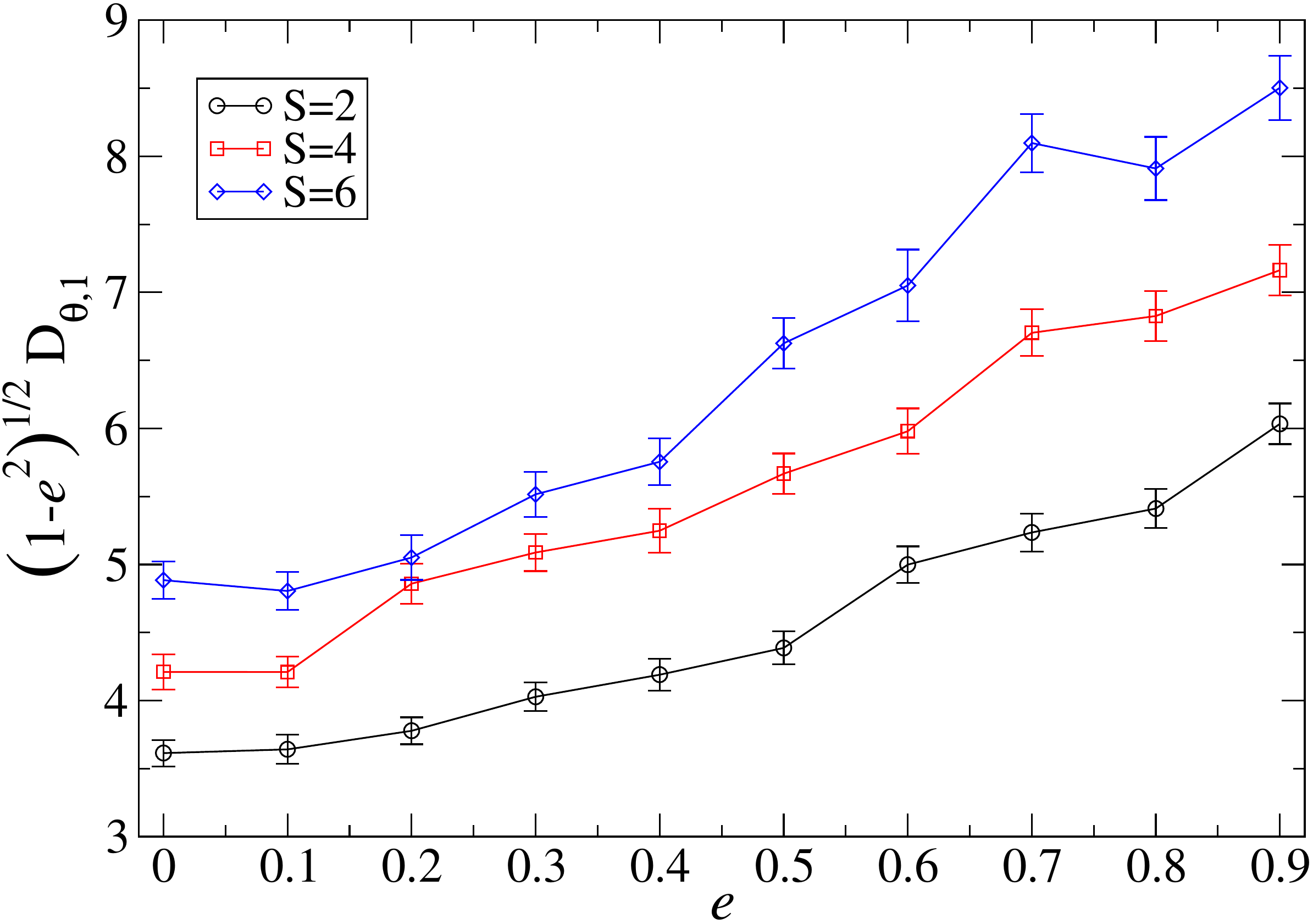}}
	\subfigure[]{\includegraphics[width=0.49\textwidth]{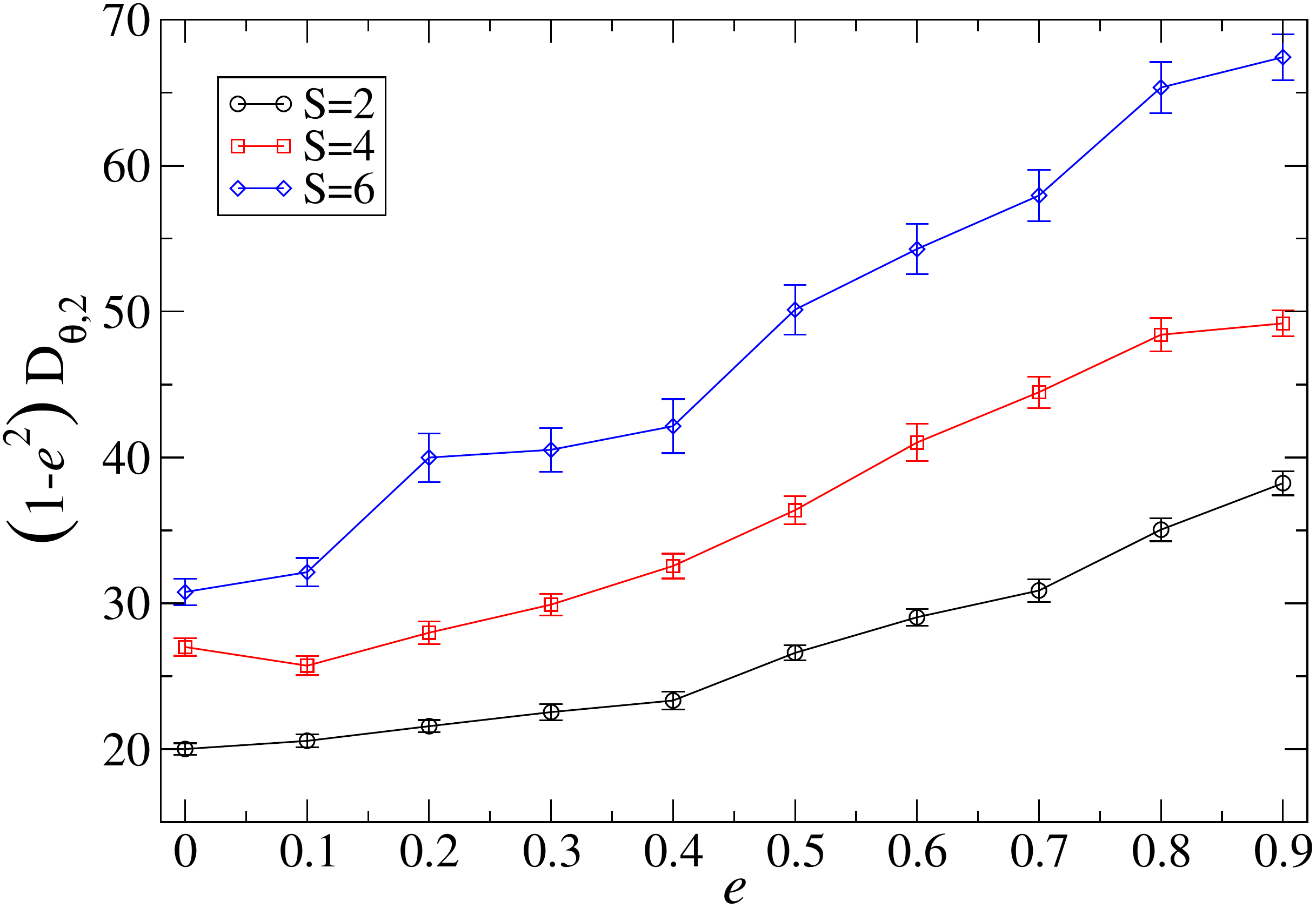}}
\caption{Continuation of Figure~\ref{Figure:D(a,eta,theta)}. The default parameter values are the same except as follows: (b) $\eta=1/2$; (d) $\theta=\pi/2$, $e=0.9$; (f) $\theta=\pi/2$. The lines on (a) and (b) are the analytical approximations given by Equation~(\ref{Equation:D(q)_fit}). The lines on (c) and (d) are $a_0+a_1\cos{2\omega}$ fits. (e) and (f) show the values averaged over the argument of periapsis $\omega$, assuming the uniform distribution of $\omega$; note that these two figures show $\sqrt{1-e^2}D_{\theta,1}$ and $(1-e^2)D_{\theta,1}$ which have finite limits at $e\rightarrow1$, so $D_{\theta,1}\sim(1-e^2)^{-1/2}$ and $D_{\theta,2}\sim(1-e^2)^{-1}$ in the high-eccentricity limit.}
\label{Figure:D(q,omega,e)}
\end{figure}

 Figures~\ref{Figure:D(a,eta,theta)} and~\ref{Figure:D(q,omega,e)} show the dependence of
 $D_{\theta,1}$ and $D_{\theta,2}$ on the various parameters.
We note the following:  
\begin{enumerate}
\item $D_{\theta,1}$ is always positive, i. e. $\langle\Delta\theta\rangle$ is always negative, 
and the angular momentum of the binary always tends to align with the rotation axis of the stellar nucleus.
\item Both $D_{\theta,1}$ and $D_{\theta,2}$ increase with increasing binary hardness (Figures~\ref{Figure:D(a,eta,theta)}a-b), reaching a maximum at $S\rightarrow\infty$, like $H$. 
The dependence is less steep than $1/a$, so that $\langle\Delta\theta\rangle$
and $\langle(\Delta\theta)^2\rangle$ are both {\it decreasing} functions of binary hardness.
\item
There is a clear trend for  $D_{\theta,2}$ to increase for decreasing mass ratio $q$, for a given $a/a_h$ (Figures~\ref{Figure:D(a,eta,theta)}b, \ref{Figure:D(q,omega,e)}b).
\item The dependence of $D_{\theta,12}$ on $\eta$ is accurately linear (Figures~\ref{Figure:D(a,eta,theta)}c-d),  consistent with definition of $\eta$ (Equation~\ref{Equation:Overline}).
\item $D_{\theta,1}(\theta)$ can be approximated as $C\sin{\theta}$ (Figure~\ref{Figure:D(a,eta,theta)}e), 
as written previously in Equation~(\ref{Equation:thetaD1D2});  the second term in that equation is zero for the scattering experiments which assume infinitesimal stellar mass. 
$D_{\theta,2}$ decreases with $\theta$, but not very dramatically: $D_{\theta,2}(0)/D_{\theta,2}(\pi)\approx1.5$ for circular binaries (Figure~\ref{Figure:D(a,eta,theta)}f).
\item For eccentric binaries, a new variable comes into play --- the argument of periapsis $\omega$. 
We define $\omega$ such that $\omega=0$ and $\omega=\pi$ correspond to the binary's major axis being perpendicular to the $z$ axis.
The dependence of $D_{\theta,12}$ on $\omega$ is shown on Figures~\ref{Figure:D(q,omega,e)}c, d. 
Both $D_{\theta,1}(\omega)$ and $D_{\theta,2}(\omega)$ can be well approximated as $C_1 + C_2\cos{2\omega}$, and for high eccentricities this dependence can be rather steep: $D_{\theta,1}(\omega=0)/D_{\theta,1}(\omega=\pi/2)\approx(1-e)^{-1}$, $D_{\theta,2}(\omega=0)/D_{\theta,2}(\omega=\pi/2)\approx1.5(1-e)^{-1}$. 
It is remarkable that the latter relation is almost independent of the degree of nuclear rotation $\eta$ 
(compare the black and red lines of Figure~\ref{Figure:D(q,omega,e)}d). 
The configurations with greatest $D_\theta$ therefore consist of eccentric binaries that are oriented 
perpendicular to the nuclear rotation axis, when changes in $\theta$ correspond to rotation of the binary orbit
about its long axis.
\item At high eccentricities, $D_{\theta,1}\sim(1-e^2)^{-1/2}$ and $D_{\theta,2}\sim(1-e^2)^{-1}$ (Figures~\ref{Figure:D(q,omega,e)}e,f). 
This is consistent with Equation~(\ref{Equation:D12}) which states that 
$D_{\theta,1}\sim1/l_b$, $D_{\theta,2}\sim1/l_b^2$.
\item $D_{\theta,1}$ and $D_{\theta,2}$ depend in rather different ways on binary mass ratio $q$
(Figures~\ref{Figure:D(q,omega,e)}a,b). 
It can be shown analytically that in the small $q$ limit, $D_{\theta,1}(q)\sim q$ and $D_{\theta,2}(q)\approx\mathrm{const}$ (see Appendix~\ref{Appendix:Large_q}). 
Accordingly, we fit the numerical values to the following simple functions:
\bsub\label{Equation:D(q)_fit}
\barr
D_{\theta,1}(q) &=& A_1\left[1+B_1\frac{(1+q)^2}{q}\right]^{-1}, \\
D_{\theta,2}(q) &=& A_2\left[1+B_2\frac{q}{(1+q)^2}\right]^{-1} .
\earr
\esub
These functions satisfy the conditions $D_{\theta,1}(q)\sim q$ and $D_{\theta,2}(q)\approx\mathrm{const}$ at small $q$ and are also invariant to the change $q\rightarrow 1/q$, appropriate given that either of the binary components can be ``first''. Figure~\ref{Figure:D(q,omega,e)}a, b verify the good fit of these analytical forms
to the data, consistent with the arguments of Appendix~\ref{Appendix:Large_q}. 
Except in the case of extreme mass ratios ($q\lesssim 10^{-2}$), an  even simpler approximation
is adequate for hard binaries: 
$D_{\theta,1}\approx\mathrm{const}$, $D_{\theta,2}\approx\sqrt{1/q}$, which works for $S\gtrsim8$.
The only other paper known to us that studied the dependence of reorientation on $q$ is \citet{Cui2014}. Their results are consistent with ours, although it is difficult to say more since they show only three points ($q=1,\,0.1,\,0.01$) with large error bars.
\end{enumerate}

We can summarize these results by writing the following, approximate expressions for the dimensionless
diffusion coefficients, which are valid in the limit of a hard binary:
\bsub\label{Equation:Dtheta12}
\barr
D_{\theta,1} &\approx& 4.5 (2\eta-1) \sqrt\frac{1+e}{1-e} \left(1+\frac{e}{2-e}\cos2\omega\right) \sin\theta, \\
D_{\theta,2} &\approx& \frac{30}{1-e}\left(1+\frac{1+2e}{5-2e}\cos2\omega\right)\sqrt{1/q}.
\earr
\esub

Or, after averaging over $\omega$,
\bsub\label{Equation:Dtheta12-omega-averaged}
\barr
D_{\theta,1} &\approx& 4.5 (2\eta-1) \sqrt\frac{1+e}{1-e} \sin\theta, \\
D_{\theta,2} &\approx& \frac{30}{1-e} \sqrt{1/q}.
\earr
\esub

Having specified the parameter dependence of the diffusion coefficients, 
we can estimate the reorientation of the binary plane in one hardening time in the diffusion-dominated 
(nonrotating nucleus) and drift-dominated (rotating nucleus) cases. 
Adopting Equation~(\ref{Equation:Definethard}) for the binary hardening time, with $H\approx 16$
(hard binary), we find for the change in inclination in one hardening time in the  
diffusion-dominated regime
\beq\label{Equation:delta-theta-2}
\delta\theta_2 \equiv \sqrt{\langle(\Delta\theta)^2\rangle t_\mathrm{hard}} 
= \sqrt{\frac{m_f}{M_{12}}\frac{D_{\theta,2}}{H}} .
\eeq
Inserting Equation~(\ref{Equation:Dtheta12-omega-averaged}b) yields
\beq\label{Equation:dthetaNonrotating}
\delta\theta_2 \approx \sqrt{\frac{m_f}{M_{12}}}\sqrt{\frac{2}{1-e}}\; (1/q)^{1/4} .
\eeq
Equation~(\ref{Equation:dthetaNonrotating}) is similar to expressions given in \citet{Merritt2002}
who considered the case $q=1, e=0$.
\citet[Equation 4.4]{Gualandris2007} presented an expression for $\delta\theta_2$ as a function 
of $q$ and $e$.
Their expression has about the same value at $q=1,\,e=0$ and the same dependence on $e$ for 
$e\rightarrow1$, although the mass ratio dependence was given by those authors as $\delta\theta_2\sim\sqrt{1/q}$.
Our expression supersedes theirs.

In the drift-dominated regime (Eq.~\ref{Equation:Dtheta12-omega-averaged}a) we find
\beq\label{Equation:dthetaRotating}
\delta\theta_1 \equiv \left|\langle\Delta\theta\rangle t_\mathrm{hard}\right| \approx 0.3(2\eta-1) 
\sqrt\frac{1+e}{1-e}  \sin\theta .
\eeq
We see that unless the corotation fraction of the nucleus is very small ($\eta-1/2 \ll 1$), $\delta\theta_1$ is of the order of $\theta$ --- a significant reorientation occurs on the hardening timescale. Comparison of (\ref{Equation:dthetaRotating}) with (\ref{Equation:dthetaNonrotating}) shows that for typical SMBH masses ($M_{1,2}=10^6...10^9\msun$) the first-order effect prevails over the second-order one even for corotation fractions as small as $\eta-1/2=0.01$ (i. e., in nuclei where only $1\%$ of all stars contribute to rotation). This is due to different dependence on the field particle mass --- first-order effects don't depend on it (only on the total number density), and second-order effects decrease as $\sqrt{m_f/M_{12}}$.

\subsection{\label{Section:D(Omega)}Diffusion coefficients for the longitude of the ascending node}

In this section the diffusion coefficients describing changes in the longitude of the binary's line of nodes,
$\Omega$, are presented. 
As shown in Figure~\ref{Figure:binary}, $\Omega$ is equivalent to the $\phi-$ coordinate
of the binary's angular momentum vector in a spherical coordinate system having the nuclear rotation
axis as reference axis.
Sections~\ref{Section:BinaryEvolveFPE} and~\ref{Section:BinaryEvolveOrientation} present relations
between $\{\Omega$, $\Delta\Omega\}$ and the ``local'' displacement variables $\Delta\Theta_\perp$ and $\Delta\Theta_\parallel$ (see Figure~\ref{Figure:binary}b and Equations~\ref{Equation:DeltaTheta}).

 From Equation~(\ref{Equation:deltaOmega}) we derive the following expressions for the first-
 and second-order diffusion coefficients, in terms of the dimensionless rates 
 $D_{\Omega,1}, D_{\Omega,2}$:
\bsub\label{Equation:DOmega12}
\barr
\langle\Delta\Omega\rangle &=& - \frac{\rho G a}{\sigma} D_{\Omega,1}\label{Equation:D1toDeltaOmega}, \\
\langle(\Delta\Omega)^2\rangle &=& \frac{m_f}{M_{12}} \frac{\rho G a}{\sigma} D_{\Omega,2} ,
\label{Equation:D2toDeltaOmega2}\\
D_{\Omega,1} &=& - \frac{1}{\nu\sqrt{1-e^2}\sin\theta}
 \frac{\overline{\overline{\delta l_\Omega}}}{aV_\mathrm{bin}},  \\
D_{\Omega,2} &=& \frac{1}{\nu^2(1-e^2)\sin^2\theta}
 \frac{\overline{\overline{\delta l^2_\Omega}}}{\left(aV_\mathrm{bin}\right)^2} .
\earr
\esub
By symmetry, none of the diffusion coefficients (either those for $\Omega$, or for the other variables presented
above) are functions of $\Omega$.
However there are no obvious constraints from symmetry that would imply the vanishing of the diffusion
coefficients in $\Omega$, at least in the case of a {\it rotating} stellar nucleus.

Immediately we see that $D_{\Omega,12} \rightarrow \infty$ at $\theta=0$ and $\theta=\pi$, which is natural since
 $\Omega$ becomes undefined when the binary orbit is aligned with the $x-y$ plane.

Our results are consistent with $D_{\Omega,1}=0$, both in nonrotating and rotating nuclei. 
This result is consistent with the results of \citet[Figure~6]{Cui2014}.

Figure~\ref{Figure:D_Omega,2} shows the dependence of $D_{\Omega,2}$ on the various parameters.
The dependences are similar to those of $D_{\theta,2}$. 
This is not surprising, since in the case of zero nuclear rotation, $D_{\Omega,2}\sin^2\theta$ at argument of periapsis $\omega$ is exactly equal to $D_{\theta,2}$ at argument of periapsis $\pi/2-\omega$, and neither
 coefficient depends strongly on the degree of nuclear rotation. From this figure, we see that $D_{\Omega,2}\sin^2\theta = 20...500$, and from this we can estimate the change in $\Omega$ on a hardening timescale by analogy with Eq.~(\ref{Equation:delta-theta-2}):
 
\beq
\delta\Omega = \sqrt{\langle(\Delta\Omega)^2\rangle t_\mathrm{hard}} 
= \sqrt{\frac{m_f}{M_{12}}\frac{D_{\Omega,2}}{H}} = 1...6 \sqrt\frac{m_f}{M_{12}\sin\theta}
\eeq

\begin{figure}[h]
	\subfigure[]{\includegraphics[width=0.49\textwidth]{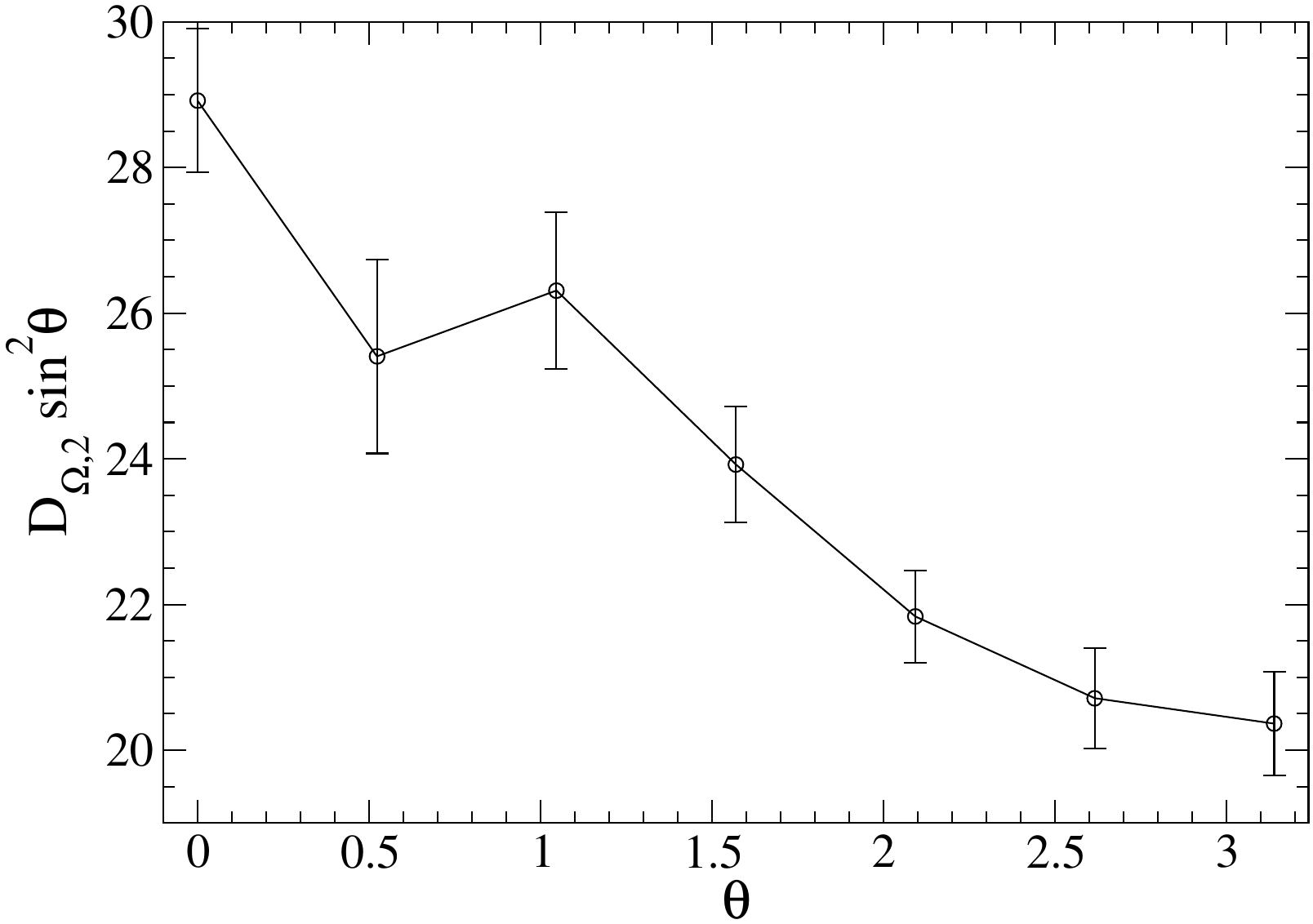}}
	\subfigure[]{\includegraphics[width=0.49\textwidth]{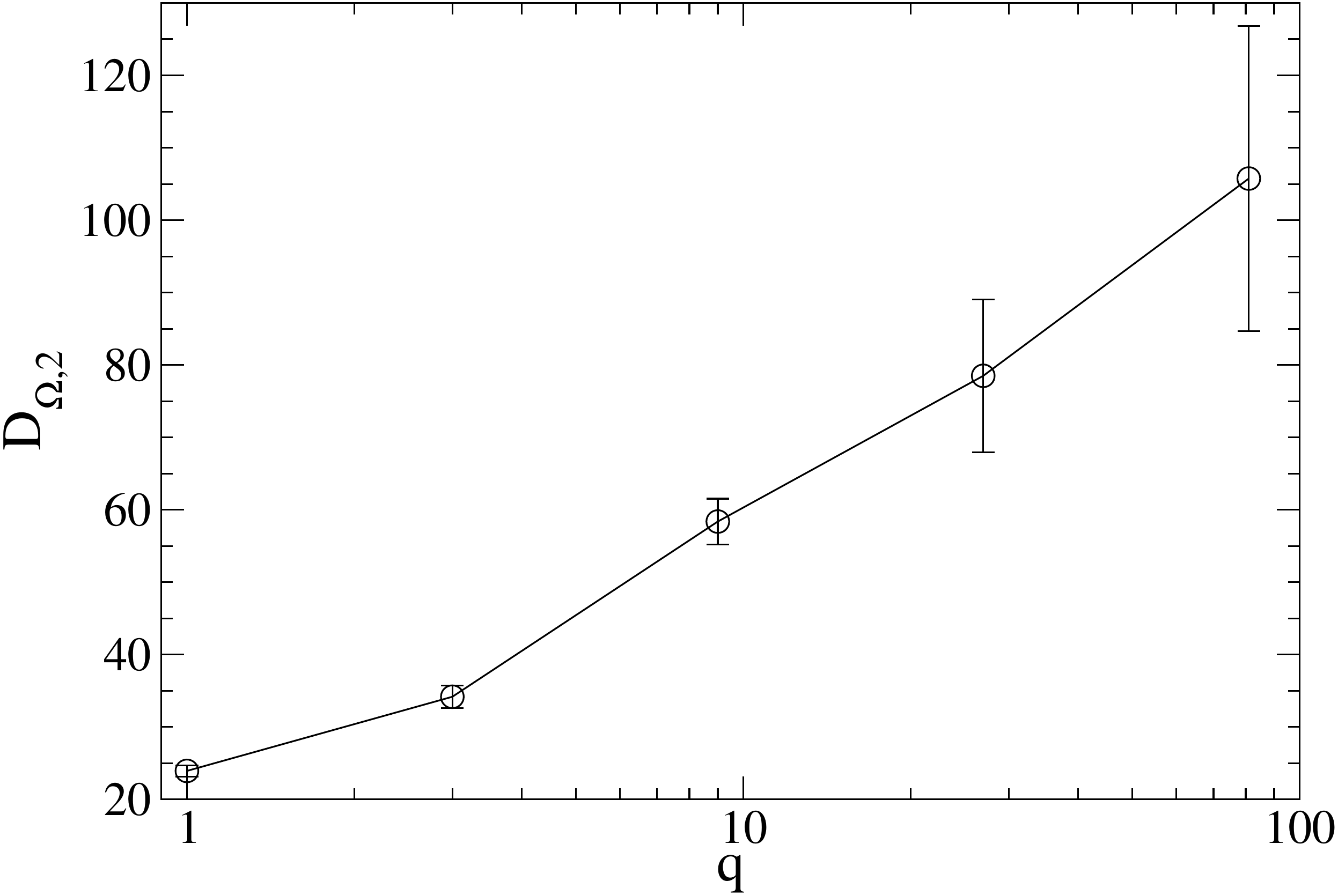}}
	\subfigure[]{\includegraphics[width=0.49\textwidth]{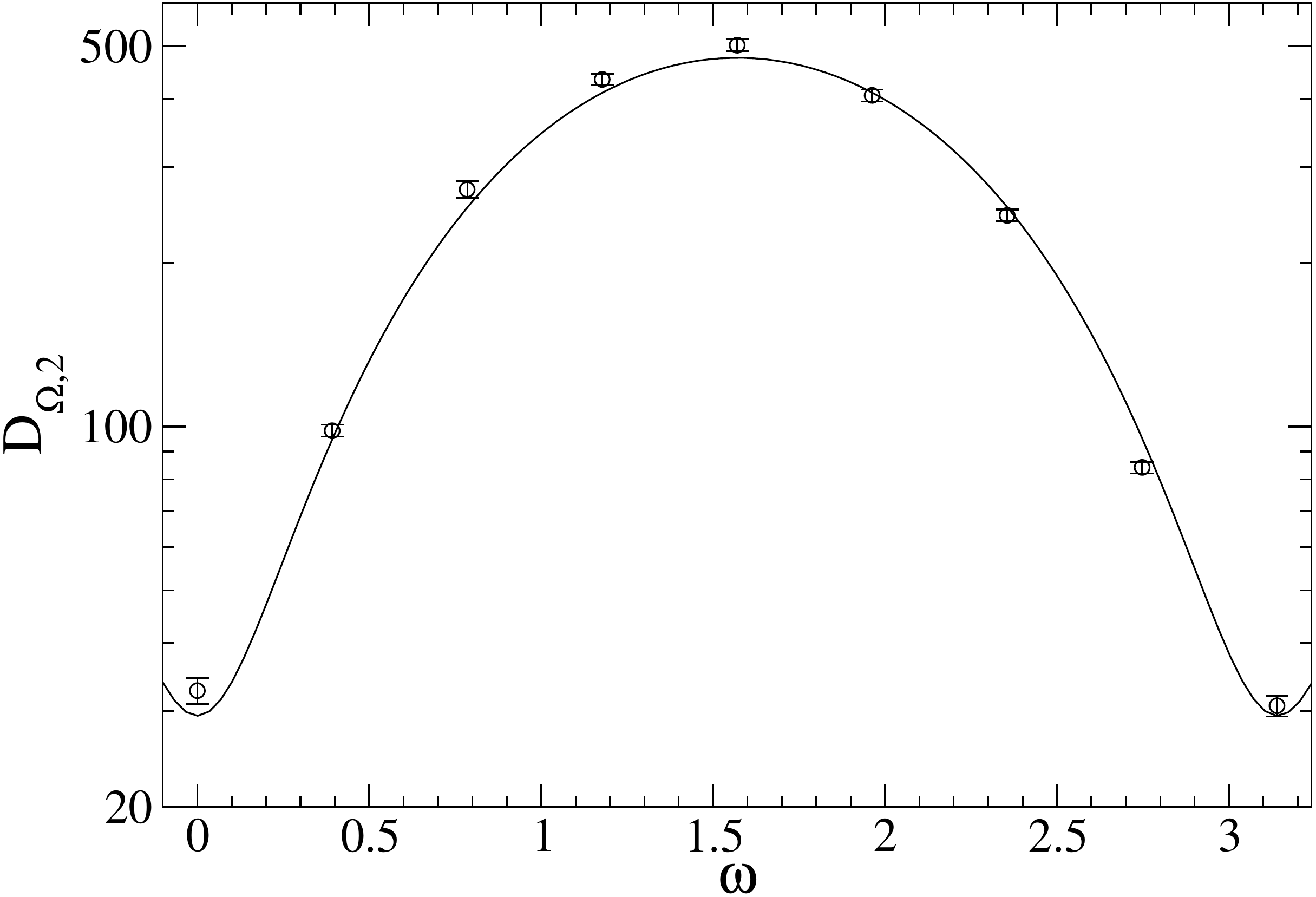}}
	\subfigure[]{\includegraphics[width=0.49\textwidth]{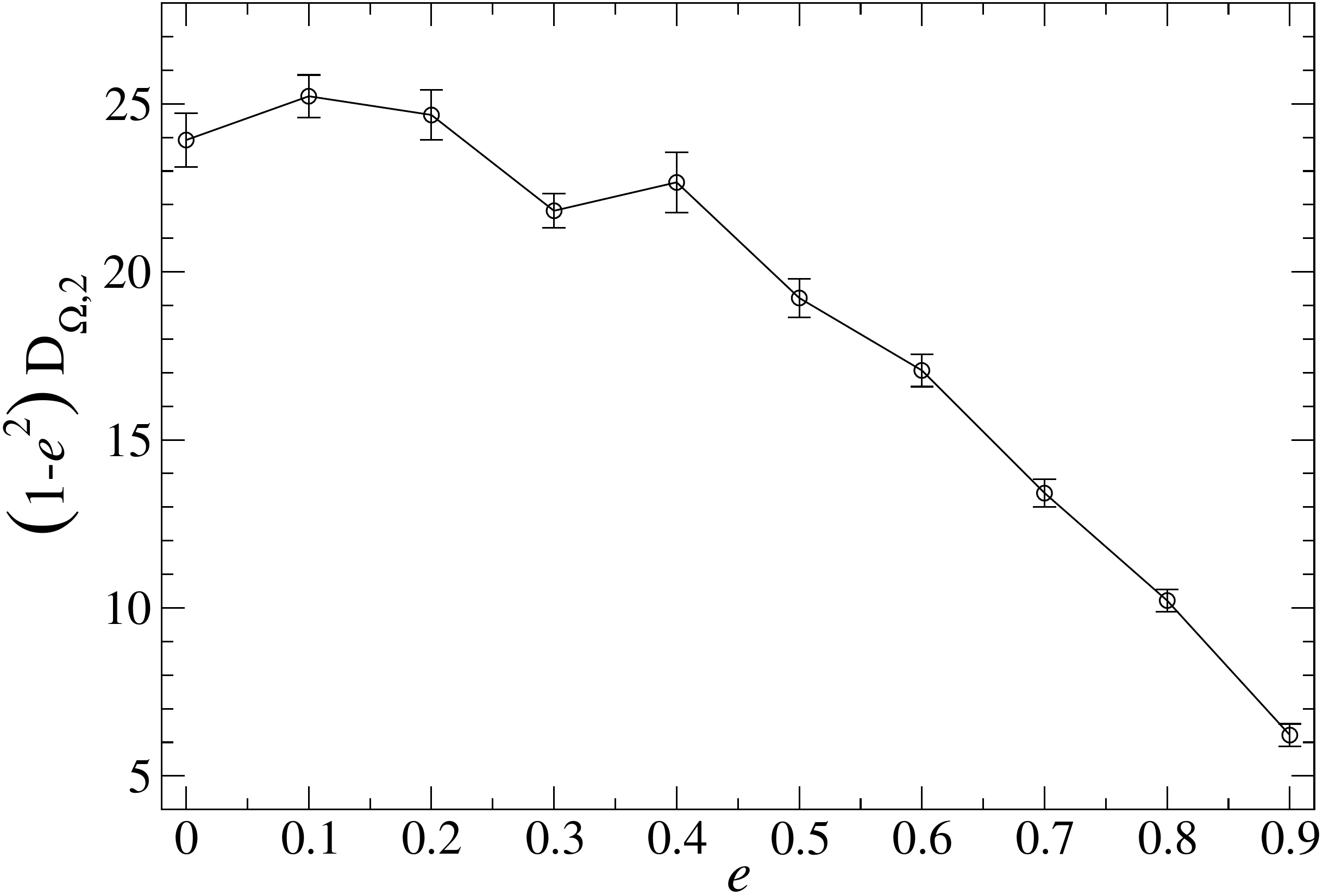}}
\caption{Dependence of $D_{\Omega,2}$ on various parameters: binary inclination $\theta$, eccentricity $e$, mass ratio $q$ and argument of periapsis $\omega$. 
Unless otherwise stated, $S=4$, $\eta=1$, $e=0$, $q=1$ and $\theta=\pi/2$, except for $e=0.9$ in (c). The line in (c) is $a_0+a_1\cos{2\omega}$ fit.}
\label{Figure:D_Omega,2}
\end{figure}

\subsection{\label{Section:D(omega)}Diffusion coefficients for the argument of periapsis}

\begin{figure}[h]
	\subfigure[]{\includegraphics[width=0.49\textwidth]{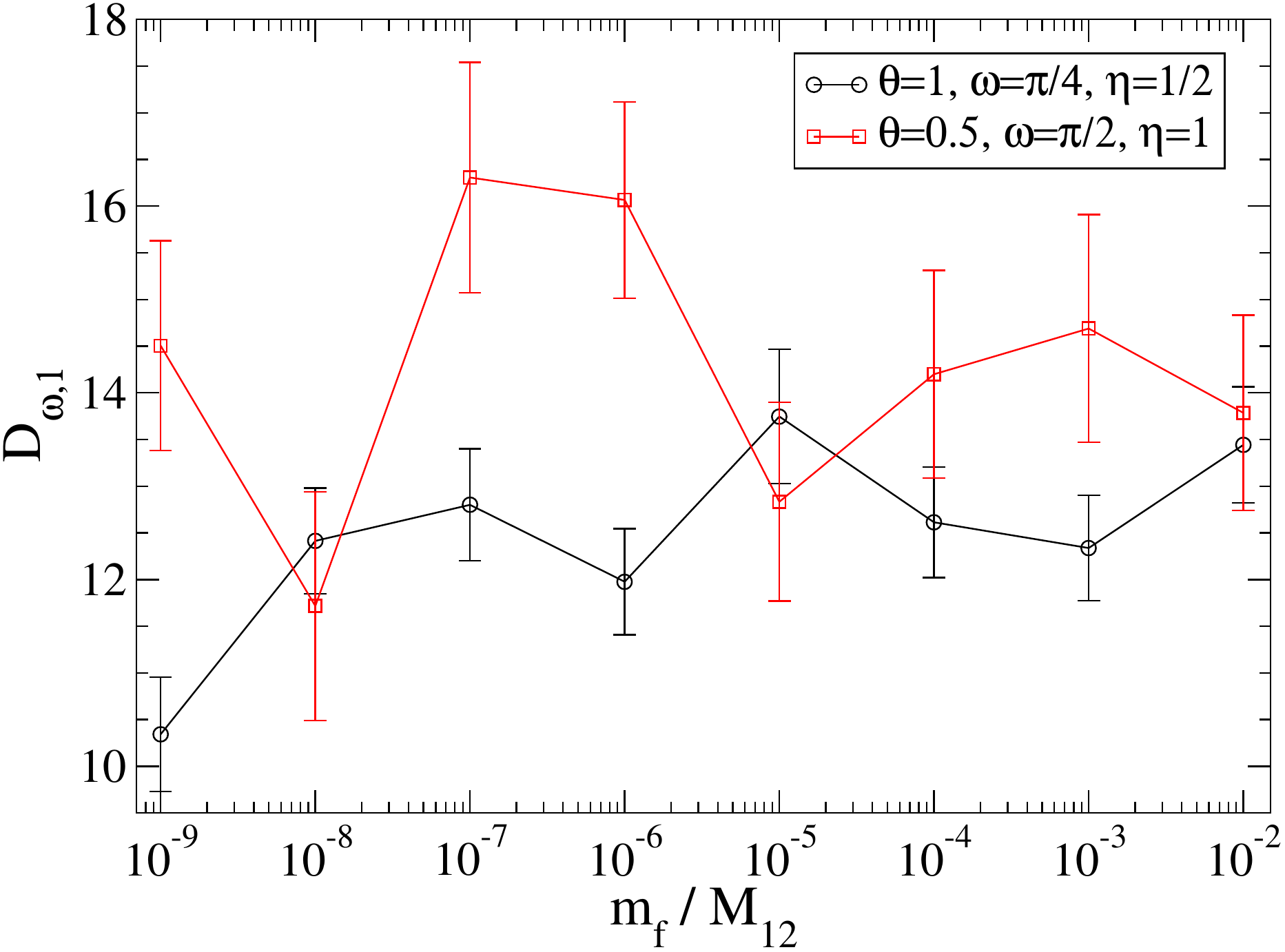}}
	\subfigure[]{\includegraphics[width=0.49\textwidth]{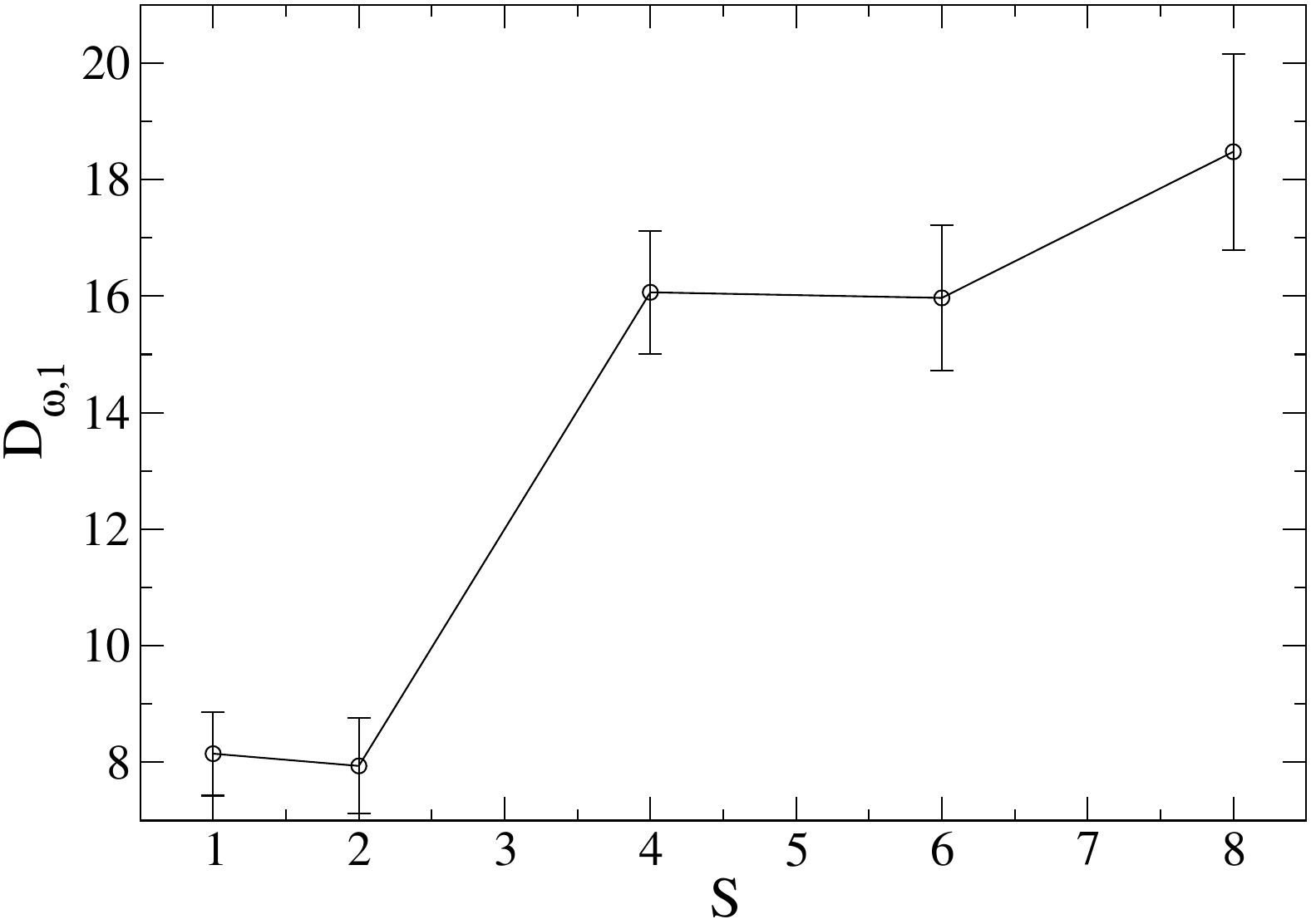}}
	\subfigure[]{\includegraphics[width=0.49\textwidth]{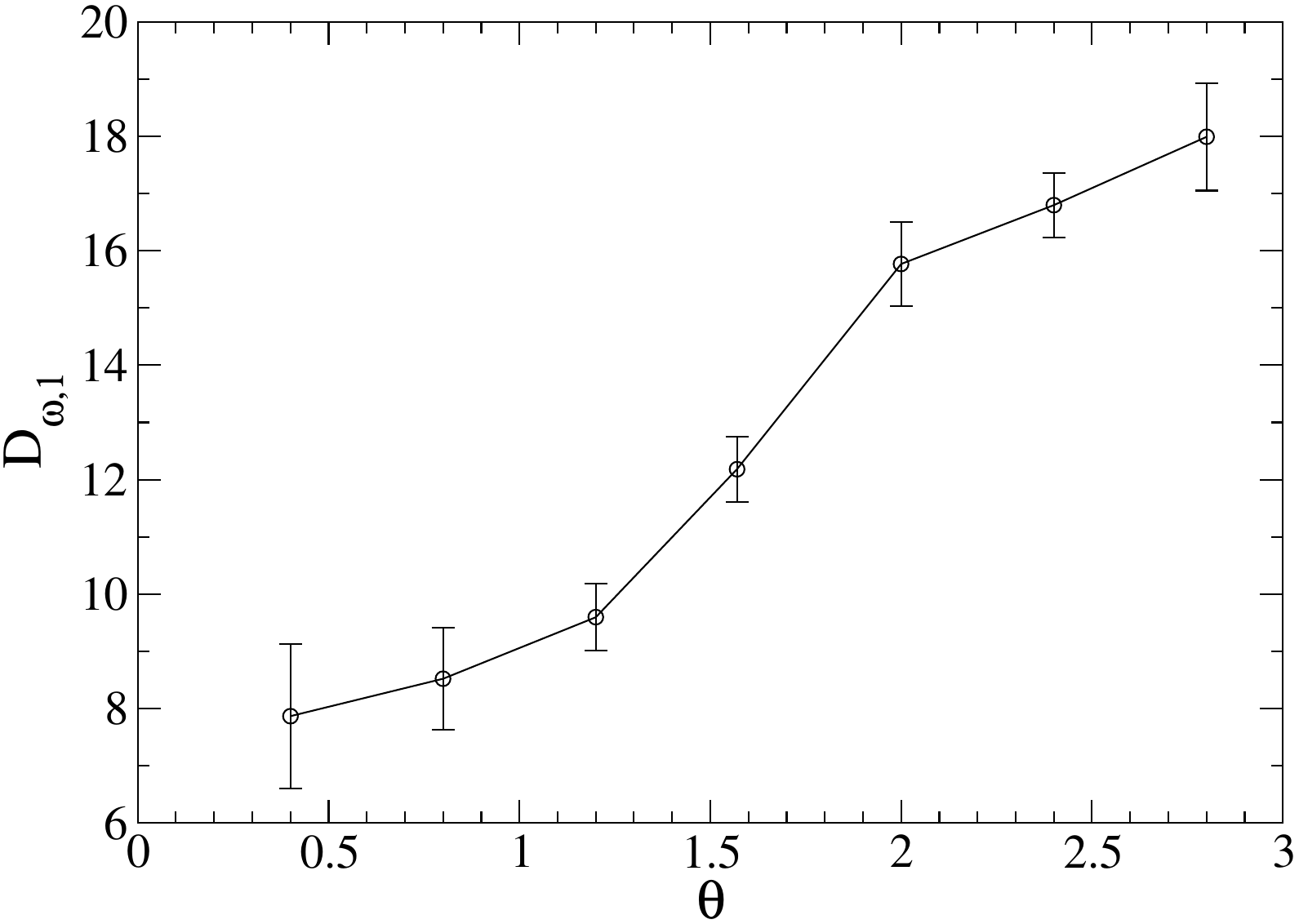}}
	\subfigure[]{\includegraphics[width=0.49\textwidth]{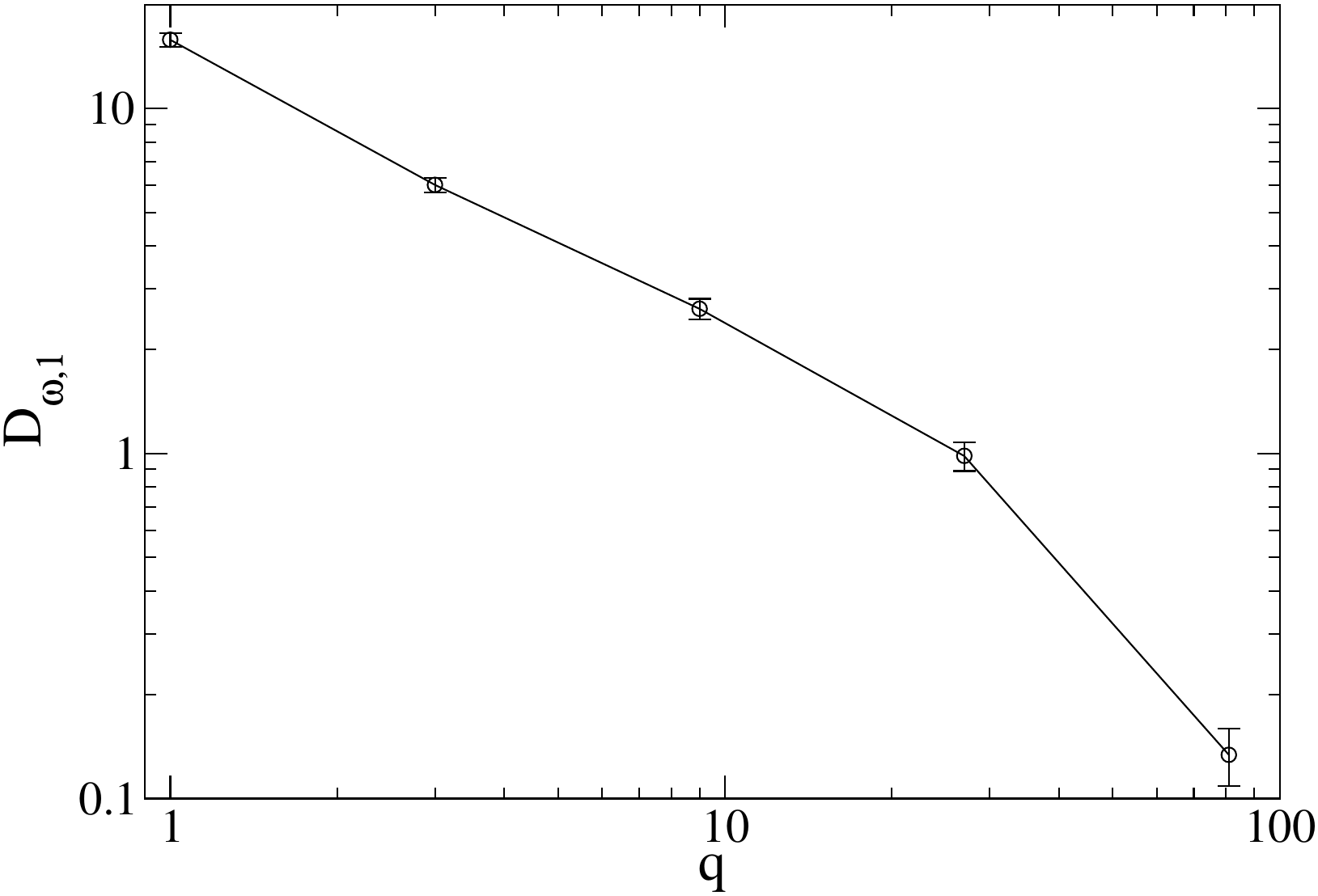}}
	\subfigure[]{\includegraphics[width=0.49\textwidth]{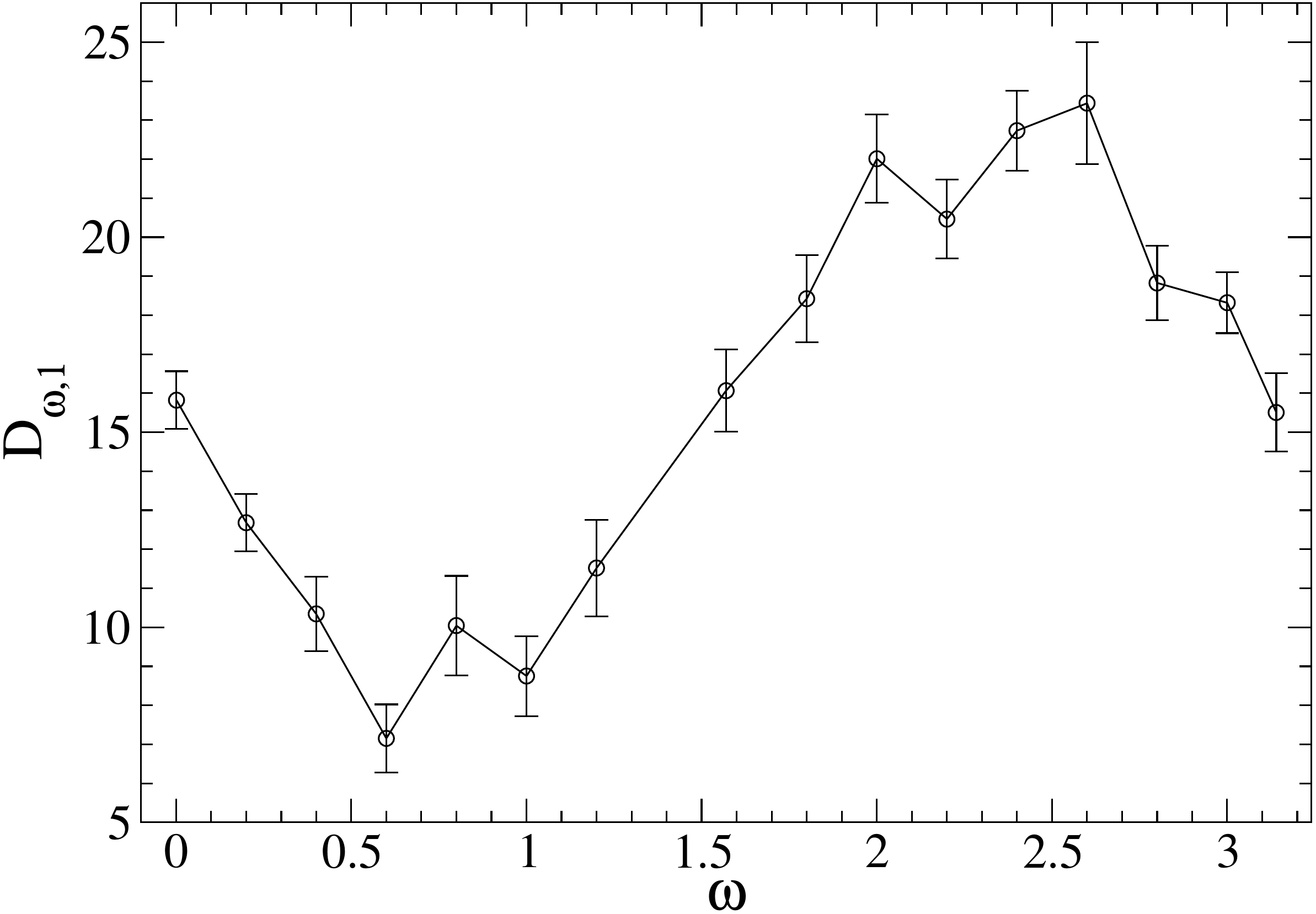}}
	\subfigure[]{\includegraphics[width=0.49\textwidth]{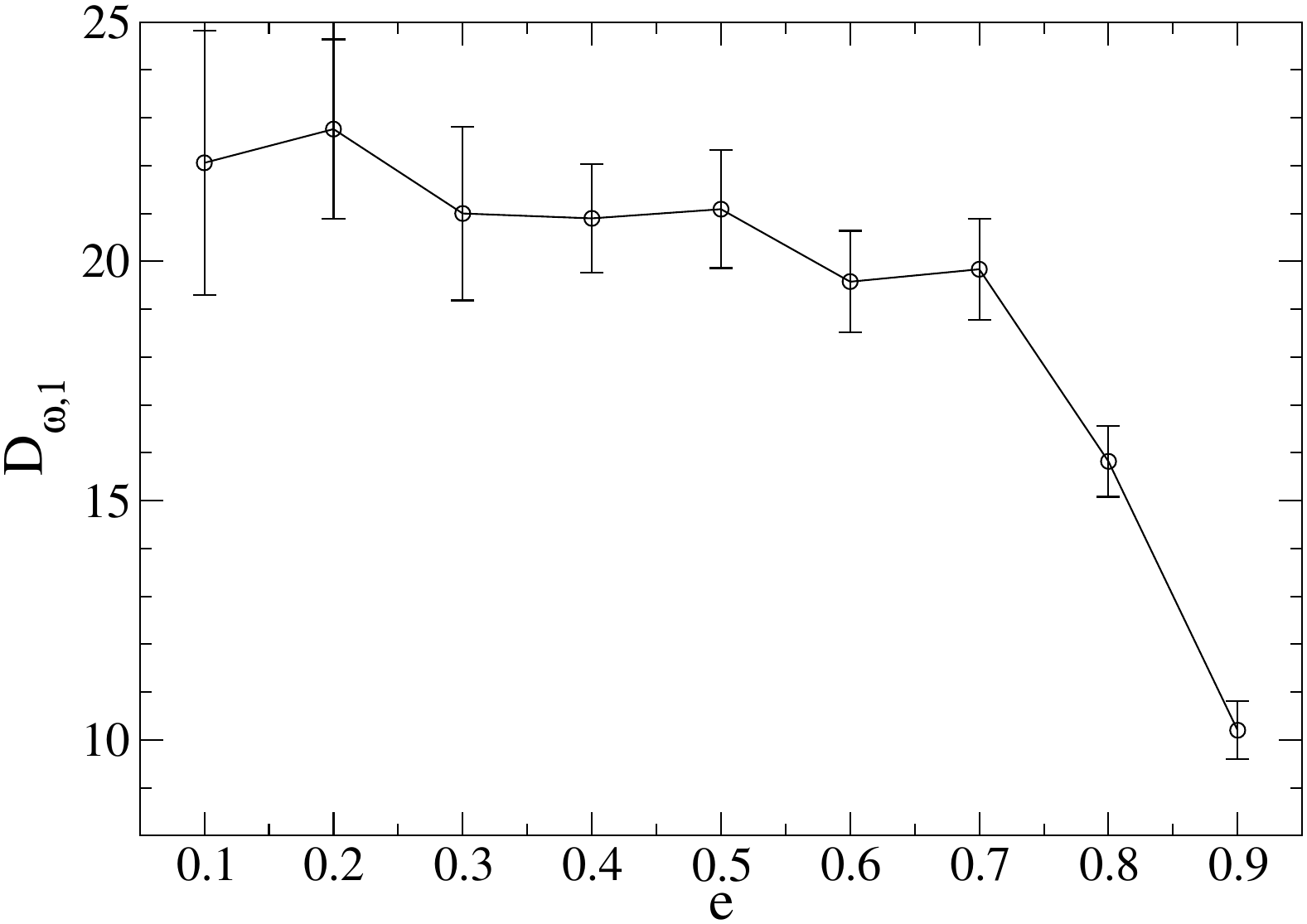}}
	\caption{
		Dependence of $D_{\omega,1}$ on various parameters. Unless otherwise stated, $m_f/M_{12}=10^{-6}$, $S=4$, $\eta=1$, $e=0.8$, $q=1$, $\theta=0.5$. (b) $\omega=\pi/2$; (c) $\omega=\pi/4$; (d)-(f) $\omega=0$.
		}
	\label{Figure:D_omega}
\end{figure}

As in the case of the angular variables $\theta$ and $\Omega$, we write the diffusion coefficients 
for the argument of periapsis, $\omega$, as:
\bsub\label{Equation:D_omega}
\barr
\langle\Delta\omega\rangle &=& - \frac{\rho G a}{\sigma} D_{\omega,1}\label{Equation:D1toDeltaomega}, \\
\langle(\Delta\omega)^2\rangle &=& \frac{m_f}{M_{12}} \frac{\rho G a}{\sigma} D_{\omega,2}\label{Equation:D2toDeltaomega2}, \\
D_{\omega,1} &=& -\overline{\overline{\delta \omega}} / m_f , \\
D_{\omega,2} &=& \overline{\overline{\delta \omega^2}} / m_f^2 .
\earr
\esub

The argument of periapsis differs from all the other orbital elements considered here,
in the sense that it is not related to the binary's energy or angular momentum.
It is therefore not possible to calculate changes in $\omega$ by means of scattering experiments with zero stellar mass. 
Instead, we carried out scattering experiments with small but nonzero stellar mass (using the same ARCHAIN integrator; see \S\ref{Section:Scattering}), 
and recorded the initial and final values of $\omega$. Because of that, we only consider the first-order coefficient below.

The minus sign in the definition of $D_{\omega,1}$ reflects the fact that $\langle\Delta\omega\rangle$ is always negative.
(Note that we define $\omega$ such that negative $\langle\Delta\omega\rangle$ means orbital precession in the direction opposite to the orbital motion of binary components.)
Figure~\ref{Figure:D_omega} shows the parameter dependences.
Figure~\ref{Figure:D_omega}a verifies that $D_{\omega,1}$ (and thus $\langle\Delta\omega\rangle$) is
independent, within the uncertainties, of the mass of the field star $m_f$ when $m_f$ is sufficiently small ($m_f \lesssim 0.01M_{12}$) as we would expect for the first-order diffusion coefficient.

Interestingly, $\langle\Delta\omega\rangle$ is significantly nonzero even in a {\it non}rotating nucleus (see the black line on Figure~\ref{Figure:D_omega}a). 
As far as we know, this source of apsidal precession has never been discussed heretofore.
We evaluate the importance of this precession by estimating how much $\omega$ changes 
in one hardening time:
\beq
\Delta\omega \equiv \left|\langle\Delta\omega\rangle\right| t_h \approx \frac{D_{\omega,1}}{H} \approx 1
\eeq
(for a binary with moderate eccentricity). 
Precession at this rate helps to justify our decision to average the diffusion coefficients in $\theta$ over $\omega$.
Below we compare changes in $\omega$ due to this mechanism with changes due to other sources
of apsidal precession, e.g., general relativity.

\section{Effect of General Relativity}
\label{Section:GR}

In the post-Newtonian approximation, the effects of general relativity (GR) on the motion can be treated
by adding terms of order $(v^2/c^2)^n$, $n=1,2,\ldots$  to the Newtonian equations of motion, where 
 $v$ are $r$ are typical velocities and separations and $m$ is the particle mass. 
At the lowest, or 1PN, order, the exact $N$-body equations of motion can be written
for arbitrary $N$: the so-called Einstein-Infeld-Hoffmann equations of motion \citep{Einstein1938}.
At higher PN orders, closed-form expressions for the accelerations only exist for 
two-particle systems.

In this section we consider the effects of GR on the orbital motion of the two SBHs.
Since $N=2$ for the binary, we are able to consider PN terms of arbitrary order.
GR also affects the motion of a star with respect to the massive binary.
We ignore those effects, partly out of convenience, but also on the grounds that the time of
interaction of a star with the massive binary is typically small compared with the time required for
GR effects to influence the star's motion.

A characteristic distance associated with the effects of GR is the gravitational radius $r_g$, 
which for a SBH of mass $M_\bullet$ is 
\begin{eqnarray}\label{Equation:r_g}
r_g \equiv \frac{GM_\bullet}{c^2} \approx 4.8\times 10^{-6}\left(\frac{\mh}{10^8\msun}\right)\,\mathrm{pc} .
\end{eqnarray}

We consider the effects of GR in PN order, from lowest to highest, and ignore for the moment spin
of the two \sbhs:

\begin{enumerate}

\item Adding the 1PN terms to the binary's equation of motion results in apsidal (in-plane) precession of the binary orbit. The time for the argument of periapsis $\omega$ to change by $\pi$ is
\begin{eqnarray}
t_\omega = \frac{1}{6}(1-e^2)\frac{a}{r_g}T   \,\label{Equation:t_omega}
\end{eqnarray}
\citep[][Eq.~4.274]{DEGN}
where $T=2\pi\sqrt{a^3/GM_{12}}$ is the binary's period. 
We can compare this time with the time for the binary orbit to precess as a result of cumulative
interactions with stars, as given by Equation~(\ref{Equation:D_omega}).
The two timescales are equal when
\bsub
\begin{eqnarray}
a = a_\omega \equiv \left[\frac{3}{(1-e^2)D_{\omega,1}}\right]^{2/7}\left(\frac{GM_{12}^3\sigma^2}{\rho^2c^4}\right)^{1/7}.
\end{eqnarray}
\esub
Due to the smallness of the exponents, we can neglect the $(1-e^2)$ factor, and we substitute $D_{\omega,1}\approx15$ (see \S\ref{Section:D(omega)}), yielding
\begin{eqnarray}\label{Equation:a_omega}
a_\omega \approx 0.36\,\mathrm{pc} \times \left(\frac{M_{12}}{10^8\msun}\right)^{3/7} \left(\frac{\sigma}{200\,\kms}\right)^{2/7} \left(\frac{\rho}{10^3\msun\mathrm{pc}^{-3}}\right)^{-2/7} .
\end{eqnarray}
This is a relatively large separation -- of order the hard-binary separation -- implying that 1PN precession 
typically dominates over three-body precession even though the precession effects themselves are small: at $a=a_\omega$, the ratio between $t_\omega$ and the orbital period $T$ is
\begin{equation}
\frac{t_\omega}{T} = \frac{1}{6}(1-e^2)\frac{a_\omega}{r_g} \approx 
1.3 \times 10^4 \,(1-e^2) \left(\frac{M_{12}}{10^8\msun}\right)^{-4/7} \left(\frac{\sigma}{200\,\kms}\right)^{2/7} \left(\frac{\rho}{10^3\msun\mathrm{pc}^{-3}}\right)^{-2/7} \gg 1
\end{equation}
As the binary orbit shrinks, this ratio becomes smaller ($t_\omega/T \propto a$); while the timescale associated with three-body interactions becomes longer ($1/\langle\Delta\omega\rangle \propto 1/a$). Thus, the overall precession rate becomes faster than $\langle\Delta\omega\rangle$, and our decision to average all the other diffusion coefficients over $\omega$ becomes more justified.
We also note that $a_\omega$ is large compared with the separation at which gravitational-wave
emission becomes important (cf. Equation~\ref{Equation:a_GR}).

\item Additional terms that appear at 2PN order imply a slightly different rate 
of apsidal precession but otherwise do not change the character of the motion
\citep[][Section 4.5.2]{DEGN}.

\item At order 2.5, the PN equations of motion become dissipative, representing the loss of energy
and angular momentum due to gravitational radiation.
The orbit-averaged rate of change of binary semimajor axis is
\bsub\label{Equation:dadt_GR}
\begin{eqnarray}
\left(\frac{da}{dt}\right)_{\rm GR} &=& - \frac{64}{5}\frac{\nu G^3M_{12}^3}{c^5a^3}f(e),\\
f(e)&\equiv&\frac{1+(73/24)e^2+(37/96)e^4}{(1-e^2)^{7/2}}
\end{eqnarray}
\esub
\citep[Eq. 4.234a]{DEGN}.
Ignoring for the moment the fact that $e$ changes, the timescale for orbital decay is
\begin{eqnarray}\label{Equation:GW_emission_time}
t_{\rm GW} &\equiv& \frac{a}{|da/dt|_{\rm GR}} = \frac{5}{64}\frac{c^5 a^4}{\nu G^3 M_{12}^3} \frac{1}{f(e)}.
\end{eqnarray}
We compare $t_\mathrm{GW}$ with $t_\mathrm{hard}$, 
the time for $a$ to change due to three-body interactions
(Eq.~\ref{Equation:Definethard}).
The two times are equal when
\begin{eqnarray}
t_{\rm GW}= \frac{\sigma}{\rho GaH} 
\end{eqnarray}
which occurs at the separation
\begin{eqnarray}
a=a_{\rm GW}\equiv\left[\frac{64}{5}\frac{\nu G^2M_{12}^3\sigma }{Hc^5\rho}f(e)\right]^{1/5}. \label{Equation:GR_regime}
\end{eqnarray}
Approximating $H=16$ at all eccentricities (a good approximation, particularly since $a_{\rm GW}\sim H^{-1/5}$), 

\bsub\label{Equation:a_GR}
\begin{eqnarray}
a_{\rm GW} &=& 0.017 
\nu^{1/5} 
\left(\frac{M_{12}}{10^8\msun}\right)^{3/5} 
\left(\frac{\sigma}{200\,\kms}\right) ^{1/5}
\left(\frac{\rho}{10^3\msun\mathrm{pc}^{-3}}\right)^{-1/5} \mathrm{pc} , \quad e=0\\
a_{\rm GW} &=& 0.071 
\nu^{1/5} 
\left(\frac{M_{12}}{10^8\msun}\right)^{3/5} 
\left(\frac{\sigma}{200\,\kms}\right) ^{1/5}
\left(\frac{\rho}{10^3\msun\mathrm{pc}^{-3}}\right)^{-1/5} \mathrm{pc} , \quad e=0.9\\
a_{\rm GW} &=& 0.35 
\nu^{1/5} 
\left(\frac{M_{12}}{10^8\msun}\right)^{3/5} 
\left(\frac{\sigma}{200\,\kms}\right) ^{1/5}
\left(\frac{\rho}{10^3\msun\mathrm{pc}^{-3}}\right)^{-1/5} \mathrm{pc}, \quad e=0.99 .
\end{eqnarray}
\esub
Except in the case of extreme eccentricities, $a_\mathrm{GW}\ll a_\mathrm{hard}$.

\item Also as a consequence of the 2.5PN terms, the binary orbit circularizes, at the rate
\begin{eqnarray}\label{Equation:dedt_GR}
\left(\frac{de}{dt}\right)_{\rm GR} = - \frac{304}{15}\frac{\nu G^3M_{12}^3}{c^5a^4} e
\frac{1+(121/304)e^2}{(1-e^2)^{5/2}}
\end{eqnarray}
 \citep[Eq. 4.234b]{DEGN}.
 As is well known, at high eccentricities changes in $a$ and $e$ tend to leave the radius of apoapsis, $r_p=a(1-e)$,
 nearly unchanged as the orbit decays, resulting in a more circular orbit \citep[Eq. 4.237]{DEGN}.

\end{enumerate}

So far we have ignored the possibility that one or both of the \sbhs\ in the binary might be spinning.
We will continue to make that assumption with regard to the equations of motion of the passing star.
But since we will later want to connect the binary orbit with the final spin
of the merged \sbhs, it is relevant to ask how the spin directions are altered due to GR effects 
before the merger occurs.

The spin angular momentum of a rotating \sbh\ is
\beq
\boldsymbol{S} = \boldsymbol{\chi} S_\mathrm{max} = \boldsymbol{\chi} \frac{G\mh^2}{c^2} 
\eeq
where $0 \le |\boldsymbol{\chi}| \le 1$ is the dimensionless spin.
The total (spin + orbital) angular momentum, $\boldsymbol{J}$, of the binary
\beq
\boldsymbol{J} = \boldsymbol{S}_1 + \boldsymbol{S}_2 + \boldsymbol{L}
\eeq
is constant; to lowest PN order, $\boldsymbol{L}$ is the Newtonian angular momentum of the binary orbit, $\boldsymbol{L}_\mathrm{N} = \mu(\boldsymbol{x}\times\boldsymbol{v})$.
Thus
\beq
\dot{\boldsymbol L} = -\left(\dot {\boldsymbol S}_1 + \dot {\boldsymbol S}_2\right) .
\eeq
The equations simplify in the case that only one of the two holes is spinning.
If the mass of the spinning hole is $M_1$, then \citep{Kidder1995}
\begin{subequations}
\begin{eqnarray}
\dot{\boldsymbol S} &=& \frac{G}{c^2r^3}\left[\frac12\left(1+3\frac{M_{12}}{M_1}\right) \boldsymbol{J}\times\boldsymbol{S}\right], \\
\dot{\boldsymbol L} &=& \frac{G}{c^2r^3}\left[\frac12\left(1+3\frac{M_{12}}{M_1}\right) \boldsymbol{J}\times\boldsymbol{L}\right] .
\end{eqnarray}
\end{subequations}
These equations imply that $\boldsymbol{L}$ and $\boldsymbol{S}$ precess about the fixed vector $\boldsymbol{J}$
at the same rate, with frequency
\begin{equation}
\Omega_p = \frac{GJ}{2c^2r^3}\left(1 + 3\frac{M_{12}}{M_1}\right) 
\end{equation}
and the magnitudes of both $\boldsymbol{S}$ and $\boldsymbol{L}$ remain fixed.
If both holes are spinning, $\boldsymbol{J}$ is still conserved; both spins precess
about a vector $\boldsymbol{\Omega}_A$ which itself precesses, leaving the two spin
magnitudes constant, although $\boldsymbol{S}=\boldsymbol{S}_1 + \boldsymbol{S}_2$ is
not constant \citep{Kidder1995}.

In the regime considered so far in this paper, $L\gg S_{1,2}$ and $\boldsymbol{J}\approx\boldsymbol{L}$.
In this regime, the two spins precess about the nearly-fixed angular momentum vector of the binary and 
the latter is hardly affected by spin-orbit torques.
The spin precession frequency in this case  (for $q=1$, $e=0$) becomes 
\barr
\Omega_{SL} \approx 3.5\frac{G}{c^2a^3}L \approx  3.5\frac{G\mu}{c^2a^3}\sqrt{GM_{12}a}.
\earr
The binary separation at which the spin precession period equals the orbital reorientation timescale 
due to three-body interactions is 
\begin{eqnarray}
a_{SL} &\approx& 2\left[\frac{G^{1/2}M^{3/2}\sigma}{c^2\rho H(2\eta-1)\sin\theta}\right]^{2/7} 
\nonumber \\
&\approx& 
0.5[(2\eta-1)\sin\theta]^{-2/7}
\left(\frac{M_{12}}{10^8\msun}\right)^{3/7} 
\left(\frac{\sigma}{200\,\kms}\right)^{2/7} 
\left(\frac{\rho}{10^3\msun\mathrm{pc}^{-3}}\right)^{-2/7} \mathrm{pc} .
\end{eqnarray}
As we see, spin-orbit precession becomes important at roughly the same separation as apsidal precession (Eq.~\ref{Equation:a_omega}), and much earlier than the binary enters the GW-dominated regime (Equation~\ref{Equation:a_GR}). This means that in a range of binary separations $a_{SL}\gtrsim a\gtrsim a_{GR}$ the spin directions are already changing due to spin-orbital effects, but the angular momentum evolution is still due to 3-body interactions. Such an interplay between the effects of GR and 3-body scattering has not been studied heretofore, and will likely be the topic of our next paper. The case $a\sim a_{SL}$, when $\boldsymbol{S}$ and $\boldsymbol{L}$ change on the same timescale, looks especially interesting since that can potentially lead to the binary being captured in one of the spin-orbit resonances identified by \citet{Schnittman2004}.

\section{Stellar capture or disruption}
\label{Section:TDE}

Stars that come sufficiently close to one of the SBHs can be tidally disrupted or captured
(i.e., continue inside the event horizon).
 Let $r_0=\Theta\, r_g$ be the distance from the center of a SBH at which capture or disruption occurs.
The value of $\Theta$ depends on the structure of the star; the mass and spin of the SBH;
and the star's orbit at the moments preceding capture (circular, radial etc.) \citep[][Section 4.6]{DEGN}.
The distribution of closest approaches to one of the binary components (for closely interacting stars) turns out to be approximately constant ($dN\sim dr$, $r=0\ldots0.5a$), so we  expect that the fraction of captured stars (the stars that come close enough to the binary's orbit) is $\alpha r_0/a \sim \alpha \Theta r_g/a$, where $\alpha$ is of the order of 1. 

\begin{figure}[b]
\includegraphics[width=0.5\textwidth]{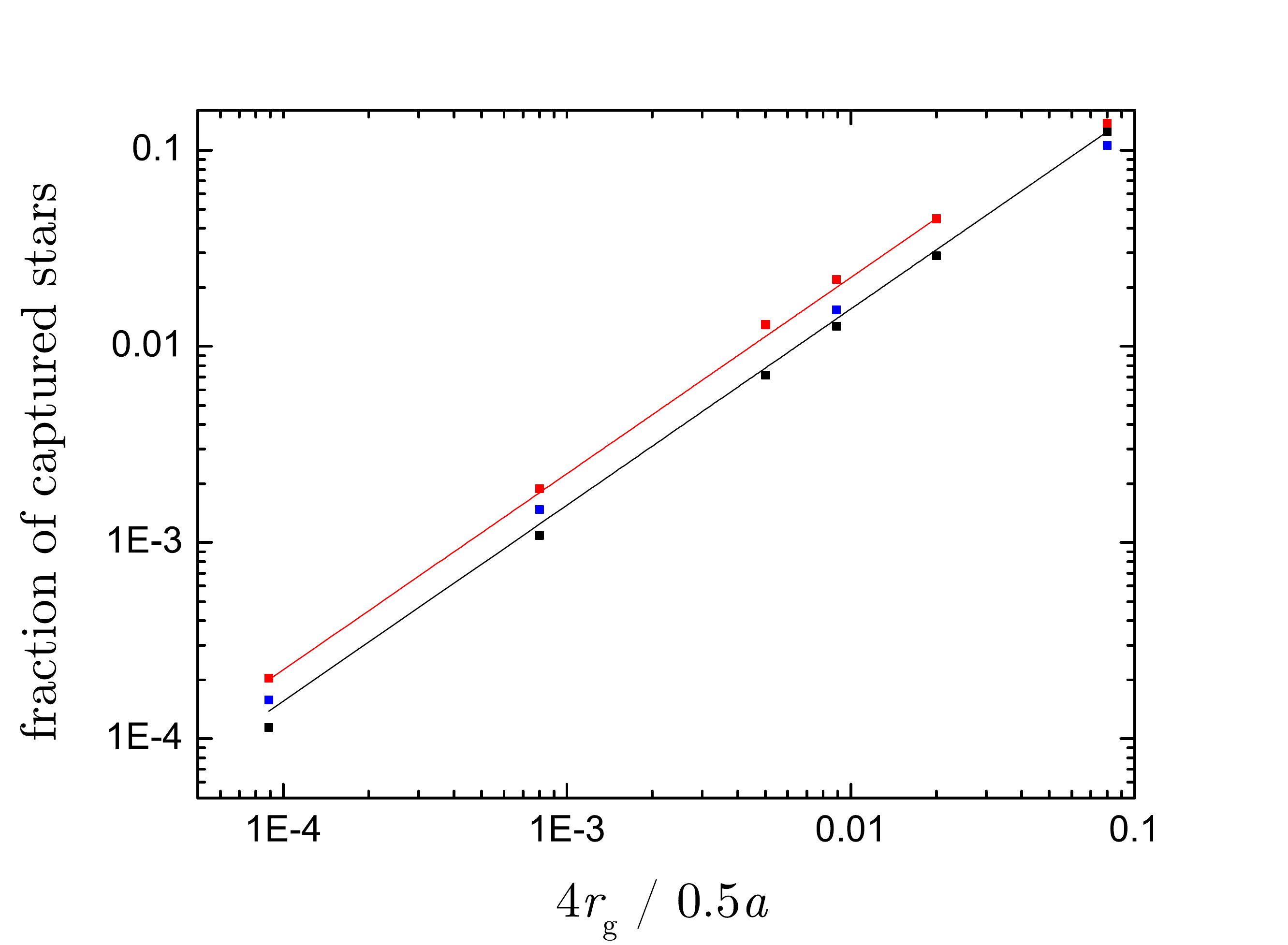}
\caption{
Fraction of stars that ended up being captured by one of the (equal mass) black holes instead of escaping to infinity after interacting with the binary. Capture radius was assumed to be $r_0=4r_g=4\cdot\frac{GM}{c^2}$. Black dots: $v=0.5v_b$, $p=0.6a..1a$; red dots: $v=200\,\mathrm{km/s}$, $p=0..\frac{\sqrt{2GMa}}{v}$ ($r_p=0..a$); blue dots: the same as red dots, but without relativistic terms in equations of motion (that would correspond to tidal disruption instead of capture). Black and red lines correspond to $\alpha=3.1$ and $\alpha=4.5$, where $\alpha$ is the ratio of fraction of captured/disrupted stars to $r_0/a$.}
\label{Figure:GR_effectiveness}
\end{figure}

Figure~\ref{Figure:GR_effectiveness} shows the fraction of captured stars in a set of scattering
experiments, assuming $\Theta = 4$. We used the same ARCHAIN code, but with post-Newtonian terms up to 2.5PN order included.

In the case of a binary SBH, even stars with large impact parameters can approach arbitrarily closely
to one of the SBHs, if their orbits carry them within a distance $\sim a$ of the binary center of mass. 
This raises the question: how much is the rate of capture
by a binary SBH enhanced compared with that of a single \sbh of the same total mass?

Consider the inflow of unbound stars with a single velocity at infinity $v$. 
In the case of a single SBH, captured stars have impact parameters less than $p_{\mathrm{capt}}=\sqrt{2GMr_0}/v$ (we assume that $r_0\ll a$). Their total number per unit time 
is
\begin{equation}\label{Equation:N1}
N_1 = nv\times \pi p_{\mathrm{capt}}^2 .
\end{equation}
In the case of a binary SBH, stars with impact parameters less than $p_{\mathrm{close}}=\sqrt{2GMa}/v$ experience close encounters with the binary, and a fraction $\alpha r_0/a$ of these are captured. The total number of captured stars per unit time 
in this case is
\begin{equation}
N_2 = nv\times \alpha \frac{r_0}{a}\times \pi p_{\mathrm{close}}^2 .
\end{equation}
The enhancement of stellar capture/TD events rate for a binary compared to a single black hole of the same mass
\begin{equation}
\frac{N_2}{N_1} 
= \alpha \frac{r_0}{a}\left(\frac{p_{\mathrm{close}}}{p_{\mathrm{capt}}}\right)^2 = \alpha
\end{equation}

Figure~\ref{Figure:GR_effectiveness} shows that $\alpha\approx3\ldots 5$, so we should expect only a few times increase of capture events rate. This result can be interpreted as follows: binary's effective capture radius ($\sim a$) is much larger than that for a single {\sbh} ($\sim r_0$); but at the same time, only a small fraction of ``effectively captured'' (closely interacting with the binary) stars get close enough to one of the binary components to get captured (almost all of them get ejected eventually rather than being captured). The fact that $\alpha\sim1$ means that these two effects almost compensate each other (within an order of magnitude) so that the total capture rate  is the same within an order of magnitude.

However, all the above results were obtained in the assumption of infinite homogenous stellar medium, which would correspond to a full loss cone approximation. In the empty
LC regime the number of stars entering the loss cone is insensitive to its size -- so that the small fraction of captured stars among those within effective LC is not compensated by the larger total number of LC stars, and the total capture rate for binaries should actually be much lower than that for single \sbh s. This a priori conclusion is confirmed by the results of \citet[Fig. 10]{Chen2008}: for realistic spherical galaxy models in a steady state (where the loss cone is empty for both single and binary \sbh) the capture rates are always a few orders of magnitude lower for binaries. However, as was shown in \citet{Chen2011}, the disruption of initially existing bound cusp by a binary \sbh results in a burst of capture/TD events with their peak rate of $\sim10^{-1}\,\mathrm{yr}^{-1}$, a few orders of magnitude higher compared to the rates for single \sbh s fed by two-body relaxation (typically $10^{-4}$ to $10^{-5}\,\mathrm{yr}^{-1}$). For a non-spherical galaxy with a non-fixed stellar distribution, capture rate is somewhere between empty- and full-LC values for both single and binary \sbh~\citep{Vasiliev2014a,Vasiliev2015} -- so, considering what was said above about these two regimes, we shouldn't expect a significant increase in capture rate compared to a single \sbh~for any galaxy.

\begin{figure}[h]
\plottwo{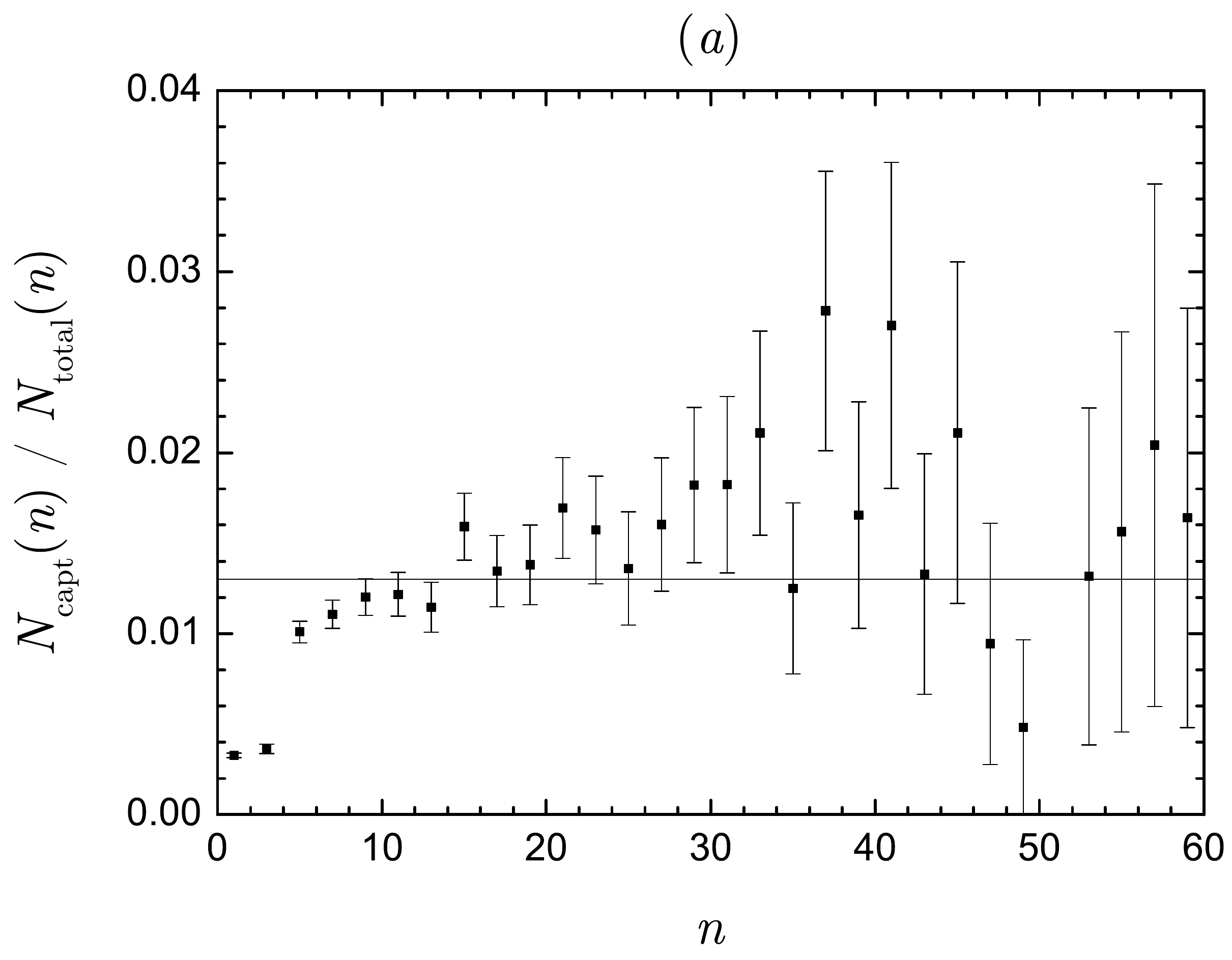}{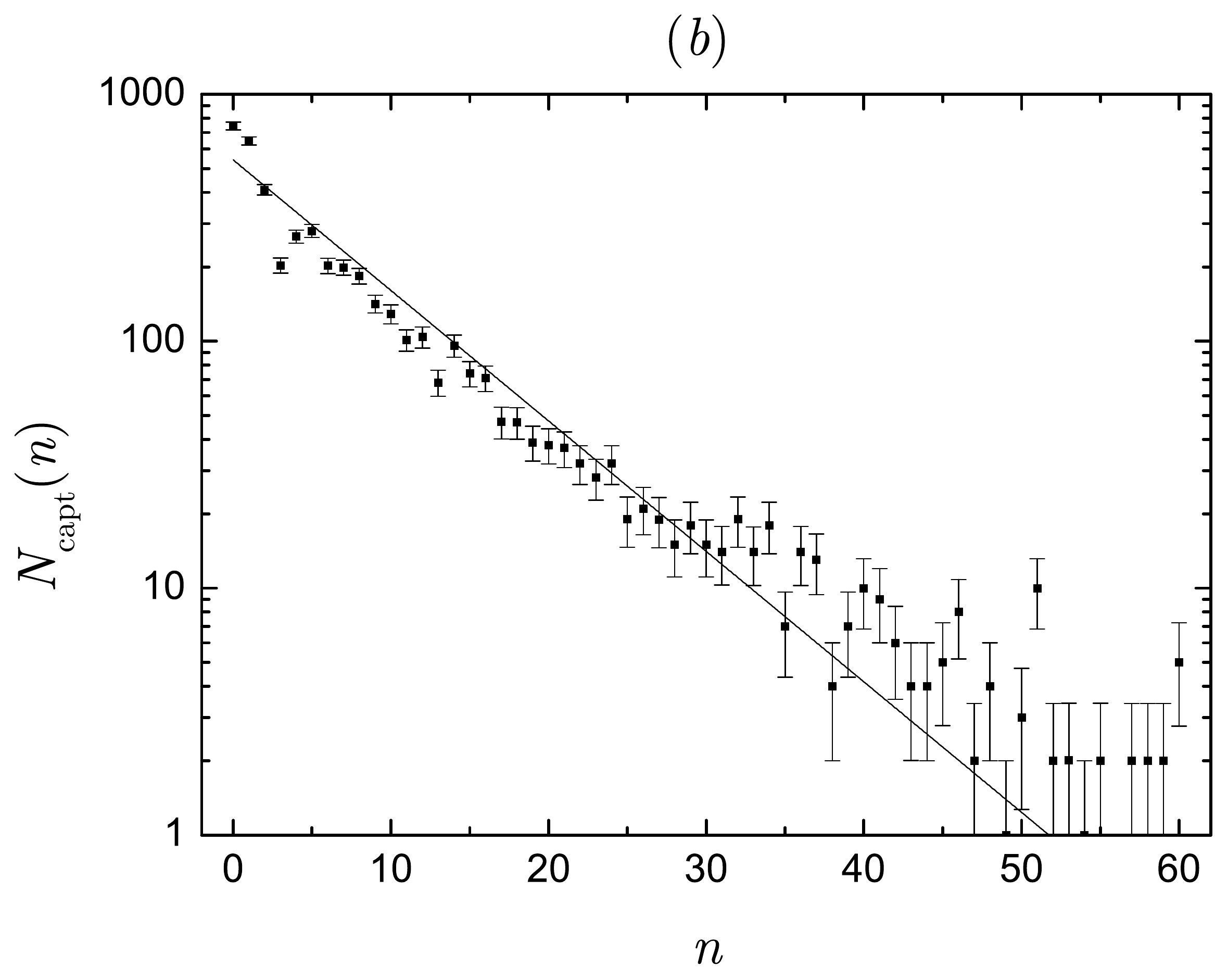}
\caption{(a) Fraction of captured stars among the ones that have experienced $n$ close interactions with the binary (or revolutions around the binary, measured as number of times when $dr_\mathrm{star}/dt=0$) for $r_0/(0.5a)=0.005$, $v=200\,\mathrm{km/s}$; only the values for odd $n$ are shown, because even $n$ already means that the star was captured. Horizontal line marks the overall fraction of captured stars.
(b) Total number of stars captured after $n$-th interaction; parameter values are the same as in Figure~\ref{Figure:Captured_vs_n}.}
\label{Figure:Captured_vs_n}
\end{figure}

Figure~\ref{Figure:Captured_vs_n}a shows the dependence between the fraction of captured stars and the number of close interactions with the binary. We see that the probability of being captured during a close interaction doesn't show any strong dependence on the number of interactions already experienced by the star --- just as one would expect assuming that the interaction between the star and the binary takes place as series of close interactions that are more or less independent from each other. Figure~\ref{Figure:Captured_vs_n}b shows the total number of stars captured after $n$-th interaction; this dependence is well fit by exponential decrease, which is, again, in agreement with aforementioned assumption about the independence of interactions.

\section{Solutions of the Fokker-Planck Equation}
\label{Section:FPESolution}

In this section, we use the analytic approximations to the diffusion coefficients derived in \S\ref{Section:DiffusionCoefficients} 
to solve the Fokker-Planck equation describing the evolution of the binary's orbital elements. 
In \S\ref{Section:FPESteadyState} - \S\ref{Section:EvolutionOfTheOrientation} we consider a one-dimensional model, 
ignoring the evolution of any orbital elements other than $\theta$ or $a$ (effectively assuming $e=0$). 
Then, in \S\ref{Section:JointEvolution}, we consider a more realistic model that accounts for changes in $\theta$, $e$ and $a$, including effects due to GR. 
It will turn out that the time dependence of $\theta$ in the latter model can be substantially different than in the simplified model.

\subsection{Steady-state orientation distribution}
\label{Section:FPESteadyState} 

We begin by considering the Fokker-Planck equation in the form of Equation~(\ref{Equation:FPtheta}),
\barr\label{Equation:f(tau)}
\frac{\partial f}{\partial \tau} &=& \frac{1}{\sin\theta}\frac{\partial}{\partial\theta}\left[\sin\theta\left(\alpha\frac{\partial f}{\partial\theta} + f\sin\theta\right)\right],
\earr
which describes changes only in the binary's orientation; 
changes in semimajor axis are incorporated into the dependence of $\tau$ on time.
Note that both first- and second-order diffusion coefficients are included.
The steady-state solution satisfies
\beq
\frac{\partial}{\partial\theta}\left[\sin\theta\left(\alpha\frac{\partial f}{\partial\theta} + f\sin\theta\right)\right] = 0
\label{Equation:SteadyStateA}
\eeq
or
\beq
\sin\theta\left(\alpha\frac{\partial f}{\partial\theta} + f\sin\theta\right) = \mathrm{constant} .
\label{Equation:SteadyStateB}
\eeq
The left hand side of Equation~(\ref{Equation:SteadyStateB}) is zero for $\theta=0$ and $\theta=\pi$,
thus the constant on the right-hand side should be zero as well:
\beq
\alpha\frac{\partial f}{\partial\theta} + f\sin\theta = 0 .
\eeq
The  solution is
\beq
f_0(\theta) = \mathrm{constant}\times\exp{\left(\frac{\cos\theta}{\alpha}\right)} .
\eeq
This distribution peaks at $\theta=0$ and declines exponentially for increasing $\theta$. 
Now it was shown in the previous section (Equation~\ref{Equation:Dtheta12}) that
\beq
\alpha \approx \frac{m_f}{M_{12}}\frac{3\sqrt{q}}{\sqrt{1-e^2}(2\eta-1)} . \nonumber
\eeq
Thus $\alpha\ll1$ for almost all reasonable parameter values, and the steady state distribution 
is substantially non-zero only for small $\theta$. 
Approximating $\cos\theta\approx1-\theta^2/2$,
\beq
f_0(\theta) \approx \mathrm{constant} \times \exp{\left(-\frac{\theta^2}{2\alpha}\right)} .
\eeq
In this approximation, the expectation value of $\theta$ in the steady state is
\barr
\theta_0 &=& \frac{\int_0^\pi \theta\; f_0(\theta) \sin\theta d\theta}{\int_0^\pi f_0(\theta) \sin\theta d\theta} 
= \frac{\int_0^\pi \theta\; \exp{\left(\frac{\cos\theta}{\alpha}\right)} \sin\theta d\theta}{\int_0^\pi \exp{\left(\frac{\cos\theta}{\alpha}\right)} \sin\theta d\theta} 
\approx \sqrt{\alpha} 
\approx \sqrt{\frac{m_f}{M_2}}(1-e^2)^{-1/4}(2\eta-1)^{-1/2}q^{1/4} .
\label{Equation:PredictedTheta0} 
\earr

\subsection{Analytical results for a Fokker–-Planck equation in the small-noise limit}
\label{Section:FPESolutionAnalytical} 

In this subsection, we consider a general one-dimensional Fokker-Planck equation:
\begin{eqnarray}
\frac{\partial f(x,t)}{\partial t} = - \frac{\partial}{\partial x}\left[K(x)f(x,t)\right] + \frac{1}{2}\frac{\partial^2}{\partial x^2}\left[D f(x,t)\right]
\end{eqnarray}
and construct approximate solutions in the limit of small diffusion term $D$.
In this limit, the time evolution of the system is mainly determined by the deterministic trajectory that corresponds to $D=0$.
Without loss of generality, $D$ is assumed constant; if it is not, it can always be made constant using the technique described in \citet[chapter 5.1]{Risken1989}. 
We begin with the zero-noise equation
\begin{eqnarray}
\frac{\partial f(x,t)}{\partial t} = - \frac{\partial}{\partial x}\left[K(x)f(x,t)\right] .
\end{eqnarray}
The corresponding deterministic equation for the position $x(t)$ of the system is easily shown to be
\begin{eqnarray}\label{Equation:dxdt}
\dot{x}(t) = K(x) .
\end{eqnarray}
Let $\overline{x}(t)$ be the solution of this equation. 
We expand the actual (stochastic) trajectory $x(t)$, in the presence of weak fluctuations,
around the deterministic path $\overline{x}(t)$. In first order of the small expansion parameter $\sqrt{D}$, we write
\begin{eqnarray}
x(t) = \overline{x}(t) + \sqrt{D}\,y(t) .
\end{eqnarray}
Then
\bsub\label{Equation:x_sigmax}
\begin{eqnarray}
\langle x\rangle &=& \overline{x}(t) + \sqrt{D}\,\langle y\rangle \label{Equation:x} , \\
\sigma_x^2 &=& D\sigma_y^2 \label{Equation:sigma_x} .
\end{eqnarray}
\esub
It is shown in \citet{Lutz2005} that
\bsub\label{Equation:y_sigmay}
\begin{eqnarray}
\frac{d\langle y\rangle}{dt} &=& K'\left(\overline{x}(t)\right) \label{Equation:y} , \\
\frac{d\sigma_y^2}{dt} &=& 2K'\left(\overline{x}(t)\right)\sigma_y^2 + 2 \label{Equation:sigma_y} .
\end{eqnarray}
\esub
To solve these differential equations, we need to set initial conditions for $\langle y\rangle$ and $\sigma_y^2$. 
They can be expressed through the initial conditions for $\langle x\rangle$ and $\sigma_x^2$ using Equations~(\ref{Equation:x_sigmax}). 
But first we should specify the initial condition for the deterministic trajectory $\overline{x}(t)$.
A natural choice is $\langle x\rangle (0) = \overline{x}(0)$, which means
\bsub\label{Equation:Initial_y_sigmay}
\begin{eqnarray}
\langle y\rangle (0) &=& 0 \label{Equation:Initial_y} , \\
\sigma_y^2 (0) &=& \frac{\sigma_x^2 (0)}{D} \label{Equation:Initial_sigma_y} .
\end{eqnarray}
\esub
Equations~(\ref{Equation:y_sigmay}) have solutions of general form
\bsub
\begin{eqnarray}
\langle y\rangle (t) &=& \mathrm{constant}\times\frac{\overline{K}(t)}{\overline{K}(0)} , \\
\sigma_y^2 (t) &=& \mathrm{constant}\times\left[\frac{\overline{K}(t)}{\overline{K}(0)}\right]^2 + 2\overline{K}(t) \int^t_0\frac{dt_1}{\overline{K}^2(t_1)} , \\
\overline{K}(t) &\equiv& K(\overline{x}(t)) .
\end{eqnarray}
\esub
Together with initial conditions (\ref{Equation:Initial_y_sigmay}), these yield
\bsub
\begin{eqnarray}
\langle y\rangle (t) &=& 0 \label{Equation:Solution_y}, \\
\sigma_y^2 (t) &=& \frac{\sigma_x^2(0)}{D}\left[\frac{\overline{K}(t)}{\overline{K}(0)}\right]^2 + 2\overline{K}(t) \int^t_0\frac{dt_1}{\overline{K}^2(t_1)} \label{Equation:Solution_sigma_y} .
\end{eqnarray}
\esub
Finally, in terms of the original variable $x$,
\bsub\label{Equation:Solution_x_sigmax}
\begin{eqnarray}
\langle x\rangle (t) &=& \overline{x}(t) \label{Equation:Solution_x} , \\
\sigma_x^2 (t) &=& \sigma_x^2(0)\left[\frac{\overline{K}(t)}{\overline{K}(0)}\right]^2 + 2D\overline{K}(t) \int^t_0\frac{dt_1}{\overline{K}^2(t_1)}\label{Equation:Solution_sigma_x} .
\end{eqnarray}
\esub

\subsection{Evolution of the orientation}
\label{Section:EvolutionOfTheOrientation}

Next we consider time-dependent solutions of the $\theta$ evolution equation (\ref{Equation:f(tau)}).
As we will see in \S\ref{Section:JointEvolution}, the predictions of such a simplified model are valid only for a 
binary that is nearly circular, and in the regime where GR effects are negligible. 
Nevertheless, the model is worth considering because it allows us to derive analytic approximations for
the mean and variance of $\theta$ and their dependence on time.

We begin by rewriting Equation~(\ref{Equation:f(tau)}) as
\bsub
\barr
\frac{\partial(f\sin\theta)}{\partial \tau} &=& - \frac{\partial}{\partial \theta}\left(K(\theta)f\sin\theta\right) + \frac{1}{2}\frac{\partial^2}{\partial \theta^2}(Df\sin\theta),\\
K(\theta) &=& -\sin{\theta} + \alpha\cot{\theta},\\ 
D &=& 2\alpha .
\earr
\esub
Since $\alpha\ll1$, we can apply the results of \S\ref{Section:FPESolutionAnalytical}:
\bsub\label{Equation:FPNumerical}
\begin{eqnarray}
\overline{\theta}(\tau) &=& \arccos[\beta \tanh[\beta (\tau + \tau_0)] - \alpha],\label{Equation:theta(tau)}\\
\beta &\equiv& \sqrt{(1+\alpha^2)},\\
\tau_0 &\equiv& \frac{1}{\beta} \arctanh\frac{\cos{\theta_0}+\alpha}{\beta} , \\
\sigma_\theta^2(\tau) &=& \sigma_\theta^2(0)\left[\frac{K(\overline{\theta}(\tau))}{K(\overline{\theta}(0))}\right]^2 + 4\alpha K(\overline{\theta}(\tau)) \int^\tau_0\frac{d\tau_1}{K^2(\overline{\theta}(\tau_1))} .
\end{eqnarray}
\esub
Substitution of $\theta=0$ or $\theta=\pi$ into Equation~(\ref{Equation:FPNumerical}) yields the boundary conditions
\beq
\frac{\partial f}{\partial\theta}\bigg|_{\theta=0} = \frac{\partial f}{\partial\theta}\bigg|_{\theta=\pi} = 0 .
\eeq

We assume a Gaussian distribution for the initial conditions: 
\beq
f (\theta, 0) = \exp\left[-\frac{(\theta-\theta_0)^2}{2\sigma_{\theta,0}^2}\right]
\eeq
and we set the mean $\theta_0 = 5\pi/6$ and the variance $\sigma_{\theta,0} = 0.03$.
Equation~(\ref{Equation:f(tau)}) was then solved numerically, setting $\alpha=0.01$, and 
the results were compared with the predictions of the approximate theory (Eq.~\ref{Equation:FPNumerical}); 
such a value of $\alpha$ is unrealistically high, but we chose it so that the second-order effects would be appreciable. 
Distribution functions $f(\theta)$ at different times are shown in Figure~\ref{Figure:f(theta,t)}. 
Comparison with the analytic approximations (for the first two moments of the distribution) is shown in Figure~\ref{Figure:FPE_theta}. 
We see that even for such a large value of the small parameter $\sqrt{D}\approx0.14$ the approximation is very good. 

\begin{figure}
\includegraphics[width=0.5\textwidth]{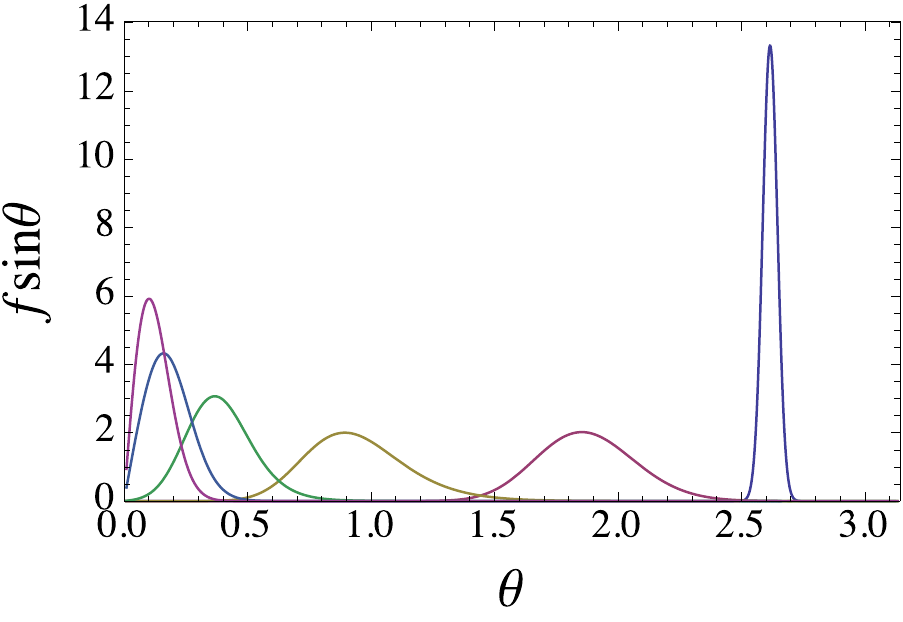}
\caption{Distribution function $f(\theta,\tau)\sin\theta$ for $\tau={0,1,2,3,4,6}$ found from numerical solution of Eq.~(\ref{Equation:f(tau)}) with $\alpha=0.01$; smaller mean values of $\theta$ correspond to later times. The steady-state distribution ($f(\theta,\tau)$ at $\tau\rightarrow\infty$) is almost indistinguishable from $f(\theta,6)$.}
\label{Figure:f(theta,t)}
\end{figure}

\begin{figure}
\plottwo{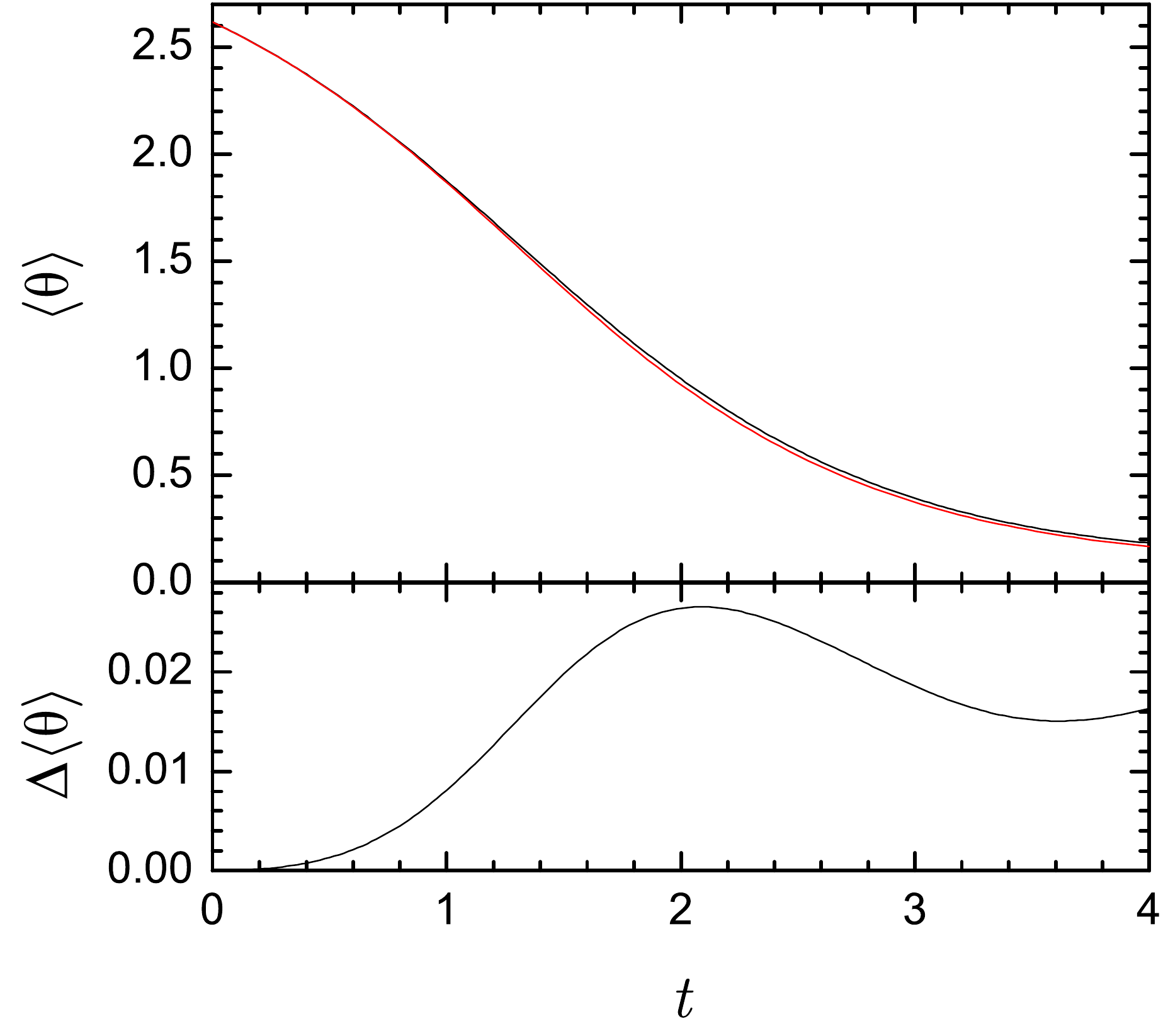}{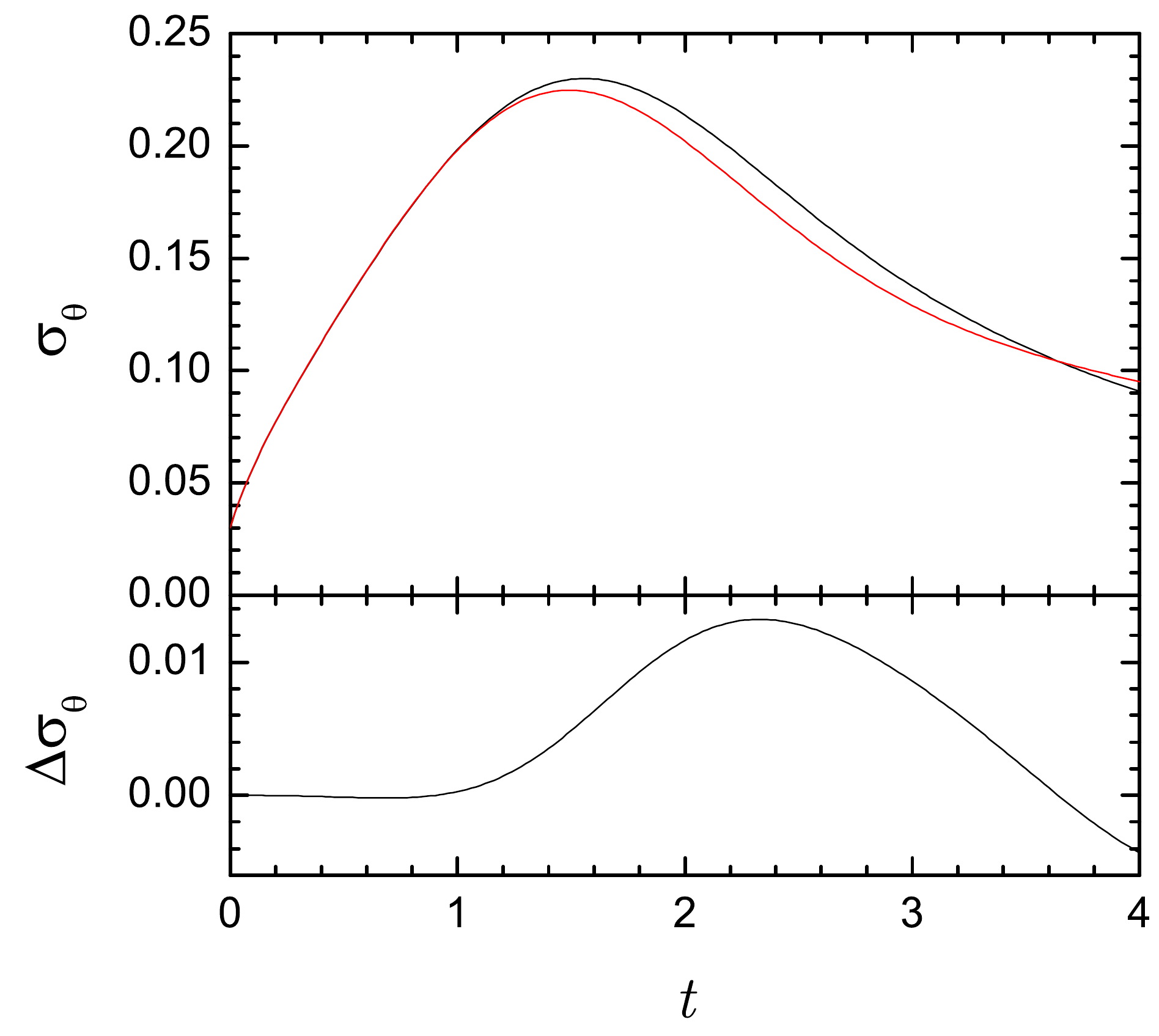}
\caption{Upper panels show the time dependence of the mean (left) or variance (right) of $\theta$, computed numerically from Eq.~(\ref{Equation:f(tau)}) (black) and analytically from Eq.~(\ref{Equation:FPNumerical}) (red). 
Lower panels show the difference between numerical and analytic solutions.
\label{Figure:FPE_theta}}
\end{figure}

Our results are in good agreement with the $N$-body simulations of \citet{Gualandris2012} and \citet{Cui2014}, who also found that reorientation of a binary's angular momentum vector always proceeds in the direction of alignment with the stellar angular momentum 
no matter what the initial conditions. 
The results of \citet{Wang2014} are seemingly in contradiction with ours: in some of their $N$-body simulations the 
binary, which is initially corotating ($\theta=0$), ends up counterrotating. 
However, most of the dramatic changes in angular momentum recorded by them take place in the early, ``unbound'' phase of dynamical evolution, when our model does not apply. After the binary components become bound, the orientation changes are consistent with our results if we take into account their low assumed degree of nuclear rotation (as shown in Figure 8 of \citet{Wang2014}, the numbers of stars with $L_z>0$ and $L_z<0$ are almost equal).

We now convert the expressions (\ref{Equation:FPNumerical}) into functions of the actual time $t$. 
As was shown in \S\ref{Section:DiffusionCoefficients}, both our drift and diffusion coefficients depend on time in the same way: in the case of a sufficiently hard binary, 
\bsub
\barr
\langle\Delta\theta\rangle &\propto& a(t),\; \langle(\Delta\theta)^2\rangle \propto a(t),\\
a(t) &=& \left(\frac{G\rho Ht}{\sigma}+\frac{1}{a_0}\right)^{-1} = \frac{a_0}{1+t/t_h},\\
t_h &=& \frac{\sigma}{G\rho a_0 H} .
\earr
\esub
Also, as we know from \S\ref{Section:BinaryEvolveOrientation}, that means 
\bsub
\barr
\tau &=& C_1\int_0^t \frac{a(t)}{a_0} dt = C_1t_h \ln\left(1+\frac{t}{t_h}\right),\\
C_1&=&\frac{\left|\langle\Delta\theta\rangle\right|}{\sin\theta} .
\earr
\esub
Then, ignoring small terms of order $\alpha$ or smaller and using Equation~(\ref{Equation:D1toDeltatheta}), 
\bsub
\barr
\theta (t) &=& \arccos\tanh\left(\tau+\arctanh\cos\theta_0\right),\\
\tau &=& \frac{D_{\theta,1}}{H} \ln\left(1+\frac{t}{t_h}\right) = \frac{D_{\theta,1}}{H} \ln\frac{a_0}{a} .
\earr
\esub
The dimensionless coefficient $D_{\theta,1} / H$ is the typical binary reorientation in one hardening time (Eq.~\ref{Equation:dthetaRotating}). It can vary depending on the parameters of the system; for a hard, equal-mass, circular binary in a maximally corotating nucleus $D_{\theta,1} \approx 5$ (about the maximum $D_{\theta,1}$ possible for a circular or mildly eccentric binary), so $D_{\theta,1} / H \approx 1/3$. For eccentric binaries it can be much higher: if we ignore the mild dependence of $H$ on eccentricity, then $D_{\theta,1} / H = 1$ for $e\approx0.85$.

\begin{figure}
\includegraphics[width=0.5\textwidth]{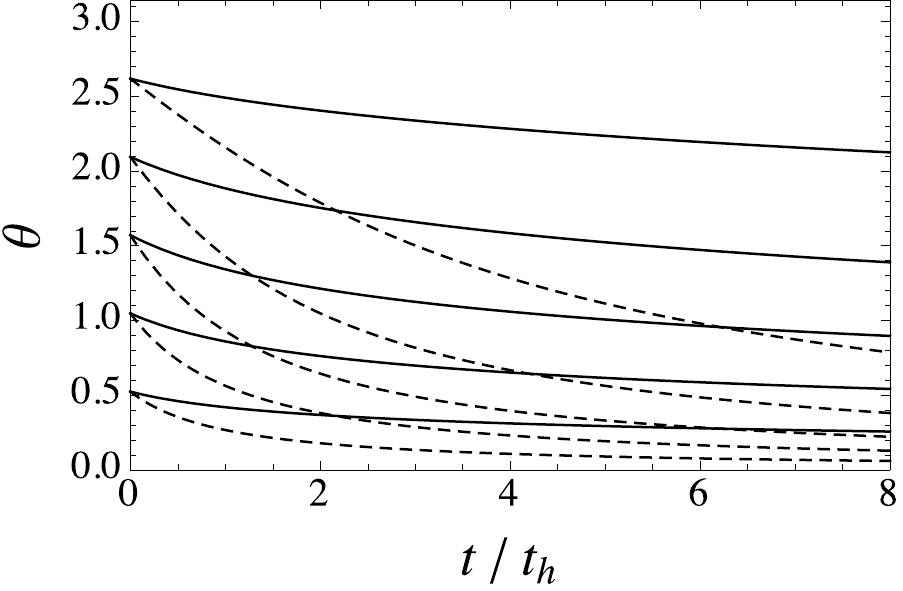}
\caption{
Evolution of binary inclination $\theta$ for $D_{\theta,1} / H = 1/3$ (solid lines) and $D_{\theta,1} / H = 1$ (dashed lines) with different initial values of $\theta$. Time is in units of the initial hardening time.}
\label{Figure:theta(t)-true}
\end{figure}

Figure~\ref{Figure:theta(t)-true} shows $\theta(t)$ for these two values of $D_{\theta,1} / H$ and different initial $\theta_0$ (the eccentricity evolution is ignored). We see that the reorientation rate declines rapidly after a few hardening times, so that the full reorientation ($\theta\ll1$) is likely not to be reached even after tens of hardening times. 
This gradual reorientation is not surprising if we recall that the energy transfer per one close encounter with a star is proportional to the binary's energy, $\delta E \sim 1/a$ \citep [chapter~8]{DEGN}, while the angular momentum transfer per  encounter is proportional
 to the binary's angular momentum, $\delta l \sim l_b$, so the inclination change per encounter $\sim \delta l / l_b$ is independent of $a$ --- it doesn't grow with hardening and, unlike energy transfer, doesn't compensate for the lowered encounter rate.

This phenomenon is another possible explanation for $\theta$ stalling at significantly nonzero value observed by \citet{Gualandris2012}, apart from the loss cone depletion proposed in their paper. Their observed reorientation would correspond to $D_{\theta,1} / H$ somewhere between $1/2$ and 1, which is consistent with the binary initially being eccentric in their simulations.

\subsection{Joint evolution of $a$, $\theta$ and $e$}
\label{Section:JointEvolution}

In previous sections we derived analytical approximations to the first-order diffusion coefficients in $a$, $\theta$ and $e$. 
We also showed that in a strongly rotating nucleus, the effects of the second-order coefficients are relatively small.  
And as demonstrated in \S\ref{Section:FPESolutionAnalytical}, if the second-order coefficients are neglected,
the evolution equations can be approximated as deterministic equations for the evolution of the average quantities,
disregarding the exact form of the distribution function (which is assumed to always remain close to a delta function).
In this approximation, we can write the joint evolution equations:
\bsub
\barr
\frac{da}{dt} &=& \langle\Delta a\rangle + \left(\frac{da}{dt}\right)_{\rm GR} = - H \frac{a^2(t)G\rho}{\sigma} - \frac{64}{5}\frac{\nu G^3M_{12}^3}{c^5a^3}f(e) , \\
\frac{de}{dt} &=& \langle\Delta e\rangle + \left(\frac{de}{dt}\right)_{\rm GR} = KH \frac{a(t)G\rho}{\sigma} - \frac{304}{15}\frac{\nu G^3M_{12}^3}{c^5a^4}g(e) , \\
\frac{d\theta}{dt} &=& \langle\Delta\theta\rangle = - D_{\theta,1} \frac{a(t)G\rho}{\sigma} , \\
f(e) &=& \frac{1+(73/24)e^2+(37/96)e^4}{(1-e^2)^{7/2}}, \\
g(e) &=& e\frac{1+(121/304)e^2}{(1-e^2)^{5/2}} .
\earr
\esub
We have included the terms that describe orbital shrinking (Eq.~\ref{Equation:dadt_GR}) 
and circularization (Eq.~\ref{Equation:dedt_GR}) due to GW emission (\S\ref{Section:GR}).
When solving these equations, we will assume the initial semimajor axis $a(0)=a_h$, which allows us to approximate 
the binary hardening rate as $H=\rm const$.
It is convenient to define a dimensionless time, expressed in initial hardening time units $t_h=\sigma/(\rho G a_h H)$, 
and a dimensionless separation, expressed in units of the hard-binary separation $a_h$:
\bsub\label{Equation:d(a,e,theta)dt-general}
\barr
\frac{d(a/a_h)}{d(t/t_h)} &=& - \left(\frac{a}{a_h}\right)^2 - \left(\frac{a_{\rm GR,0}}{a_h}\right)^5 \left(\frac{a}{a_h}\right)^{-3} f(e), \\
\frac{de}{d(t/t_h)} &=& K\frac{a}{a_h} - \frac{19}{12} \left(\frac{a_{\rm GR,0}}{a_h}\right)^5 \left(\frac{a}{a_h}\right)^{-4} g(e), \\
\frac{d\theta}{d(t/t_h)} &=& -\frac{D_{\theta,1}}{H}\frac{a}{a_h}, \\
a_{\rm GR,0} &\equiv& \frac{a_{\rm GR}}{f^{1/5}(e)} = \left(\frac{64}{5}\frac{\nu G^2M_{12}^3\sigma}{Hc^5\rho}\right)^{1/5} .
\earr
\esub
Since $a<a_h$, we can use the analytic approximations to $K$ and $D_{\theta,1}$ derived earlier 
(Eqs.~\ref{Equation:K(e,theta)} and \ref{Equation:Dtheta12-omega-averaged}a):
\bsub\label{Equation:d(a,e,theta)dt}
\barr
\frac{d(a/a_h)}{d(t/t_h)} &=& - \left(\frac{a}{a_h}\right)^2 - \left(\frac{a_{\rm GR,0}}{a_h}\right)^5 \left(\frac{a}{a_h}\right)^{-3} f(e), \\
\frac{de}{d(t/t_h)} &=& 1.5\,e\, (1-e^2)^{0.7}\, [0.15-(2\eta-1)\cos\theta]\,\frac{a}{a_h} - \frac{19}{12} \left(\frac{a_{\rm GR,0}}{a_h}\right)^5 \left(\frac{a}{a_h}\right)^{-4} g(e), \\
\frac{d\theta}{d(t/t_h)} &=& - 0.3\,(2\eta-1)\sin{\theta}\sqrt\frac{1+e}{1-e}\frac{a}{a_h} .
\earr
\esub
Equations~(\ref{Equation:d(a,e,theta)dt}) comprise a closed system of ordinary differential equations which we can solve given
initial values of $e$ and $\theta$ (assuming $a(0)=a_h$). 
Since these equations include terms describing the effects of GR, they are valid for $a_h > a \gg r_g$. 
From Equations~(\ref{Equation:r_g}) and (\ref{Equation:a_h}) we know that 
\barr\label{Equation:ahrg}
\frac{a_h}{r_g} = \nu \frac{c^2}{4\sigma^2} = 6.9 \times 10^5 \,\nu \left(\frac{M_{12}}{10^8\msun}\right)^{-2/5} .
\earr
For the second equality we have used the $M-\sigma$ relation \citep[Eq. 2.33]{DEGN}:
\beq
\frac{\sigma}{200\,\kms} \approx 0.90 \left(\frac{M_{12}}{10^8 \msun}\right)^{1/5} .
\eeq
In what follows, we are going to consider $a_h / a \leq 10^3$, 
which is well below the limit given by Equation~(\ref{Equation:ahrg}), so the condition $a\gg r_g$ is always satisfied.
We also know from \S\ref{Section:GR} that effects due to GR become important when $a \lesssim a_{\rm GR}$, where
\bsub
\barr
\frac{a_{\rm GR}}{a_h} &=& 6.3\times 10^{-3} \, f^{1/5}(e) 
\,\nu^{-4/5} 
\left(\frac{M_{12}}{10^8 \msun}\right)^{-2/5} 
\left(\frac{\sigma}{200\,\kms}\right) ^{11/5}
\left(\frac{\rho}{10^3\msun\mathrm{pc}^{-3}}\right)^{-1/5} \\
&=& 4.9\times 10^{-3} \, f^{1/5}(e) 
\,\nu^{-4/5} 
\left(\frac{M_{12}}{10^8 \msun}\right)^{1/25} 
\left(\frac{\rho}{10^3\msun\mathrm{pc}^{-3}}\right)^{-1/5} .
\label{Equation:t_final-eccentric}
\earr
\esub
To eliminate $\rho$ from this equation, we use an expression from \citet{Vasiliev2015} that gives the hardening rate 
in terms of the radius of influence $r_{\rm infl}$:
\barr
\frac{d}{dt}\left(\frac{1}{a}\right) = \frac{HG\rho}{\sigma} \approx 4\sqrt\frac{GM_{12}}{r_{\rm infl}^5} .
\earr
Combining this with the definition of the radius of influence, $r_{\rm infl}=GM_{12}/\sigma^2$, 
\barr
\rho \approx \frac{4}{H}\frac{\sigma}{G}\sqrt{\frac{GM_{12}}{r_{\rm infl}^5}} = 1.16\times 10^4 \msun\mathrm{pc}^{-3} \left(\frac{M_{12}}{10^8 \msun}\right)^{-4/5} ,
\earr
and Equation~(\ref{Equation:t_final-eccentric}) becomes
\barr\label{Equation:agrah}
\frac{a_{\rm GR}}{a_h} = 3.0\times 10^{-3} \, f^{1/5}(e) \,\nu^{-4/5} 
\left(\frac{M_{12}}{10^8 \msun}\right)^{1/5}  .
\earr

Solutions to Equations~(\ref{Equation:d(a,e,theta)dt}) are shown in Figures~\ref{Figure:theta,e(a)}, \ref{Figure:theta,e(a),eta=08} and \ref{Figure:theta,e(a),eta=06} for $\eta=1$, 0.8 and 0.6 respectively; $\theta$ and $1-e$ are plotted {\it vs}. $a_h / a$. 
Since $a(t)$ is always a decreasing function of time, $a_h / a(t)$ can be used as a dimensionless proxy for time. 
As expected, the reorientation always proceeds in the direction $\theta\rightarrow0$, but at a much faster rate for highly eccentric binaries. Because of that, and because of the rapid eccentricity increase for counterrotating binaries, binaries with initial $\theta$ close to $\pi$
(e. g. $\theta=5\pi/6$) may end up more nearly corotating than those with lower initial $\theta$; this can be see in Figure~\ref{Figure:theta,e(a)} as well as Figure~3 of \citet{Gualandris2012}. 

When the binary enters the GW regime ($a=a_{\rm GR}$ given by Eq.~\ref{Equation:GR_regime} or \ref{Equation:agrah}) 
it may seem that $\theta$ (plotted vs. $a_h/a$) has stopped changing.
The reason is that $da/dt$ increases dramatically so that $d\theta/da\rightarrow0$.

The eccentricity is either always decreasing with time if the binary is initially corotating or, if it is counterrotating, it increases at first, 
but then reaches its maximum when $\theta\approx\pi/2$ or the binary enters the GW regime (whichever happens first), 
and then decreases to zero. 
Of particular importance is the eccentricity at the moment when the binary enters the GW-dominated regime, $e_{\rm GR}$, since it determines $a_{\rm GR}/a_h$ and hence the coalescence timescale; also, as shown in \citet{PaperII}, the higher $e_{\rm GR}$ for a population of binaries, the more their stochastic GW background spectrum is attenuated compared to that for circular binaries.
The border between the GW-dominated and the stelar-encounter dominated regimes is plotted as the red curve on Figures \ref{Figure:theta,e(a)}-\ref{Figure:e(a)-nonrotating}, so that the binary's trajectory in $(a, e)$ space crosses this line at $(e_{\rm GR}, a_{\rm GR})$. The defining equation for that red curve is $a=a_{\rm GR}(e)$, or, if we take the definition of $a_{\rm GR}$ from Eq.~(\ref{Equation:agrah}),
\barr\label{Equation:e_GR}
\frac{a}{a_h} = 3.0\times 10^{-3} \, f^{1/5}(e) \,\nu^{-4/5} \left(\frac{M_{12}}{10^8 \msun}\right)^{1/5} .
\earr
The quantity $e_{\rm GR}$ generally increases as we decrease $\eta$ from 1 to $1/2$ as both reorientation and circularization become less pronounced (compare Figures \ref{Figure:theta,e(a)} and \ref{Figure:theta,e(a),eta=06}). 
This trend can be seen more clearly in Figure~\ref{Figure:e_GW} which shows $e_{\rm GR}$ for all  possible combinations of the initial parameters $(e_0,\theta_0)$. 
One other noteworthy detail is that for $\eta\gtrsim0.8$ there exists a certain ``critical'' value of $\theta_0$ at which $e_{\rm GR}$ dramatically increases above $\sim0.99$. 
This happens due to the strong effect of eccentricity increase for counterrotating binaries (large $K$ at $\eta\sim1$ and $\theta\sim\pi$) that is normally cancelled by quick reorientation except for this case of almost exactly counterrotating binaries when reorientation is slow enough ($d\theta/dt\propto\sin\theta$).  

\begin{figure}
	\centering
	\subfigure[]{\includegraphics[width=0.49\textwidth]{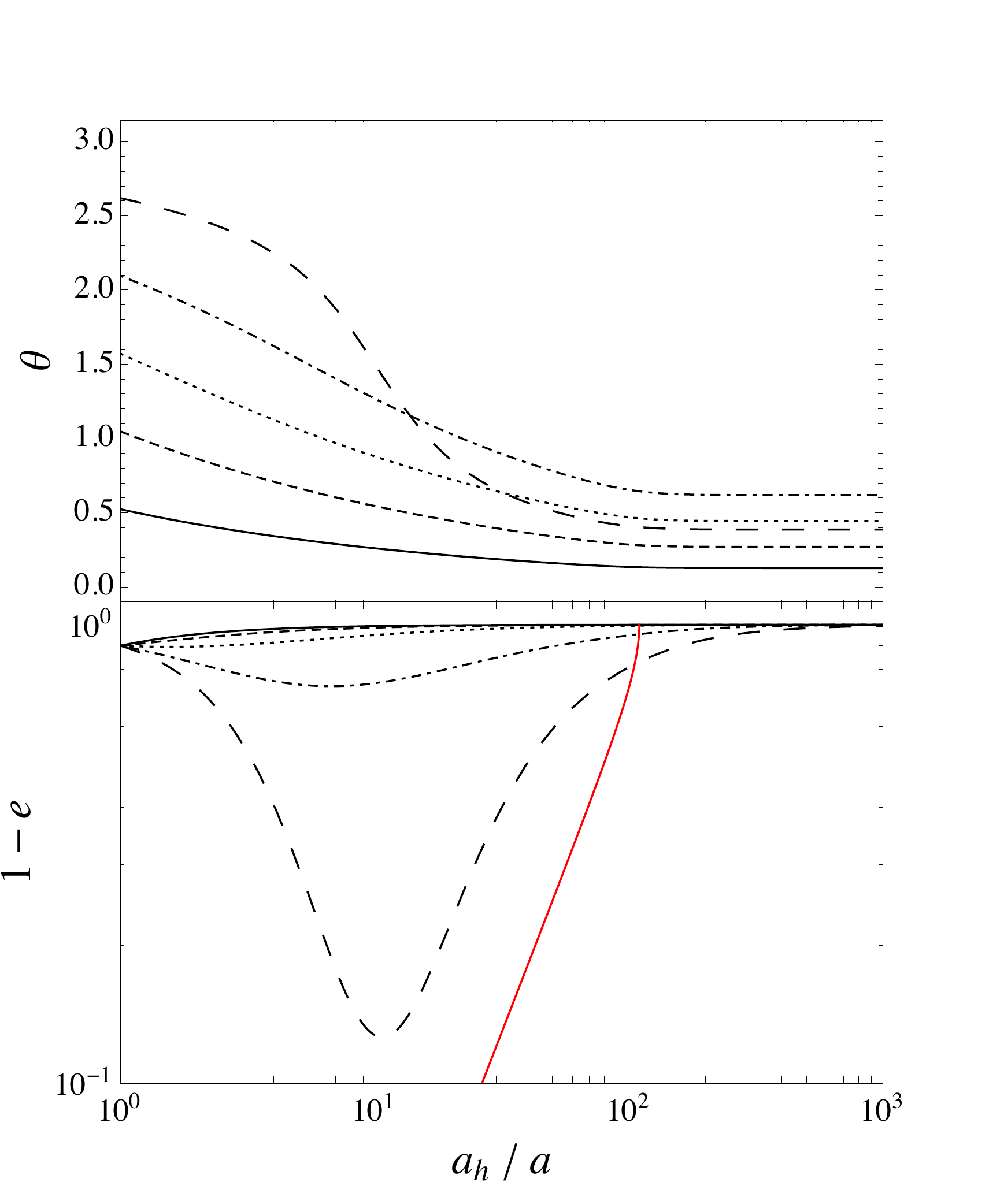}}
	\subfigure[]{\includegraphics[width=0.49\textwidth]{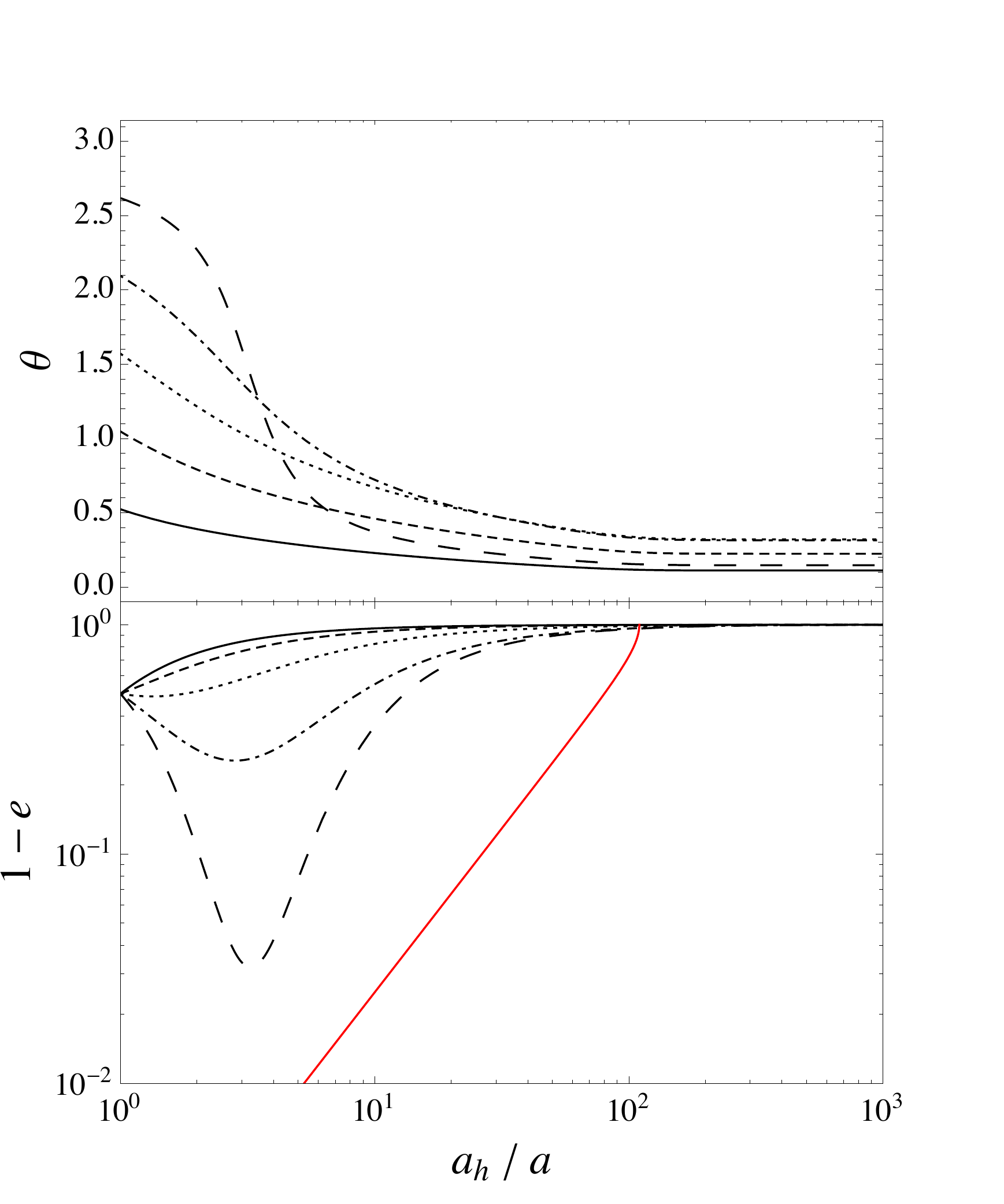}}
	\subfigure[]{\includegraphics[width=0.49\textwidth]{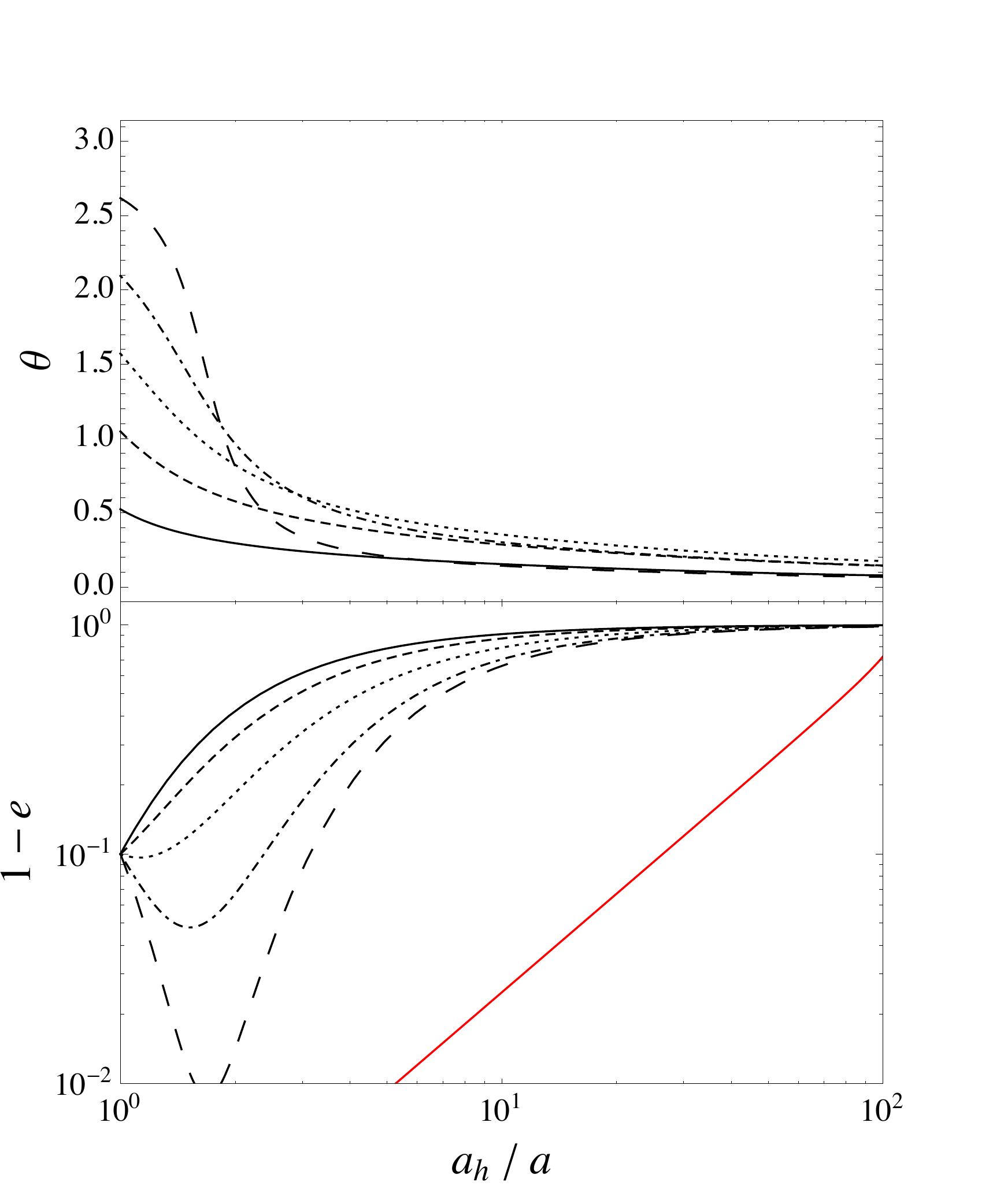}}
	\caption{
		Evolution of orbital inclination $\theta$ and eccentricity $e$ of a binary with $M_{12}=10^8\msun$ and $q=1$ in a maximally corotating nucleus ($\eta=1$), according to Equations~(\ref{Equation:d(a,e,theta)dt}) and (\ref{Equation:agrah}). 
Different line styles correspond to different initial values of $\theta$. 
The initial eccentricity is (a) 0.1, (b) 0.5, (c) 0.9. 
The red curve separates the regimes where the hardening of the binary is dominated by stellar encounters (to the left) and GW emission (to the right); its equation is $a=a_{\rm GR}$ (see Eq.~\ref{Equation:agrah}). 
Note the use of a different scale for different plots. 
		}
	\label{Figure:theta,e(a)}
\end{figure}

\begin{figure}
	\centering
	\subfigure[]{\includegraphics[width=0.49\textwidth]{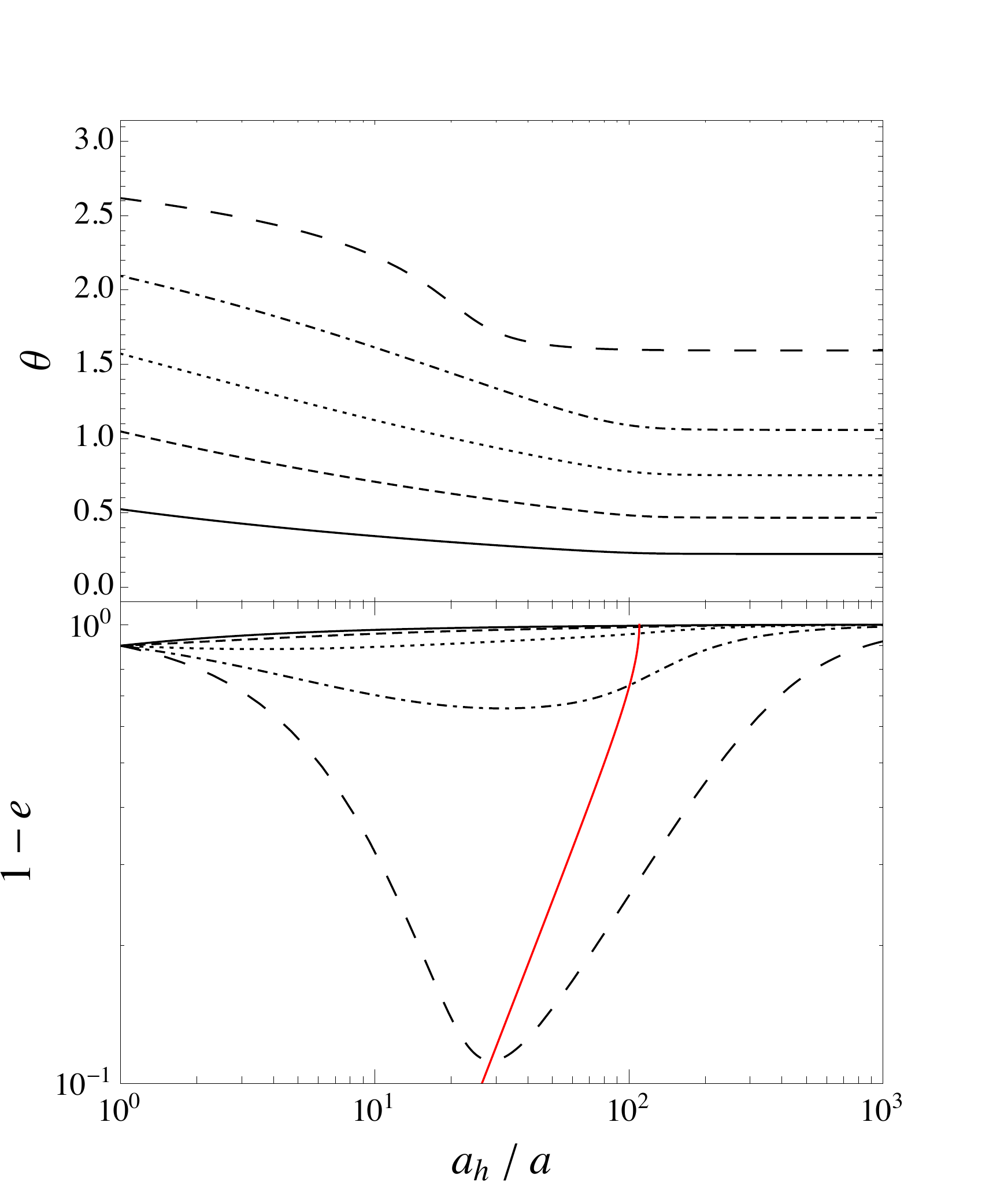}}
	\subfigure[]{\includegraphics[width=0.49\textwidth]{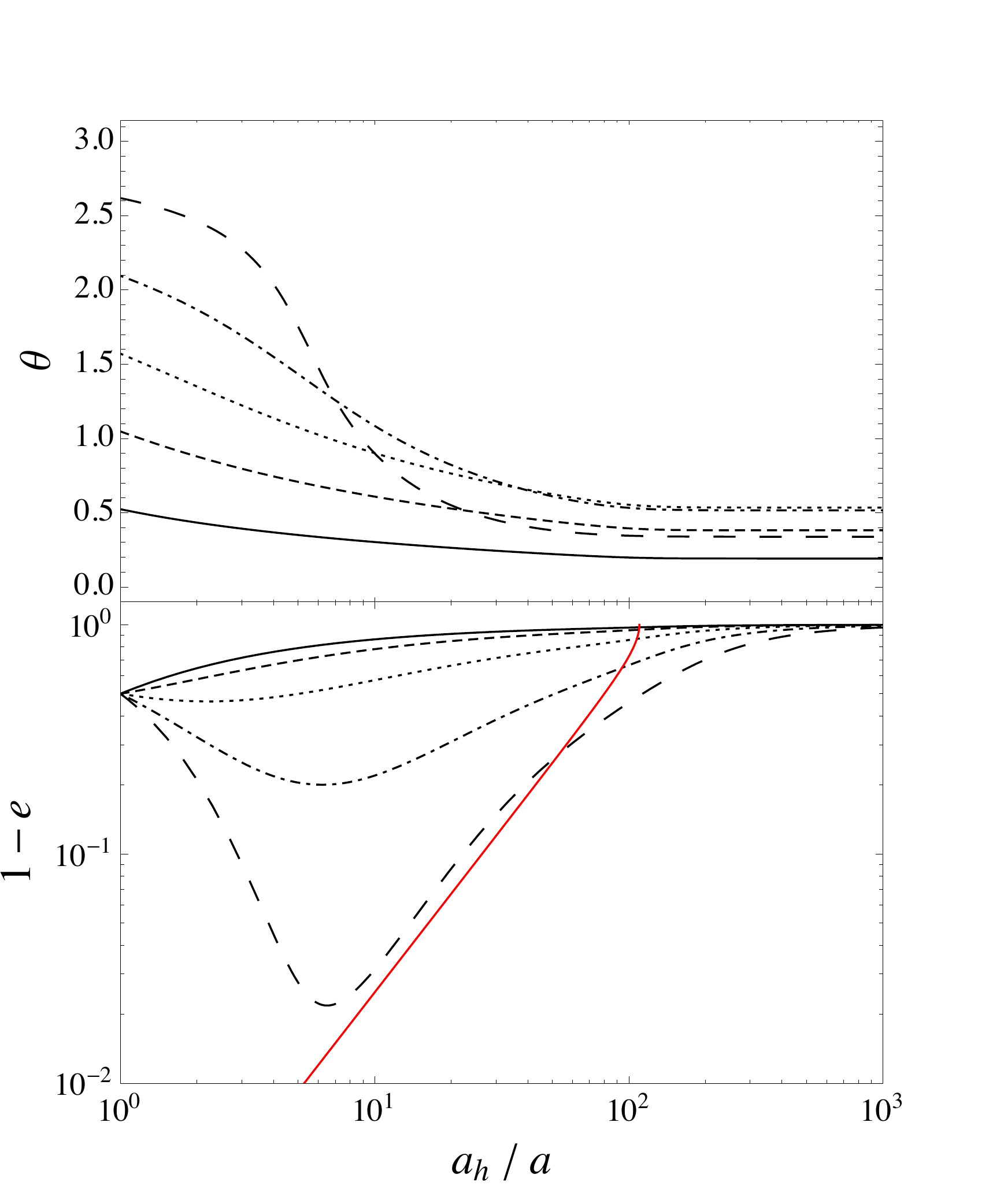}}
	\subfigure[]{\includegraphics[width=0.49\textwidth]{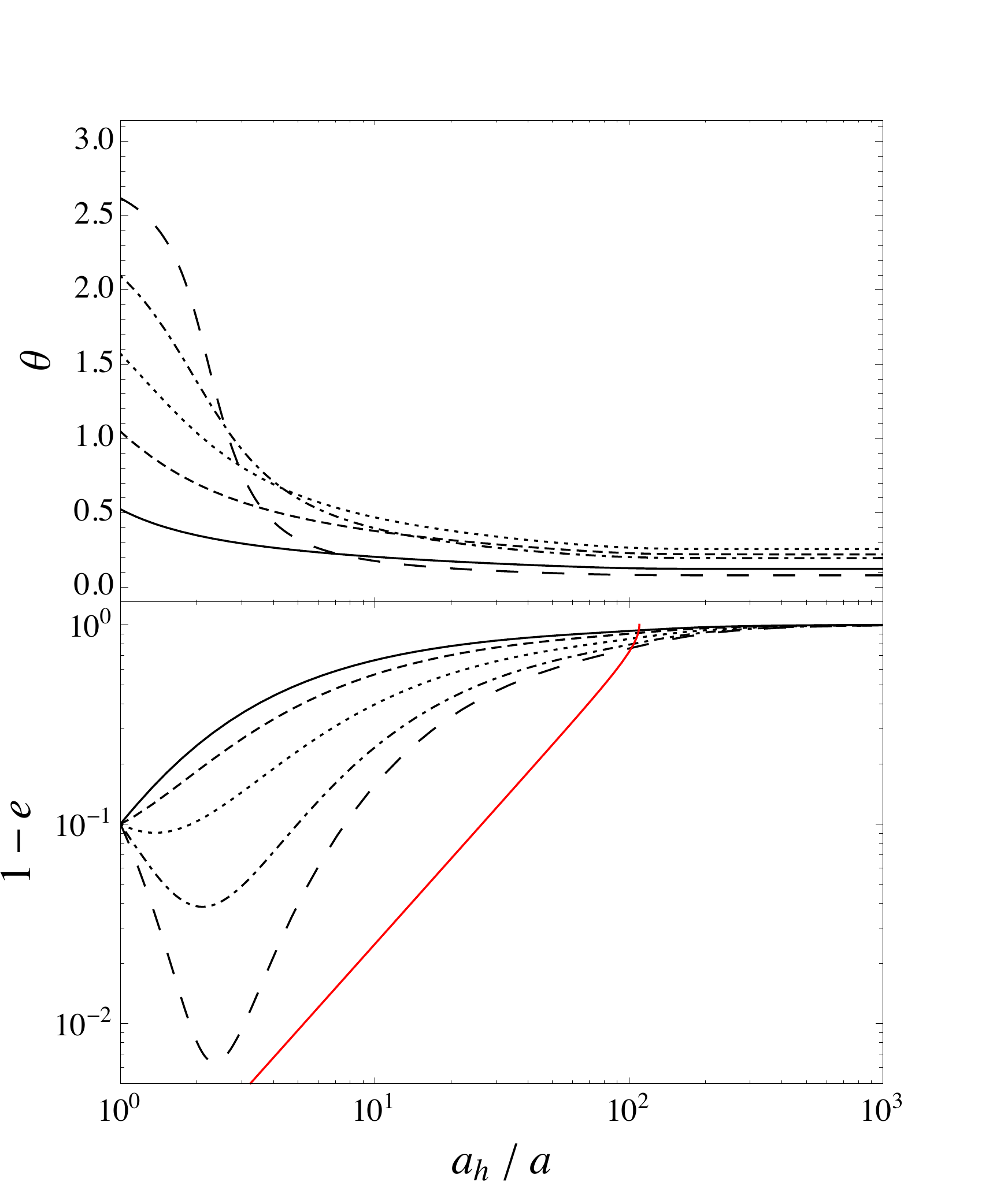}}
	\caption{
		The same as Figure~\ref{Figure:theta,e(a)} but for $\eta=0.8$.
		}
	\label{Figure:theta,e(a),eta=08}
\end{figure}

\begin{figure}
	\centering
	\subfigure[]{\includegraphics[width=0.49\textwidth]{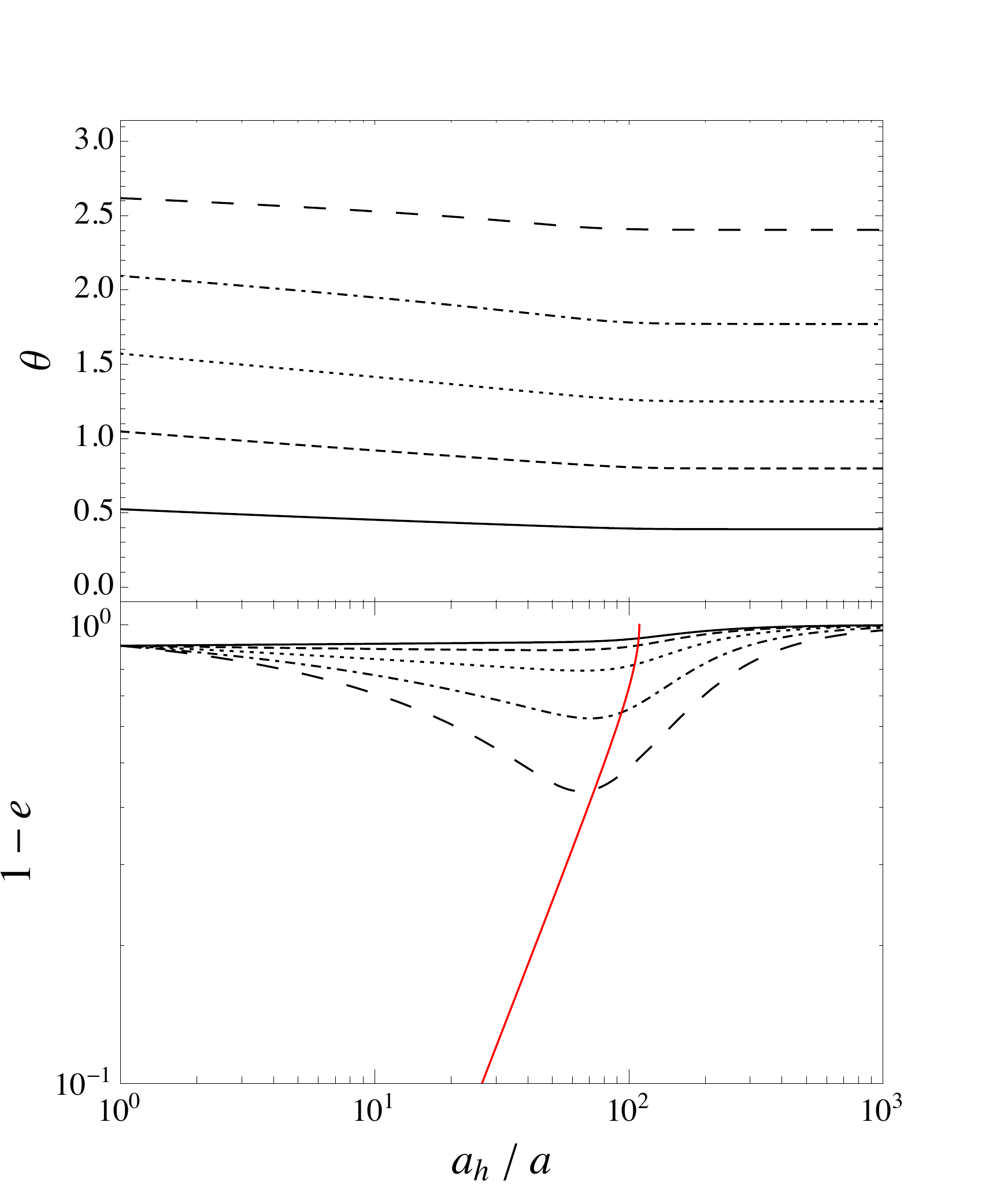}}
	\subfigure[]{\includegraphics[width=0.49\textwidth]{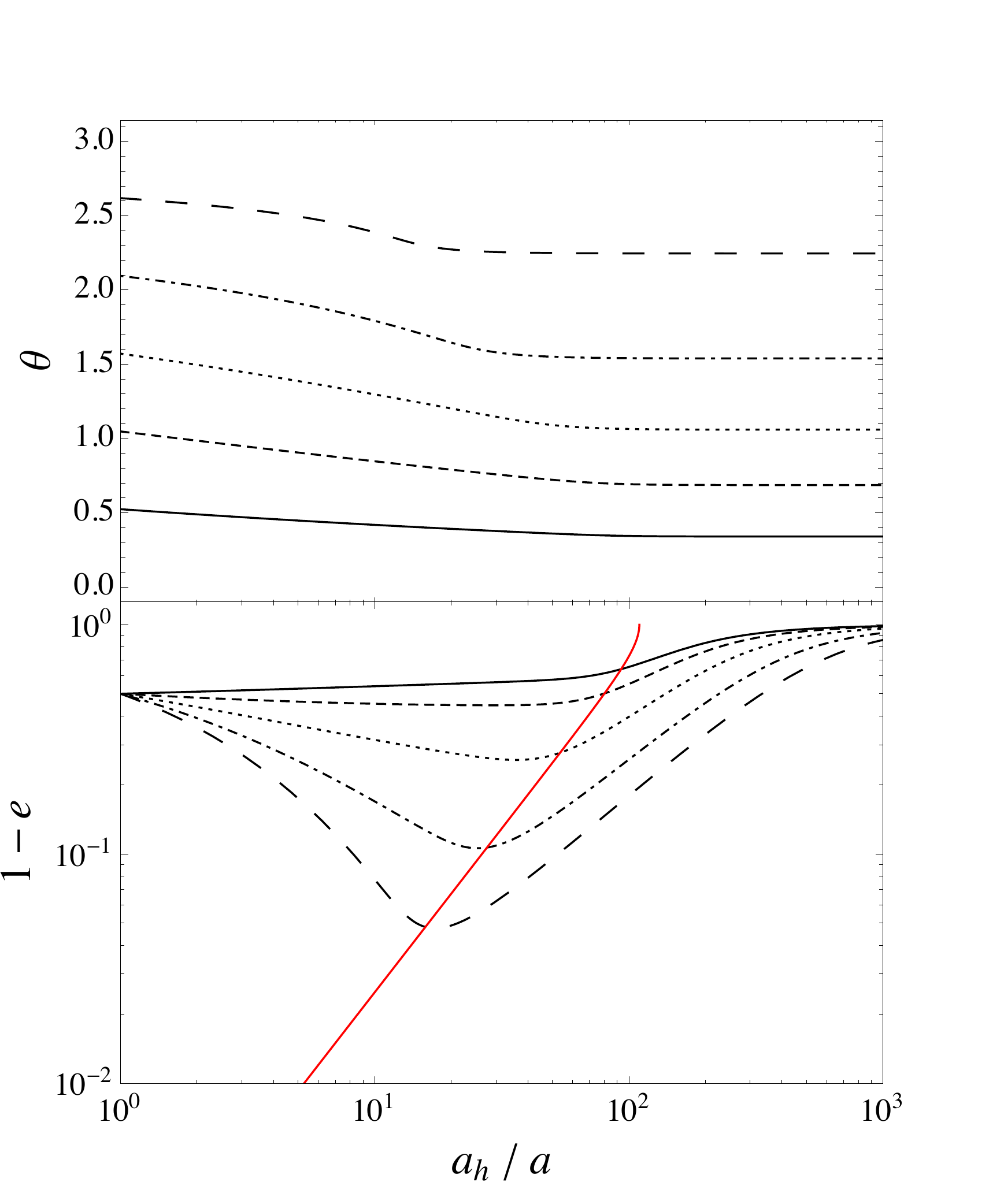}}
	\subfigure[]{\includegraphics[width=0.49\textwidth]{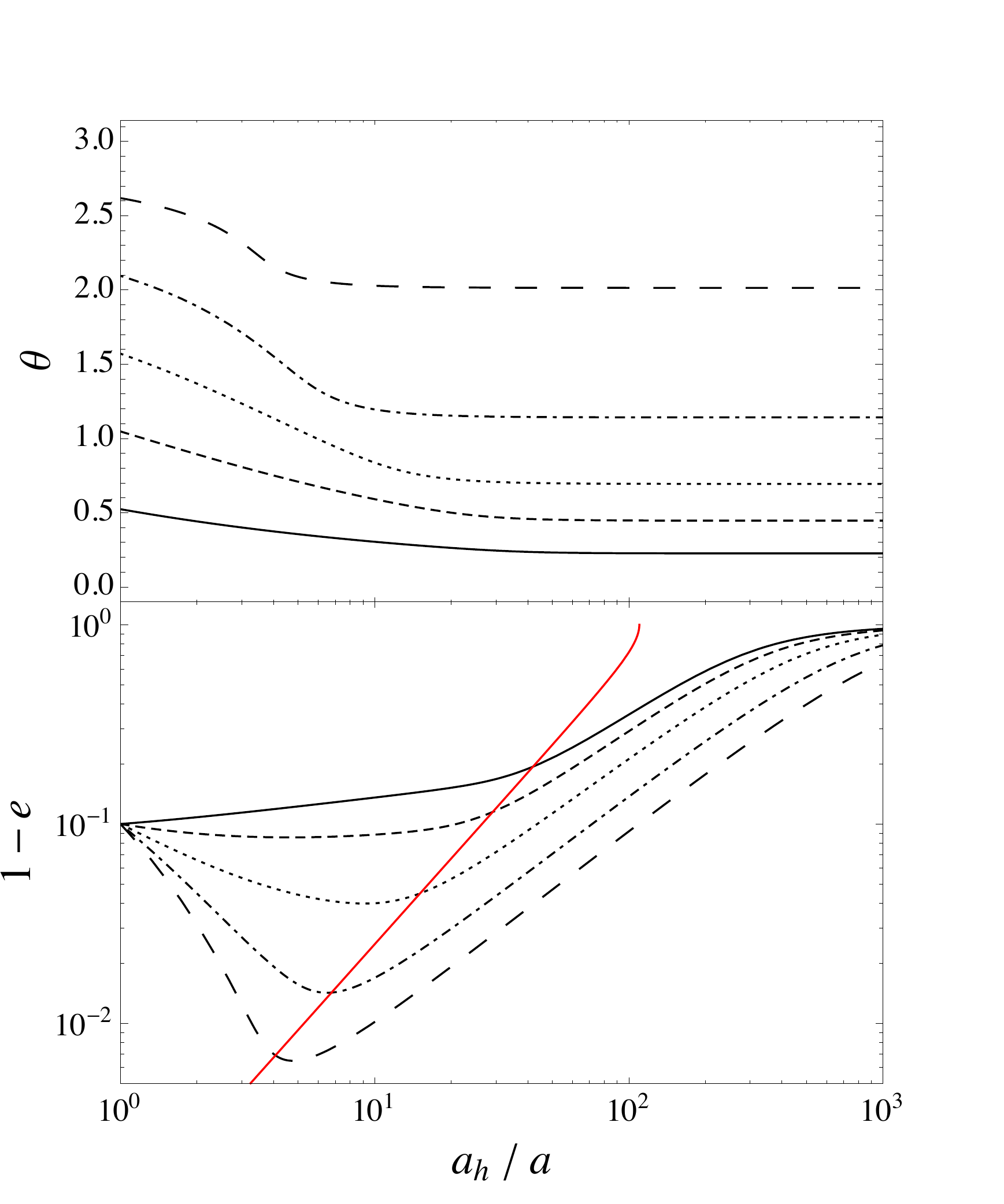}}
	\caption{
		The same as Figure~\ref{Figure:theta,e(a)} but for $\eta=0.6$.
		}
	\label{Figure:theta,e(a),eta=06}
\end{figure}

\begin{figure}[h]
\includegraphics[width=0.5\textwidth]{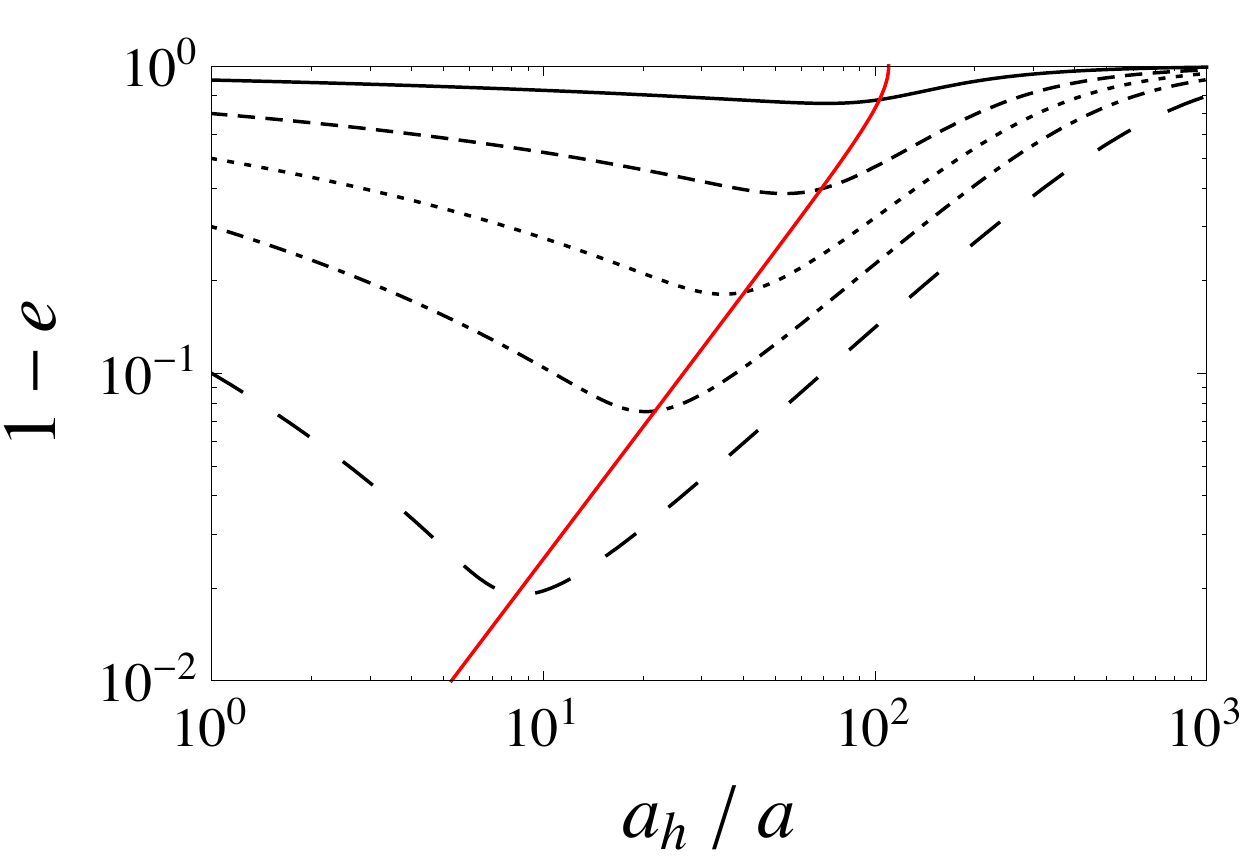}
\caption{The same as Figure~\ref{Figure:theta,e(a)} but for a nonrotating nucleus ($\eta=0.5$) and initial eccentricities 0.1, 0.3, 0.5, 0.7 and 0.9. The evolution of $\theta$ is not shown because $\theta=\mathrm{const}$.}
\label{Figure:e(a)-nonrotating}
\end{figure}

\begin{figure}
	\centering
	\subfigure{\includegraphics[width=0.49\textwidth]{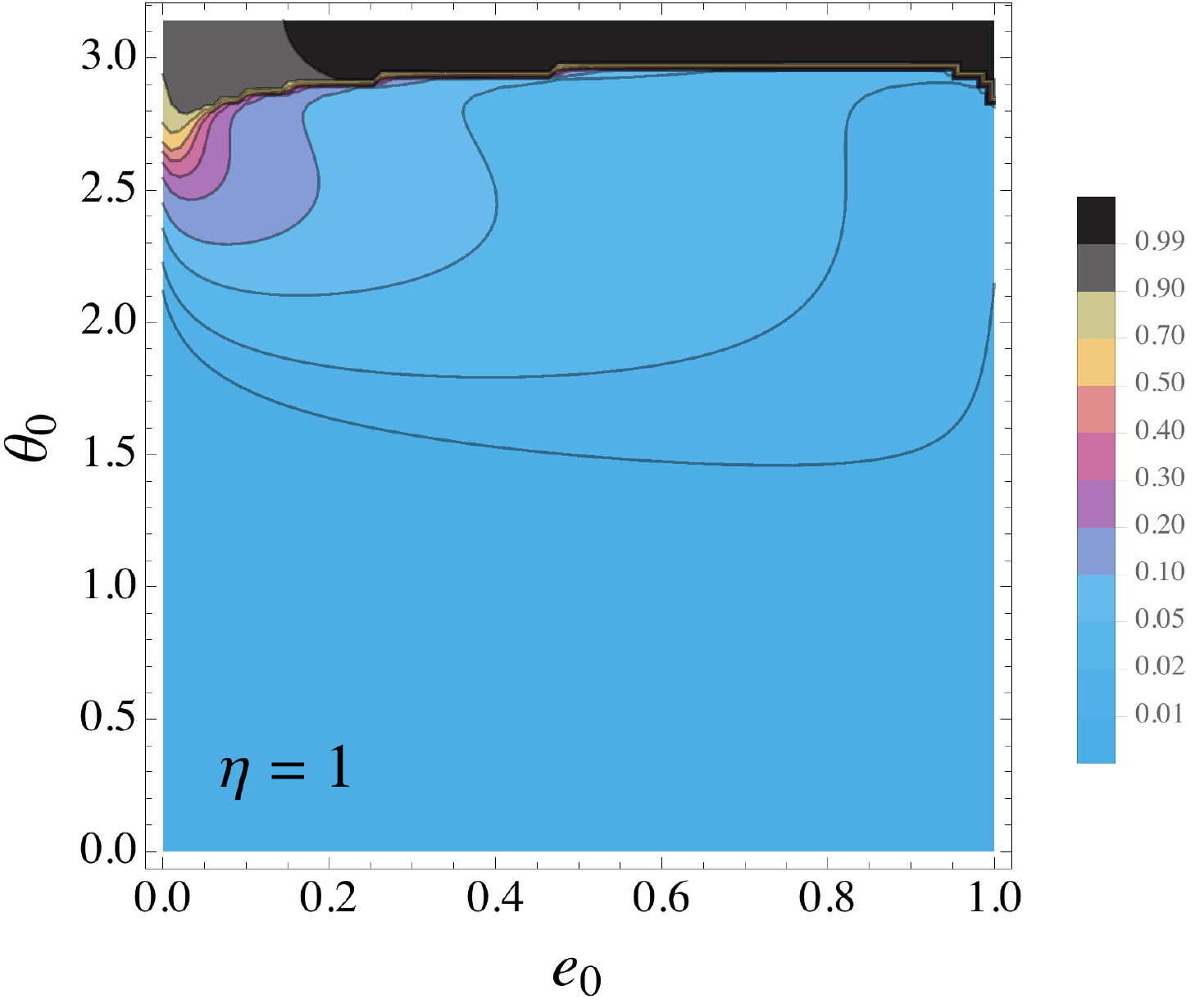}}
	\subfigure{\includegraphics[width=0.49\textwidth]{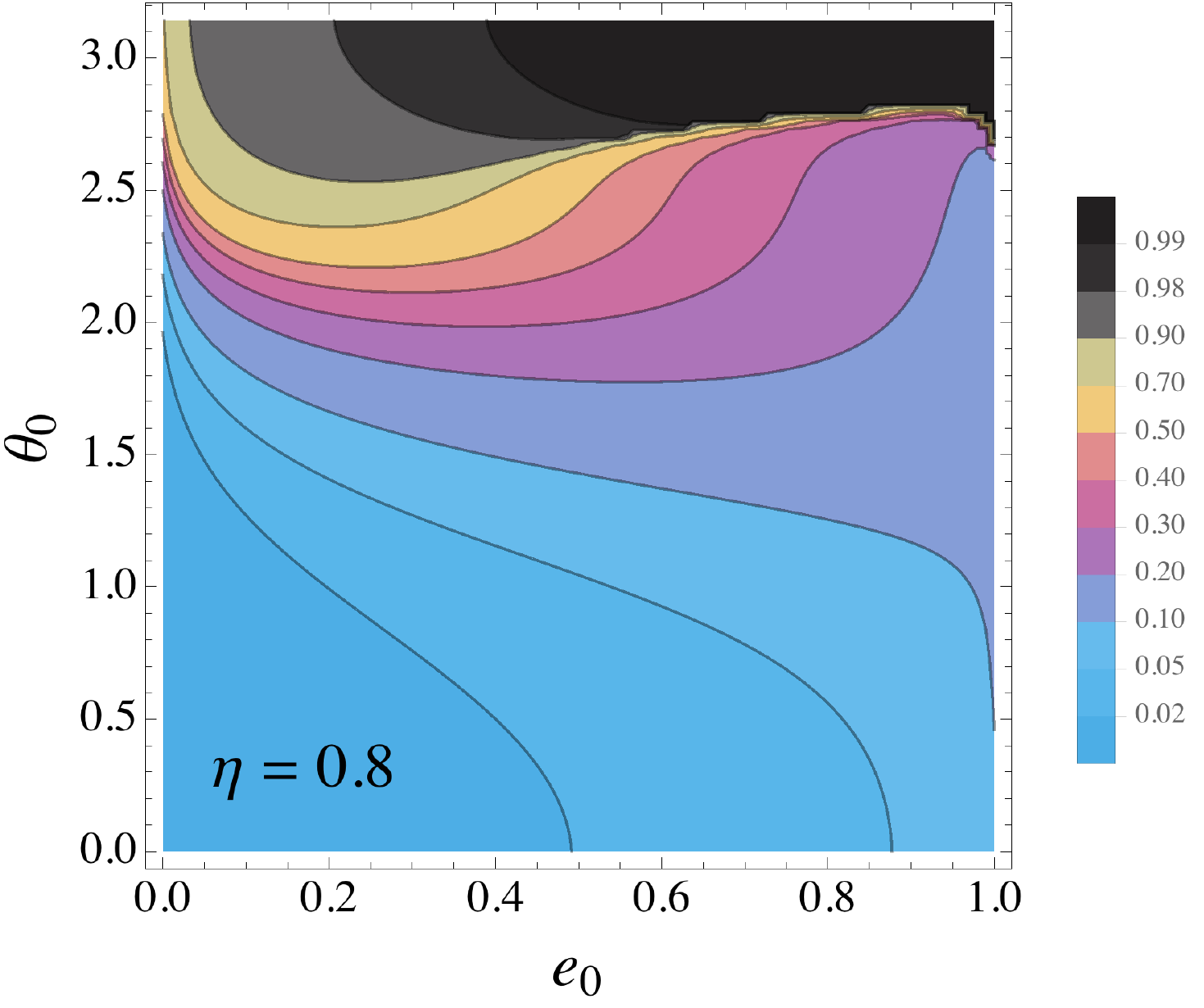}}
	\subfigure{\includegraphics[width=0.49\textwidth]{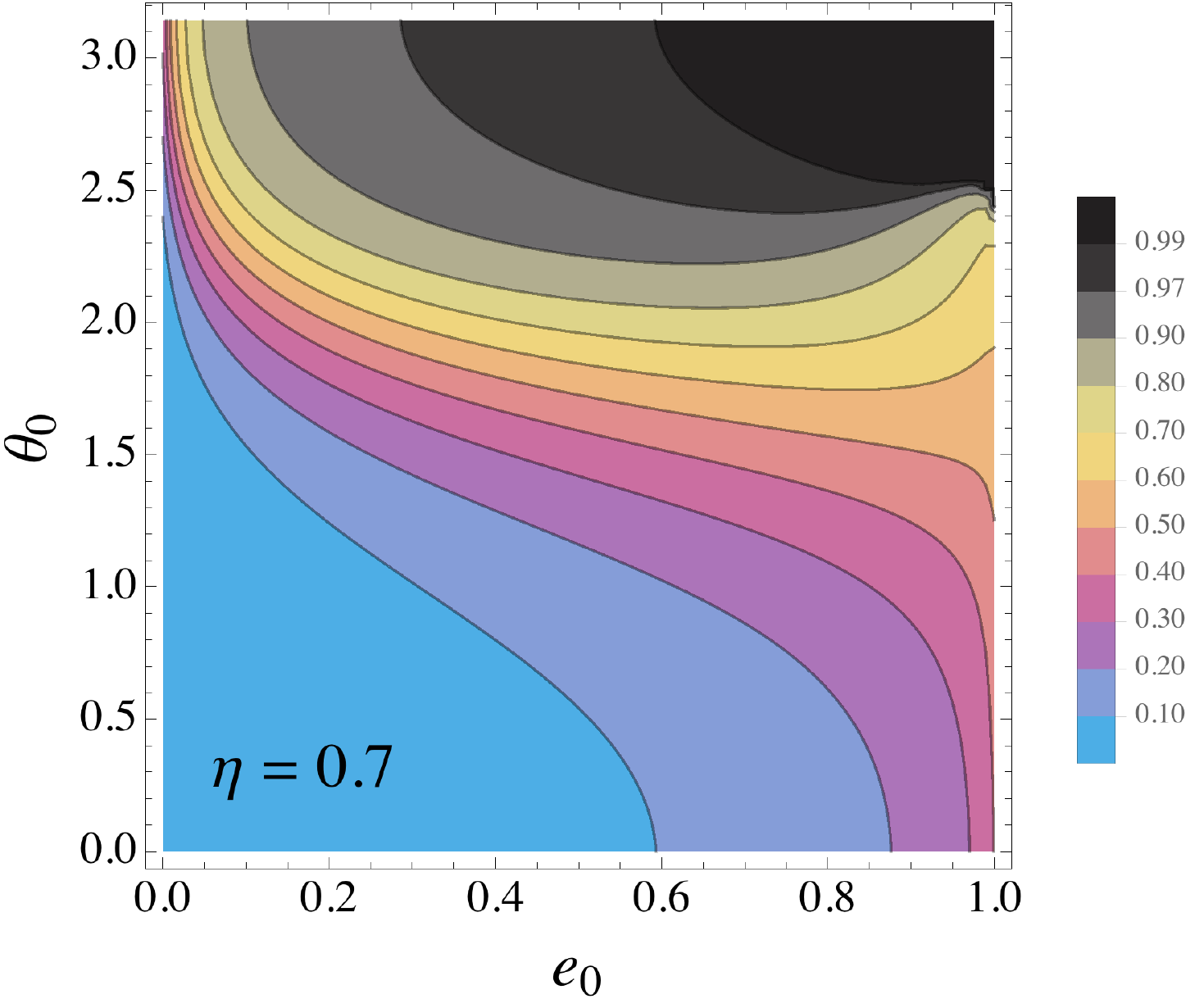}}
	\subfigure{\includegraphics[width=0.49\textwidth]{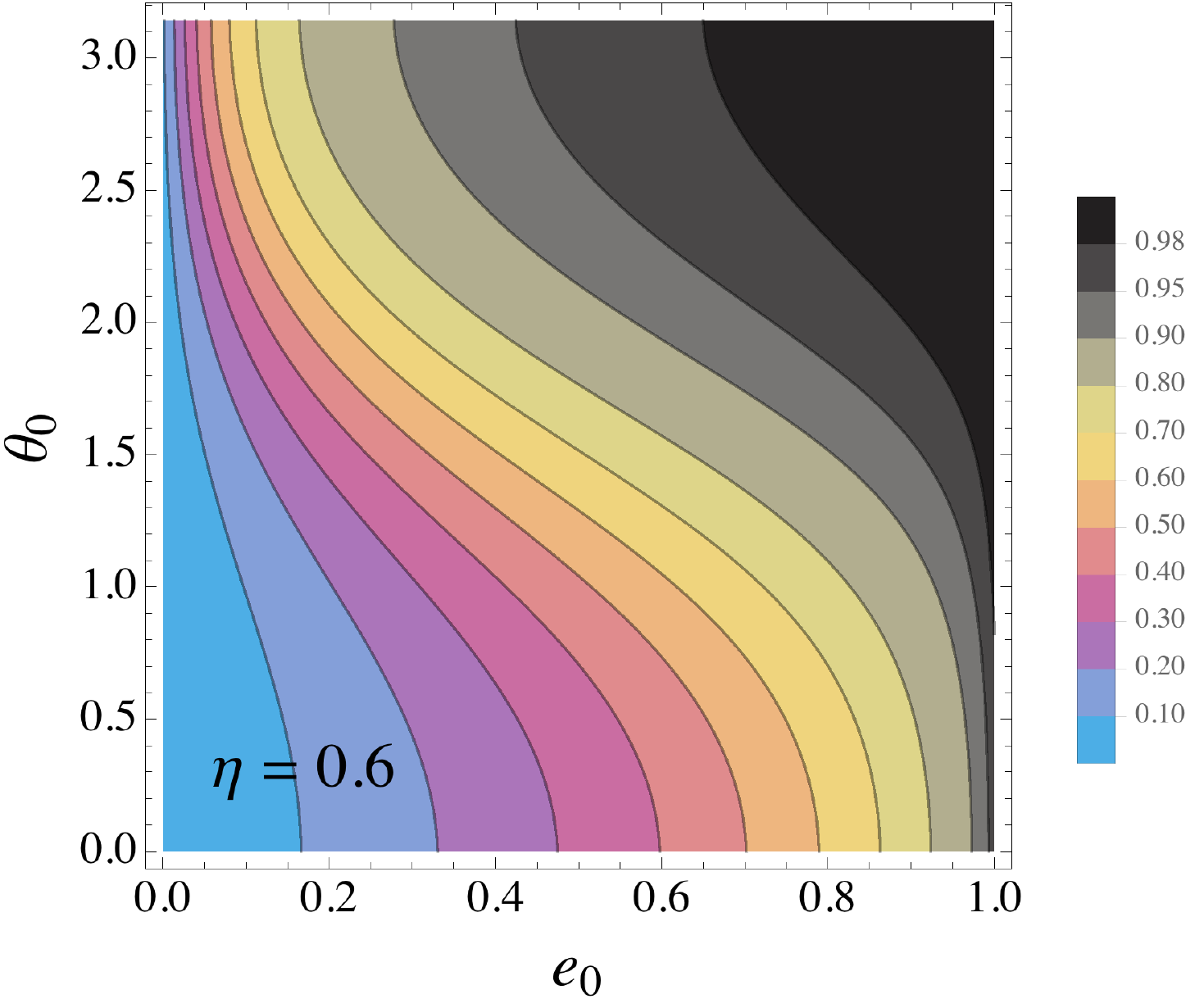}}
	\caption{
		Contour plots of $e_{\rm GR}$ (Eq.~\ref{Equation:e_GR}) in the $(e_0,\theta_0)$ plane (initial eccentricity and initial inclination at $a/a_h=1$) for $M_{12}=10^8\msun$, $q=1$ and four different corotation fractions.
		}
	\label{Figure:e_GW}
\end{figure}

\subsection{Loss-cone depletion}
\label{Section:LossConeDepletion}

So far we have assumed that the distribution of stars in the nucleus is unchanging. 
But in real galaxies, only a finite number of stars are on orbits that carry them close to the massive binary,
and the ejection of such stars leads to a gradual ``loss-cone depletion''. 
In a precisely spherical galaxy, the number of stars on orbits that intersect the binary will be small; if in addition 
the two-body relaxation time is long, repopulation of depleted orbits would be extremely slow,
and the binary separation would be expected to ``stall'' at a separation $a\sim a_h$ \citep[][chapter 8]{DEGN}.
But rates of loss-cone repopulation can be much higher in nonspherical galaxies, due to the combined effects
of gravitational encounters, and changes in orbital eccentricity due to torques from the large-scale potential
\citep{MerrittVasiliev2011}.

\citet{Vasiliev2015}  studied this phenomenon quantitatively using a  Monte-Carlo technique that properly accounts for dynamical
 relaxation even when the number of particles in a simulation is much lower than in a real galaxy. 
 Vasiliev et al. suggested the following expressions for the binary hardening rate in galaxies with different morphologies:
\bsub\label{Equation:galaxyType}
\barr
\frac{da}{dt} &=& k \left(\frac{da}{dt}\right)_\mathrm{full} \left(\frac{a}{a_h}\right)^\alpha,\\
k&=&0.4, \;\;\alpha = 0.3 \;\; {\rm for\;triaxial\;nuclei},\\
k&=&(N_\star/10^5)^{-1/2}, \;\;\alpha = 0 \;\; {\rm for\;axisymmetric\;nuclei},\\
k&=&(N_\star/10^5)^{-1}, \;\;\alpha = 0 \;\; {\rm for\;spherical\;nuclei} .
\earr
\esub
In these expressions, $N_\star$ is the number of stars in the galaxy and  $(da/dt)_\mathrm{full}$ is the hardening rate 
calculated under the ``full-loss-cone'' assumption -- the same expression that we have been using until now.
\citet{Vasiliev2015} studied only the hardening rate, but since all of our diffusion coefficients are proportional to the stellar encounter rate, 
it is reasonable to assume that their dependence on galaxy morphology is the same as for the hardening rate. 
In so doing, we ignore the possibility that loss-cone depletion has a systematic effect on the  change in any orbital parameter per encounter; such an assumption is justified considering the chaotic nature of a binary-star interaction where the final velocity and orbital angular momentum of a star are weakly correlated with their initial values.

The $N_\star$-dependence in Equations~(\ref{Equation:galaxyType}c,d) reflects the fact that in the
spherical and axisymmetric geometries, conservation of angular momentum (spherical symmetry) 
or its component along the symmetry axis (axisymmetry)
fixes the minimum periapsis distance accessible to a star.
Once all the stars on an orbit with given periapsis have been removed, 
continued supply of stars to the binary is only possible after new stars have been scattered onto the orbit by gravitational encounters, at rates that are $N_\star-$dependent. 
In triaxial galaxies, much of the phase space corresponds to orbits with no minimum periapsis;
the time for a star on such an orbit to reach the binary depends much more on torques from the
large-scale mass distribution than on two-body relaxation, hence the lack of an appreciable $N_\star$
dependence in the expression for the ``triaxial'' hardening rate.

Applying the corrections implied by Equations~(\ref{Equation:galaxyType}) to 
Equations~(\ref{Equation:d(a,e,theta)dt-general}), the new  evolution equations are
\bsub\label{Equation:d(a,e,theta)dt-new}
\barr
\frac{d(a/a_h)}{d(t/t_h)} &=& - k \left(\frac{a}{a_h}\right)^{2+\alpha} - \left(\frac{a_{\rm GR,0}}{a_h}\right)^5 \left(\frac{a}{a_h}\right)^{-3} f(e), \\
\frac{de}{d(t/t_h)} &=& k K \left(\frac{a}{a_h}\right)^{1+\alpha} - \frac{19}{12} \left(\frac{a_{\rm GR,0}}{a_h}\right)^5 \left(\frac{a}{a_h}\right)^{-4} g(e), \\
\frac{d\theta}{d(t/t_h)} &=& - k \frac{D_{\theta,1}}{H} \left(\frac{a}{a_h}\right)^{1+\alpha}
\earr
\esub
where $H$, $K$, $D_{\theta,1}$, $a_h$ and $a_{\rm GR, 0}$ are the same as before.

\begin{figure}
	\centering
	\subfigure{\includegraphics[width=0.49\textwidth]{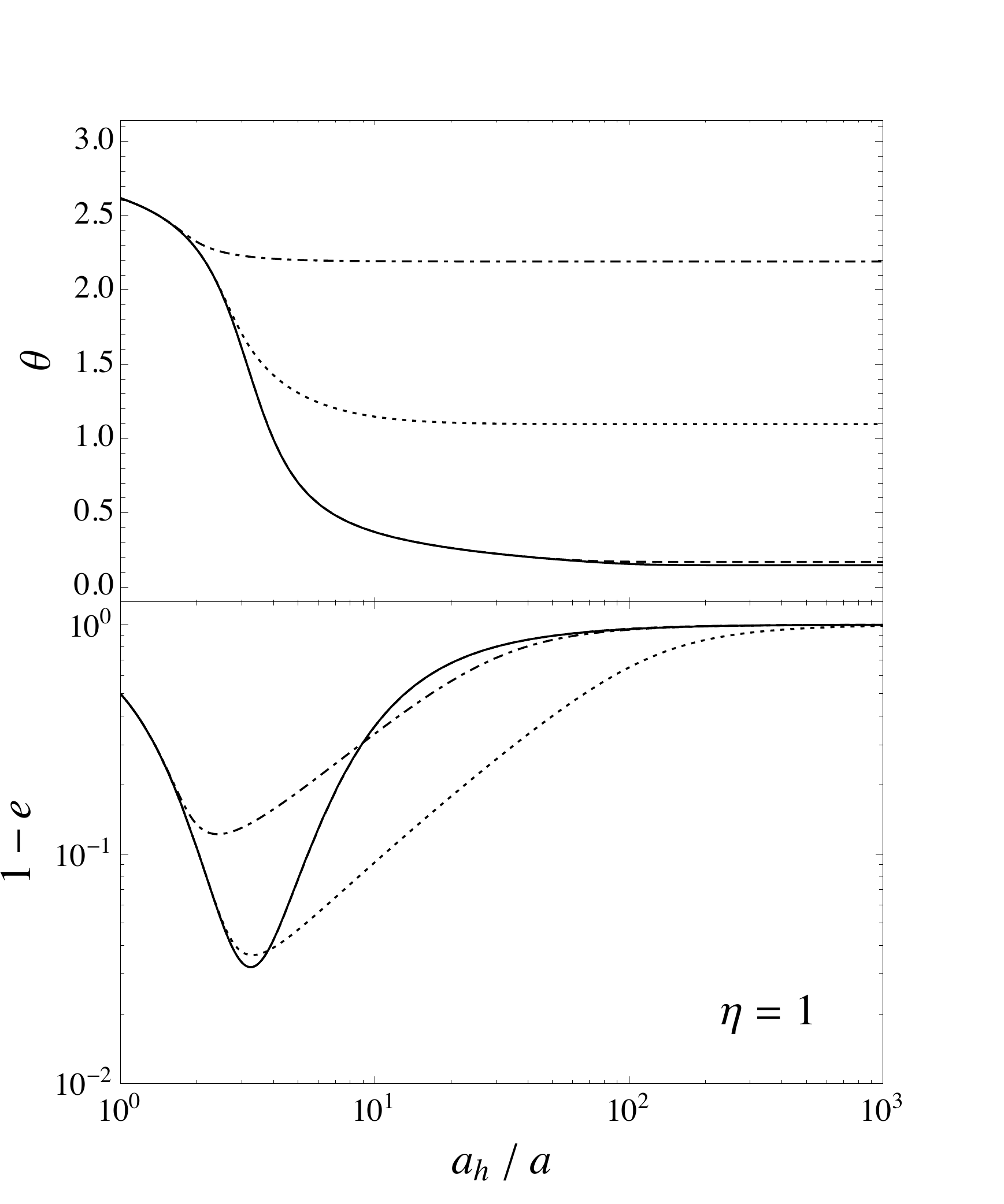}}
	\subfigure{\includegraphics[width=0.49\textwidth]{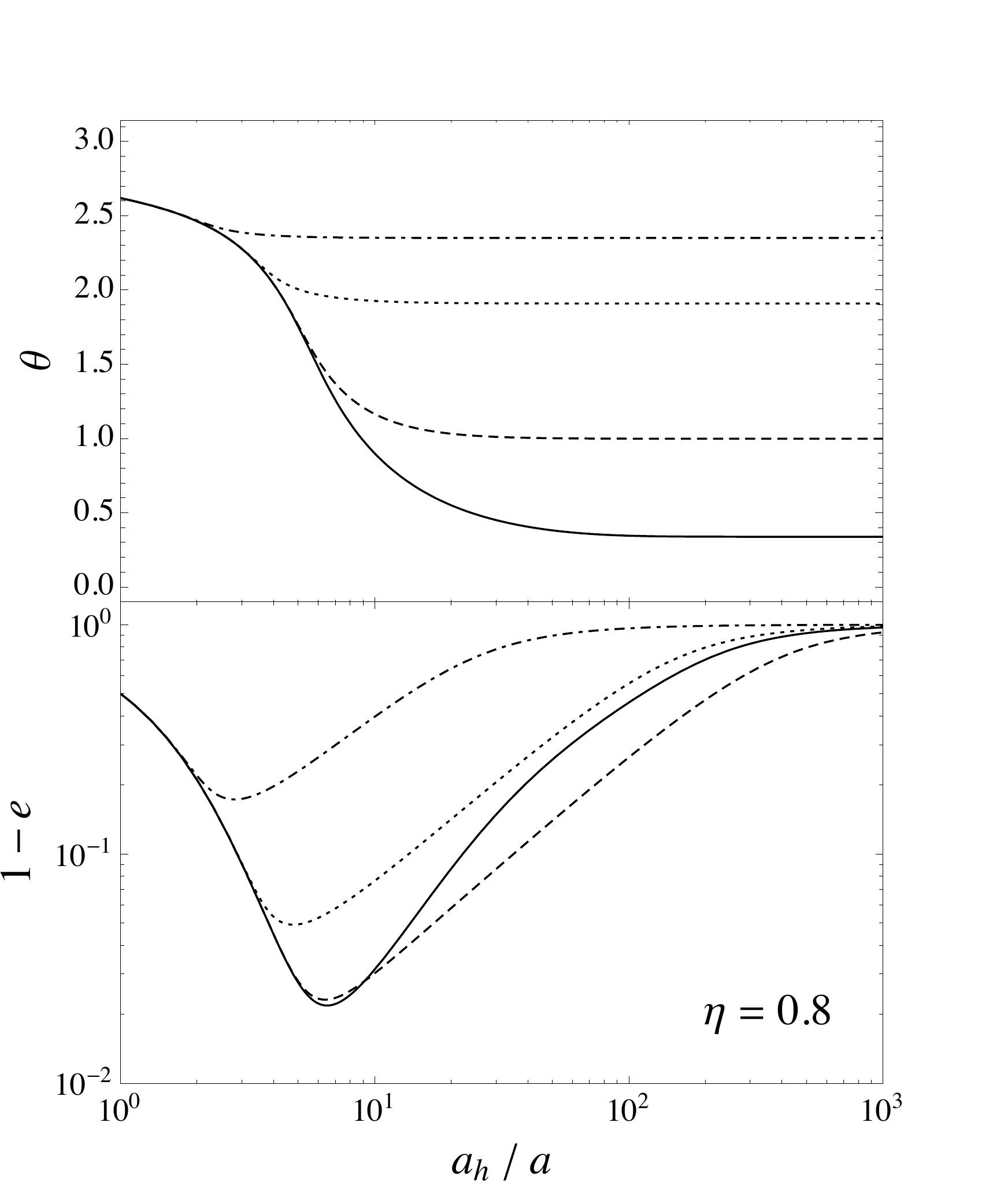}}
	\subfigure{\includegraphics[width=0.49\textwidth]{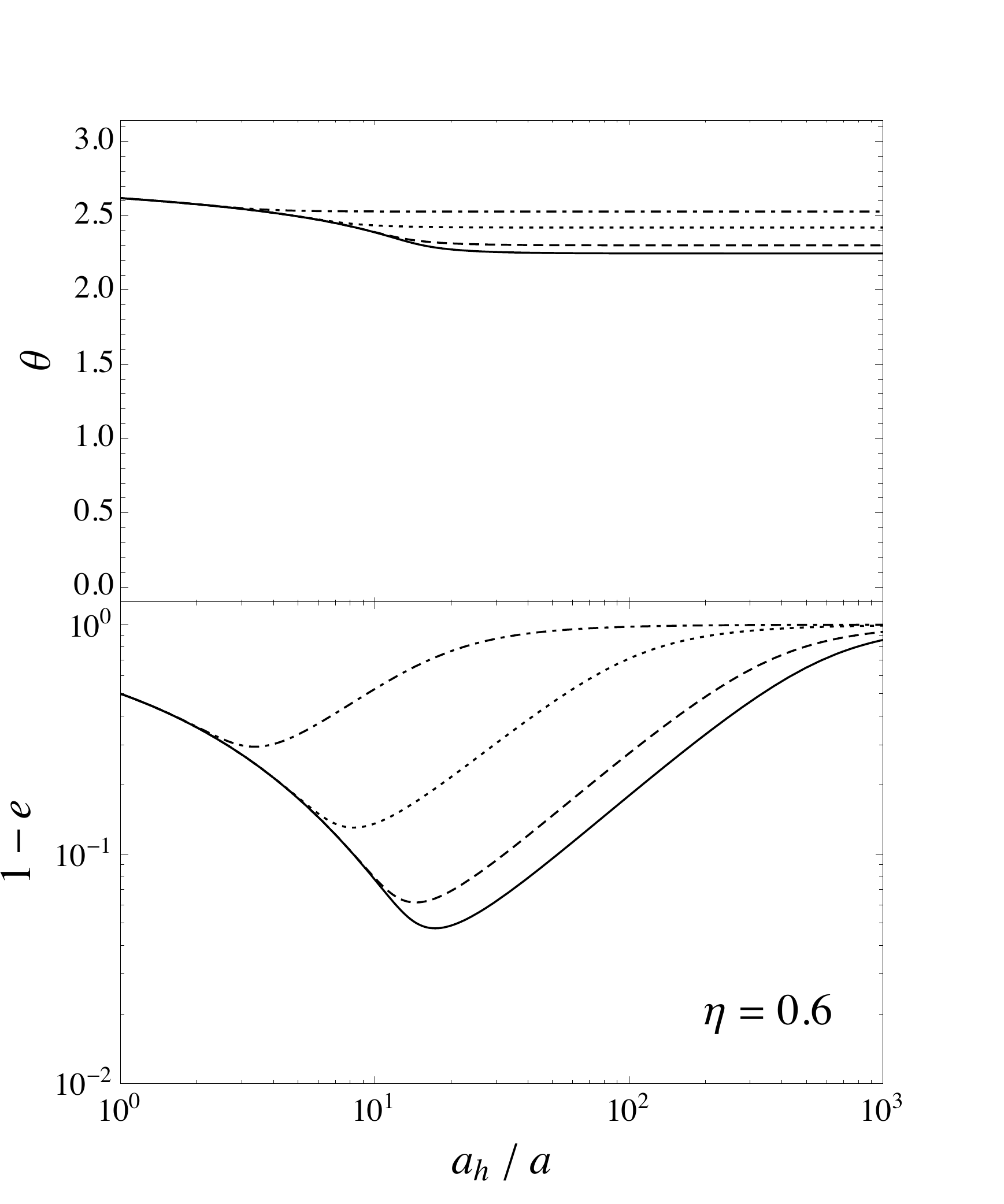}}
	\subfigure{\includegraphics[width=0.49\textwidth]{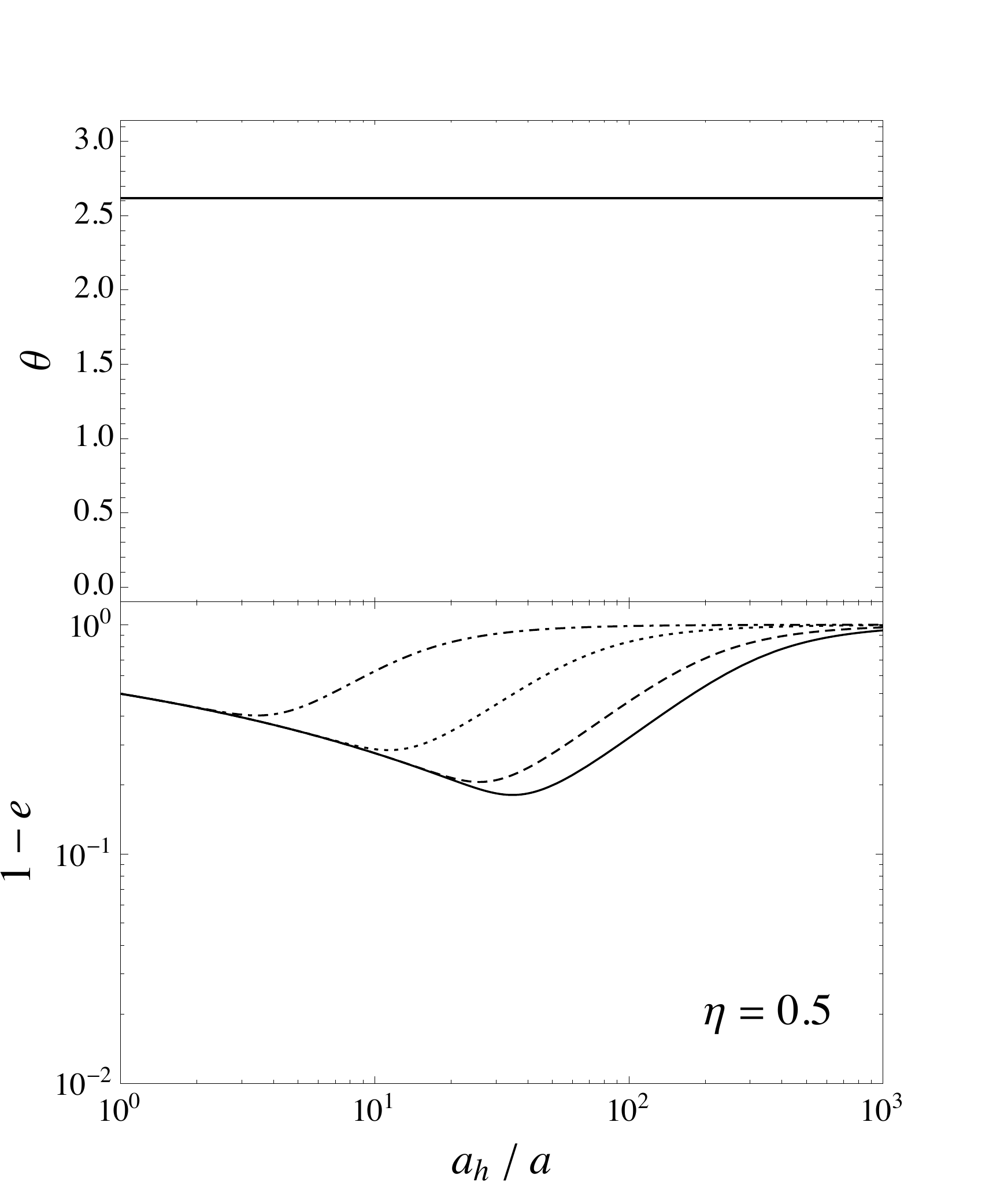}}
	\caption{
		Evolution of orbital inclination $\theta$ and eccentricity $e$ of a binary with $M_{12}=10^8\msun$, $q=1$, $e_0=0.5$ and $\theta=5\pi/6$ at different degrees of corotation, integrated using Equations~(\ref{Equation:galaxyType}) and (\ref{Equation:d(a,e,theta)dt-new}) for triaxial (dashed), axisymmetric (dotted) and spherical (dot-dashed) galaxies as well as in the full-loss-cone approximation (solid).
		}
	\label{Figure:geometry}
\end{figure}

\begin{figure}
	\centering
	\subfigure{\includegraphics[width=0.33\textwidth]{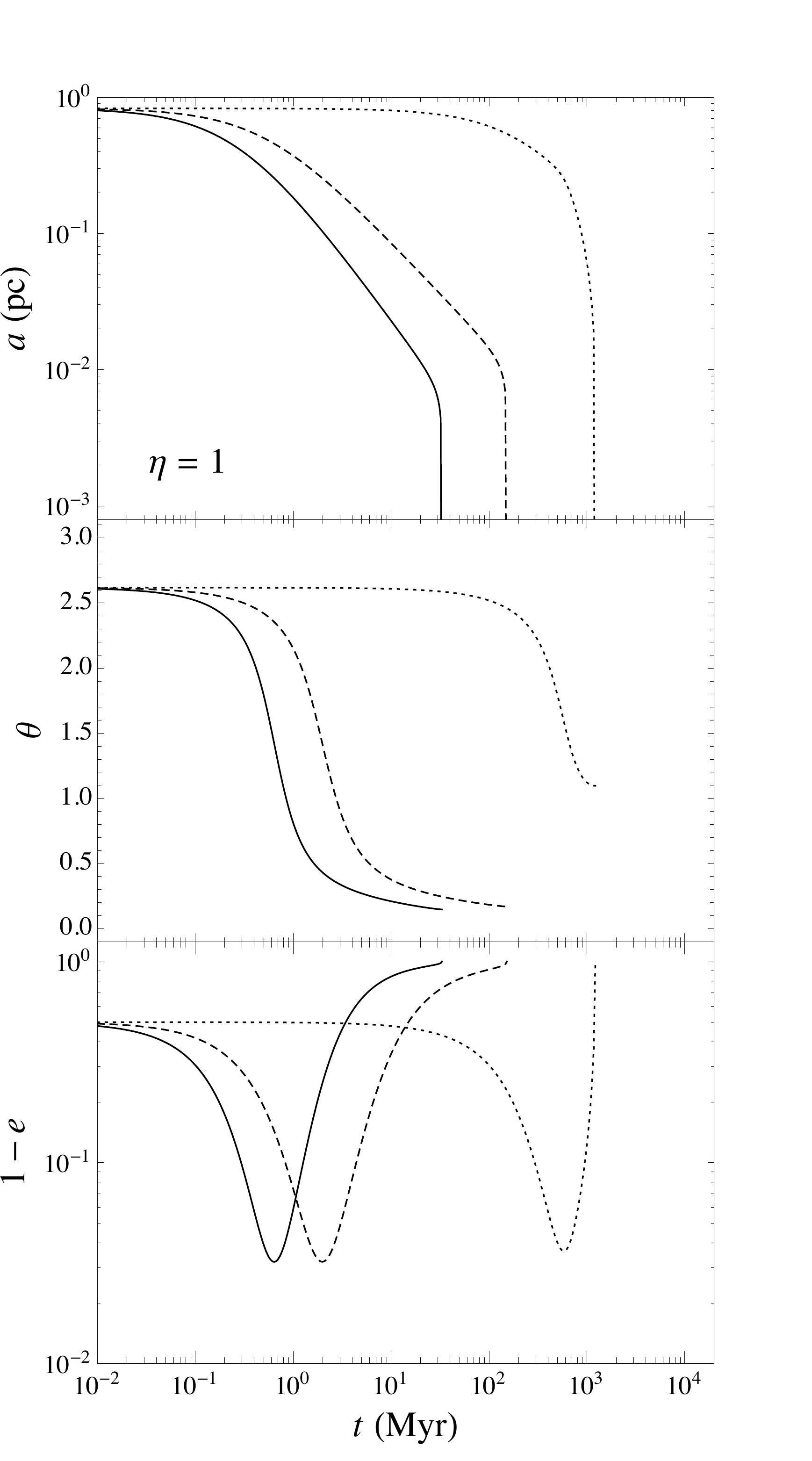}}
	\subfigure{\includegraphics[width=0.33\textwidth]{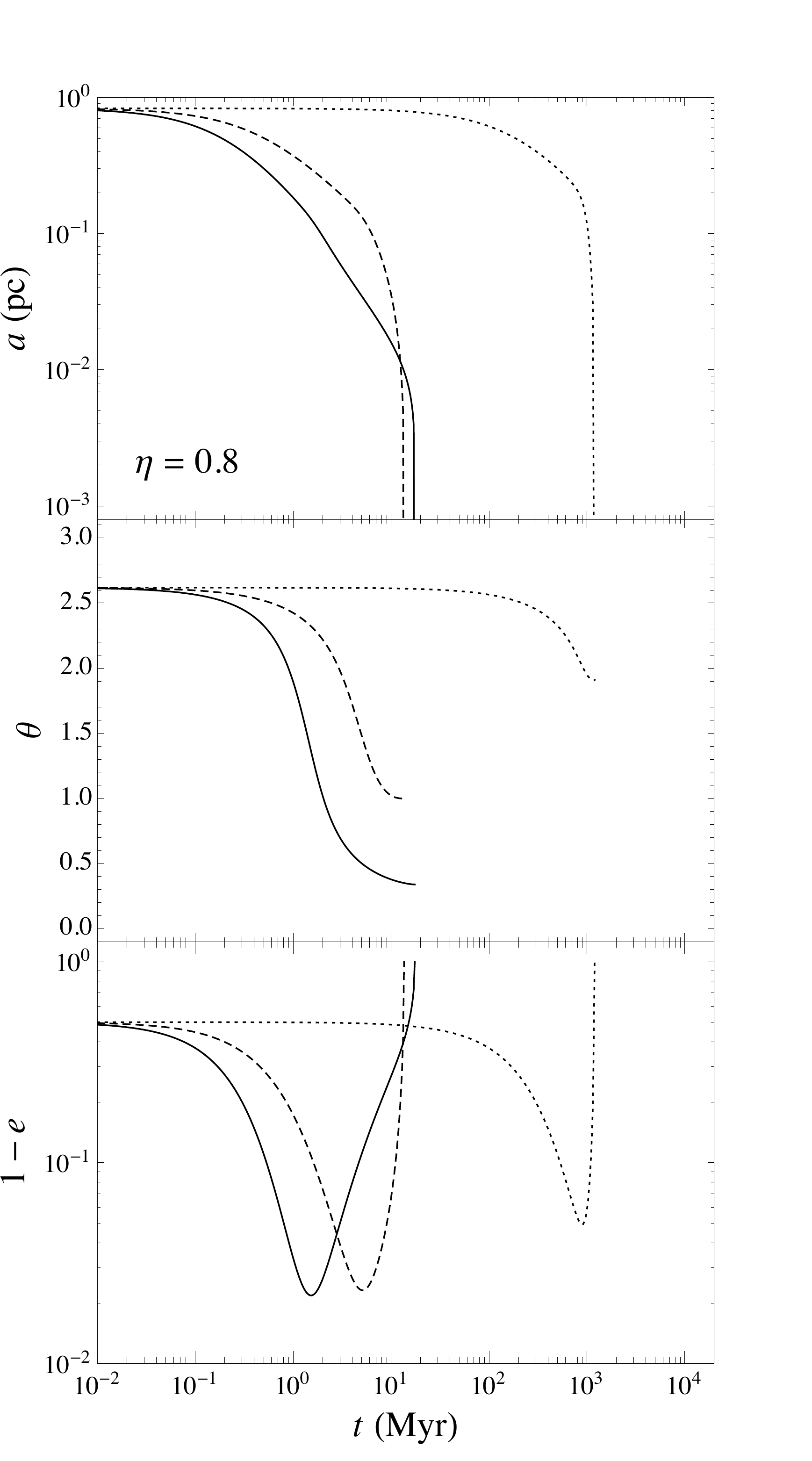}}
	\subfigure{\includegraphics[width=0.33\textwidth]{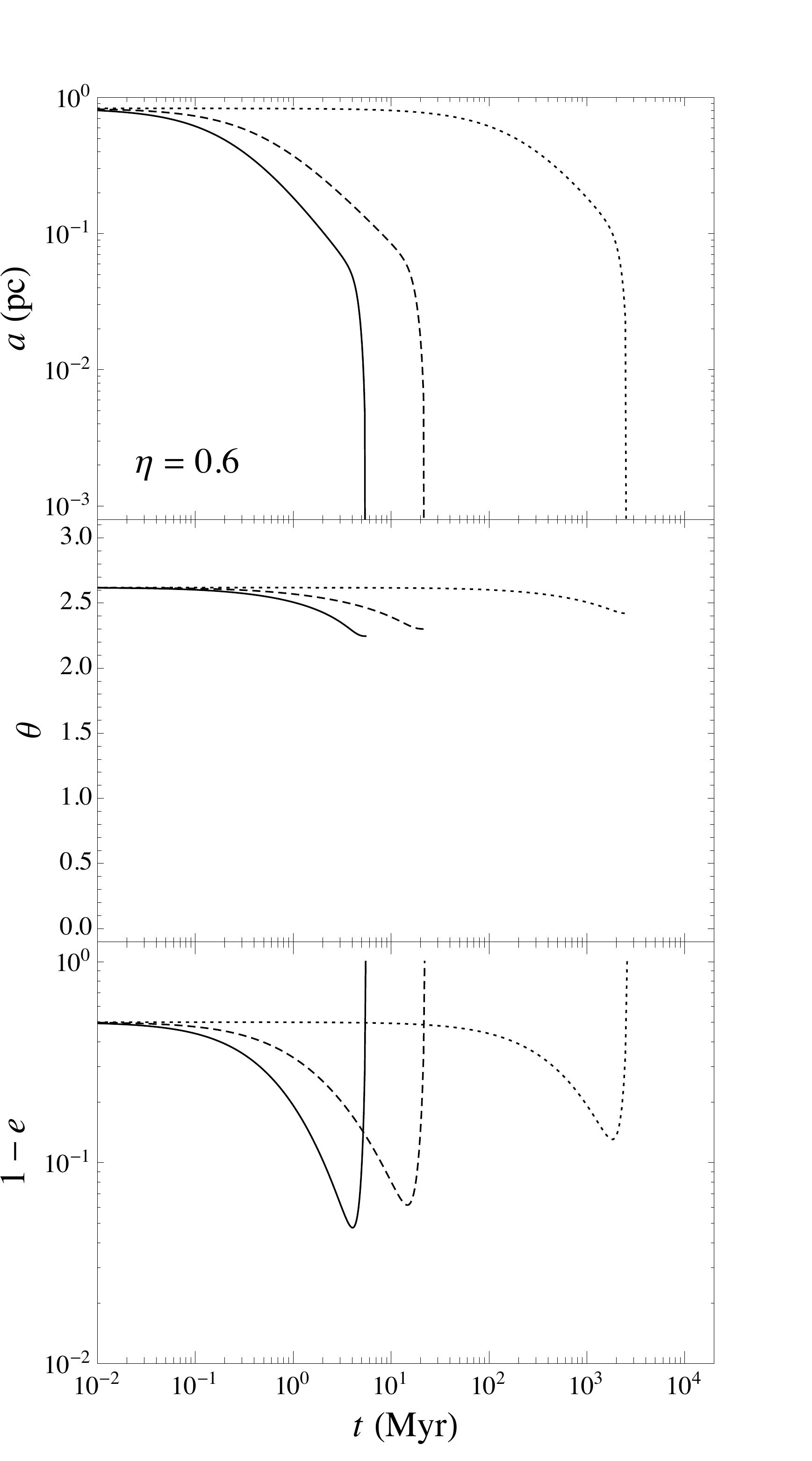}}
	\caption{
		Time dependence of orbital parameters of a binary with $M_{12}=10^8\msun$, $q=1$, $e_0=0.5$ and $\theta=5\pi/6$ at different degrees of corotation, integrated using Equations~(\ref{Equation:galaxyType}) and (\ref{Equation:d(a,e,theta)dt-new}) for triaxial (dashed), axisymmetric (dotted) and spherical (dot-dashed) galaxies as well as in the full-loss-cone approximation (solid).
		}
	\label{Figure:realtime}
\end{figure}

Some illustrative solutions to these equations are shown in Figure~\ref{Figure:geometry} (with $a/a_h$ as a proxy for time) 
and Figure~\ref{Figure:realtime} (in physical time units). 
Galaxy geometry can have an enormous influence on the coalescence timescale.
The latter is comparable to the full-loss-cone case for triaxial galaxies;
1-2 orders of magnitude longer in the axisymmetric geometry;
and extremely long (longer than the Hubble time) for spherical galaxies. 
At the same time, lower hardening rates for these three ``depleted loss cone'' models mean that binaries enter the 
GW-dominated regime earlier and $e_{\rm GR}$ for them is higher than determined by Equation~(\ref{Equation:e_GR}). 
A more detailed analysis of coalescence timescales in different geometries can be found in \citet{PaperII}.

\section{Conclusions}
\label{Section:Conclusions} 

We derived a Fokker-Planck equation describing the evolution of the orbital elements of a binary supermassive black hole
(SBH) due to interacting stars, and applied it to the case of a binary in a rotating stellar nucleus. 
First- and second-order diffusion coefficients for the binary's orbital parameters ($a,e,i,\Omega,\omega$) were calculated by means of scattering experiments. 
Excepting the case of a nucleus with very low rotation, the first-order (drift) terms almost always dominate over the second-order 
(stochastic) terms due to large ratio between the mass of a single star and the binary SBH. 
In particular, changes in the binary's orbital inclination (with respect to the axis of rotation of the nucleus) are almost always determined 
by the drift term, which is always negative, i. e. the inclination tends to decrease, toward a configuration in which the binary's 
angular momentum is aligned with that of the nucleus.
The first-order coefficient describing changes in eccentricity was found to depend strongly on inclination: 
eccentricity decreases for co-rotating binaries and increases for counterrotating ones. 
The inclination drift term, in turn, is an increasing function of binary eccentricity, so that evolution of the eccentricity and inclination are interdependent. 
These results are in agreement with previous numerical studies \citep{Sesana2011,Gualandris2012}. 

Invoking the smallness of the second-order terms, we derived a system of deterministic differential equations that describe 
the time evolution of a binary's eccentricity, $e$, and inclination, $\theta$. 
Included were the effects of gravitational wave emission, which become important for small semimajor axis and/or large eccentricity.  
Eccentricity evolution was found to depend strongly on the initial $\theta$.
For initially co-rotating binaries ($\theta_0\lesssim\pi/2$), the eccentricity decreases to zero fairly quickly,
while for counterrotating binaries ($\theta_0\gtrsim\pi/2$), $e$ increases initially but then decreases 
due either to binary reorientation or to the effects of GW emission. 
Counterrotating binaries can reach high eccentricities ($e>0.9$), but in nuclei with a high degree of rotation,
eccentricity decreases again to low values due to fast reorientation, so that the binary enters the final, GW-dominated, 
stage of its evolution with an almost circular orbit. 

We were able to take into account, in an approximate way, depletion of the binary's ``loss-cone''
by rescaling the diffusion coefficients according to the results of \citet{Vasiliev2015}, who derived expressions for the rate
of loss-cone repopulation in galaxies with various geometries.
The main result of this correction was found to be a longer evolution timescale compared with the full-loss-cone approximation:
 a few times longer for triaxial nuclei, about two orders of magnitude longer for axisymmetric nuclei, and many orders of magnitude longer (typically, longer than the Hubble time) for spherically symmetric nuclei. 
 Another consequence is that the transition to the GW-dominated regime happens at larger semimajor axes.

One of the important applications of our work is to the production of GWs by binary SBHs, and the generation of a stochastic GW
background by a population of massive binaries.
In the low-frequency regime accessible to pulsar timing arrays (PTAs), much of the signal would be produced by binaries
at large separations, where the main source of evolution is likely to be interaction with ambient stars \citep[e.g.][]{Sesana2013}.
Evolution of binary eccentricity is of crucial importance: circular-orbit binaries emit GWs at only one frequency -- twice the orbital frequency -- while eccentric binaries radiate at all harmonics \citep{PM}. 
 As we have shown, eccentricity of the binary in the GW-dominated regime is determined by the initial values ($e_0, \theta_0$)
 of $e$ and $\theta$ and by the degree of nuclear rotation ($\eta$). 
 We would therefore expect GW emission to be strongly affected by those parameters as well. 
 Upper limits inferred from the lack of detection by PTA observations have already excluded some of models of 
 binary \sbh\ evolution \citep{Shannon2015}.
 Existing models include neither the effects of nuclear rotation, nor loss-cone depletion.
 It is therefore important to calculate the stochastic background spectrum for different assumed distributions of 
 $e_0$, $\theta_0$ and $\eta$ and to test which are consistent with current (or possible future) observational limits.
 These questions are addressed in detail in Paper II \citep{PaperII}.

Not just the eccentricity evolution, but the orbital plane reorientation itself may also have significant observational implications. 
It was shown in post-Newtonian numerical simulations \citep{MerrittEkers2002,Gergely2009,Kesden2010} that the spin direction of the coalescence product of two black holes is usually in the same direction as their orbital angular momentum at the beginning of GW-driven phase, except in the case where the binary mass
ratio is extreme and the spin of the primary \sbh\ is almost exactly counter-aligned with the orbital angular momentum. 
And as the spin direction, in turn, is believed to define the jet direction in active galactic nuclei, we infer that in rotating nuclei, 
the jet should be preferentially aligned with the stellar rotation axis. 
There is, indeed, some observational evidence for that: \citet{Battye2009} found preferential alignment of major radio and minor optical axes in relatively radio-quiet galaxies (which they identify with fast-rotating axisymmetric ellipticals) and the absence of such alignment in more radio-loud galaxies (which they identify with slowly rotating triaxial ellipticals). 
\citet{Middleton2016} found a similar bimodality in accretion disk orientations.
 \citet{Lagos2011} studied the orientation angles of Type I and II AGN hosts, and their results also imply significant alignment between AGN components (torus and accretion disk) and galaxy rotation axes. 
However, a number of other studies have failed to find strong evidence for the aforementioned correlations \citep[e. g.][]{Kinney2000,Gallimore2006}. All these results should be interpreted carefully since it is possible for \sbh\ spin directions
 to change due to accretion of gas having angular momentum that is misaligned with the spin \citep{Dotti2013}.

\acknowledgments
We thank Alessia Gualandris, Richard O'Shaughnessy, Jeremy Schnittman and Eugene Vasiliev for useful conversations.
This work was supported by the National Science Foundation under grant no. AST 1211602 
and by the National Aeronautics and Space Administration under grant no. NNX13AG92G.
\appendix
\section{$(E,L,\mu,\phi)$ diffusion coefficients}
   \label{Appendix:DiffCoefs}
Using equation~(\ref{Equation:Delta}), we construct expressions for the diffusion coefficients describing changes in the binary's energy and angular momentum defined via the variables
\beq
x_1 = L, \ \ \ x_2 = \mu = \cos\theta = L_z/L, \ \ \ x_3 = \phi, \ \ \ x_4 = E 
\eeq
(equation~\ref{Equation:NewVariables}),
in terms of diffusion coefficients based on the variables $E$ and $L_i, i=1, 2, 3$.
The results are as follows:
\begin{eqnarray}
\langle\Delta L\rangle &=& \displaystyle\sum_i\frac{L_i}{L}\langle\Delta L_i\rangle
+ \frac{1}{2L}\displaystyle\sum_{i,j}\left(\delta_{ij}
-\frac{L_i L_j}{L^2}\right)\langle\Delta L_i\Delta L_j\rangle, \\
\langle\Delta L^2\rangle &=& \frac{1}{L^2}\displaystyle\sum_{i,j}L_i L_j\langle\Delta L_i\Delta L_j\rangle
\nonumber , \\
\langle\Delta \mu\rangle &=& \frac{1}{L}\langle\Delta L_z\rangle - \frac{L_z}{L^3}(L_x\langle\Delta L_x\rangle+L_y\langle\Delta L_y\rangle) -
 \frac{L_z}{2L^3}\displaystyle\sum_i\langle\Delta L_i\rangle +
 \frac{3L_z}{2L^5}(L_x^2\langle\Delta L_x^2\rangle+L_y^2\langle\Delta L_y^2\rangle+2L_x L_y\langle\Delta L_x\Delta L_y\rangle) \nonumber , \\ \nonumber \\
\langle\Delta \mu^2\rangle &=& \frac{1}{L^2}\langle\Delta L_z^2\rangle
- \frac{2L_z}{L^4}(L_x\langle\Delta L_x\Delta L_z\rangle+L_y\langle\Delta L_y\Delta L_z\rangle)
+ \frac{L_z^2}{L^6}(L_x^2\langle\Delta L_x^2\rangle+L_y^2\langle\Delta L_y^2\rangle+2L_x L_y\langle\Delta L_x\Delta L_y\rangle) \nonumber , \\ \nonumber \\
\langle\Delta L\Delta \mu\rangle &=& \frac{L_z}{L^2}\langle\Delta L_z^2\rangle -
\frac{L_z}{L^4}(L_x^2\langle\Delta L_x^2\rangle+L_y^2\langle\Delta L_y^2\rangle+2L_x L_y\langle\Delta L_x\Delta L_y\rangle) + \frac{L_x^2+L_y^2}{L^4}(L_x\langle\Delta L_x\Delta L_z\rangle+L_y\langle\Delta L_y\Delta L_z\rangle) \nonumber , \\ \nonumber \\
\langle\Delta \phi\rangle &=& \frac{1}{L^2}(-L_y\langle\Delta L_x\rangle+L_x\langle\Delta L_y\rangle)+\frac{1}{L^4}(L_xL_y(\langle\Delta L_x^2\rangle-\langle\Delta L_y^2\rangle)+(L_y^2-L_x^2)\langle\Delta L_x\Delta L_y\rangle) \nonumber , \\
\langle\Delta \phi^2\rangle &=& \frac{1}{L^4}(L_y^2\langle\Delta L_x^2\rangle+L_x^2\langle\Delta L_y^2\rangle-2L_xL_y\langle\Delta L_x\Delta L_y\rangle)\nonumber , \\
\langle\Delta L\Delta \phi\rangle &=& \frac{1}{L^3}(L_xL_y(\langle\Delta L_y^2\rangle-\langle\Delta L_x^2\rangle)+ (L_x^2-L_y^2)\langle\Delta L_x\Delta L_y\rangle-L_yL_z\langle\Delta L_x\Delta L_z\rangle+ L_xL_z\langle\Delta L_y\Delta L_z\rangle) \nonumber , \\
\langle\Delta \mu\Delta \phi\rangle &=& \frac{1}{L^3}(L_x\langle\Delta L_y\Delta L_z\rangle-L_y\langle\Delta L_x\Delta L_z\rangle)- \frac{L_z}{L^5}(L_xL_y(\langle\Delta L_y^2\rangle-\langle\Delta L_x^2\rangle)+(L_x^2-L_y^2)\langle\Delta L_x\Delta L_y\rangle) \nonumber , \\
\langle\Delta E\Delta L\rangle &=& \displaystyle\sum_i\frac{L_i}{L}\langle\Delta E\Delta L_i\rangle
\nonumber , \\
\langle\Delta E\Delta \mu\rangle &=& \frac{1}{L}\langle\Delta E\Delta L_z\rangle
- \frac{L_zL_x}{L^3}\langle\Delta E\Delta L_x\rangle
- \frac{L_zL_y}{L^3} \langle\Delta E\Delta L_y\rangle \nonumber , \\
\langle\Delta E\Delta \phi\rangle &=&
- \frac{L_y}{L^2}\langle\Delta E\Delta L_x\rangle
+ \frac{L_x}{L^2} \langle\Delta E\Delta L_y\rangle \nonumber .
\end{eqnarray}

\section{Fokker-Planck equation in terms of $\Theta_\parallel, \Theta_\perp$}
\label{Appendix:Debye}

The sine rule from spherical trigonometry states
\begin{eqnarray}
\frac{\sin{\Delta\phi}}{\sin\Theta} = \frac{\sin\xi}{\sin\theta'}
\label{sine rule} .
\end{eqnarray}
Following Debye, we write
\begin{eqnarray}
\Delta\phi = \chi\Theta + \beta\Theta^2 + \ldots 
\label{phi series}
\end{eqnarray}
and we assume that $\Theta$ is small.
Also, we already know that
\begin{eqnarray}
\sin{\theta'} = \sin\theta + \cos\theta\cdot\Delta\theta + \mathcal{O}(\Theta^2) = \sin\theta - \cos\theta\cdot\Theta\cos\xi + \mathcal{O}(\Theta^2)
\end{eqnarray}
so
\begin{eqnarray}
\frac{1}{\sin{\theta'}} = \frac{1}{\sin\theta (1 - \cot\theta\cdot\Theta\cos\xi + \mathcal{O}(\Theta^2))} = \frac{1}{\sin\theta} + \frac{\cos\theta}{\sin^2\theta}\Theta\cos\xi + \mathcal{O}(\Theta^2) 
\label{1sintheta series} .
\end{eqnarray}
Substitution of (\ref{phi series}) and (\ref{1sintheta series}) into (\ref{sine rule}) yields
\begin{eqnarray}
\Delta\phi = \chi\Theta + \beta\Theta^2 + \ldots = \frac{\sin\xi}{\sin\theta}\Theta + \frac{\cos\theta}{\sin^2\theta}\sin\xi\cos\xi\cdot\Theta^2 + \ldots
\end{eqnarray}
and finally
\begin{eqnarray} 
\label{Equation:LocalDiffusionCoefficients}
\sin\theta\langle\Delta\phi\rangle &=& \langle\Delta\Theta_\perp\rangle + \cot\theta\langle\Delta\Theta_\parallel\Delta\Theta_\perp\rangle ,\nonumber \\
\sin^2\theta\langle\Delta\phi^2\rangle &=& \langle(\Delta\Theta_\perp)^2\rangle ,\nonumber \\
\sin\theta\langle\Delta\phi\Delta\theta\rangle &=& -\langle\Delta\Theta_\parallel\Delta\Theta_\perp\rangle .
\end{eqnarray}
The inverse relations are
\begin{eqnarray}
\label{Equation:DeltaTheta}
\langle\Delta\Theta_\perp\rangle &=&  \sin\theta\langle\Delta\phi\rangle + \cos\theta\langle\Delta\phi\Delta\theta\rangle ,\nonumber \\
\langle\Delta\Theta_\parallel\rangle &=& -\langle\Delta\theta\rangle + \frac{1}{2}\sin\theta\cos\theta\langle\Delta\phi^2\rangle ,\nonumber \\
\langle(\Delta\Theta_\perp)^2\rangle &=& \sin^2\theta\langle\Delta\phi^2\rangle ,\nonumber \\
\langle(\Delta\Theta_\parallel)^2\rangle &=& \langle(\Delta\theta)^2\rangle, \nonumber \\
\langle\Delta\Theta_\parallel\Delta\Theta_\perp\rangle &=& - \sin\theta\langle\Delta\phi\Delta\theta\rangle .
\end{eqnarray}
In terms of $(\theta,\phi)$,
the Fokker-Planck equation for the angular part of the probability density is
\begin{eqnarray}
\label{Equation:FPLocalThetaPhi}
\frac{\partial g}{\partial t} = - \frac{\partial}{\partial\theta} \left(g\langle\Delta\theta\rangle\right) - \frac{\partial}{\partial\phi}\left(g\langle\Delta\phi\rangle\right) +
+ \frac{1}{2}\frac{\partial^2}{\partial\theta^2}\left(g\langle(\Delta\theta)^2\rangle\right) + \frac{\partial^2}{\partial\phi\partial\theta}\left(g\langle\Delta\theta\Delta\phi\rangle\right) + \frac{1}{2}\frac{\partial^2}{\partial\phi^2}\left(g\langle\Delta\phi^2\rangle\right) .
\end{eqnarray}
Substitution of Equations
(\ref{Equation:LocalDiffusionCoefficientsInTheta})
and
(\ref{Equation:LocalDiffusionCoefficients})
into Equation (\ref{Equation:FPLocalThetaPhi})
results in
\begin{eqnarray}
\frac{\partial g}{\partial t} &=& - \frac{\partial}{\partial\theta} \left[g\left(-\langle\Delta\Theta_\parallel\rangle + \frac{1}{2}\cot\theta\langle(\Delta\Theta_\perp)^2\rangle\right)\right] - \frac{\partial}{\partial\phi}\left[g\left(\frac{1}{\sin\theta}\langle\Delta\Theta_\perp\rangle + \frac{\cos\theta}{\sin^2\theta} \langle\Delta\Theta_\parallel\Delta\Theta_\perp\rangle\right)\right]
\nonumber\\
&+& \frac12\frac{\partial^2}{\partial\theta^2}\left[g\langle(\Delta\Theta_\parallel)^2\rangle\right] -
\frac{\partial^2}{\partial\phi\partial\theta}\left[g\frac{1}{\sin\theta}\langle\Delta\Theta_\parallel\Delta\Theta_\perp\rangle\right] + \frac12\frac{\partial^2}{\partial\phi^2}\left[g\frac{1}{\sin^2\theta}\langle(\Delta\Theta_\perp)^2\rangle\right]
\end{eqnarray}
where $g=f\sin\theta$.  Alternatively,
\begin{eqnarray}
\sin\theta\frac{\partial f}{\partial t} &=& - \frac{\partial}{\partial\theta} \left[f\left(-\sin\theta\langle\Delta\Theta_\parallel\rangle + \frac12\cos\theta\langle(\Delta\Theta_\perp)^2\rangle\right)\right] -
\frac{\partial}{\partial\phi}\left[f\left(\langle\Delta\Theta_\perp\rangle + \cot\theta\langle\Delta\Theta_\parallel\Delta\Theta_\perp\rangle\right)\right] +\nonumber\\
&+& \frac12\frac{\partial^2}{\partial\theta^2}\left[f\sin\theta\langle(\Delta\Theta_\parallel)^2\rangle\right] - \frac{\partial^2}{\partial\phi\partial\theta}\left[f\langle\Delta\Theta_\parallel\Delta\Theta_\perp\rangle\right] + \frac{1}{2}\frac{\partial^2}{\partial\phi^2}\left[f\frac{1}{\sin\theta}\langle(\Delta\Theta_\perp)^2\rangle\right] .
\end{eqnarray}

\section{Number density and velocity dispersion values in the integral expression for diffusion coefficients} 
\label{Appendix:n(r)}

We model the stellar density  as
\begin{eqnarray}\label{Equation:DehnenModel}
\rho(r)=\frac{(3-\gamma)}{4\pi} \frac{M_\mathrm{gal}}{r_b^3}
\left(\frac{r}{r_b}\right)^{-\gamma}
\left(1+\frac{r}{r_b}\right)^{\gamma-4} ,
\end{eqnarray}
a ``Dehnen model'' \citep{Dehnen1993}, 
where $M_\mathrm{gal}$ is the total galaxy mass and $r_b$ is a ``break radius'' or ``core radius''.
We expect the latter to be determined by the binary itself during its formation and
to be of order the gravitational influence radius of the binary,
$r_\mathrm{infl}$, defined as the radius where
\beq\label{Equation:r_infl}
M_\star(r<r_\mathrm{infl}) = 2 M_{12}
\eeq
\citep[][Sect. 8.2]{DEGN}.
The same process of binary formation is expected to
result in a shallow central density profile, $\gamma\lesssim 1$.
In fact, for any $\gamma<2$, the contribution to the gravitational potential from the
stars in this model is finite at all radii.
Now, it is only stars with $r_p\lesssim a$ that contribute appreciably to the integral
(\ref{Equation:DiffCoeffinQ}).
If we assume a hard binary, $a\ll r_b$, then $r_p\ll r_b$ and $\Phi_\star(r_p)\approx\Phi_\star(0)$.
In this limit, the field-star energy (\ref{Equation:Eofpva}) is given approximately by
\begin{eqnarray}
E \approx \Phi_\star(0) + \frac{v_\infty^2}{2} .
\end{eqnarray}
Substituting this expression for $E$ into equation (\ref{Equation:DiffCoeffinQ})
and again assuming $f_\star=f_\star(E)$ yields
\begin{equation}
\langle\Delta Q\rangle \approx \int_0^{\sqrt{-2\Phi_\star(0)}} \int_0^{p_\mathrm{max}} \overline{\delta Q}\,n'\times 2\pi p\,dp\, v_\infty \times 4\pi v^2 f^\prime_v(v) dv
\end{equation}
where
\begin{equation}
n' \equiv \int_0^{\sqrt{-2\Phi_\star(0)}} 4\pi\, dv\,v^2
f\left(\Phi_\star(0)+\frac{v^2}{2}\right),
\end{equation}
the number density at the radius $r_n$, defined such that 
$\Phi(r_n)=-GM_{12}/r_n + \Phi_\star(r_n) = \Phi_\star(0)$,
and $f'_v(v)$ is the normalized velocity distribution at this radius.

\begin{figure}[h]
\includegraphics[width=0.65\textwidth]{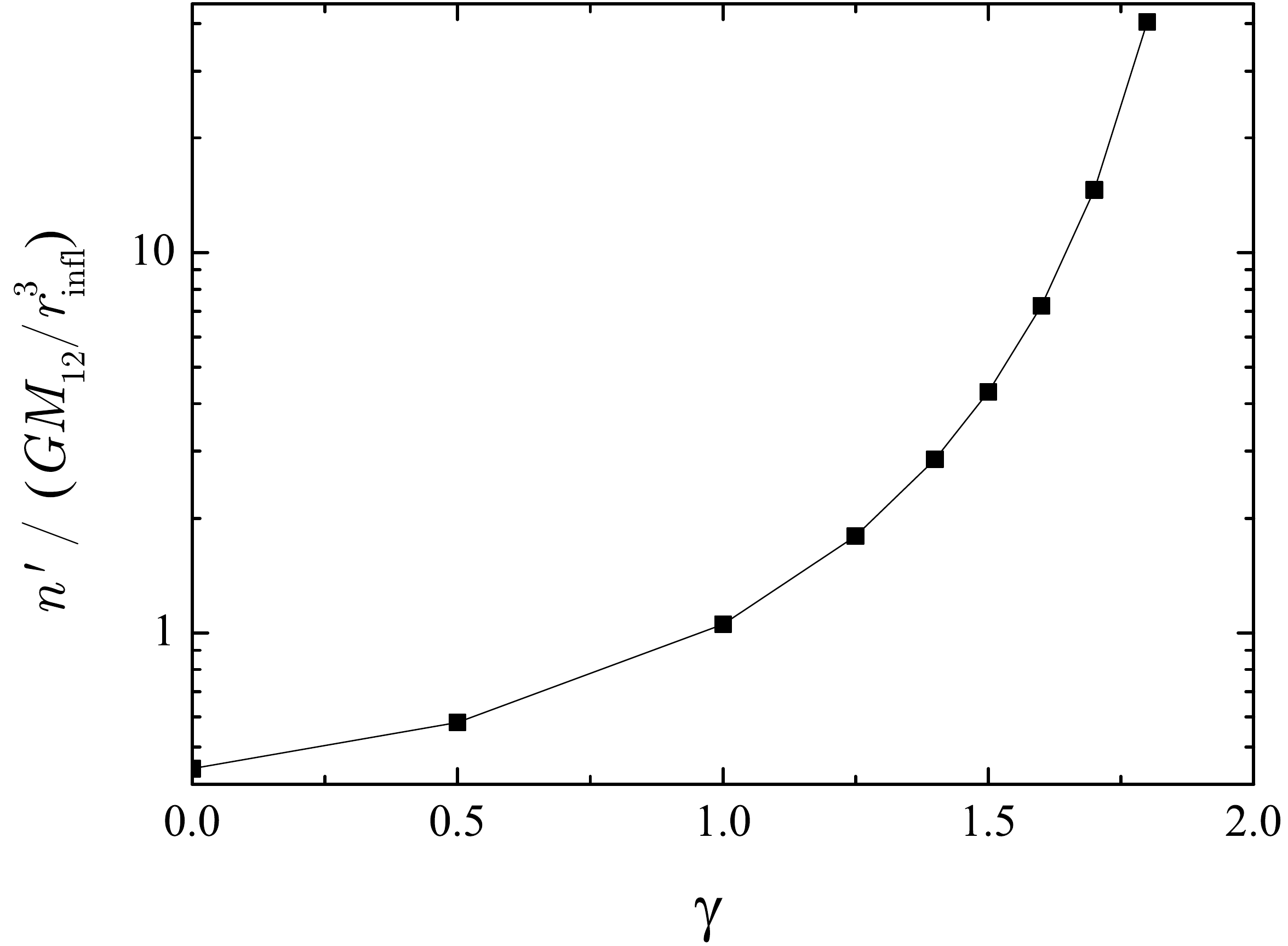}
\caption{The effective stellar number density $n'$ as a function of $\gamma$.}
\label{Figure:n_vs_gamma}
\end{figure}

Since
\beq
\Phi_\star(r_n) - \Phi_\star(0) = \frac{GM_{12}}{r_n} ,
\eeq
it is clear that $r_n$ is similar to $r_\mathrm{infl}$ and hence to $r_b$.
For instance, setting $\gamma=1$ in equation (\ref{Equation:DehnenModel}),
one finds $r_n\approx0.42r_b$.
Furthermore $n'\approx M_{12}/r_\mathrm{infl}^3$ with some leading coefficient that depends on the density slope $\gamma$; this coefficient is plotted as a function of $\gamma$ in Figure~\ref{Figure:n_vs_gamma}.
When $\gamma\lesssim1$, $n'\approx M_{12}/r_\mathrm{infl}^3$, and $n'\rightarrow\infty$ as $\gamma\rightarrow2$.
It turns out that the velocity distribution $f'_v(v)$ can be well approximated for all $\gamma$ by a  Maxwellian distribution $f_v\sim e^{-v^2/2\sigma^2}$
with  $\sigma^2\approx3GM_{12}/r_\mathrm{infl}$ \citep[cf. Fig. 3.8 of][]{DEGN}. The exact number doesn't matter when our binary is sufficiently hard $\left(V_{\mathrm{bin}}\gg GM_{12}/r_\mathrm{infl},\,V_{\mathrm{bin}}\equiv\sqrt{\frac{GM_{12}}{a}}\right)$. 
The role of binary hardness is discussed in \S\ref{Section:DiffusionCoefficients} where we calculate the diffusion coefficients.

\section{Diffusion coefficients in large mass ratio limit} 
\label{Appendix:Large_q}

Let $\lbin=\Lbin/\mu=\sqrt{GM_{12}a(1-e^2)}$ and $\lstar=\Lstar/m_f$ be the angular momentum per unit mass of the binary and the star, interacting with the binary, respectively. The diffusion coefficients can then be estimated as
\bsub
\begin{eqnarray}
\langle\Delta\theta\rangle &\sim& \frac{dn_\mathrm{enc}}{dt}\frac{\delta l_\mathrm{bin}}{\lbin}\\
\langle(\Delta\theta)^2\rangle &\sim& \frac{dn_\mathrm{enc}}{dt}\left(\frac{\delta l_\mathrm{bin}}{\lbin}\right)^2
\end{eqnarray}
\esub
where $dn_\mathrm{enc}/dt$ is the encounter rate and $\delta l_\mathrm{bin}$ is the change in $\lbin$ in one interaction. Due to angular momentum conservation
\beq
\delta l_\mathrm{bin} \sim \frac{m_f}{\mu}\delta\lstar
\eeq
Since only the close encounters with one of the binary components matter, the average change per encounter in the stellar angular momentum per unit mass is of the order of the binary angular momentum per unit mass (which is unity in the dimensionless units we use in our scattering experiments):
\beq
\delta l_\mathrm{star} \sim \lbin
\eeq
Combined with the previous equation, this gives us
\beq
\delta l_\mathrm{bin} \sim \frac{m_f}{\mu}\lbin \sim q\frac{m_f}{M_1}\lbin
\eeq
In the very large mass ratio assumption ($M_2\ll M_1$ or $q\gg1$) the encounter rate can be estimated as follows. The motion of stars is mainly determined by the potential of the primary component, but only the stars that experience a close encounter with the secondary contribute to the angular momentum exchange (for them $\delta\lstar\sim\lbin$). This means the encounter rate is actually the rate of close encounters with the secondary. Only the stars passing closer than $\lesssim a$ to the primary can experience a close interaction with the secondary; according to Eq.~(\ref{Equation:r_p}), this corresponds to the maximum impact parameter $p_{\rm max} = \sqrt{2GM_1a}/\sigma$. Only a small fraction of them actually do, because the radius of influence of the secondary $R_{\mathrm{infl},2} = GM_2/v^2_\mathrm{rel}$ is small compared to $a$. This fraction $\xi$ can be estimated as the probability of a particle crossing the sphere with radius $a$ to cross the sphere of radius $R_{\mathrm{infl},2}$, which is in a random point inside the larger sphere, i. e. $\xi \sim R_{\mathrm{infl},2}^2 / a^2$. That makes the following estimate of encounter rate:
\beq
\frac{dn_\mathrm{enc}}{dt} = n\sigma \cdot \pi p_{\rm max}^2 \cdot \xi \sim \frac{2\pi n GM_1 a}{\sigma} \cdot \frac{1}{q^2}
\eeq
Finally, for the diffusion coefficients we have
\bsub
\begin{eqnarray}
\langle\Delta\theta\rangle &\sim& \frac{2\pi n GM_1 a}{\sigma}\frac{m_f}{M_1}\cdot\frac{1}{q}\\
\langle(\Delta\theta)^2\rangle &\sim& \frac{2\pi n GM_1 a}{\sigma}\left(\frac{m_f}{M_1}\right)^2
\end{eqnarray}
\esub
At large mass ratios $\langle\Delta\theta\rangle$ decreases as $1/q$ and $\langle(\Delta\theta)^2\rangle$ is independent of $q$. However, as Fig.~\ref{Figure:D(q,omega,e)}a-b show that such an approximation probably works only at rather large mass ratios $q\gtrsim100$, though it strongly depends on hardness: this approximation starts working for smaller values of $q$ for softer binaries.


\end{document}